\documentclass[12 pt,twoside,aps,prd,amsmath,amssymb,
tightenlines,showpacs,showkeys,eqsecnum]{revtex4}
\usepackage{mathptmx}
\usepackage{bm}
\usepackage{graphicx}
\usepackage[%
breaklinks,colorlinks,
naturalnames,
unicode]{hyperref}
\newcommand{\bs}{http://wwwinfo.jinr.ru/~bijan/my_papers}

\renewcommand{\cal}{\mathcal}

\newcommand {\ve}{\varepsilon}

\newcommand {\prm}{\prime}
\newcommand {\cG}{\cal G}
\newcommand {\cD}{\cal D}
\newcommand {\cL}{\cal L}

\newcommand {\bp}{\bar \psi}

\newcommand {\p}{\psi}
\newcommand {\vf}{\varphi}

\def \myfigures #1#2#3#4#5#6#7#8
{\begin{figure}[ht]
    \begin{center}
        \includegraphics[width=#2 \textwidth, angle=-90]{#1.eps}
        \hfill
        \includegraphics[width=#6 \textwidth, angle=-90]{#5.eps}
        \parbox[t]{#4\textwidth}{\caption {#3}\label{#1}}
        \hfill
        \parbox[t]{#8\textwidth}{\caption {#7}\label{#5}}
    \end{center}
\end{figure} }

\def \myfiguress #1#2#3#4#5#6#7#8
{\begin{figure}[ht]
    \begin{center}
        \includegraphics[width=#2 \textwidth]{#1.eps}
        \hfill
        \includegraphics[width=#6 \textwidth]{#5.eps}
        \parbox[t]{#4\textwidth}{\caption {#3}\label{#1}}
        \hfill
        \parbox[t]{#8\textwidth}{\caption {#7}\label{#5}}
    \end{center}
\end{figure} }

\def \myf #1#2#3#4#5#6#7#8#9
{\begin{figure}[ht]
    \begin{center}
        \includegraphics[width=#2 \textwidth]{#1.eps}
        \hfill
        \includegraphics[width=#5 \textwidth]{#4.eps}
        \hfill
        \includegraphics[width=#8 \textwidth]{#7.eps}\\
        \parbox[t]{#2\textwidth}{\caption {#3}\label{#1}}
        \hfill
        \parbox[t]{#5\textwidth}{\caption {#6}\label{#4}}
        \hfill
        \parbox[t]{#8\textwidth}{\caption {#9}\label{#7}}
    \end{center}
\end{figure} }

\def \myff #1#2#3#4#5#6#7#8
{\begin{figure}[ht]
    \begin{center}
        \includegraphics[width=#2 \textwidth]{#1.eps}
        \hfill
        \includegraphics[width=#4 \textwidth]{#3.eps}
        \hfill
        \includegraphics[width=#6 \textwidth]{#5.eps}\\
        \parbox[t]{#7\textwidth}{\caption {#8}}
     \end{center}
\end{figure} }

\def\myfigure #1#2#3#4
{\begin{figure}[ht]\begin{center}
\includegraphics[width=#2 \textwidth, angle=-90]{#1.eps}
\parbox[t]{#4\textwidth}{\caption{#3}\label{#1}}
\end{center}\end{figure}}

\def\myfigured #1#2#3#4
{\begin{figure}[ht]\begin{center}
\includegraphics[width=#2 \textwidth]{#1.eps}\\
\vskip 2 cm
\parbox[t]{#4\textwidth}{\caption{#3}\label{#1}}
\end{center}\end{figure}}

\date{\today}
\begin{document}
\title{Anisotropic cosmological models with spinor and scalar fields
and viscous fluid in presence of a $\Lambda$ term: qualitative solutions}
\author{Bijan Saha and Victor Rikhvitsky}
\affiliation{Laboratory of Information Technologies\\
Joint Institute for Nuclear Research, Dubna\\
141980 Dubna, Moscow region, Russia} \email{bijan@jinr.ru}
\homepage{http://wwwinfo.jinr.ru/~bijan/}

\begin{abstract}

The study of a self-consistent system of interacting spinor and
scalar fields within the scope of a Bianchi type I (BI) gravitational
field in presence of a viscous fluid and $\Lambda$ term has been
carried out. The system of equations defining the evolution of the
volume scale of BI universe, energy density and corresponding Hubble
constant has been derived. The system in question has been thoroughly
studied qualitatively. Corresponding solutions are graphically
illustrated. The system in question is also studied from the view point
of blow up. It has been shown that the blow up takes place only in
presence of viscosity.

\end{abstract}



\maketitle

\bigskip


\section{Introduction}

The problem of an initial singularity still remains at the center of
modern day cosmology. Though the Big Bang theory is deeply rooted
among the scientists dealing with the cosmology of the early
Universe, it is natural to reconsider models of a universe free from
initial singularities. Another problem that the modern day cosmology
deals with is the accelerated mode of expansion. In order to answer
to these questions a number of theories has been proposed by
cosmologists. It has been shown that the introduction of a nonlinear
spinor field or an interacting spinor and scalar fields depending on
some special choice of nonlinearity can give rise to singularity free
solutions in one hand \cite{sahajmp,sahagrg,sahaprd,sahaecaa}, on the
other hand they may exploited to explain the late time acceleration
\cite{sahaprd06,kramer}.

Why study a nonlinear spinor field? It is well known that the nonlinear
generalization of classical field theory remains one possible way to overcome
the difficulties of a theory that considers elementary particles as
mathematical points. In this approach elementary particles are modelled by
regular (solitonlike) solutions of the corresponding nonlinear equations. The
gravitational field equation is nonlinear by nature and the field itself is
universal and unscreenable. These properties lead to a definite physical
interest in the gravitational field that goes with these matter fields. We
prefer a spinor field to scalar or electromagnetic fields, as the spinor field
is the most sensitive to the gravitational field.

Why study an anisotropic universe? Though spatially homogeneous and isotropic,
Friedmann-Robertson-Walker (FRW) models are widely considered as a good
approximation of the present and early stages of the Universe. However, the
large scale matter distribution in the observable Universe, largely manifested
in the form of discrete structures, does not exhibit a high degree of
homogeneity. Recent space investigations detect anisotropy in the cosmic
microwave background. The Cosmic Background Explorer's differential radiometer
has detected and measured cosmic microwave background anisotropies at different
angular scales.

These anisotropies are supposed to contain in their fold the entire
history of cosmic evolution dating back to the recombination era and
are being considered as indicative of the geometry and the content of
the Universe. More information about cosmic microwave background
anisotropy is expected to be uncovered by the investigations of the
microwave anisotropy probe. There is widespread consensus among
cosmologists that cosmic microwave background anisotropies at small
angular scales are the key to the formation of discrete structures.
The theoretical arguments~\cite{mis1} and recent experimental data
that support the existence of an anisotropic phase that approaches an
isotropic phase leads one to consider universe models with an
anisotropic background.

Why study a system with viscous fluid? The investigation of
relativistic cosmological models usually has the energy momentum
tensor of matter generated by a perfect fluid. To consider more
realistic models one must take into account the viscosity mechanisms,
which have already attracted the attention of many researchers.
Misner \cite{mis1,mis2} suggested that strong dissipative due to the
neutrino viscosity may considerably reduce the anisotropy of the
black-body radiation. Viscosity mechanism in cosmology can explain
the anomalously high entropy per baryon in the present universe
\cite{wein,weinb}. Bulk viscosity associated with the
grand-unified-theory phase transition \cite{lang} may lead to an
inflationary scenario \cite{waga,pacher,guth}.

A uniform cosmological model filled with fluid which possesses
pressure and second (bulk) viscosity was developed by Murphy
\cite{murphy}. The nature of cosmological solutions for homogeneous
Bianchi type I (BI) model was investigated by Belinskii and
Khalatnikov \cite{belin} by taking into account dissipative process
due to viscosity. They showed that viscosity cannot remove the
cosmological singularity but results in a qualitatively new behavior
of the solutions near singularity. They found the remarkable property
that during the time of the \textit{big bang} matter is created by
the gravitational field.

Given the importance of both viscous mechanism and nonlinear spinor
field we have recently studied the system in question from various
aspects. In \cite{Vismpl05} we have studied the evolution of a BI
universe filled with viscous fluid in presence of a $\Lambda$ term.
Exact solutions to the corresponding system of equations were found
for some special choice of viscosity parameters. This study was
further developed in \cite{Visrykh04}, where the system was studied
qualitatively. Introduction of a nonlinear spinor field into the
system considerably changes the situation giving rise to some
unexpected results such as Big Rip without phantom dark energy. The
system in question was studied analytically in
\cite{saharrp,visnlsrev2} and generalized in \cite{visspqual}
employing both numerical and qualitative methods. Since the
interacting system of spinor and scalar fields gives rise to a
induced nonlinearity of the spinor field that can change the picture
drastically, we plan to consider this system as well. Some exact
solutions to the system of equations were obtained in \cite{visspsc}.
Here we thoroughly study the interacting spinor and scalar fields
within the framework of a BI gravitational field in presence of a
viscous fluid and $\Lambda$ term. In doing so we will exploit both
numerical and qualitative methods.

\section{Basic equations}

We consider a self-consistent system of interacting nonlinear spinor
and scalar fields within the scope of a Bianchi type-I (BI)
gravitational field filled with a viscous fluid in presence of a
cosmological term. Corresponding Lagrangian takes the form:
\begin{equation}
{\cal L}_{ss}=\frac{i}{2} \biggl[ \bp \gamma^{\mu} \nabla_{\mu} \p-
\nabla_{\mu} \bp \gamma^{\mu} \p \biggr] - m\bp \p +
\frac{1}{2}\vf_{,\alpha}\vf^{,\alpha}(1 + \lambda F), \label{nlspin}
\end{equation}
Here $m$ is the spinor mass, $\lambda$ is the coupling constant and
$F = F(I,J)$ with $I = S^ 2= (\bar \psi \psi)^2$ and $J = P^2 = (i
\bar \psi \gamma^5 \psi)^2$. According to the Pauli-Fierz theorem
among the five invariants only $I$ and $J$ are independent as all
other can be expressed by them: $I_V = - I_A = I + J$ and $I_Q = I -
J.$ Therefore, the choice $F = F(I, J)$, describes the nonlinearity
in the most general of its form \cite{sahaprd}. Note that setting
$\lambda = 0$ in \eqref{nlspin} we come to the case with minimal
coupling.

The gravitational field in our case is given by a Bianchi type I (BI)
metric
\begin{equation}
ds^2 = dt^2 - a^2 dx^2 - b^2 dy^2 - c^2 dz^2, \label{BI}
\end{equation}
with $a,\, b,\, c$ being the functions of time $t$ only. Here the
speed of light is taken to be unity.

For the BI space-time \eqref{BI} on account of the $\Lambda$ term
this system has the form
\begin{subequations}
\label{BID}
\begin{eqnarray}
\frac{\ddot b}{b} +\frac{\ddot c}{c} + \frac{\dot b}{b}\frac{\dot
c}{c}&=&  \kappa T_{1}^{1} +\Lambda,\label{11}\\
\frac{\ddot c}{c} +\frac{\ddot a}{a} + \frac{\dot c}{c}\frac{\dot
a}{a}&=&  \kappa T_{2}^{2} + \Lambda,\label{22}\\
\frac{\ddot a}{a} +\frac{\ddot b}{b} + \frac{\dot a}{a}\frac{\dot
b}{b}&=&  \kappa T_{3}^{3} + \Lambda,\label{33}\\
\frac{\dot a}{a}\frac{\dot b}{b} +\frac{\dot b}{b}\frac{\dot c}{c}
+\frac{\dot c}{c}\frac{\dot a}{a}&=&  \kappa T_{0}^{0} + \Lambda,
\label{00}
\end{eqnarray}
\end{subequations}
where over dot means differentiation with respect to $t$ and
$T_{\nu}^{\mu}$ is the energy-momentum tensor of the material field
given by
\begin{eqnarray}
T_{\mu}^{\rho} &=& \frac{i}{4} g^{\rho\nu} \biggl(\bp \gamma_\mu
\nabla_\nu \psi + \bp \gamma_\nu \nabla_\mu \psi - \nabla_\mu \bar
\psi \gamma_\nu \psi - \nabla_\nu \bp \gamma_\mu \psi \biggr)  \label{tem}\\
& & + (1 - \lambda F)  \vf_{,\mu}\vf^{,\rho} - \delta_{\mu}^{\rho}
{\cL} + T_{\mu\,{\rm m}}^{\,\,\,\nu}. \nonumber
\end{eqnarray}
Here $T_{\mu\,{\rm m}}^{\nu}$ is the energy-momentum tensor of a
viscous fluid having the form
\begin{equation}
T_{\mu\,{\rm m}}^{\nu} = (\ve + p^\prm) u_\mu u^\nu - p^{\prm}
\delta_\mu^\nu + \eta g^{\nu \beta} [u_{\mu;\beta}+u_{\beta:\mu}
-u_\mu u^\alpha u_{\beta;\alpha} - u_\beta u^\alpha u_{\mu;\alpha}],
\label{imper}
\end{equation}
where
\begin{equation}
p^{\prm} = p - (\xi - \frac{2}{3} \eta) u^\mu_{;\mu}. \label{ppr}
\end{equation}
Here $\ve$ is the energy density, $p$ - pressure, $\eta$ and $\xi$
are the coefficients of shear and bulk viscosity, respectively. In a
comoving system of reference such that $u^\mu = (1,\,0,\,0,\,0)$ we
have
\begin{subequations}
\begin{eqnarray}
T_{0\,{\rm m}}^{0} &=& \ve, \\
T_{1\,{\rm m}}^{1} &=& - p^{\prm} + 2 \eta \frac{\dot a}{a}, \\
T_{2\,{\rm m}}^{2} &=& - p^{\prm} + 2 \eta \frac{\dot b}{b}, \\
T_{3\,{\rm m}}^{3} &=& - p^{\prm} + 2 \eta \frac{\dot c}{c}.
\end{eqnarray}
\end{subequations}
WE consider the case when both the spinor and the scalar fields
depend on $t$ only. We also define a new function
\begin{equation}
\tau = a b c, \label{taudef}
\end{equation}
which is indeed the volume scale of the BI space-time. It was shown
in \cite{saharrp,visnlsrev2,visspsc} that the solutions of the spinor
and scalar field equations can be expressed in terms of $\tau$. Then
for the components of the energy-momentum tensor we find
\begin{subequations} \label{total}
\begin{eqnarray}
T_{0}^{0} &=& m S + \frac{C^2}{2\tau^2(1+\lambda F)} + \ve \equiv \tilde{T}_{0}^{0},\\
T_{1}^{1} &=& {\cD} S + {\cG} P - \frac{C^2}{2\tau^2(1+\lambda F)}
- p^{\prm} + 2 \eta \frac{\dot a}{a} \equiv \tilde{T}_{1}^{1} + 2 \eta \frac{\dot a}{a}, \\
T_{2}^{2} &=& {\cD} S + {\cG} P - \frac{C^2}{2\tau^2(1+\lambda F)}
- p^{\prm} + 2 \eta \frac{\dot b}{b} \equiv \tilde{T}_{1}^{1} + 2 \eta \frac{\dot b}{b},, \\
T_{3}^{3} &=& {\cD} S + {\cG} P - \frac{C^2}{2\tau^2(1+\lambda F)} -
p^{\prm} + 2 \eta \frac{\dot c}{c} \equiv \tilde{T}_{1}^{1} + 2 \eta
\frac{\dot c}{c},.
\end{eqnarray}
\end{subequations}
In account of \eqref{total} from \eqref{BID} we find the metric
functions \cite{sahaprd}
\begin{subequations}
\label{abc}
\begin{eqnarray}
a(t) &=& Y_1 \tau^{1/3}\exp \biggl[\frac{X_1}{3} \int\,\frac{e^{-2
\kappa \int \eta dt}}{\tau (t)} dt \biggr],
\label{a} \\
b(t) &=& Y_2\tau^{1/3}\exp \biggl[\frac{X_2}{3} \int\,\frac{e^{-2
\kappa \int \eta dt} }{\tau (t)} dt \biggr],
\label{b} \\
c(t) &=& Y_3\tau^{1/3}\exp \biggl[\frac{X_3}{3} \int\,\frac{e^{-2
\kappa \int \eta dt}}{\tau (t)} dt \biggr], \label{c}
\end{eqnarray}
\end{subequations}
with the constants $Y_i$ and $X_i$ obeying
$$Y_1Y_2Y_3 = 1, \qquad X_1 + X_2 + X_3 = 0.$$

As one sees from \eqref{a}, \eqref{b} and \eqref{c}, for $\tau = t^n$
with $n > 1$ the exponent tends to unity at large $t$, and the
anisotropic model becomes isotropic one.

So one needs to find the function $\tau$, explicitly. Corresponding
equation can be derived from Einstein equations and Bianchi identity
[a detailed description of this procedure can be found in
\cite{saharrp,visnlsrev2,visspsc}]. For convenience, we also define
the generalized Hubble constant. The system then reads
\cite{visspsc}:

\begin{subequations}
\label{HVespsc}
\begin{eqnarray}
\dot {\tau } &=& 3 H \tau, \label{tauss}\\
\dot {H} &=& \frac{\kappa}{2}\bigl(3 \xi H - \omega\bigr) - \bigl(3 H^2 - \kappa \ve - \Lambda
\bigr) + \frac{\kappa}{2}
\bigl(\frac{m}{\tau} + \frac{n \tau^{n-2}}{2(\lambda + \tau^n)^2}\bigr),  \label{Hss}\\
\dot {\ve} &=& 3 H\bigl(3 \xi H - \omega\bigr) + 4 \eta \bigl(3 H^2 - \kappa \ve - \Lambda\bigr) -
4 \eta \bigl[ \kappa \bigl( \frac{m}{\tau} + \frac{\tau^{n-2}}{2(\lambda +\tau^n)}\bigr)\bigr].
\label{Vess}
\end{eqnarray}
\end{subequations}

Here $\kappa$ is the Einstein's gravitational constant, $\Lambda$
is the cosmological constant, $\lambda $ is the self-coupling
constant, $m$ is the spinor mass and $n$ is the power of
nonlinearity of the spinor field (here we consider only power law
nonlinearity). In \eqref{HVespsc} $\eta$ and $\xi$ are the bulk
and shear viscosity, respectively and they are both positively
definite, i.e.,
\begin{equation}
\eta > 0, \quad \xi > 0.
\end{equation}
They may be either constant or function of time or energy. We
consider the case when
\begin{equation}
\eta = A \ve^{\alpha}, \quad \xi = B \ve^{\beta}, \label{etaxi}
\end{equation}
with $A$ and $B$ being some positive quantities. For $p$ we set as
in perfect fluid,
\begin{equation}
p = \zeta \ve, \quad \zeta \in (0, 1]. \label{pzeta}
\end{equation}Vismpl05
Note that in this case $\zeta \ne 0$, since for dust pressure,
hence temperature is zero, that results in vanishing viscosity.
Note that a system in absence of spinor field has been studied in
\cite{Vismpl05,Visrykh04}. In that case the corresponding system is
analogical to the one given in \eqref{HVespsc} without the third
terms in \eqref{Hss} and \eqref{Vess}.

\section{Qualitative analysis}

The study of the behavior of dynamic system given by a system of
ordinary differential equations implies the survey of all possible
scenarios of development for different values of the problem
parameters. It is necessary to understand at least how the process
of evolution comes to an end if it does so at infinitively large
time for a given set of initial conditions which can be given
anywhere.

So, under the specific behavior of the system we understand the
phase portrait of the system, i.e., the family of integral curves,
covering the total phase space. It is easy to imagine as far as
any point of the space can be declared as the initial one and at
least one integral curve will pass through it (or it will be fixed
point).

Certainly, it is difficult to imagine such a set of curves. In
many cases, close (and not only) curves transform into each other
at some diffeomorphism of space. These curves are known as
topologically equivalent. The differences between them are not
very important for our study. They all behave in the same manner.
This relation - "the relation of equivalence" - divides the family
of curves into the classes of equivalence. For graphical
demonstration it will be convenient to present at least one
representative of each class.

The change of the value of problem parameters not always results
in significant change of the phase portrait. Repeating this method, we
say that one family of integral curves (covering the total space)
for the given set of parameters is equivalent to the other for
another set of parameters, if there exists a diffeomorphism of
space transforming the first family into the second. It is clear
that there occurs the division into the classes of equivalence, and
we are not very interested in differences between equivalent
families. We argue that the corresponding changes in parameters
do not alter anything on principle. So it is sufficient to
demonstrate only one phase portrait for a given set of parameters
underlining the features of the given class.

However, for some critical relations between the parameters there
occurs significant changes. These are the boundary relations of
parameters, dividing, as usual, parameter space into regions of
similar behavior. Thus accomplishes the qualitative classification
of the mode of evolution of dynamic system. Now, giving the
concrete value of parameters, we can define which region of
parameters they correspond to, thus define the type of behavior.
Moreover, given the specific initial conditions, we can answer
the question to which region of phase space the evolution of the
system lead in time.

In our cosmological model, numerical parameters $A$, $\alpha$, $B$,
$\beta$ are related to the viscosity, while $\lambda$ and
$\Lambda$ are the (self)-coupling and cosmological constants.

Initially, we consider the system of Einstein and Dirac equations.
Solving these equations, we find the components of the spinor
field and metric functions $a,\,b,\,c$ in terms of volume scale
$\tau = abc$ of the BI universe. Finally, in order to find $\tau$
from Einstein equations and Bianchi identity, we deduce three
first order ordinary differential equations. Further for
convenience we introduce a new function $\nu$ inverse to $\tau$,
i.e., $\nu = 1/\tau$.

The fact that the system has the dimension greater than 2,
strongly complicates qualitative analysis. Note that well known
Lorentz system of three ordinary differential equations with
polynomial right hand side with degree less or equal to 2,
possesses in some region of parameter space chaotic behavior known
as a strange attractor and in that region there do not exist first
integrals (i.e., globally defined invariants). Though the set of
singularities is very simple, there exist only three singular
(fixed) points: two focus and one saddle. The presence of such
example does not allow us to make an optimistic conclusion on the
basis of simple construction of our system (with polynomials in
the right hand side and absence of singular points the in region
of space we are interest in, which is even dynamically closed.

Nevertheless, on the boundary of the the space $\epsilon=0$, as
well as $\nu=0$ ($\tau=+\infty$), which are dynamically closed
themselves, the complete classification has been done. The
dynamical closeness of these planes simultaneously as an obstacle
for penetration from positive octant $\epsilon>0$ $\land$ $\nu>0$
to the region with negative values. But, there are no
singularities, fixed points (there are fixed points on the
boundary) in the positive octant, we were not able to prove the
simplicity of its behavior, e.g., presence of first integrals, as
well as their absence.

Thus let us go back to the system \eqref{HVespsc} in details. As
it was already mentioned, tt is convenient to define a new
function $\nu = 1/\tau$. In this case the obvious singularity that
occurs at $\tau = 0$ vanishes and $\nu = 0$ corresponds to $\tau =
\infty$ while $\nu = \infty$ to $\tau = 0$. The system
\eqref{HVespsc} on account of \eqref{etaxi} takes the form:

\begin{subequations}
\label{HVespscnu}
\begin{eqnarray}
\dot {\nu} &=& -3 H \nu, \label{taussnu}\\
\dot {H} &=& \frac{\kappa}{2}\bigl(3 \xi H - \omega\bigr) - \bigl(3 H^2 - \kappa \ve - \Lambda
\bigr) + \frac{\kappa}{2}
\bigl(m\nu + \frac{n \nu^{2-n}}{2(\lambda + \nu^{-n})^2}\bigr),  \label{Hssnu}\\
\dot {\ve} &=& 3 H\bigl(3 \xi H - \omega\bigr) + 4 \eta \bigl(3 H^2 - \kappa \ve - \Lambda\bigr) -
4 \eta \bigl[ \kappa \bigl( m\nu + \frac{\nu^{2-n}}{2(\lambda +  \nu^{-n})}\bigr)\bigr].
\label{Vessnu}
\end{eqnarray}
\end{subequations}

Let us now study the foregoing system of equations in details.

\subsection{Behavior of the solutions on $\nu = 0$ plane}

As one can see, in this case the system \eqref{HVespscnu} takes the form:
\begin{subequations}
\label{HVespscnu=0}
\begin{eqnarray}
\dot {H} &=& \frac{\kappa}{2}\bigl(3 \xi H - \omega\bigr) - \bigl(3 H^2 - \kappa \ve - \Lambda
\bigr),  \label{Hssnu=0}\\
\dot {\ve} &=& 3 H\bigl(3 \xi H - \omega\bigr) + 4 \eta \bigl(3 H^2 - \kappa \ve - \Lambda\bigr).
\label{Vessnu=0}
\end{eqnarray}
\end{subequations}
This system of equations completely coincides with the one when the
BI universe is filled with viscous fluid only. The system in question
was thoroughly studied in \cite{Visrykh04}, hence we skip this study
in the present report.

\subsection{Behavior of the solutions on $\ve = 0$ plane}

The plane $\ve = 0$ is dynamic invariant, since $\dot
\ve\bigl|_{\ve = 0} = 0$. Depending on the sign of $H$ this plane
is either attractive or repulsive, namely, for $H > 0$ it is
attractive and for $H < 0$ it is repulsive, since
$$\frac{\partial \dot \ve}{\partial \ve} = -3H(1 + \zeta) < 0.$$

In presence of of spinor and scalar fields the system
\eqref{HVespscnu} at $\ve = 0$ has the form

\begin{subequations}
\label{HVessnu2}
\begin{eqnarray}
\dot \nu &=& - 3 H \nu, \label{tauH22}\\
\dot {H} &=&  - 3 H^2 + \Lambda + \frac{1}{2} \Bigl( m\nu + \frac{ n
\nu^{2-n}}{2(\lambda+\nu^{-n})^2} \Bigr). \label{Hn22}
\end{eqnarray}
\end{subequations}
The system \eqref{HVessnu2} has the following integral curves
\begin{subequations}
\begin{eqnarray}
6H^2 &=& 2\Lambda+2m\nu+C\nu^2-\frac{\nu^{2+n}}{(\lambda \nu^n+1)}
\end{eqnarray}
\end{subequations}
where $C$ is some arbitrary constant.

The characteristic equation of nontrivial singular points on $\ve
= 0$ plane for the system \eqref{HVespsc} takes the form
\begin{equation}
2m\lambda^2\nu^{2n+1}+4\Lambda\lambda^2\nu^{2n}+n\nu^{n+2} +
4m\lambda\nu^{n+1}+8\Lambda\lambda\nu^{n}+2m\nu+4\Lambda = 0.
\end{equation}

Depending on changes of signs in the sequence of $\lambda$, $m$,
$\Lambda$ it has one, two or no solutions.

In Tables A1, B1, C1, D1 we illustrated the phase-portrait on $\ve =
0$ plane for a positive and a negative $\Lambda$, respectively for
$n=1,2,3,4$ and $\lambda<0$. In Tables A2, B2, C2, D2 we illustrated
the phase-portrait on $\ve = 0$ plane for a positive and a negative
$\Lambda$, respectively for $n=1,2,3,4$ and $\lambda=0$. In Tables
A3, B3, C3, D3 we illustrated the phase-portrait on $\ve = 0$ plane
for a positive and a negative $\Lambda$, respectively for $n=1,2,3,4$
and $\lambda>0$.

As it was mentioned earlier, here we deal with the multi-parametric
system of ordinary nonlinear differential equation. In doing so we
consider all possible variants independent to their physical
validity. Therefore, we demonstrate the results obtained for a
negative spinor mass ($m < 0$).

The singular point around which the oscillation takes place has $H =
0$, and therefore, the trajectory of oscillation partially passes in
the region which is attractive to the plane $\ve = 0$ and partially
in the region that is repulsive. In the long run in the repulsive
region at some moment the growth of $\ve$ becomes dominant. It
results in the fact that $\ve$ becomes infinity within a finite range
of time.

\newpage
\begin{center}
\begin{tabular}{|c|c|c|c|}
  \hline
     & $\Lambda<0$ & $\Lambda=0$ & $\Lambda>0$ \\
  \hline
   $m<0$ &
   \begin{tabular}{c} \includegraphics[width=0.15 \textwidth]{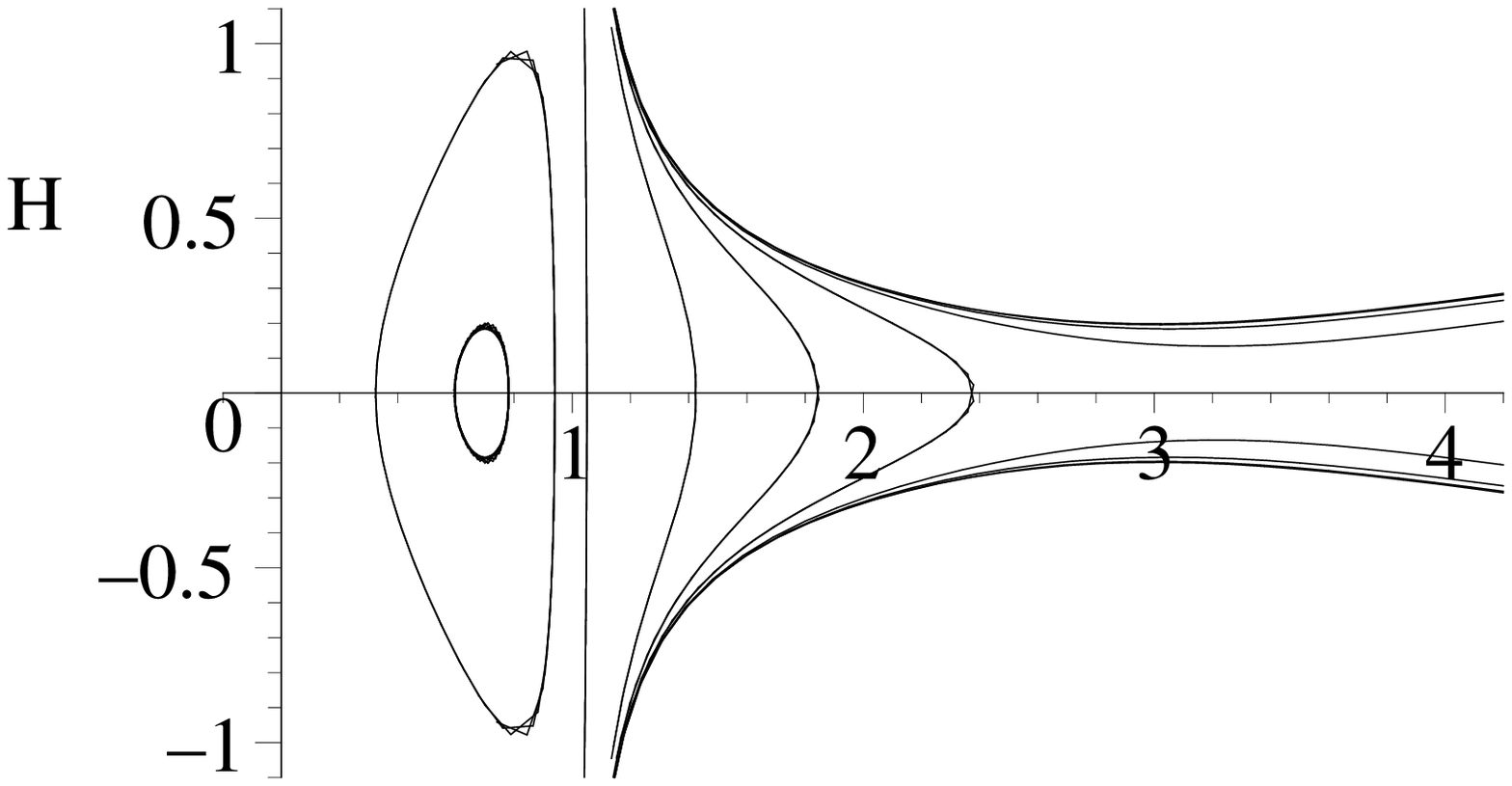} \\ a  \end{tabular} &
   \begin{tabular}{c} \includegraphics[width=0.15 \textwidth]{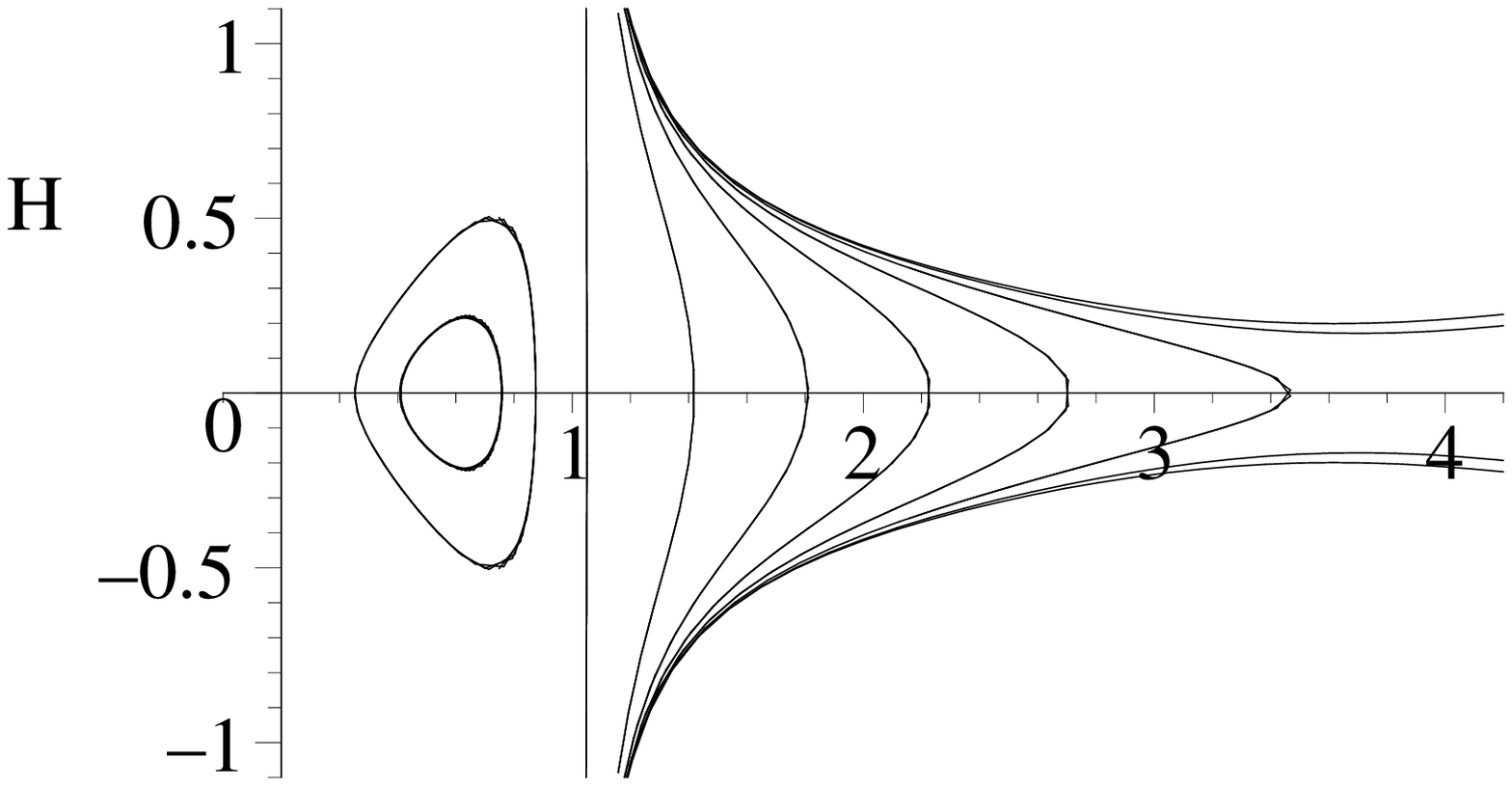} \\ b \end{tabular} &
   \begin{tabular}{c} \includegraphics[width=0.15 \textwidth]{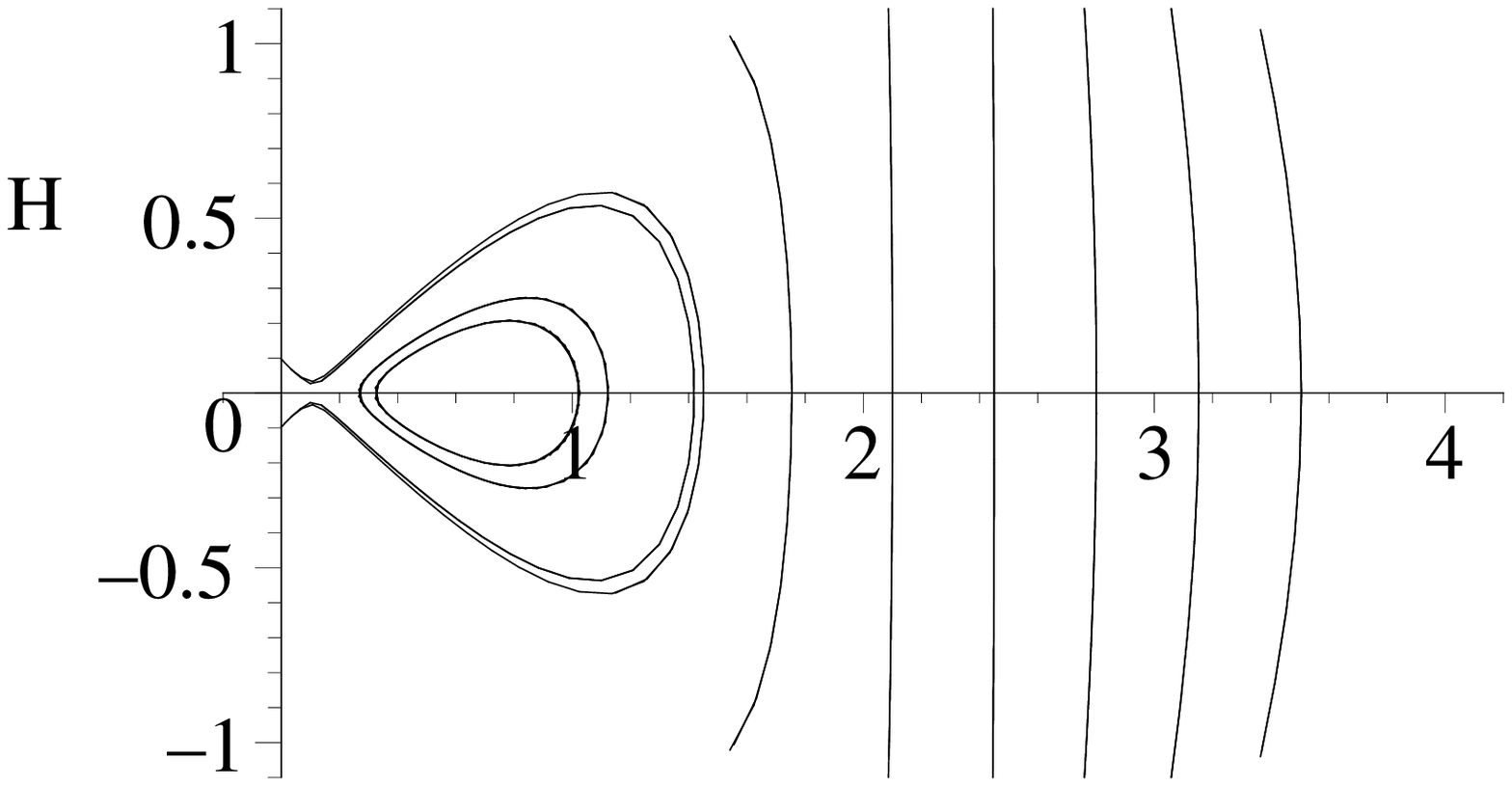} \\ c \end{tabular} \\
  \hline
   $m=0$ &
   \begin{tabular}{c} \includegraphics[width=0.15 \textwidth]{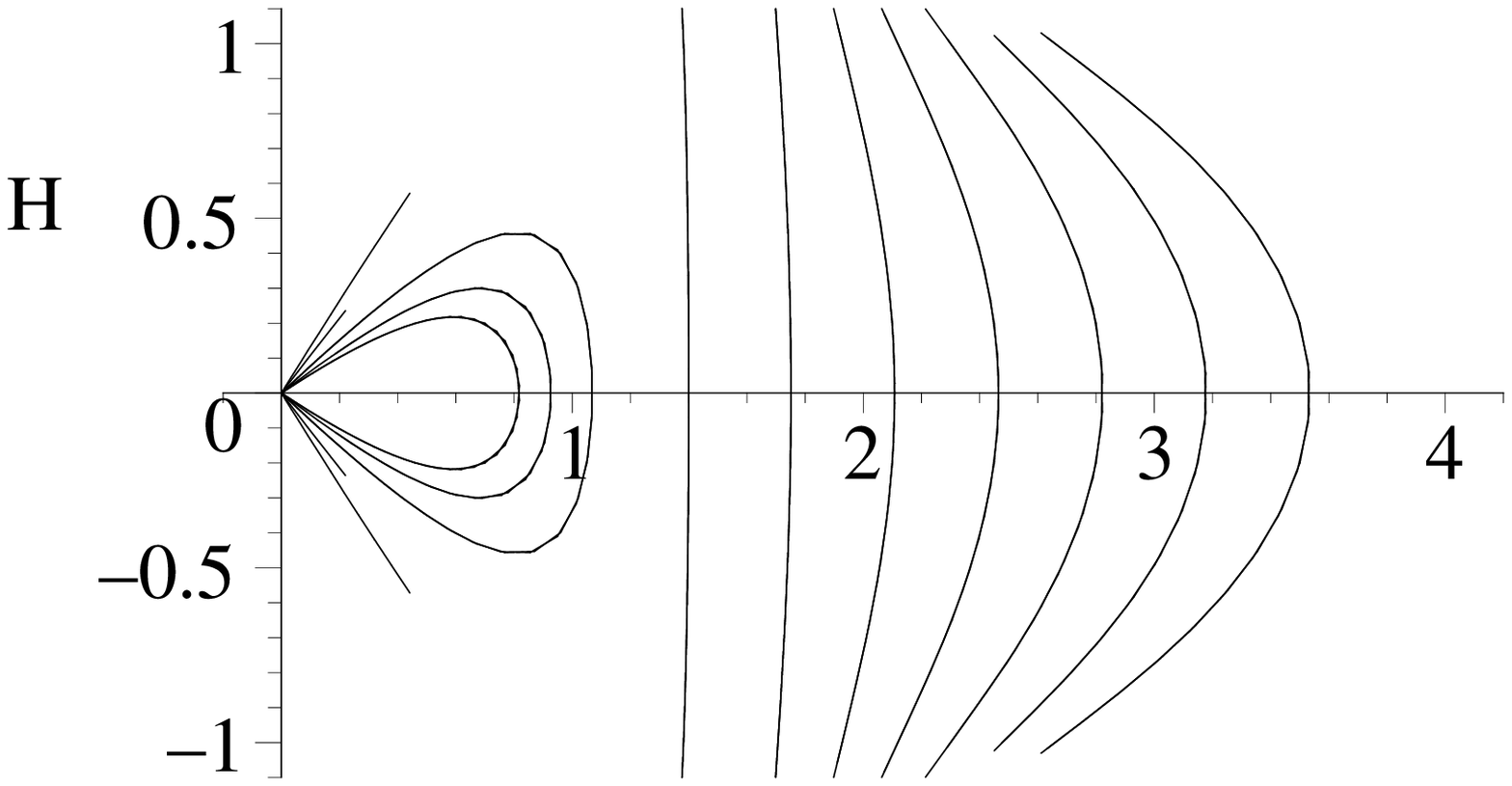} \\ d \end{tabular} &
   \begin{tabular}{c} \includegraphics[width=0.15 \textwidth]{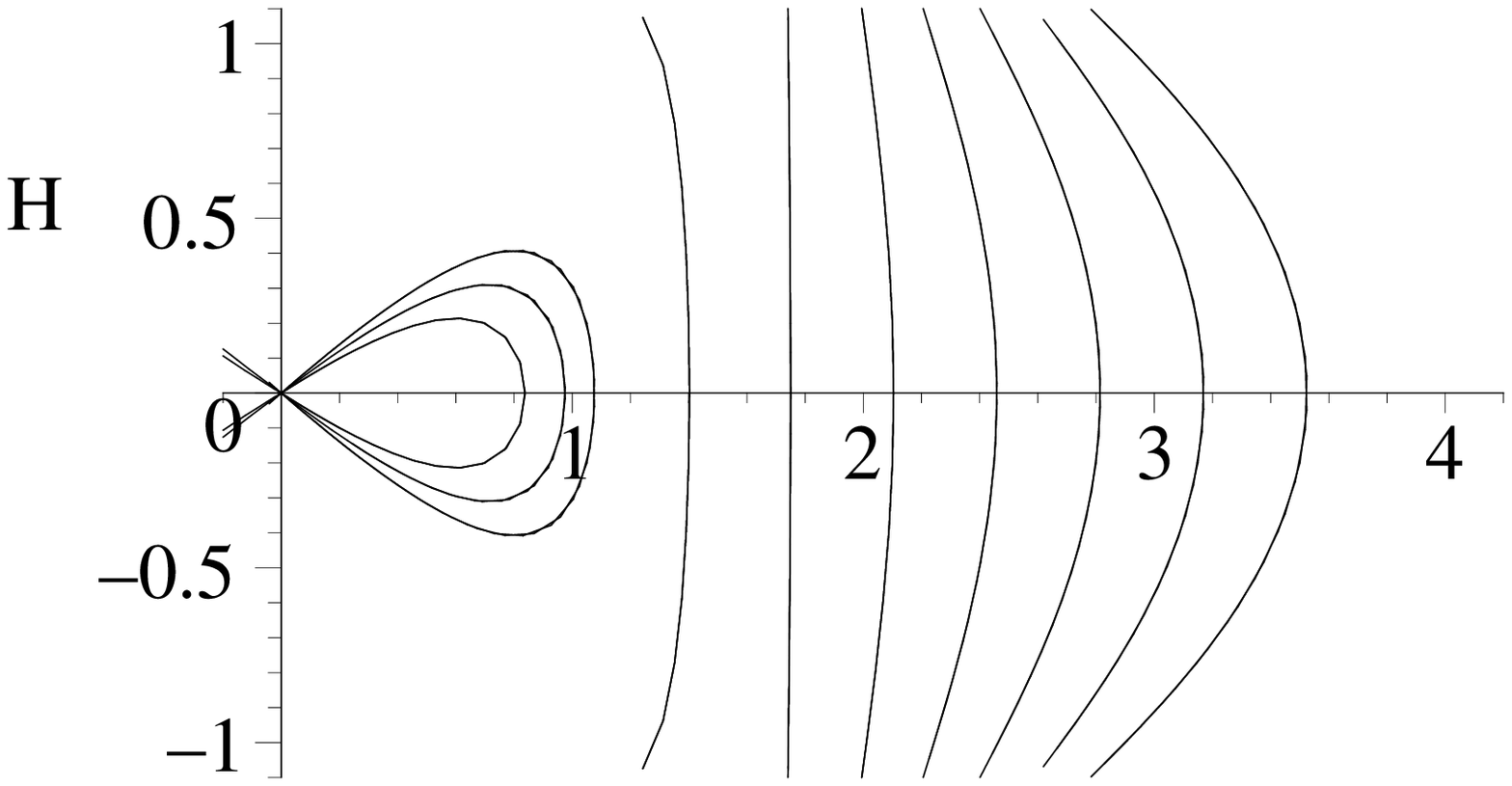} \\ e \end{tabular} &
   \begin{tabular}{c} \includegraphics[width=0.15 \textwidth]{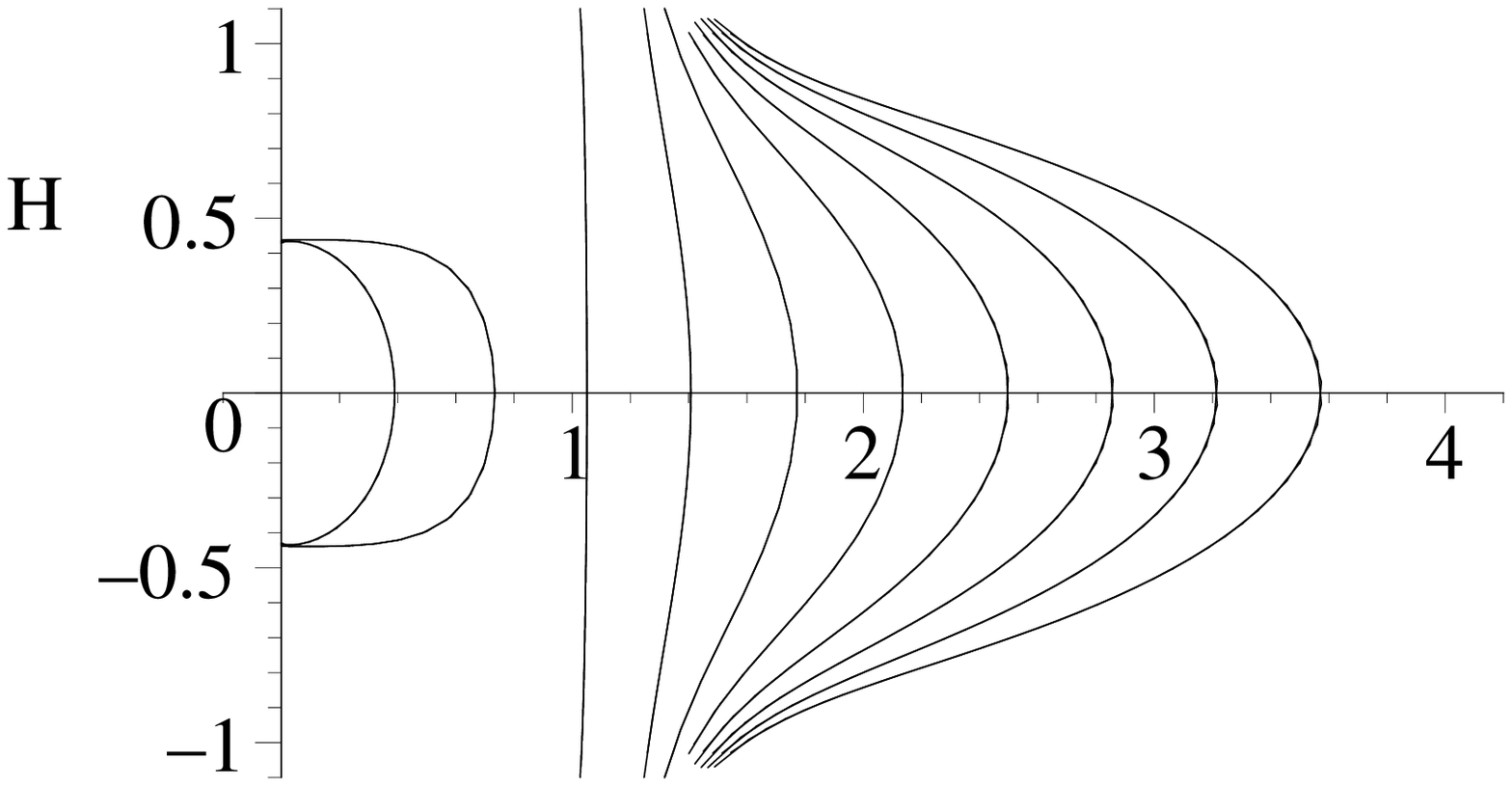} \\ f \end{tabular} \\
  \hline
   $m>0$ &
   \begin{tabular}{c} \includegraphics[width=0.15 \textwidth]{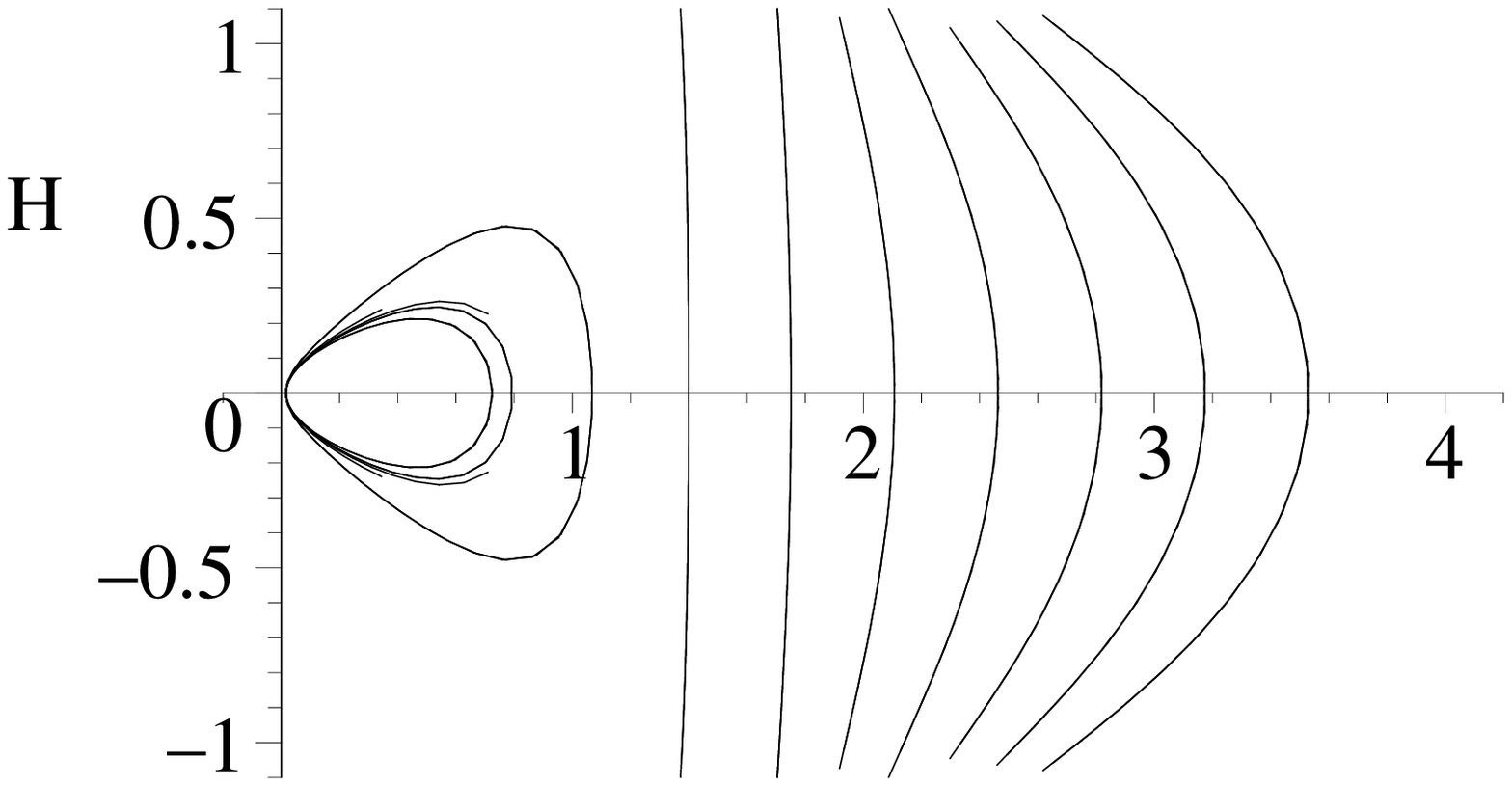} \\ g \end{tabular} &
   \begin{tabular}{c} \includegraphics[width=0.15 \textwidth]{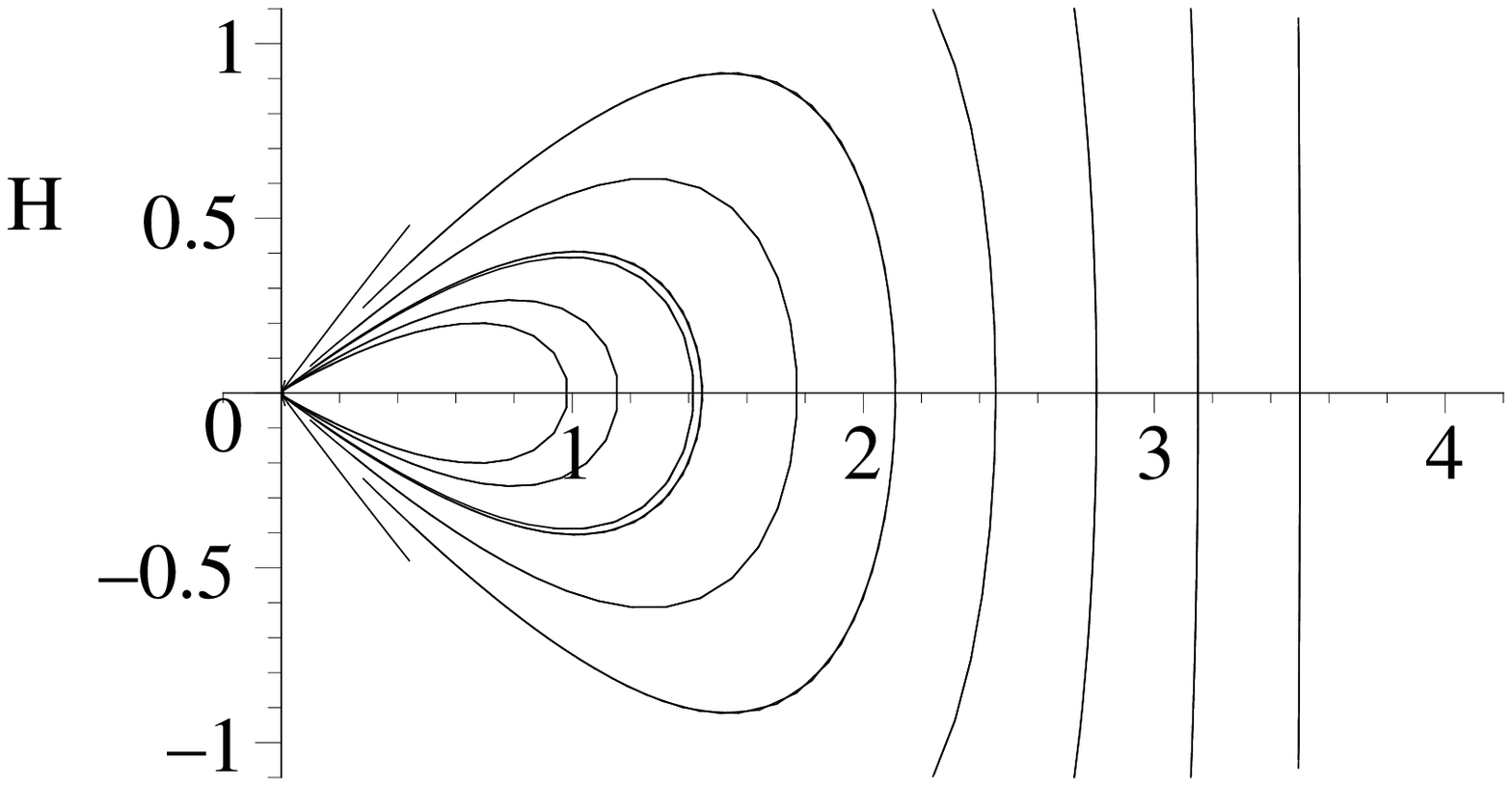} \\ h \end{tabular} &
   \begin{tabular}{c} \includegraphics[width=0.15 \textwidth]{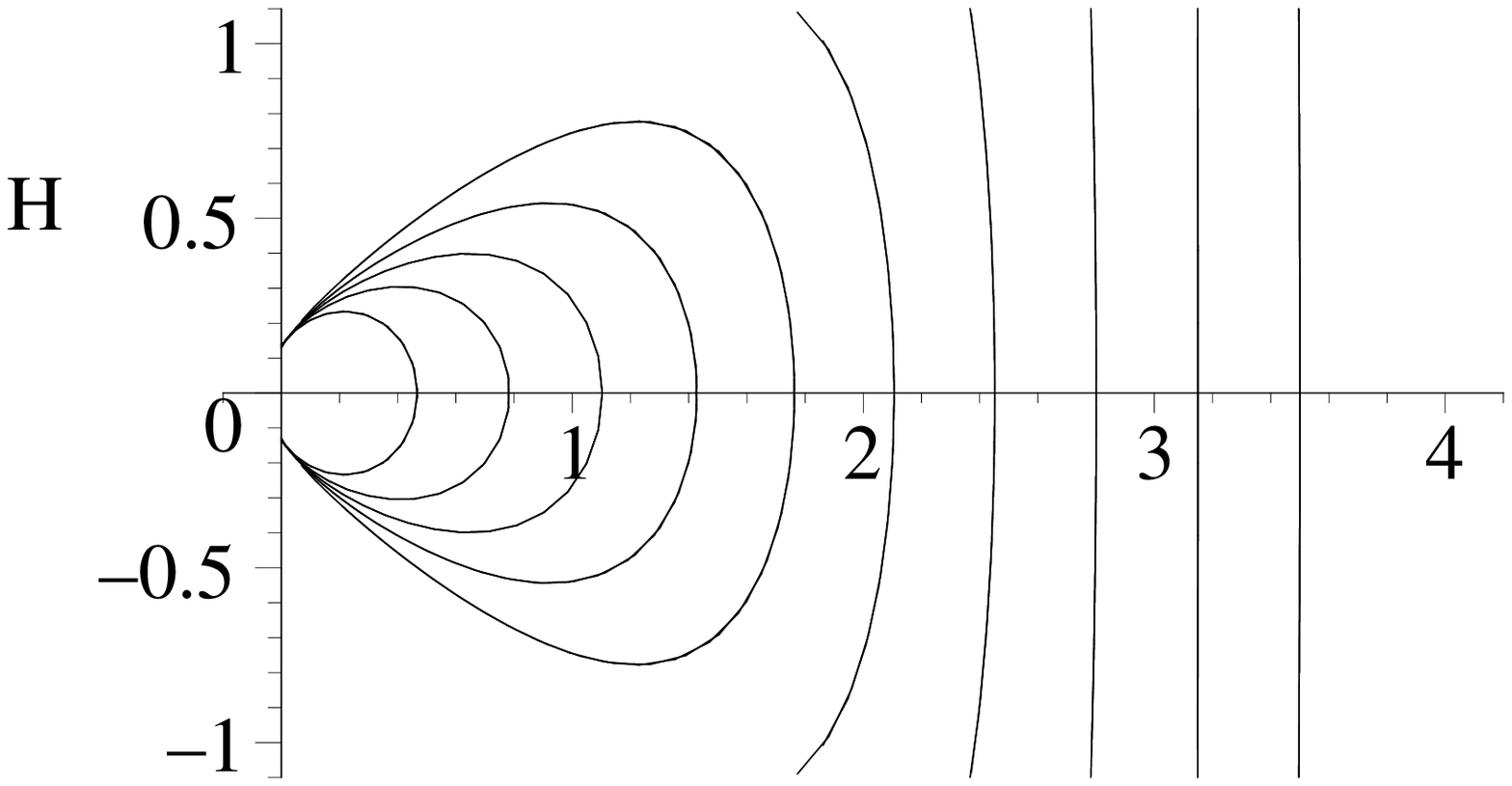} \\ i \end{tabular} \\
  \hline
\end{tabular}//
\vskip 2 mm
Table A1. Case with  $\ve=0$, $n=1$ and $\lambda<0$. \\
\vskip 2 mm

\begin{tabular}{|c|c|c|c|}
  \hline
     & $\Lambda<0$ & $\Lambda=0$ & $\Lambda>0$ \\
  \hline
   $m<0$ &
   \begin{tabular}{c} \includegraphics[width=0.15 \textwidth]{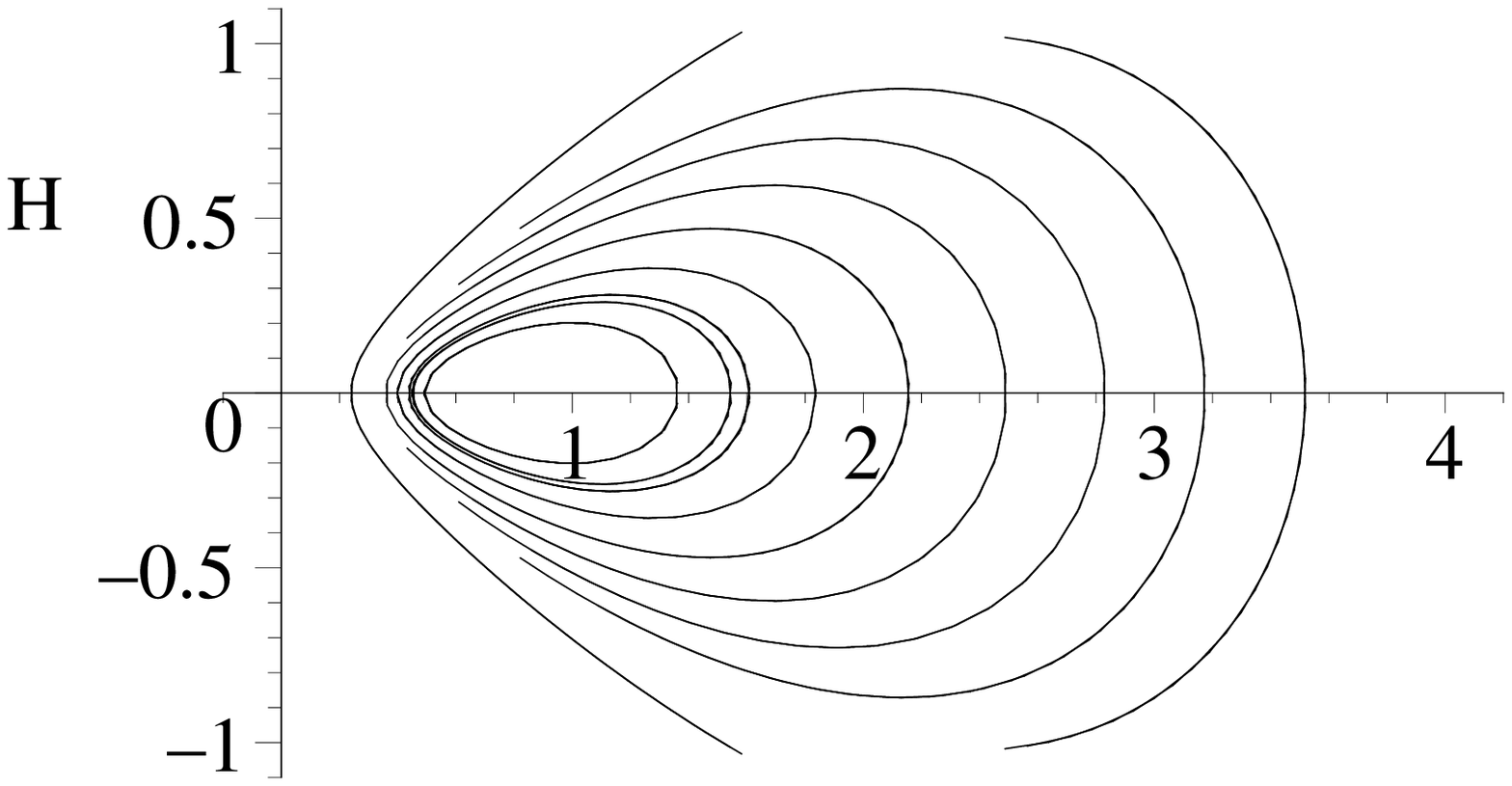} \\ a  \end{tabular} &
   \begin{tabular}{c} \includegraphics[width=0.15 \textwidth]{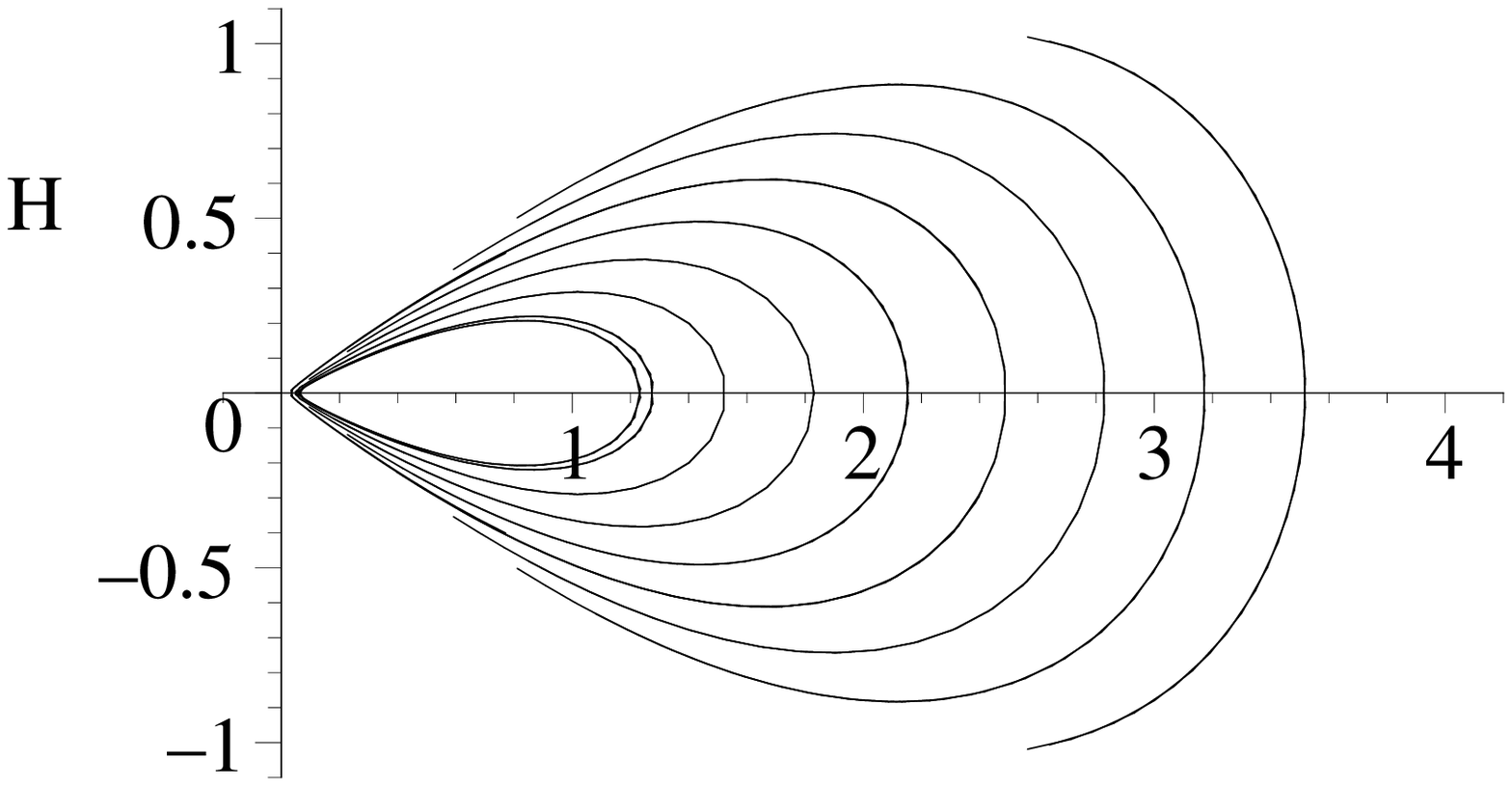} \\ b \end{tabular} &
   \begin{tabular}{c} \includegraphics[width=0.15 \textwidth]{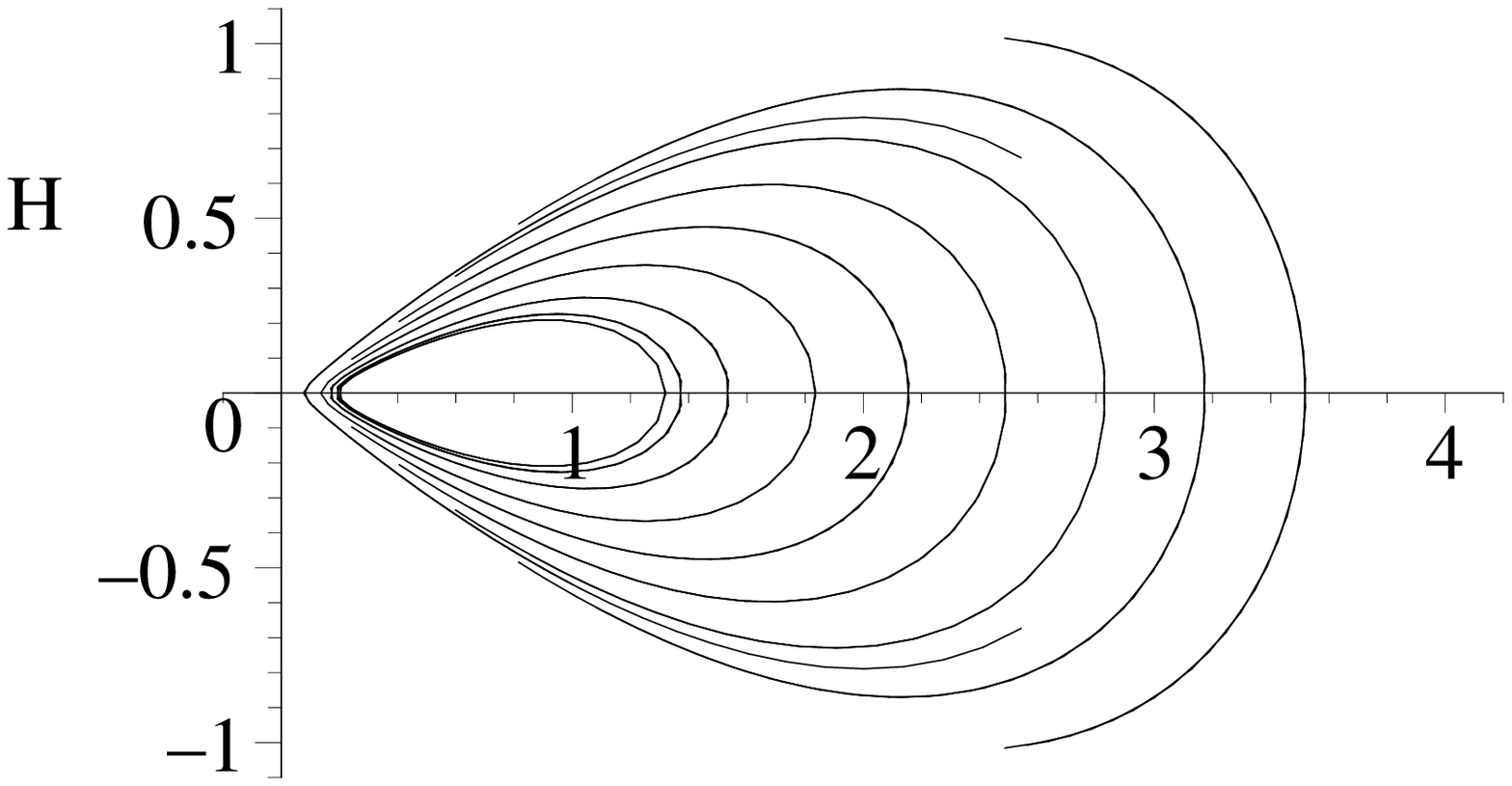} \\ c \end{tabular} \\
  \hline
   $m=0$ &
   \begin{tabular}{c} \includegraphics[width=0.15 \textwidth]{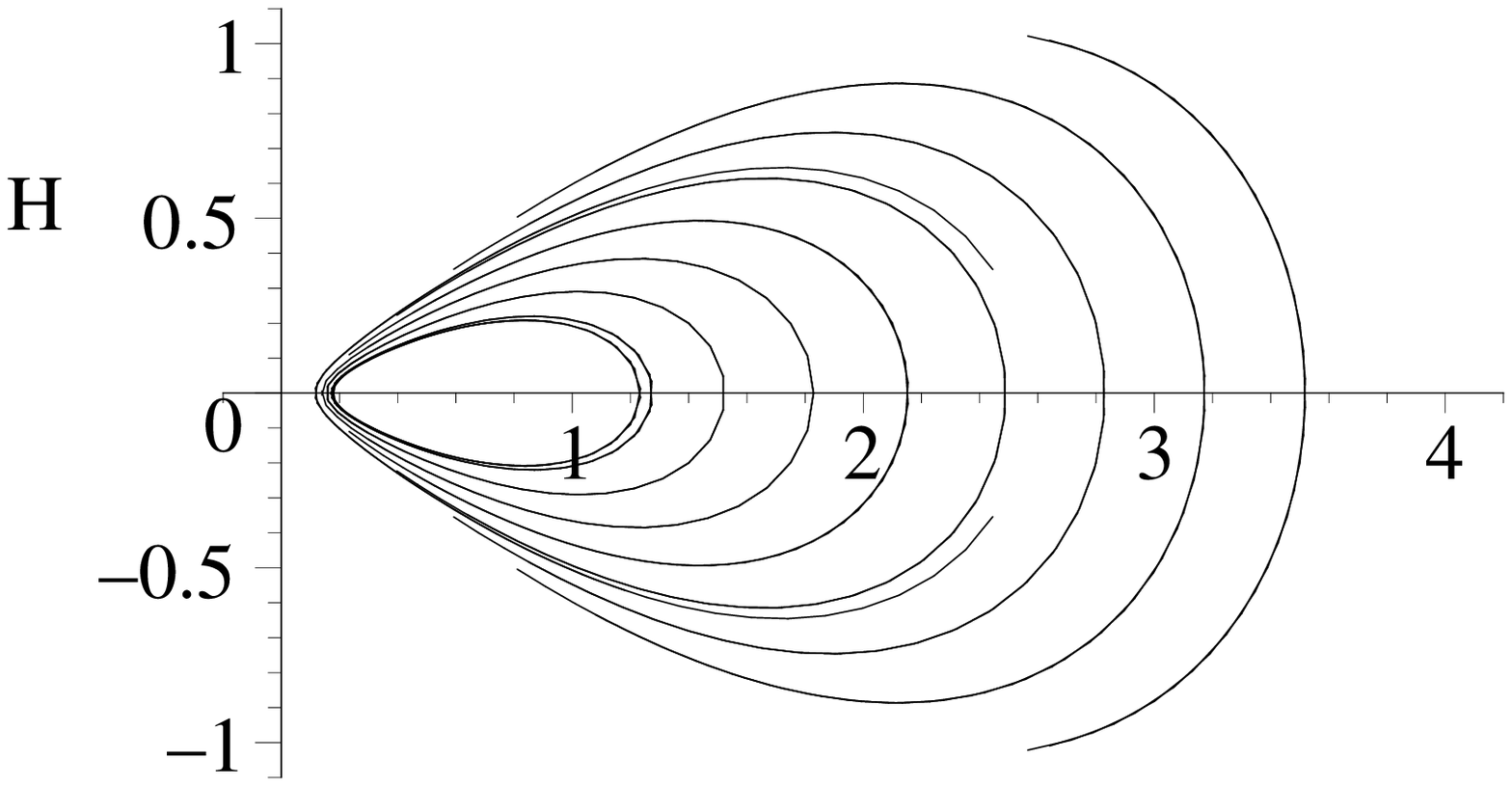} \\ d \end{tabular} &
   \begin{tabular}{c} \includegraphics[width=0.15 \textwidth]{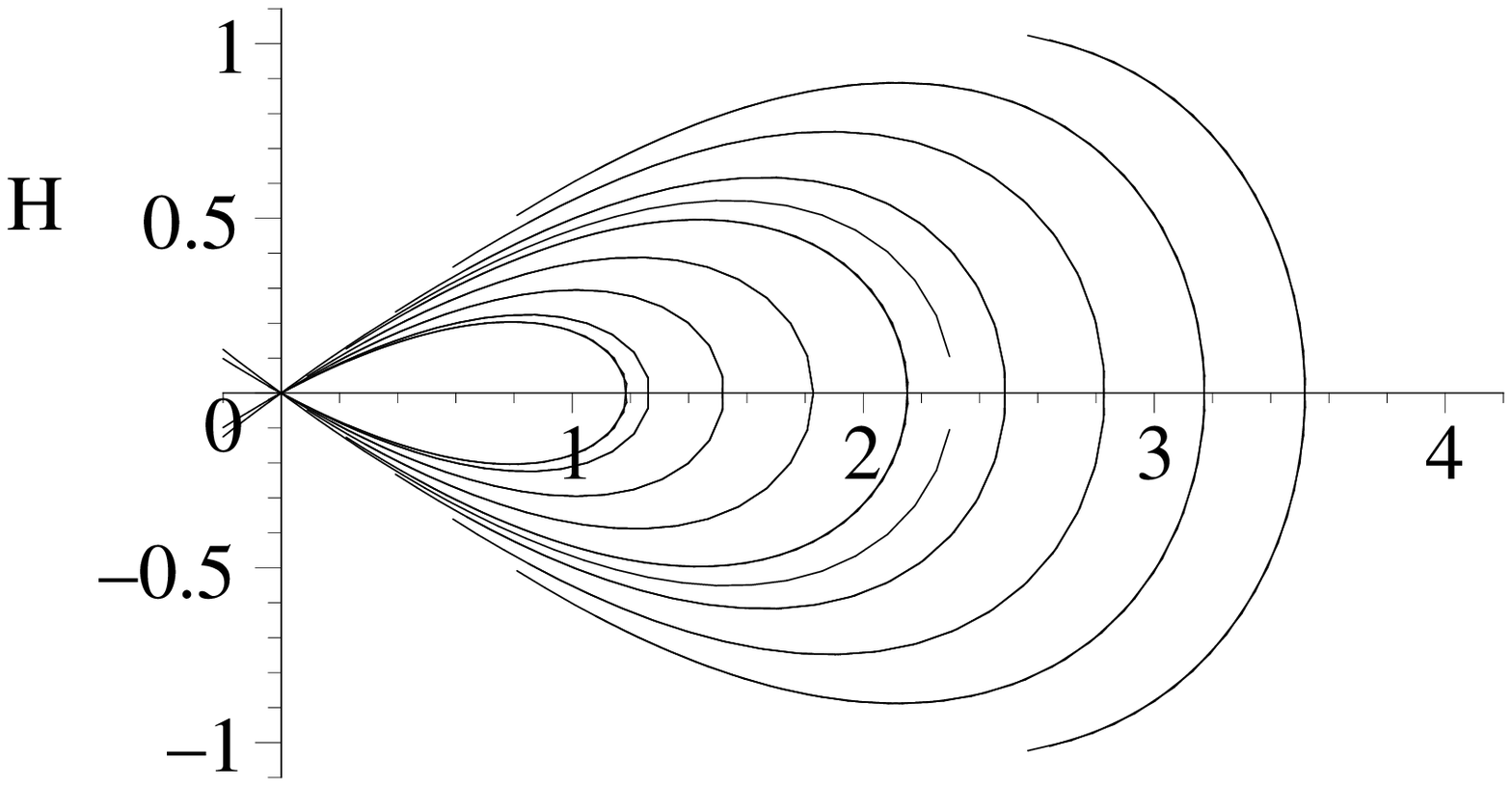} \\ e \end{tabular} &
   \begin{tabular}{c} \includegraphics[width=0.15 \textwidth]{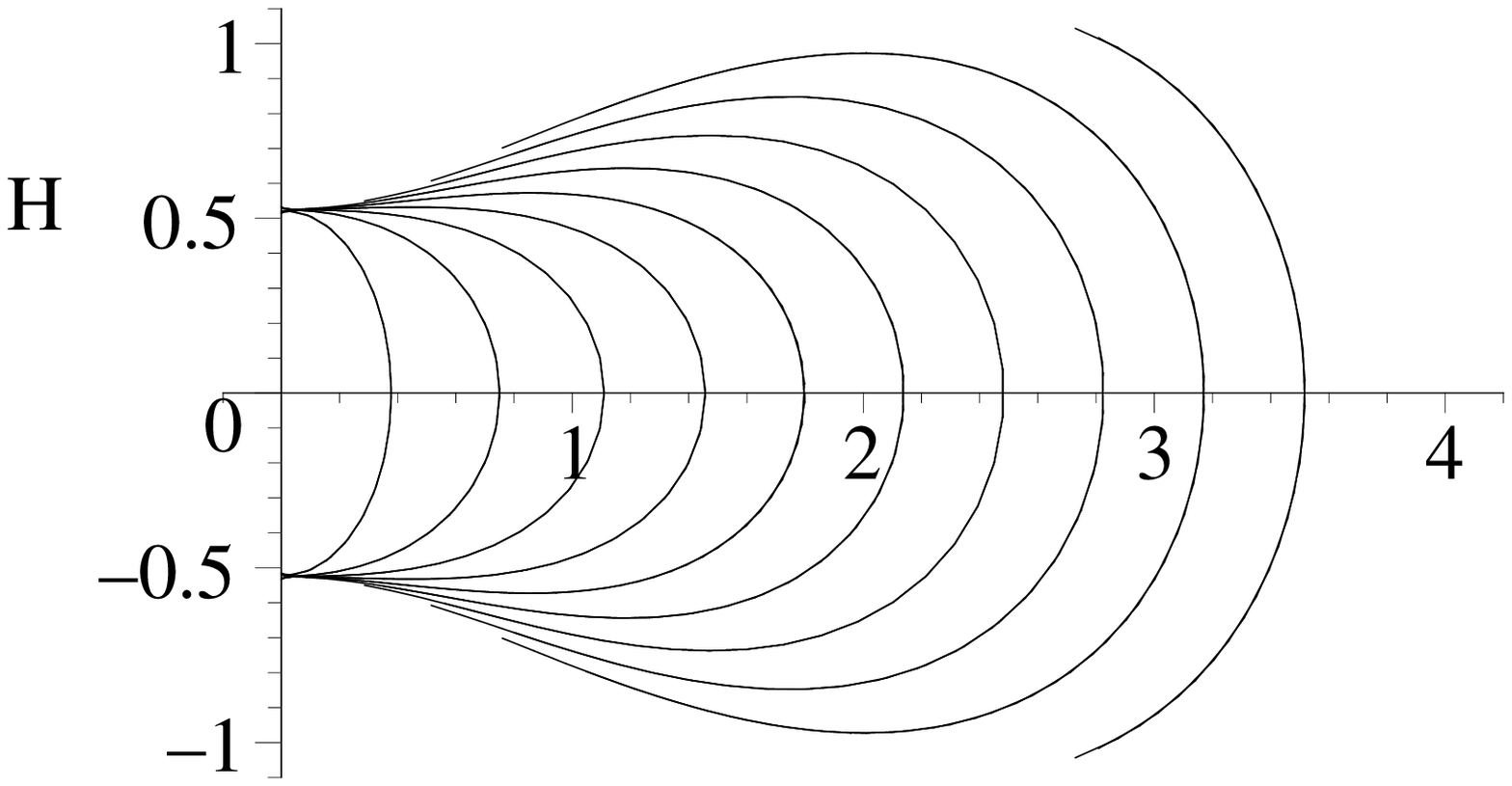} \\ f \end{tabular} \\
  \hline
   $m>0$ &
   \begin{tabular}{c} \includegraphics[width=0.15 \textwidth]{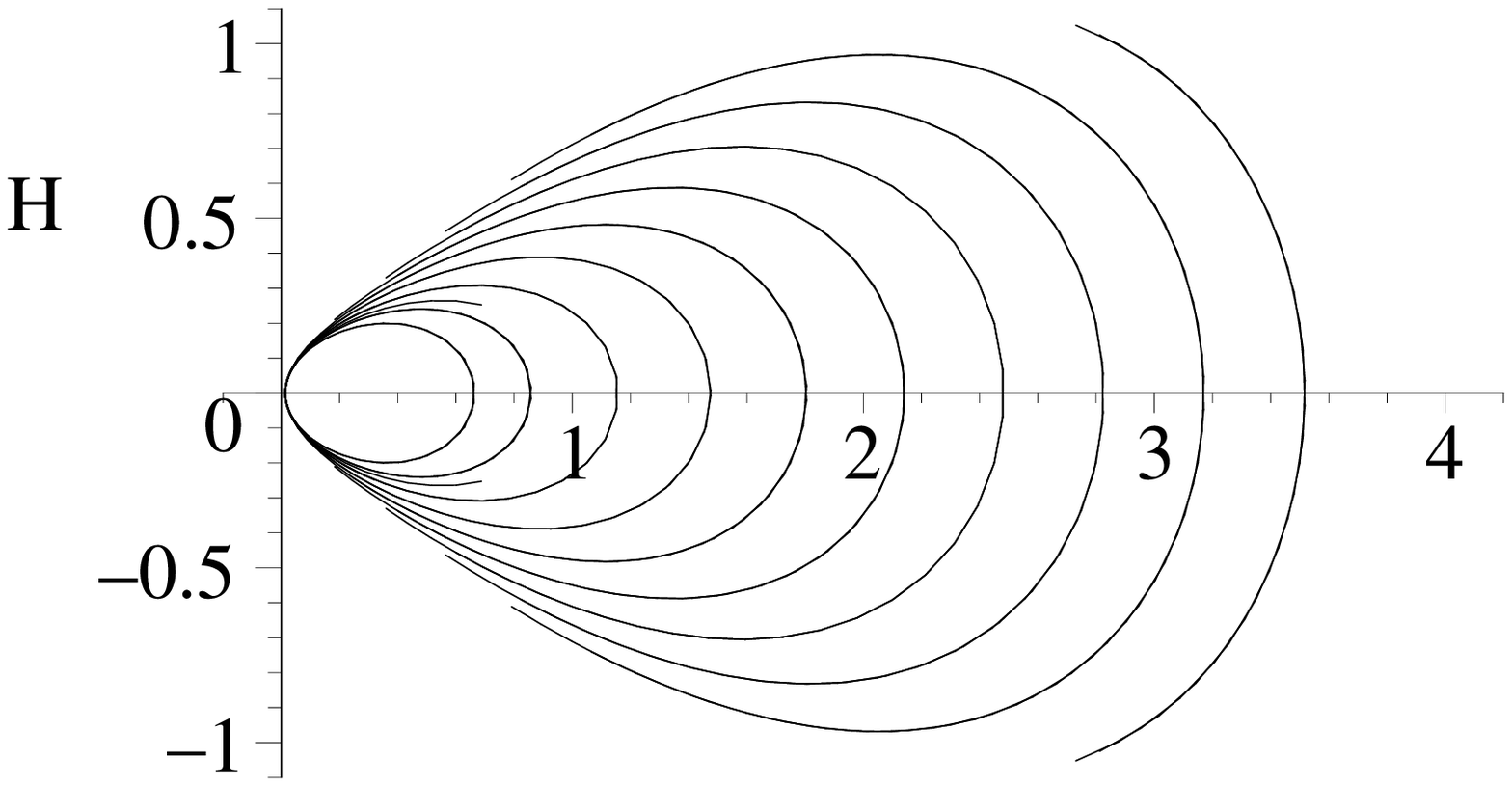} \\ g \end{tabular} &
   \begin{tabular}{c} \includegraphics[width=0.15 \textwidth]{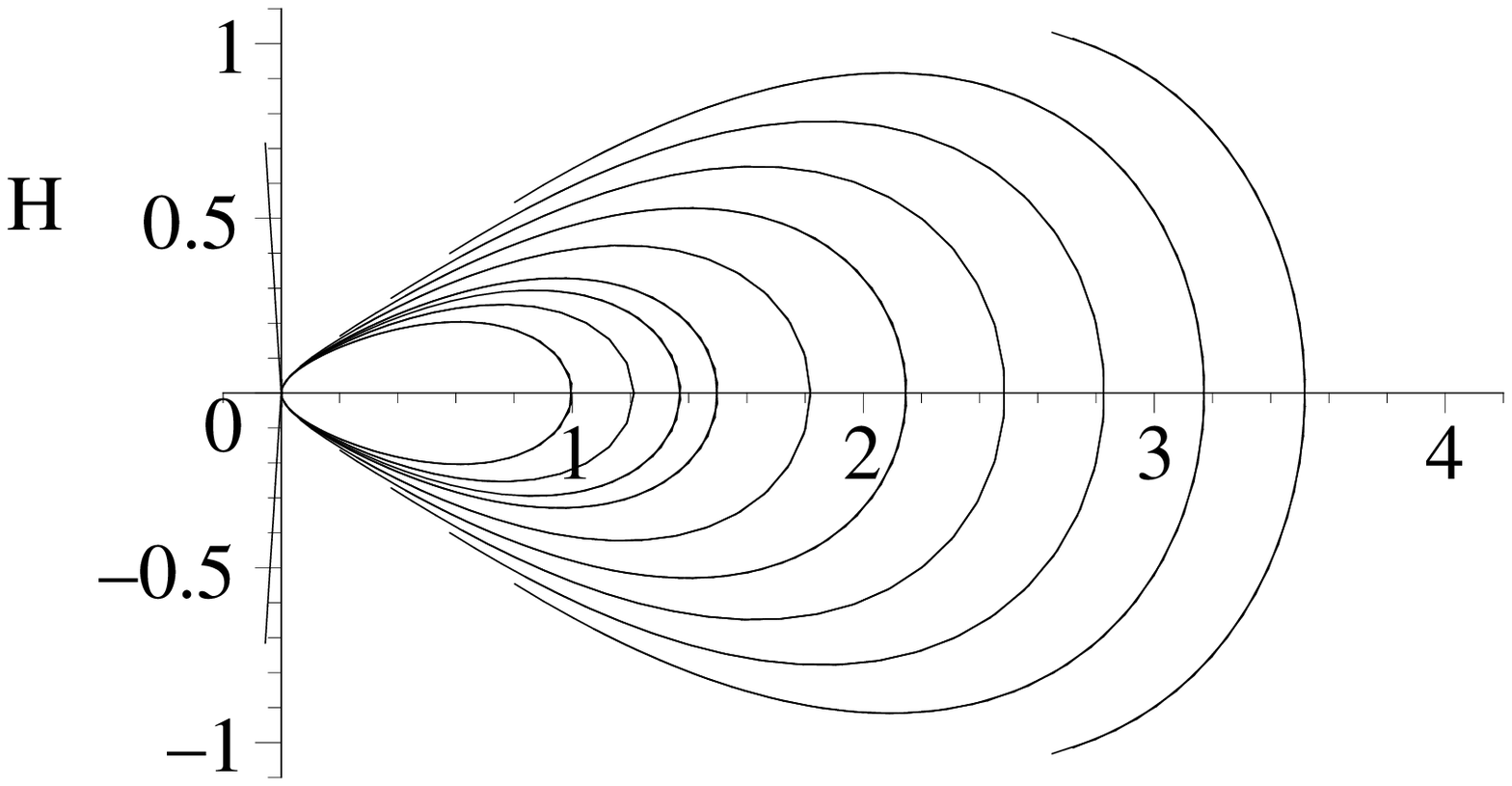} \\ h \end{tabular} &
   \begin{tabular}{c} \includegraphics[width=0.15 \textwidth]{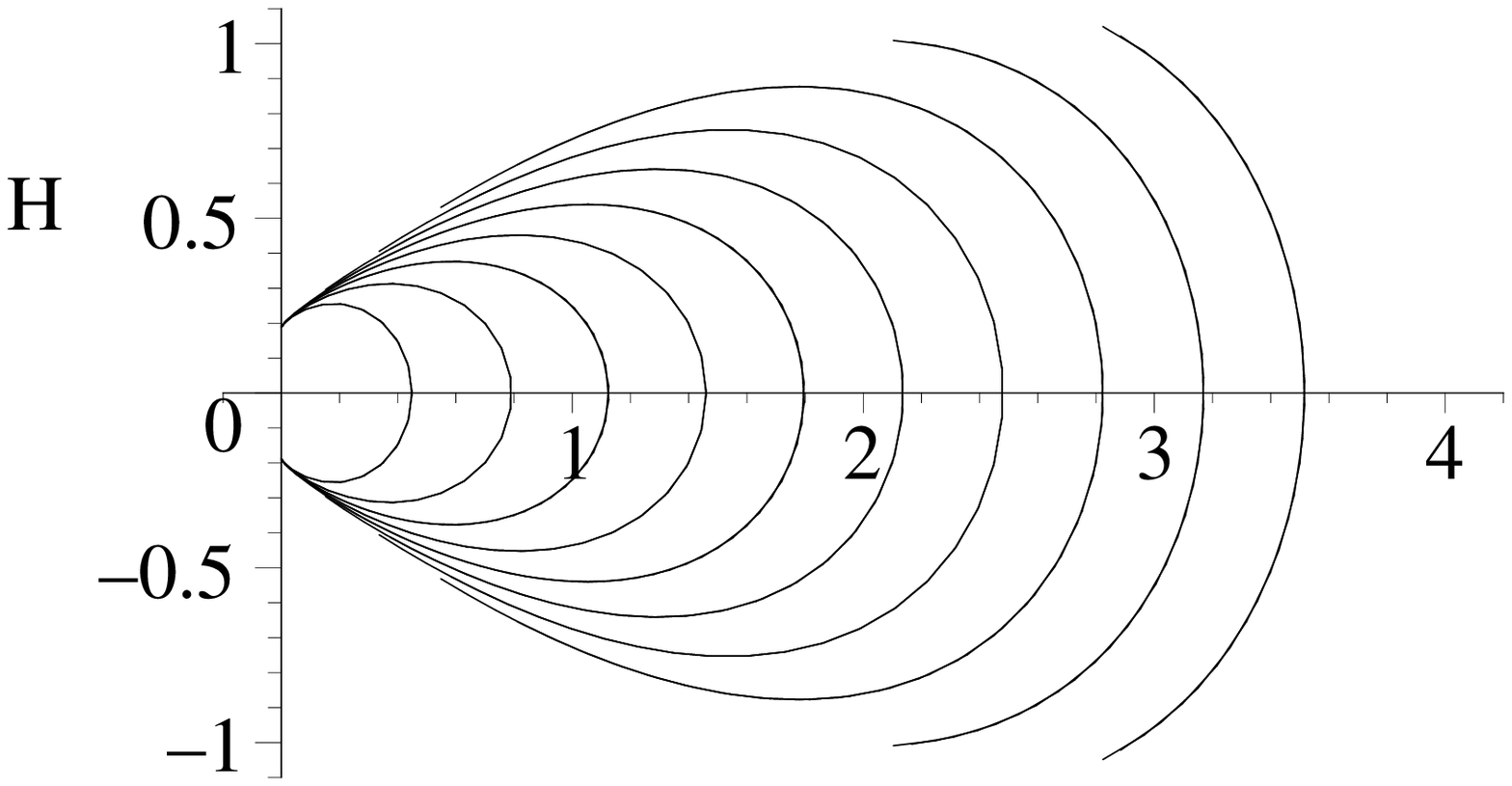} \\ i \end{tabular} \\
  \hline
\end{tabular}//
\vskip 2 mm
Table E1. Case with  $\ve=0$, $n=1$ and $\lambda=0$. \\

\begin{tabular}{|c|c|c|c|}
  \hline
     & $\Lambda<0$ & $\Lambda=0$ & $\Lambda>0$ \\
  \hline
   $m<0$ &
   \begin{tabular}{c} \includegraphics[width=0.15 \textwidth]{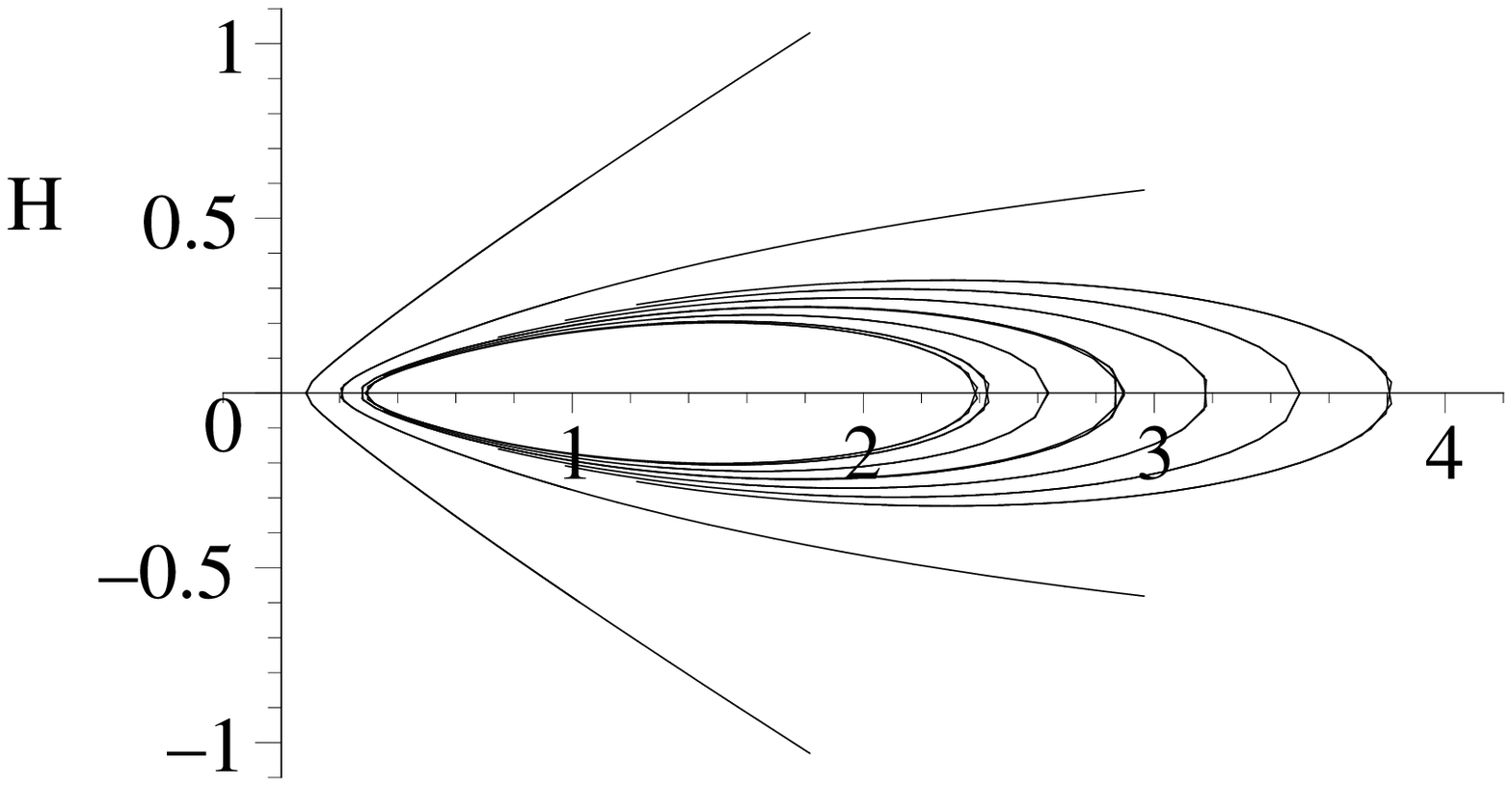} \\ a  \end{tabular} &
   \begin{tabular}{c} \includegraphics[width=0.15 \textwidth]{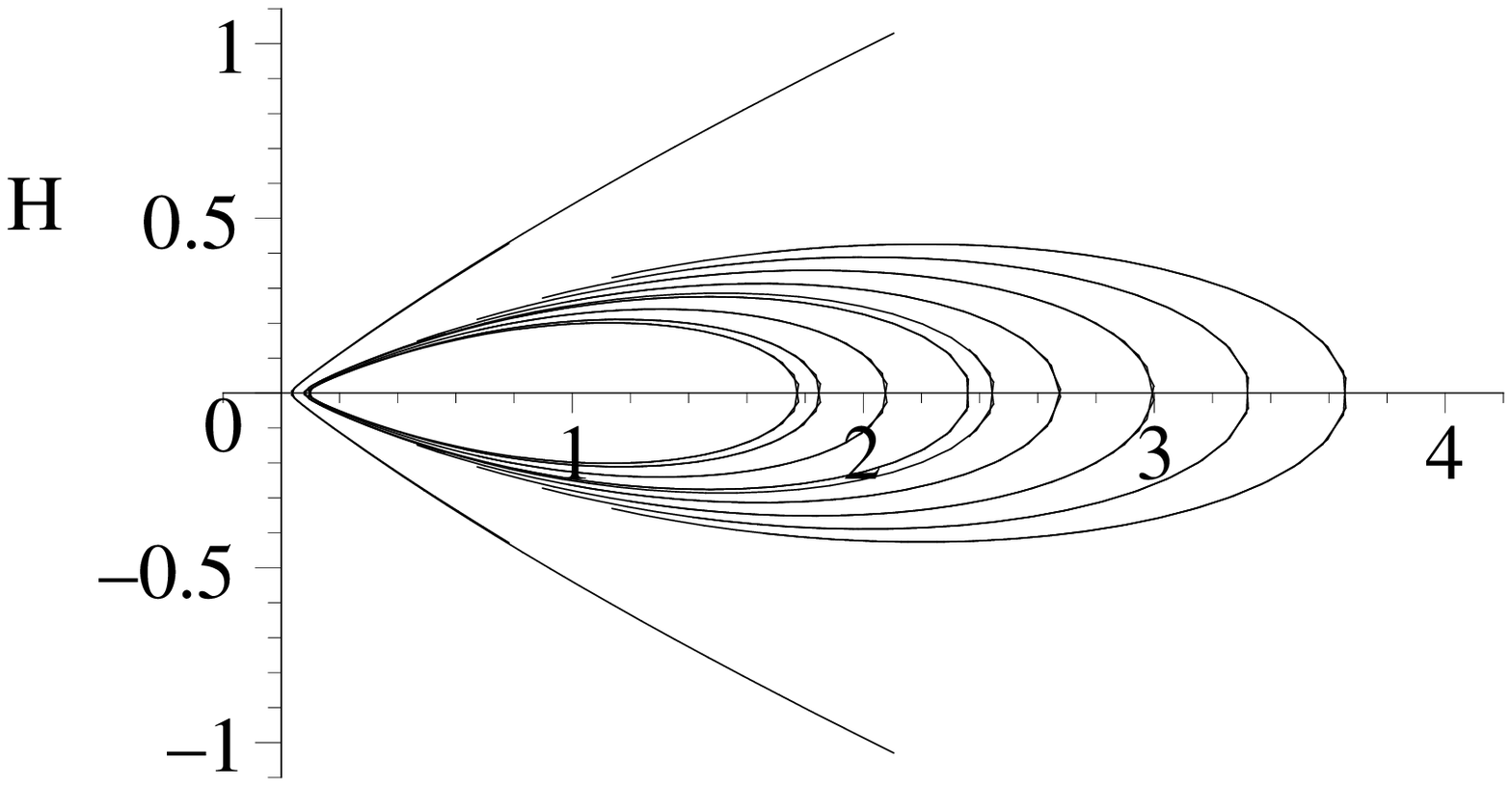} \\ b \end{tabular} &
   \begin{tabular}{c} \includegraphics[width=0.15 \textwidth]{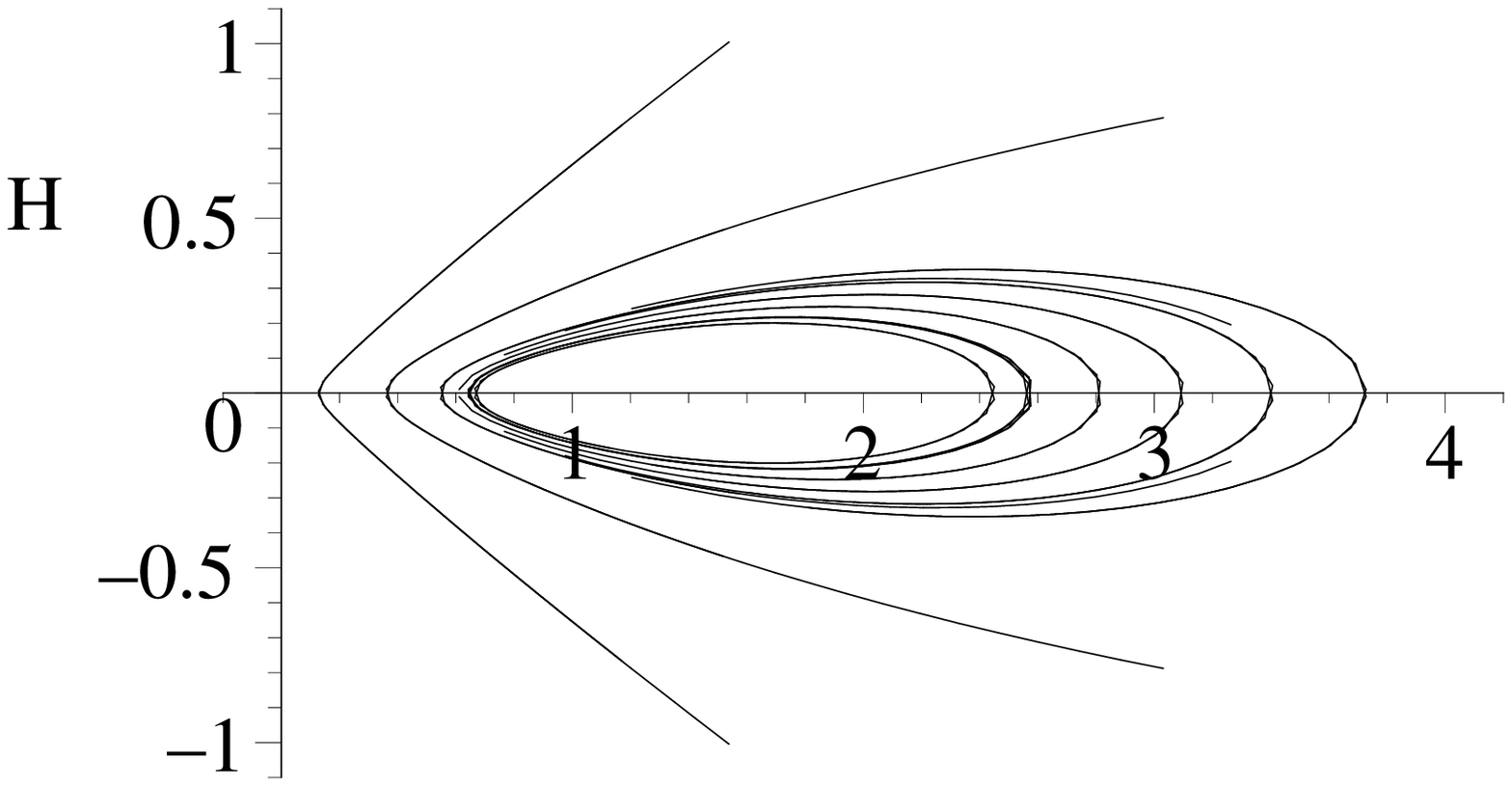} \\ c \end{tabular} \\
  \hline
   $m=0$ &
   \begin{tabular}{c} \includegraphics[width=0.15 \textwidth]{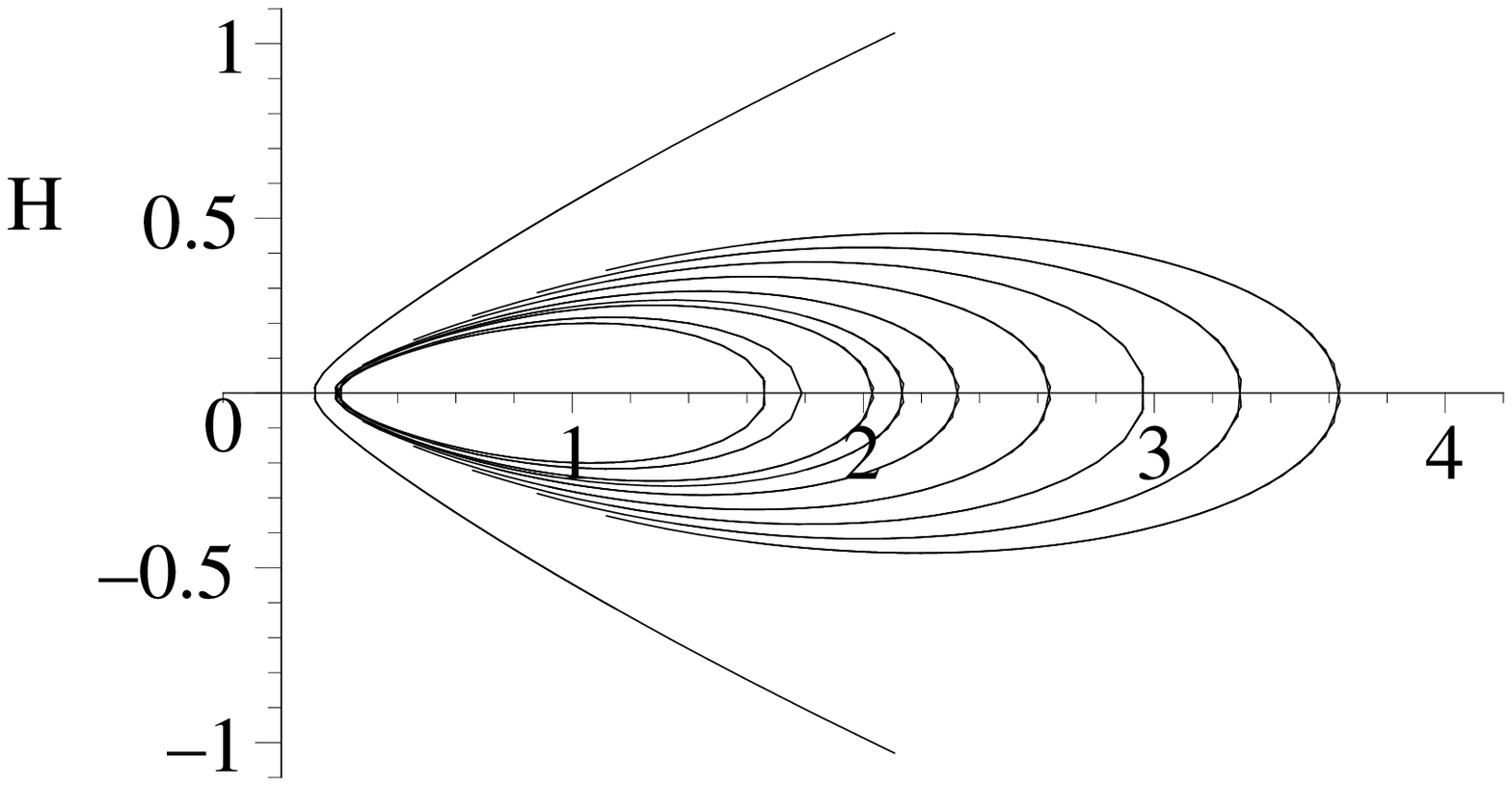} \\ d \end{tabular} &
   \begin{tabular}{c} \includegraphics[width=0.15 \textwidth]{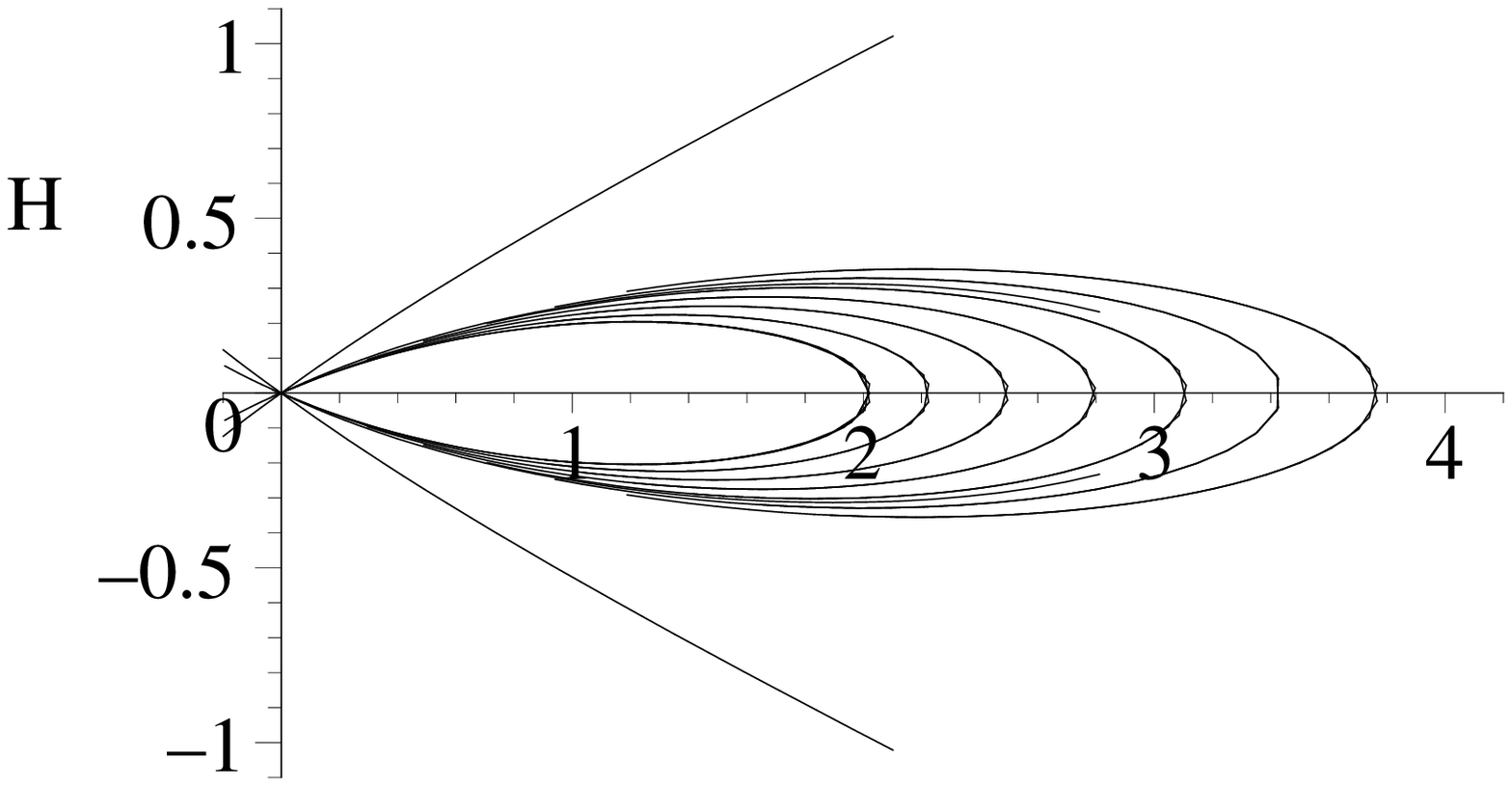} \\ e \end{tabular} &
   \begin{tabular}{c} \includegraphics[width=0.15 \textwidth]{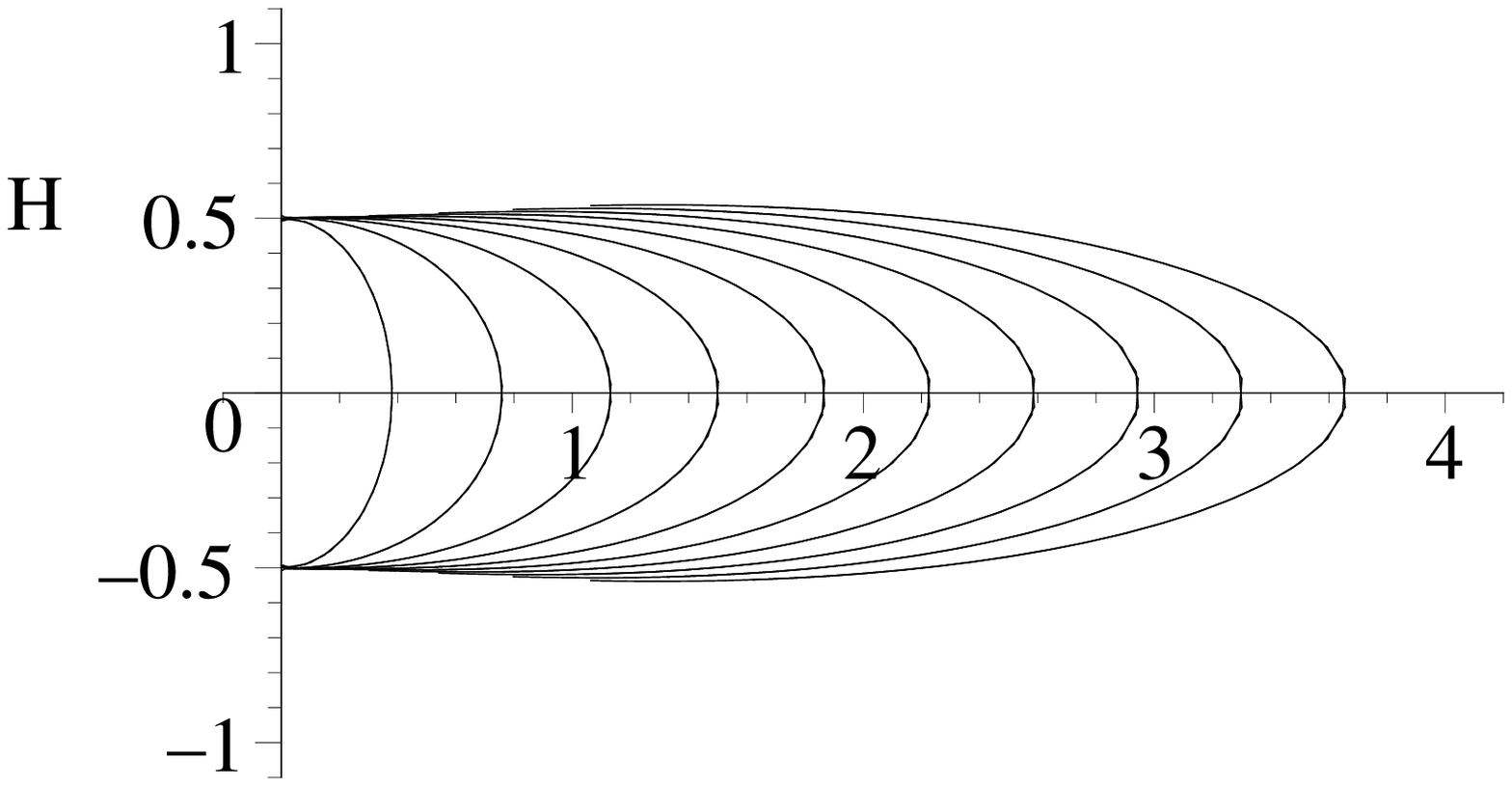} \\ f \end{tabular} \\
  \hline
   $m>0$ &
   \begin{tabular}{c} \includegraphics[width=0.15 \textwidth]{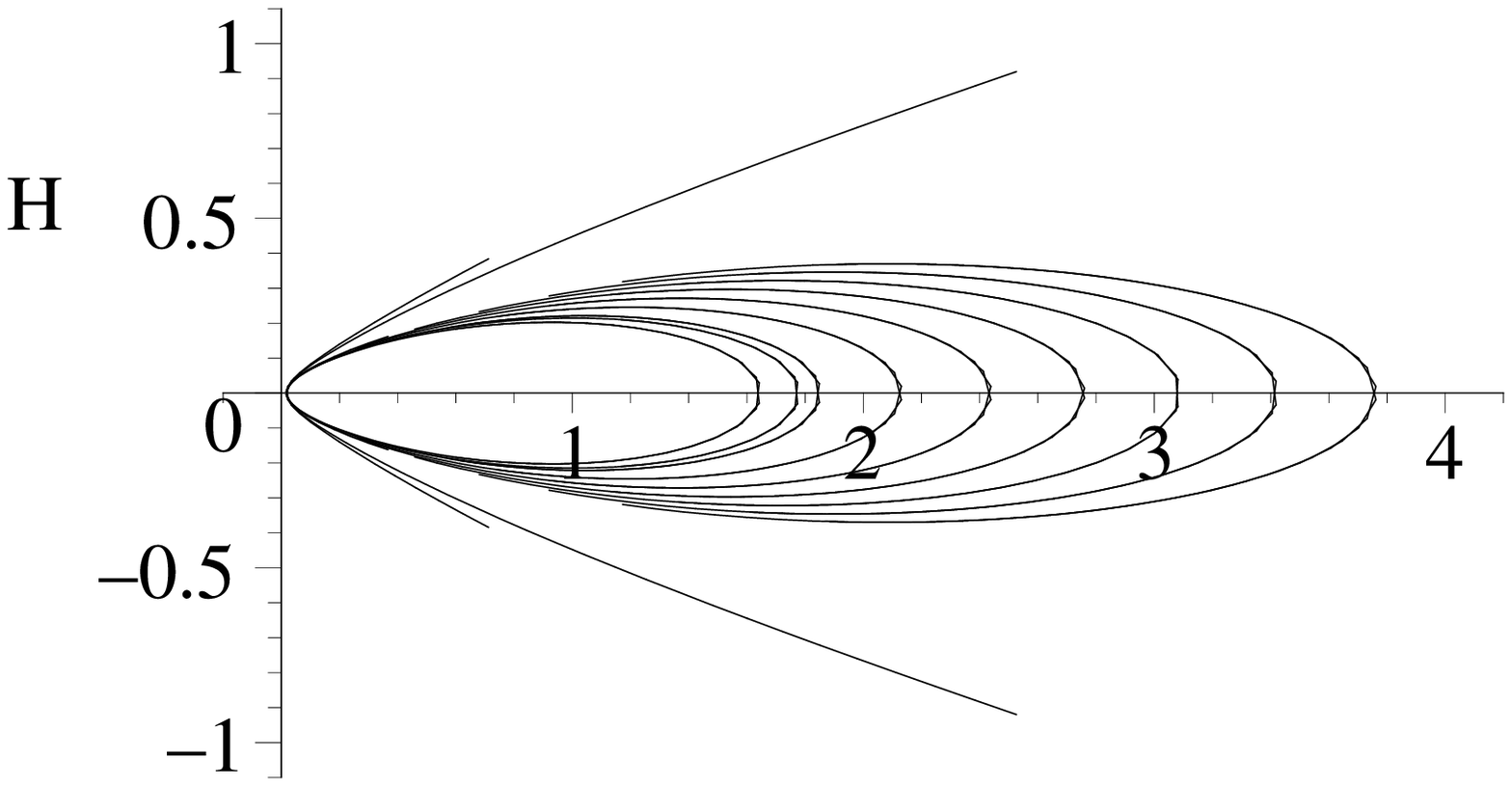} \\ g \end{tabular} &
   \begin{tabular}{c} \includegraphics[width=0.15 \textwidth]{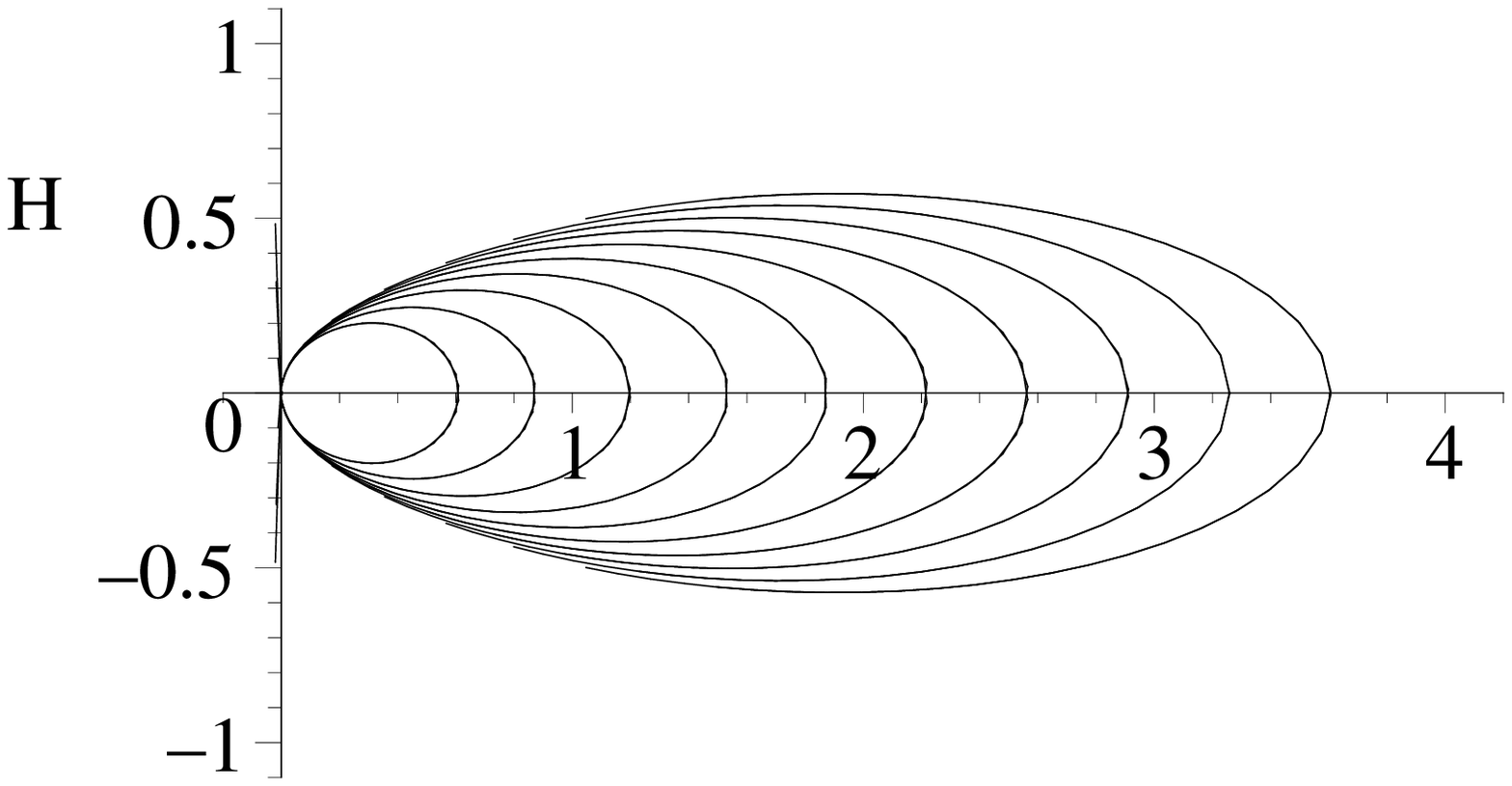} \\ h \end{tabular} &
   \begin{tabular}{c} \includegraphics[width=0.15 \textwidth]{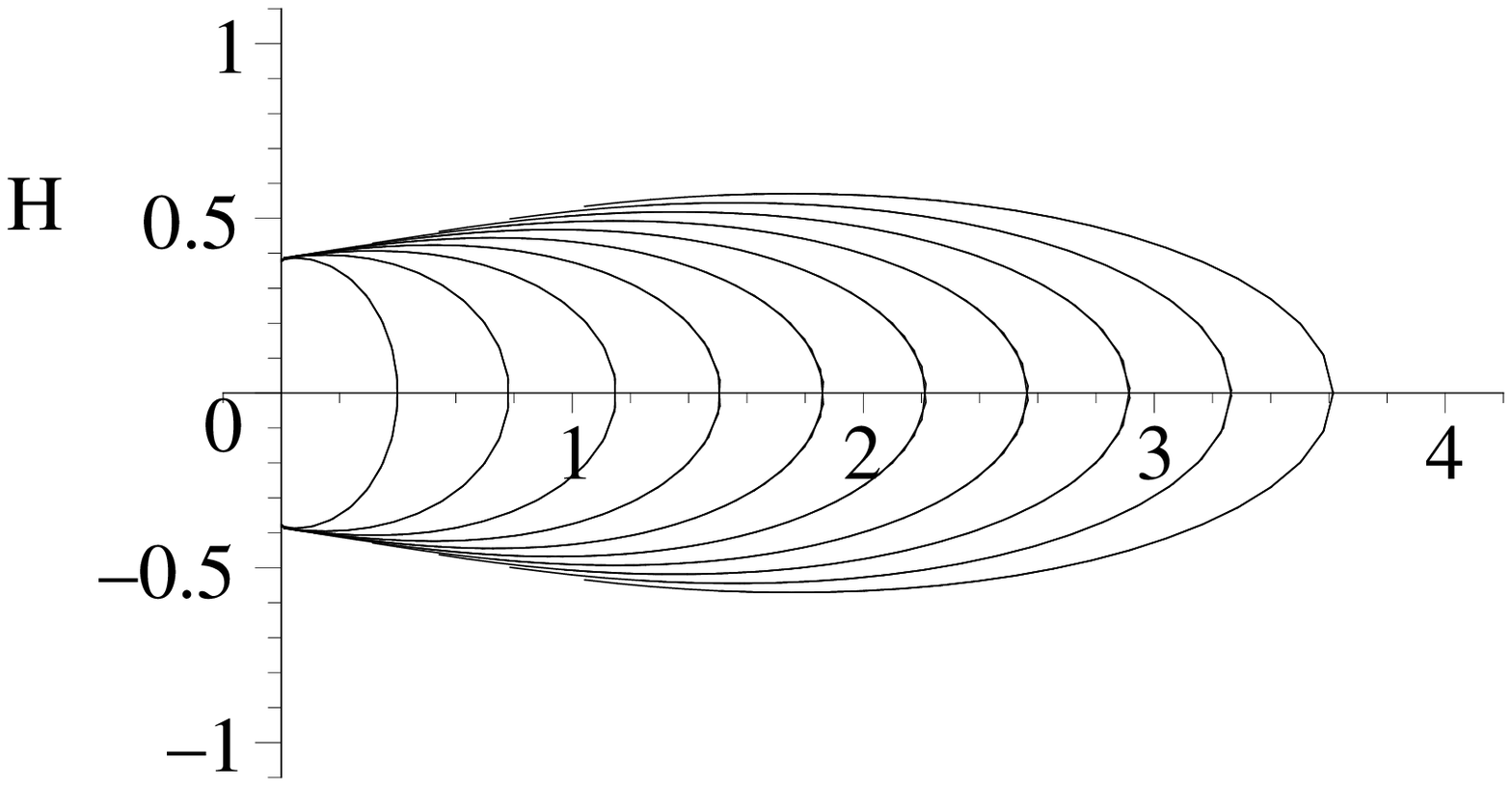} \\ i \end{tabular} \\
  \hline
\end{tabular}//
\vskip 2 mm
Table A2. Case with  $\ve=0$, $n=1$ and $\lambda>0$. \\
\vskip 2 mm

\begin{tabular}{|c|c|c|c|}
  \hline
     & $\Lambda<0$ & $\Lambda=0$ & $\Lambda>0$ \\
  \hline
   $m<0$ &
   \begin{tabular}{c} \includegraphics[width=0.15 \textwidth]{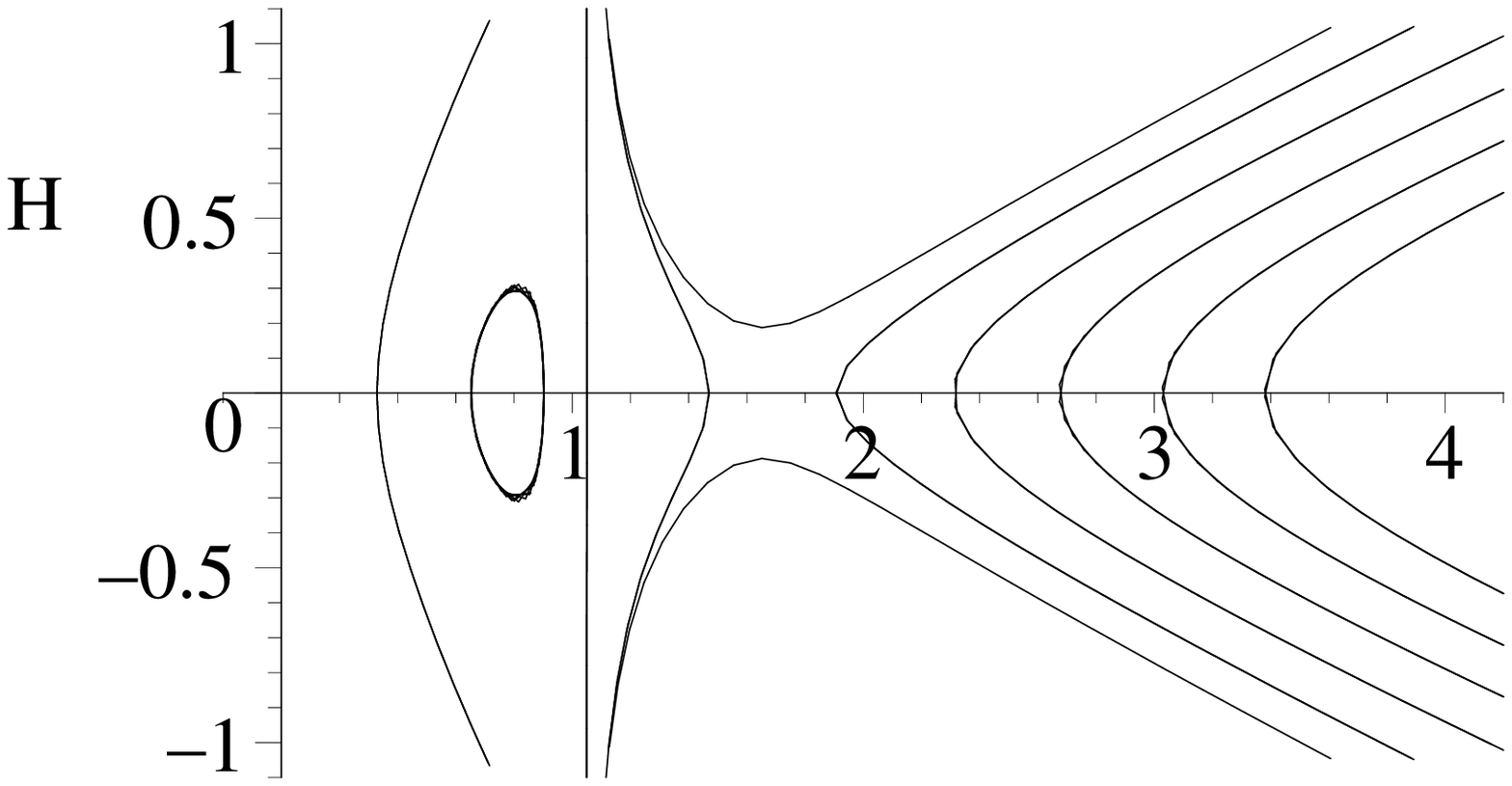} \\ a  \end{tabular} &
   \begin{tabular}{c} \includegraphics[width=0.15 \textwidth]{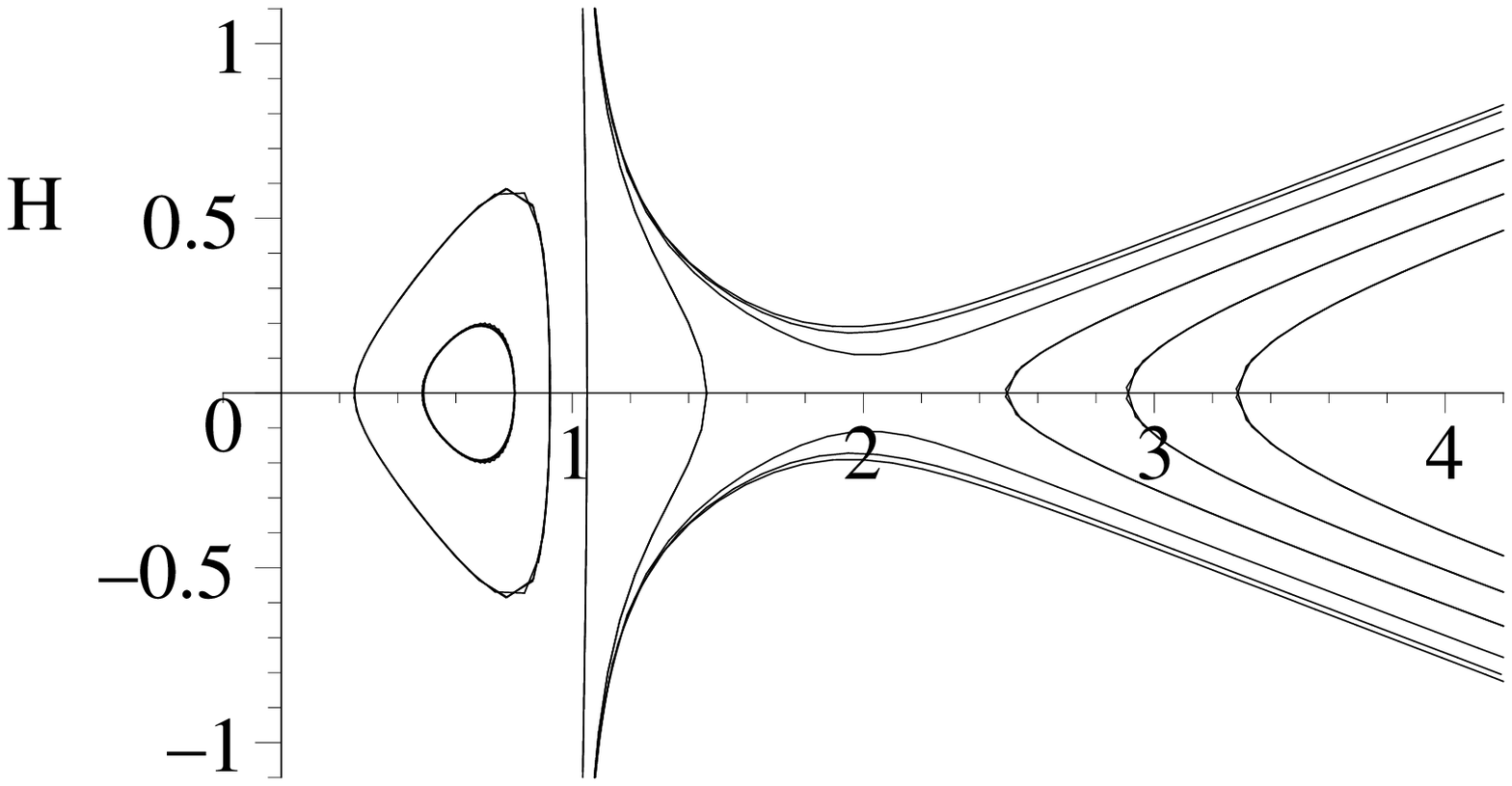} \\ b \end{tabular} &
   \begin{tabular}{c} \includegraphics[width=0.15 \textwidth]{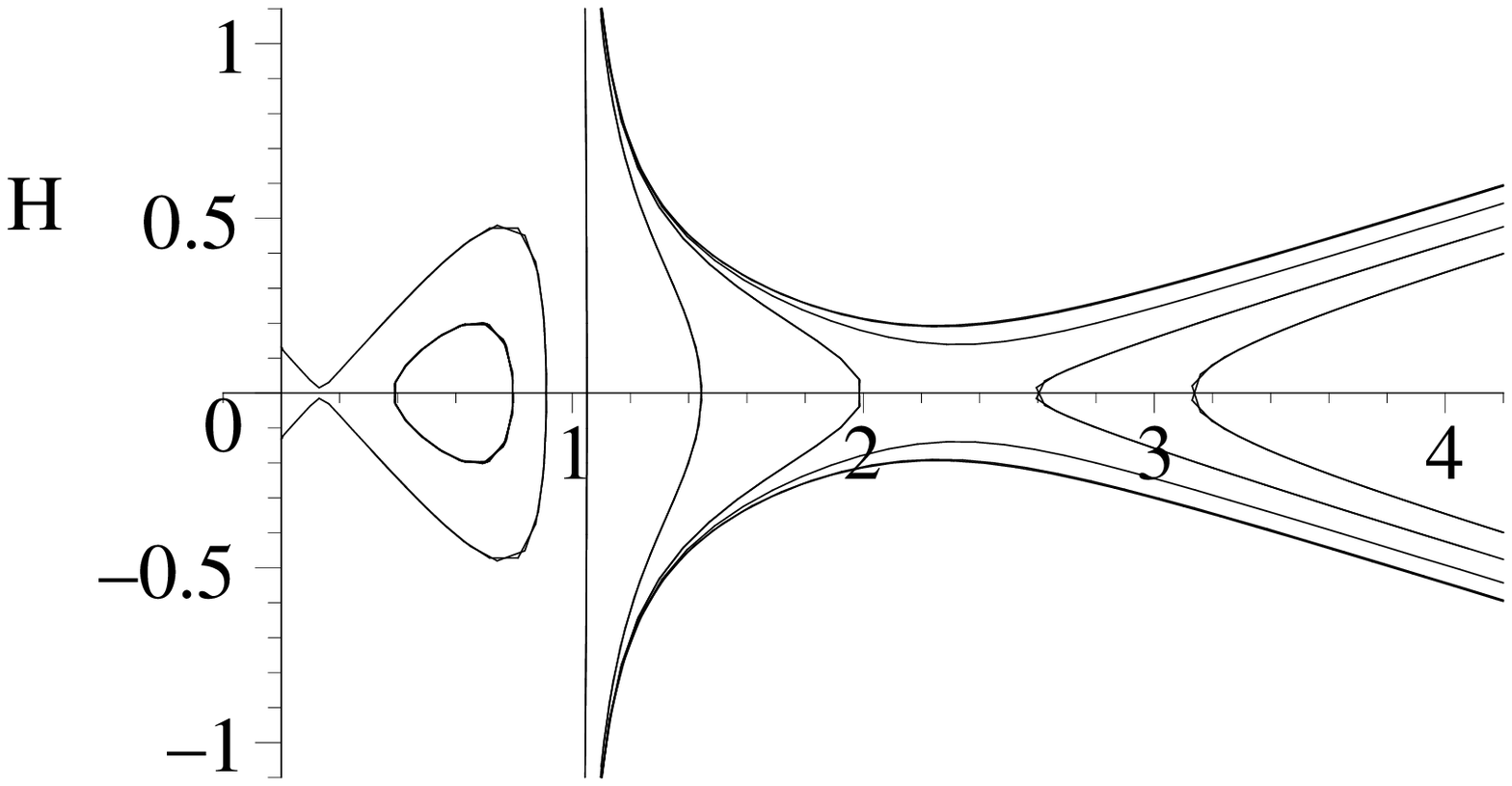} \\ c \end{tabular} \\
  \hline
   $m=0$ &
   \begin{tabular}{c} \includegraphics[width=0.15 \textwidth]{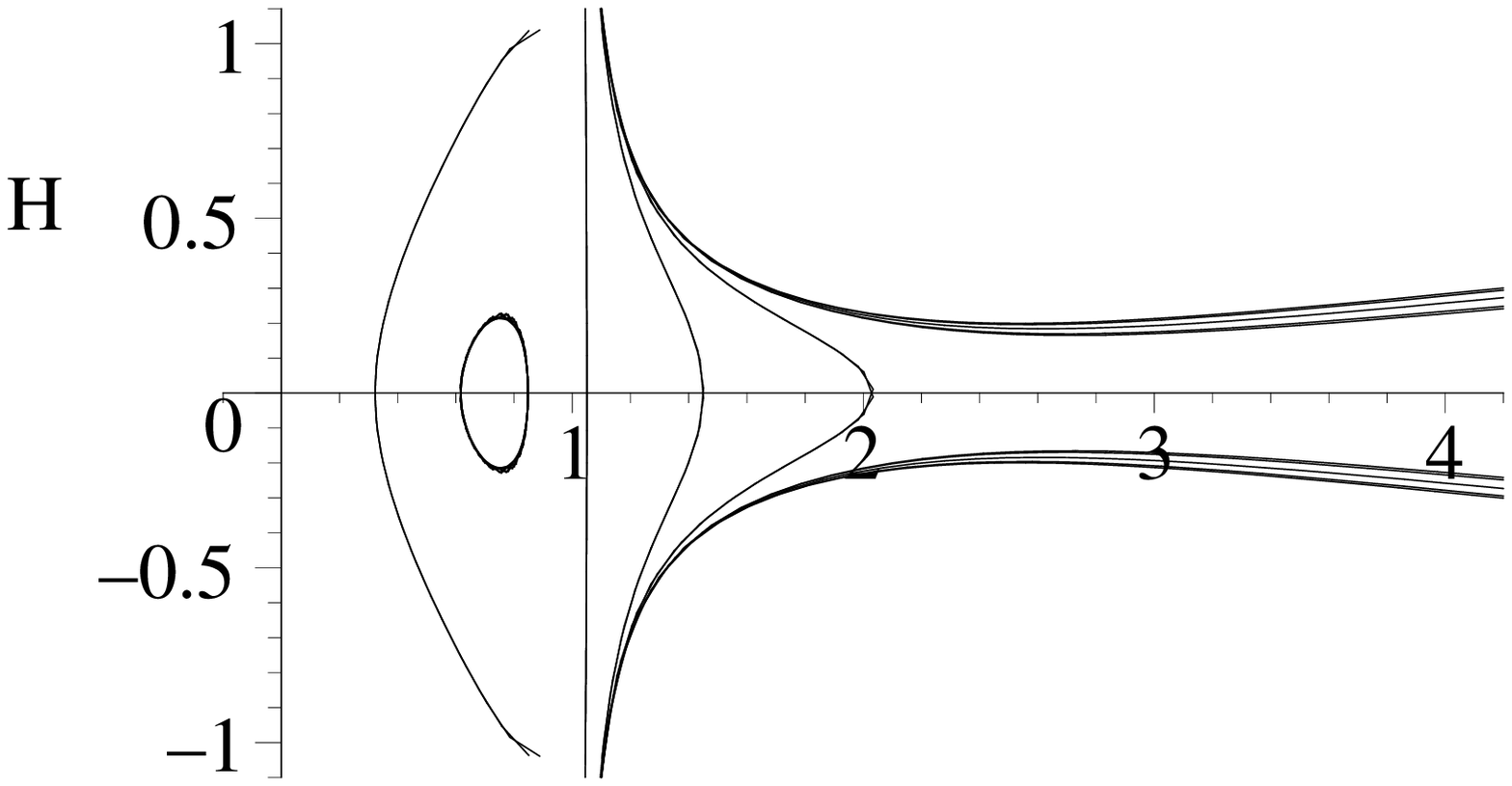} \\ d \end{tabular} &
   \begin{tabular}{c} \includegraphics[width=0.15 \textwidth]{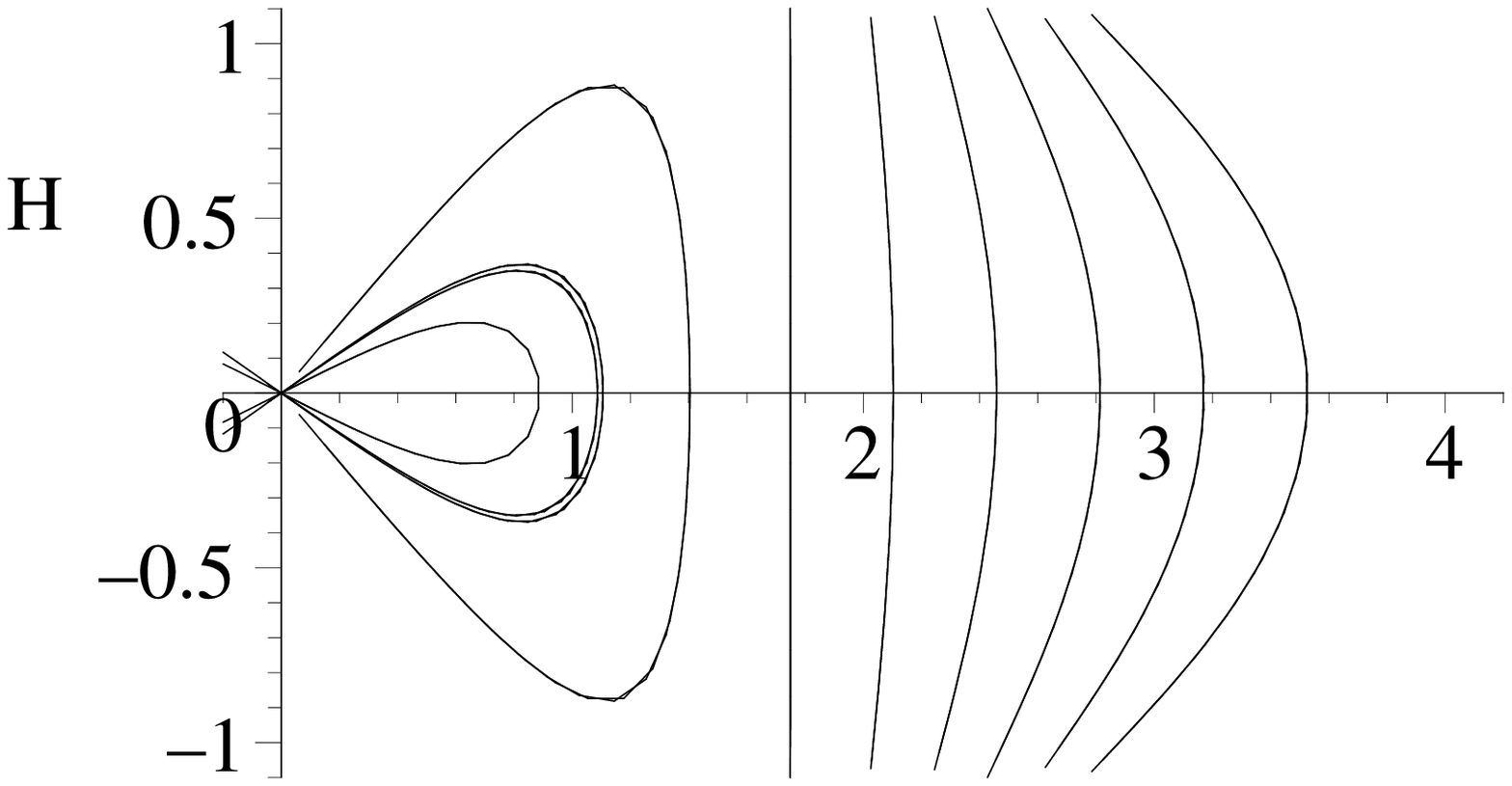} \\ e \end{tabular} &
   \begin{tabular}{c} \includegraphics[width=0.15 \textwidth]{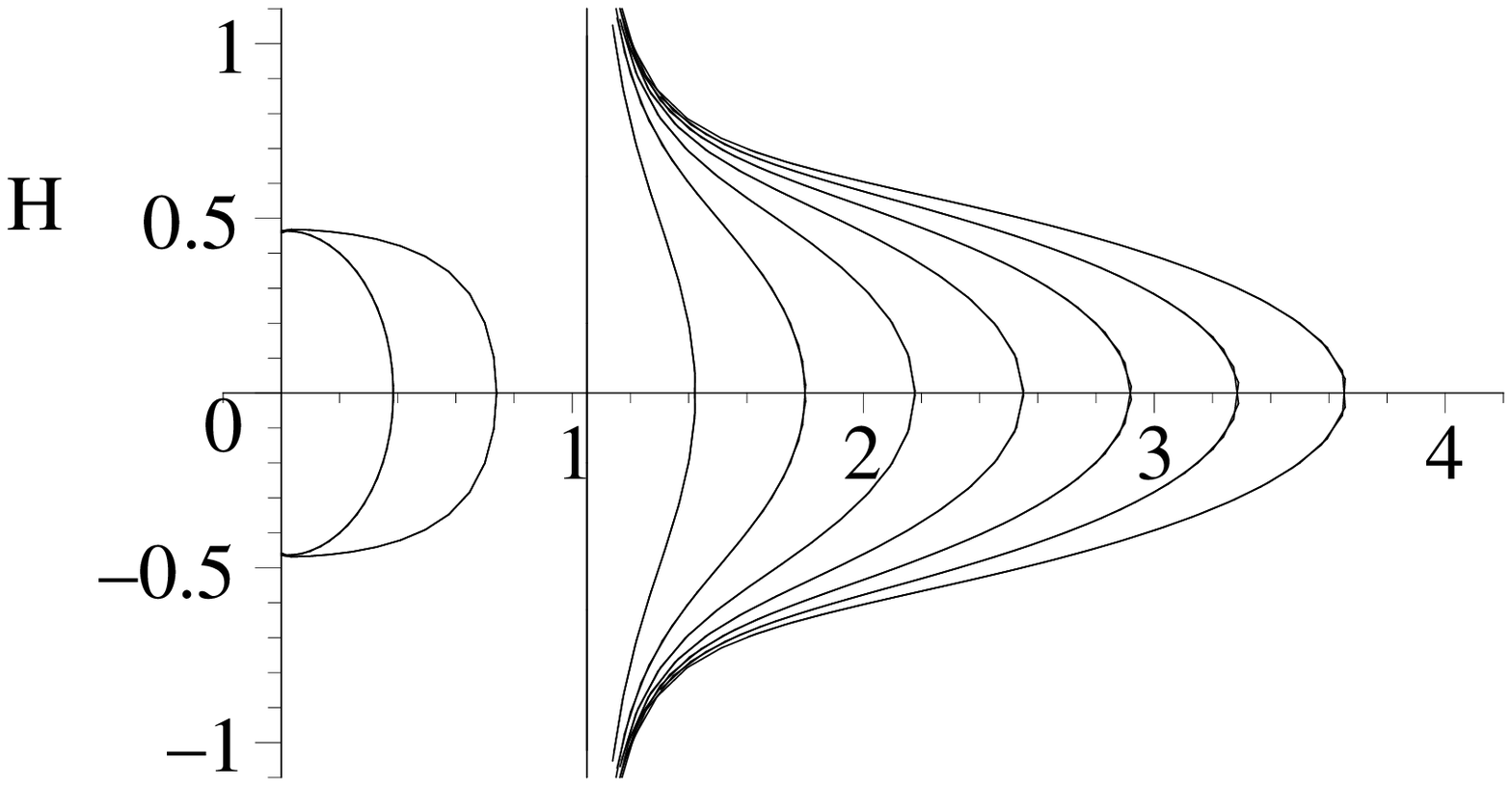} \\ f \end{tabular} \\
  \hline
   $m>0$ &
   \begin{tabular}{c} \includegraphics[width=0.15 \textwidth]{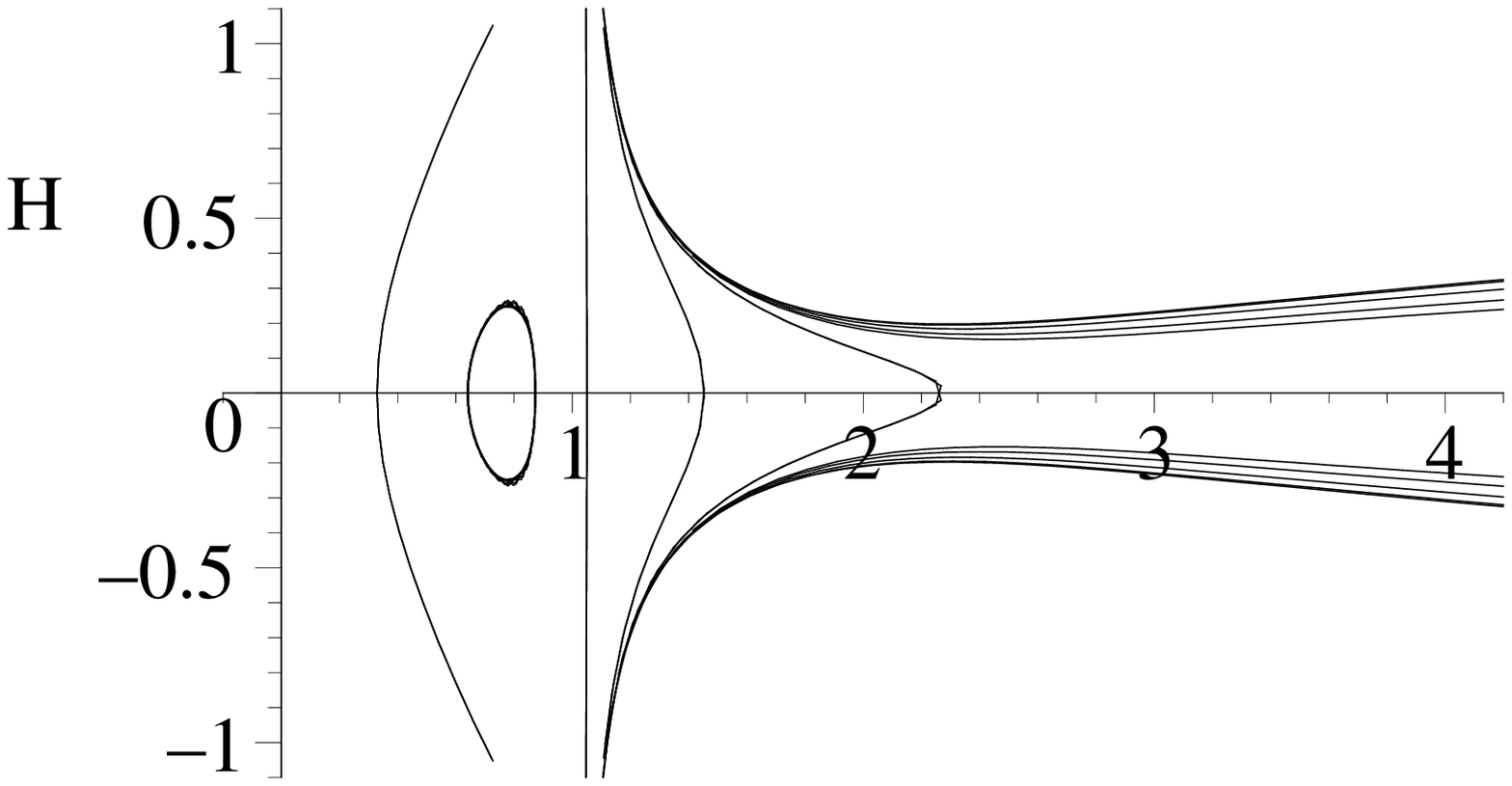} \\ g \end{tabular} &
   \begin{tabular}{c} \includegraphics[width=0.15 \textwidth]{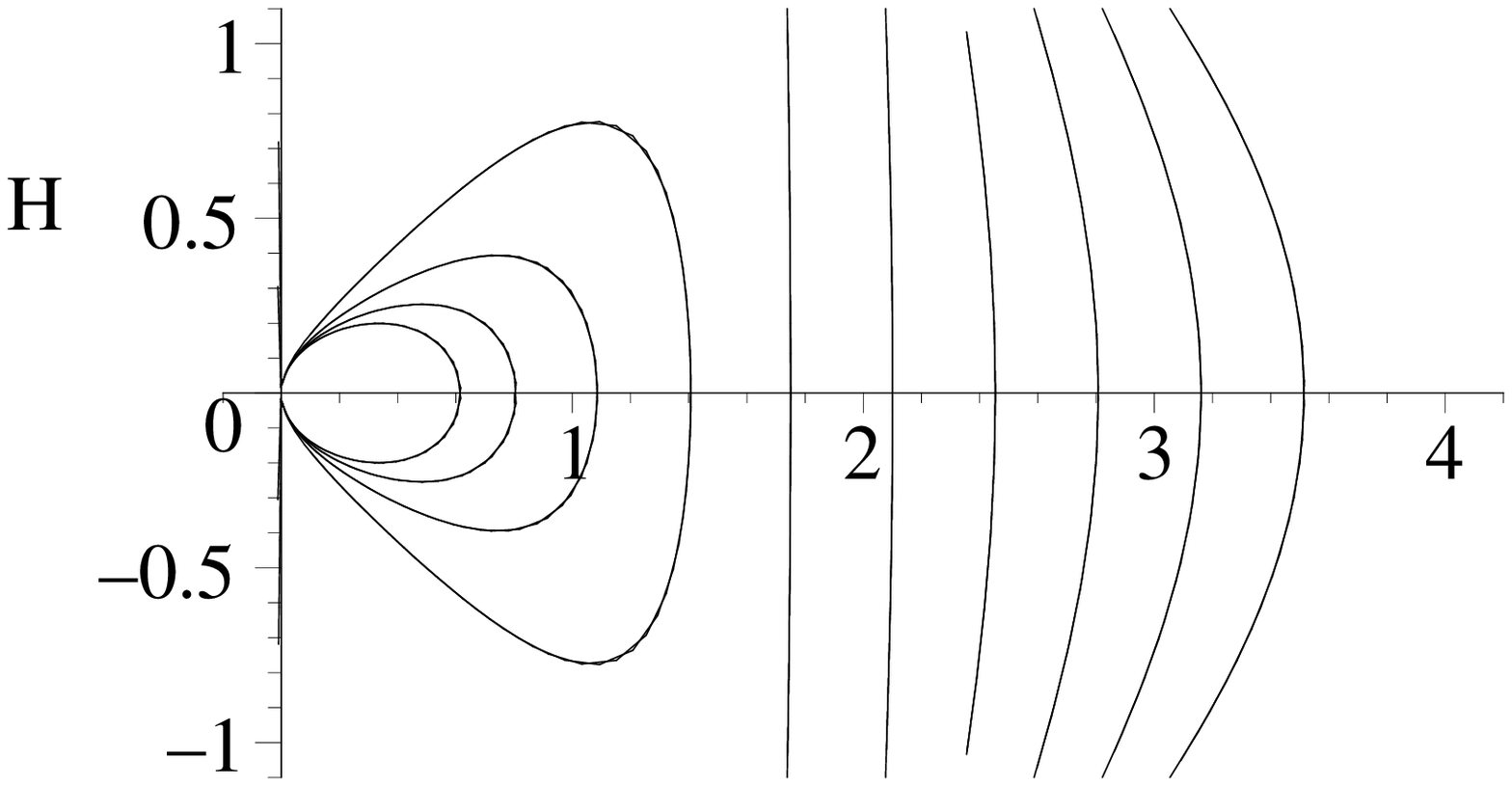} \\ h \end{tabular} &
   \begin{tabular}{c} \includegraphics[width=0.15 \textwidth]{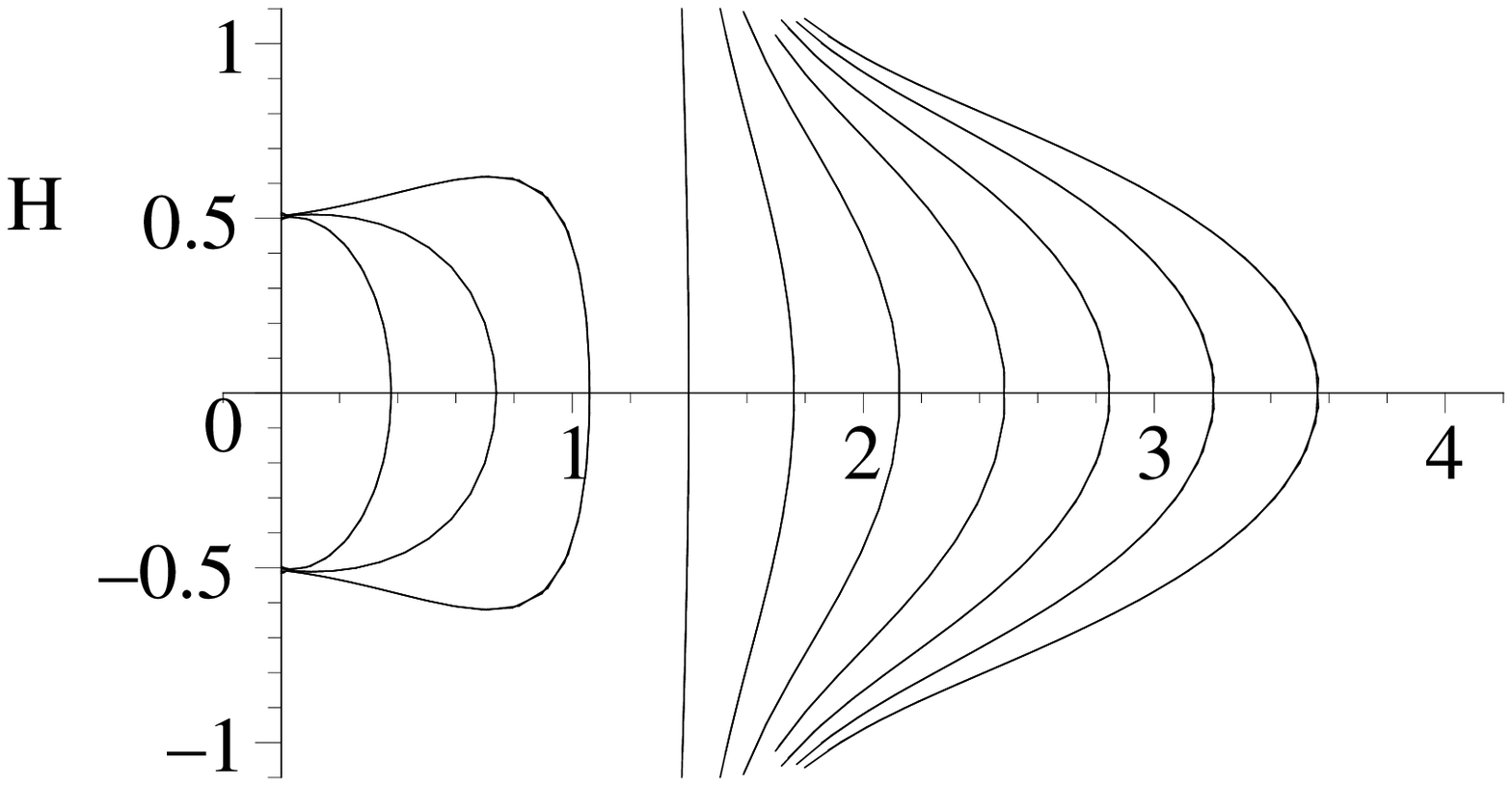} \\ i \end{tabular} \\
  \hline
\end{tabular}//
\vskip 2 mm
Table B1. Case with  $\ve=0$, $n=2$ and $\lambda<0$. \\
\vskip 2 mm

\begin{tabular}{|c|c|c|c|}
  \hline
     & $\Lambda<0$ & $\Lambda=0$ & $\Lambda>0$ \\
  \hline
   $m<0$ &
   \begin{tabular}{c} \includegraphics[width=0.15 \textwidth]{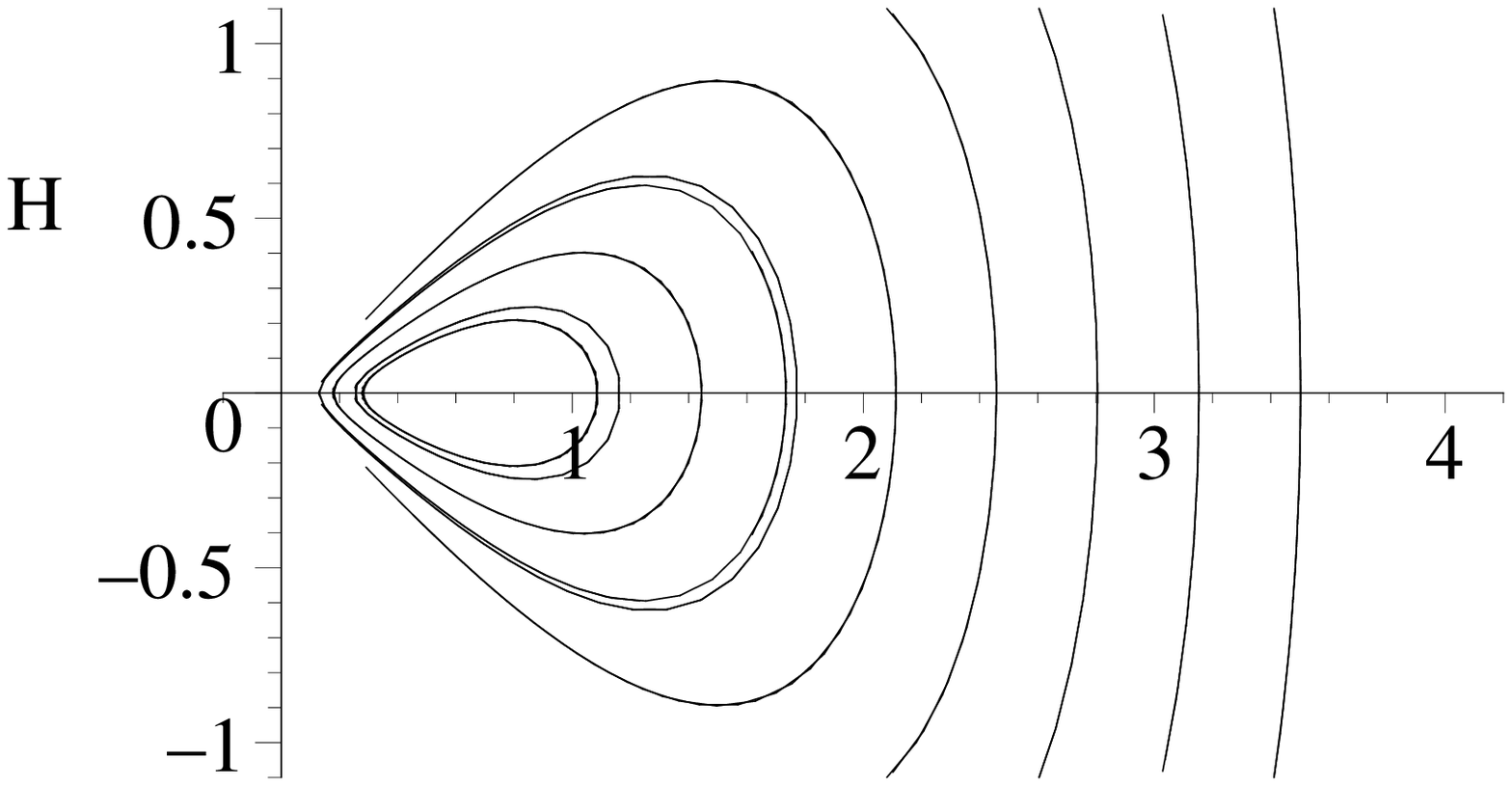} \\ a  \end{tabular} &
   \begin{tabular}{c} \includegraphics[width=0.15 \textwidth]{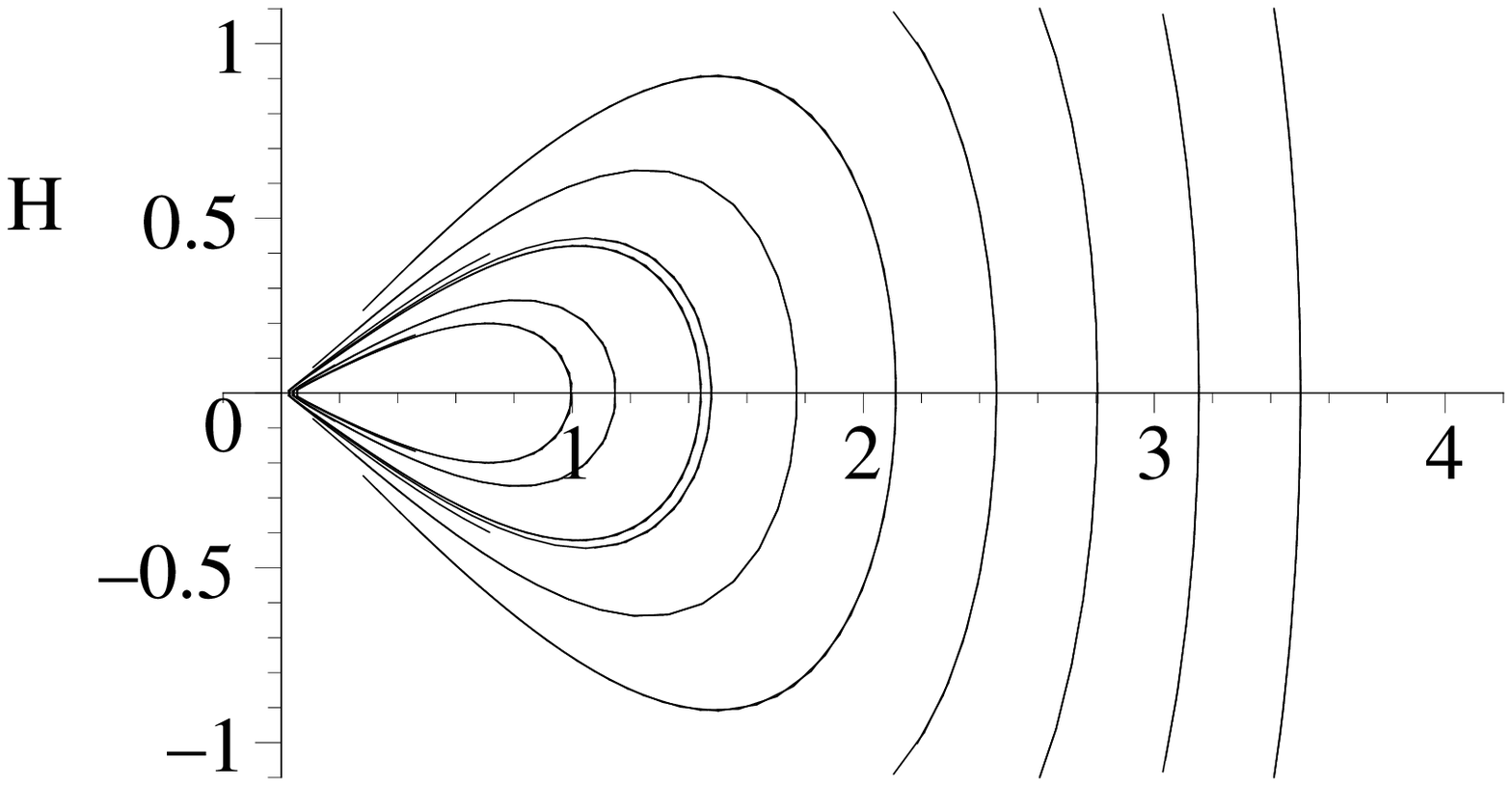} \\ b \end{tabular} &
   \begin{tabular}{c} \includegraphics[width=0.15 \textwidth]{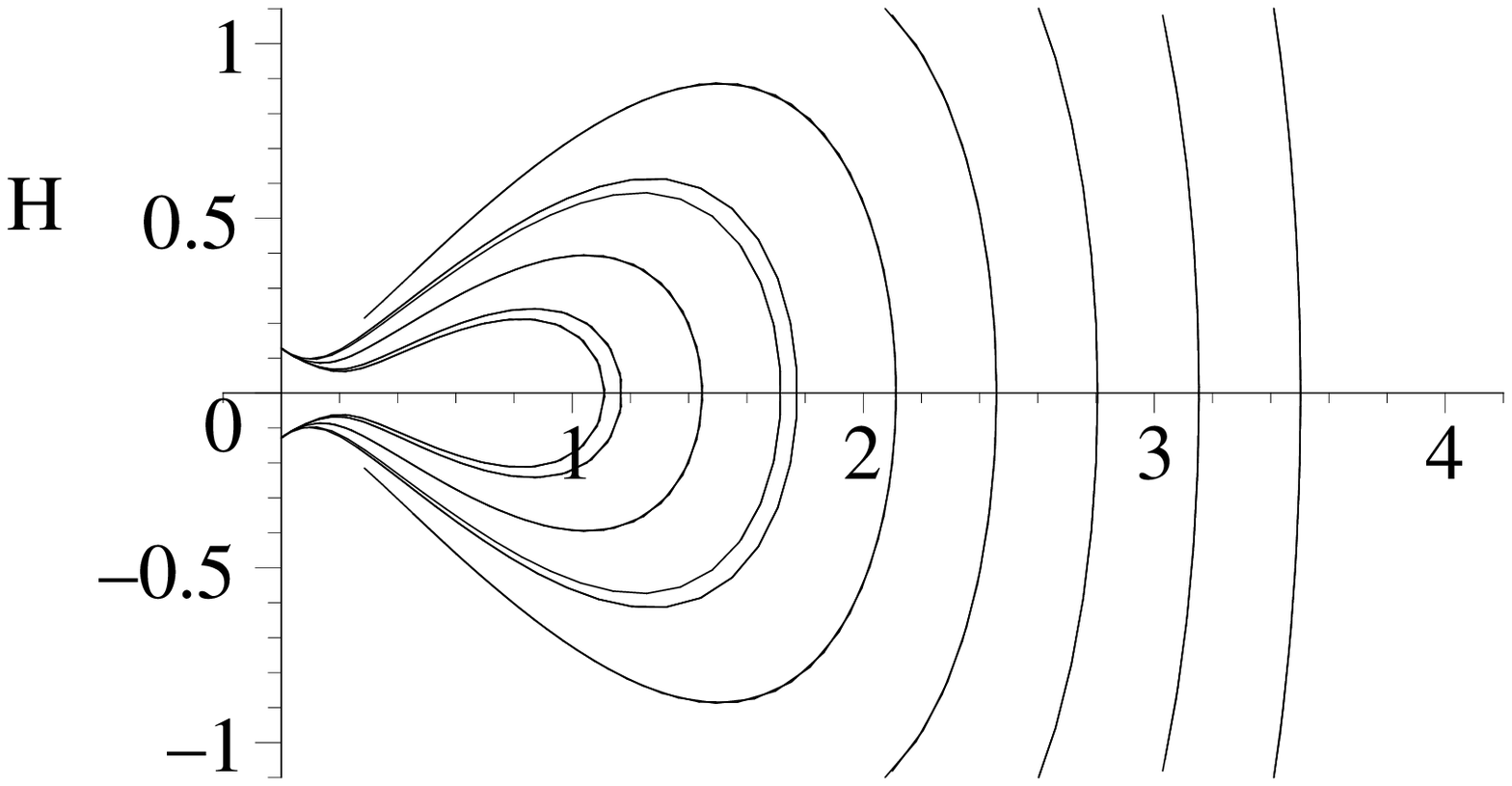} \\ c \end{tabular} \\
  \hline
   $m=0$ &
   \begin{tabular}{c} \includegraphics[width=0.15 \textwidth]{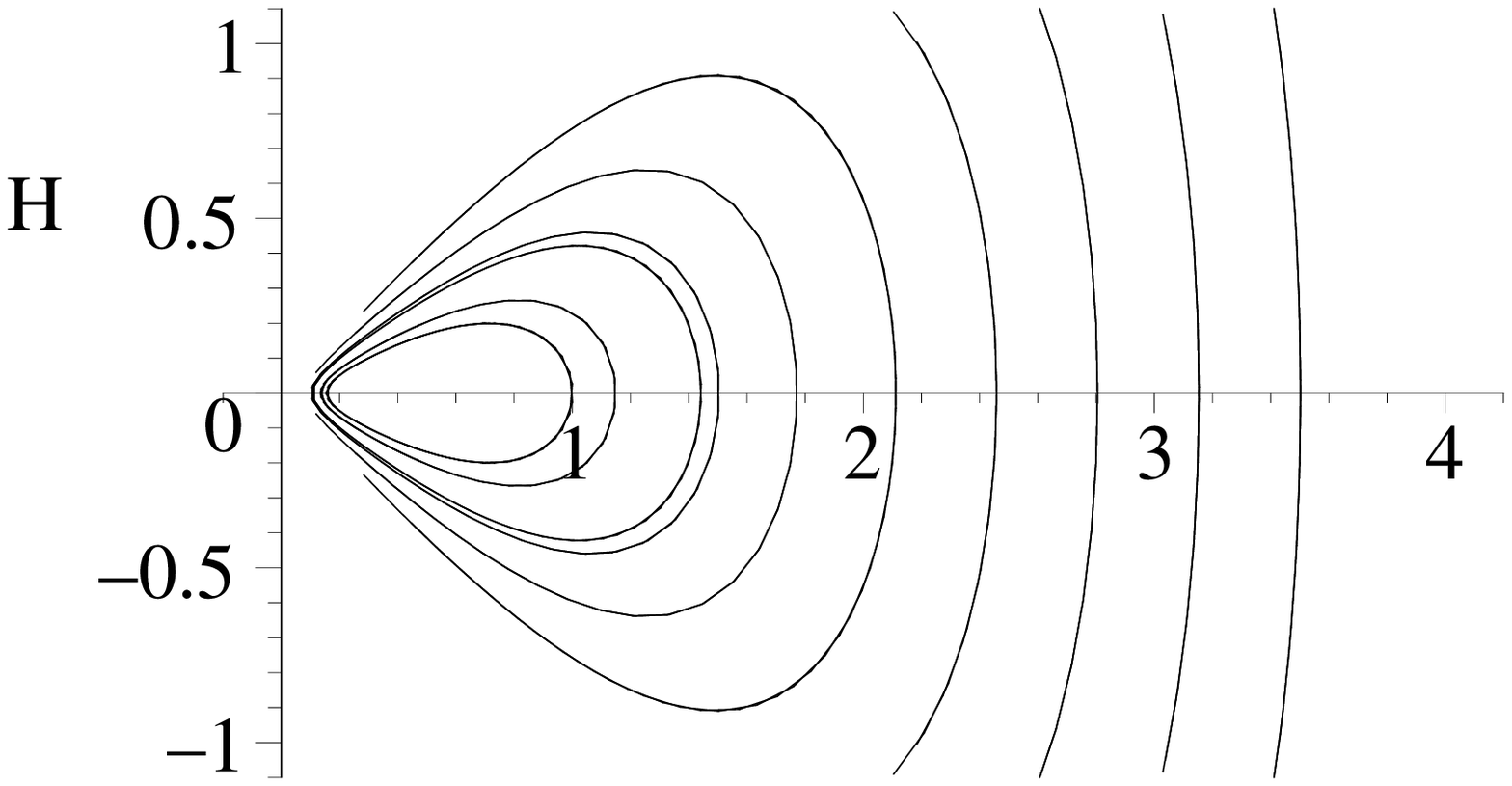} \\ d \end{tabular} &
   \begin{tabular}{c} \includegraphics[width=0.15 \textwidth]{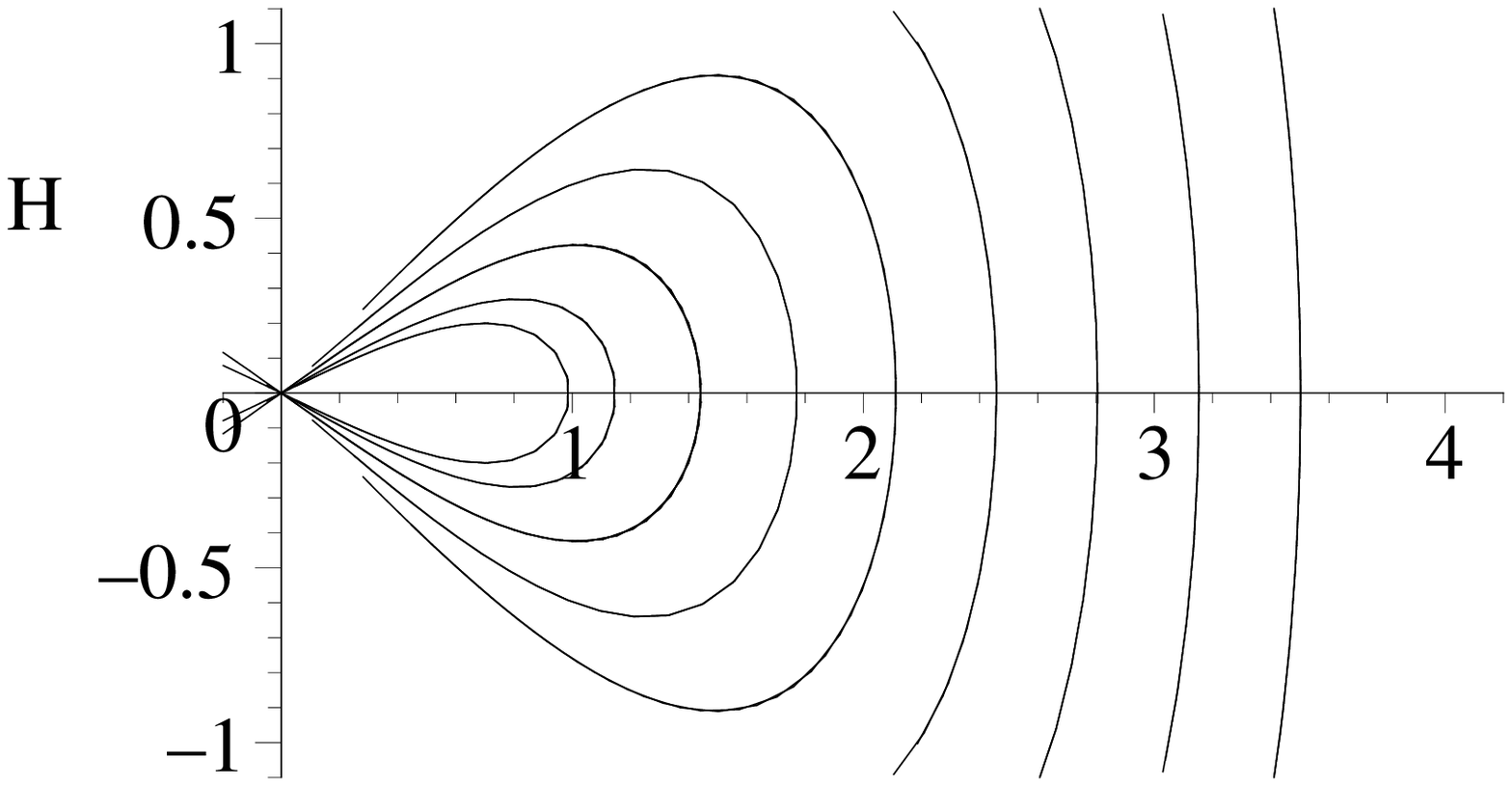} \\ e \end{tabular} &
   \begin{tabular}{c} \includegraphics[width=0.15 \textwidth]{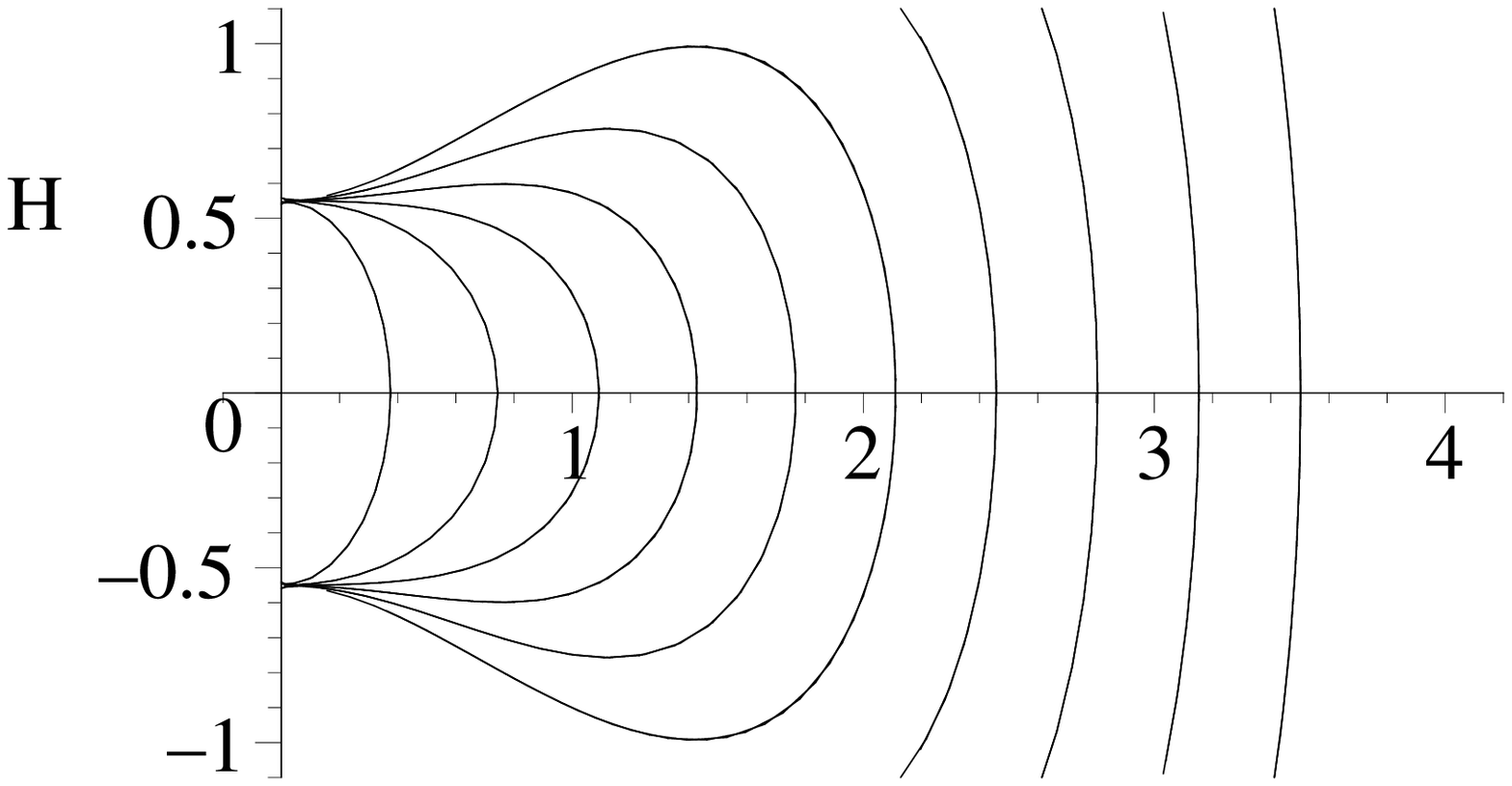} \\ f \end{tabular} \\
  \hline
   $m>0$ &
   \begin{tabular}{c} \includegraphics[width=0.15 \textwidth]{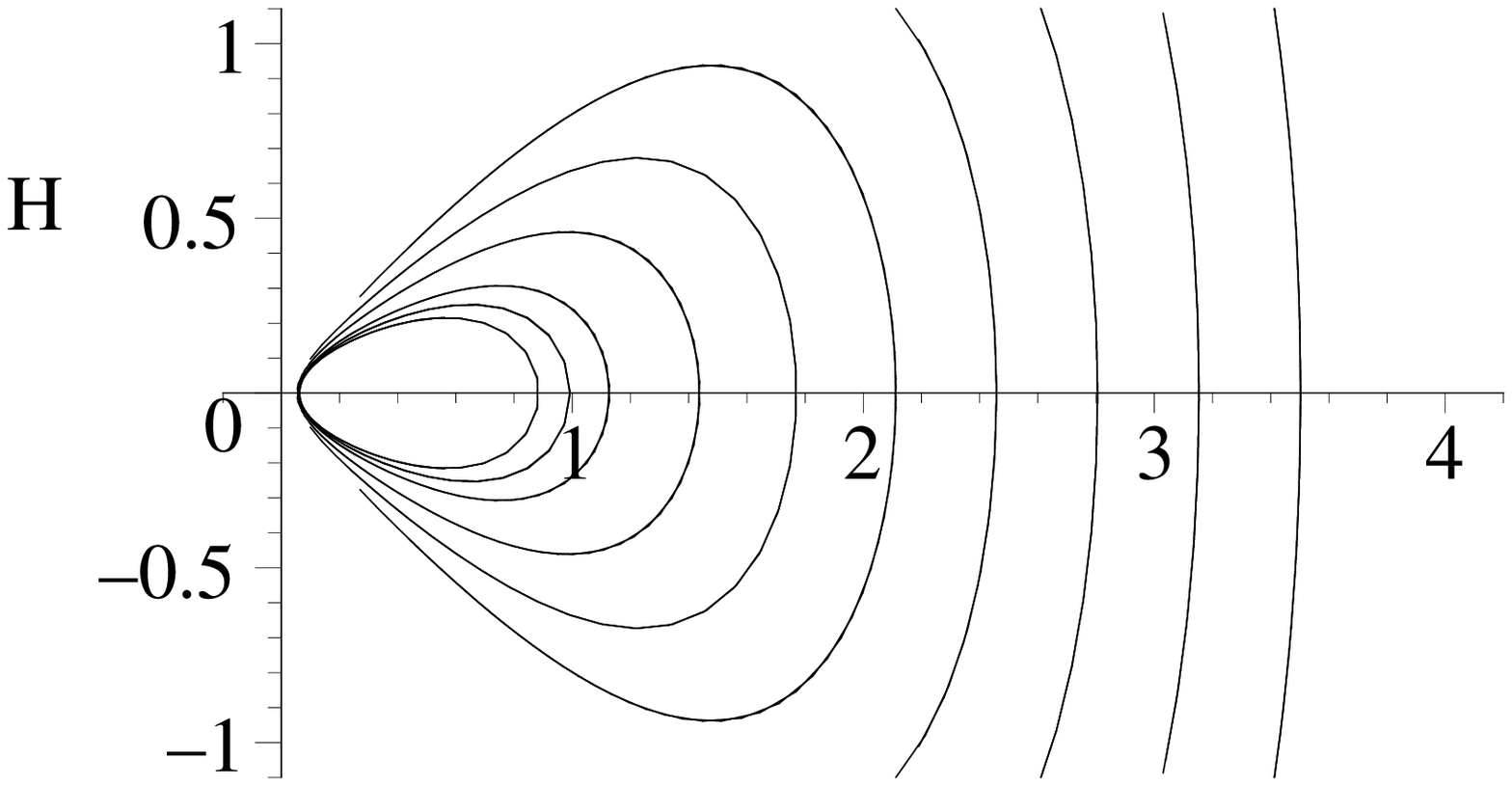} \\ g \end{tabular} &
   \begin{tabular}{c} \includegraphics[width=0.15 \textwidth]{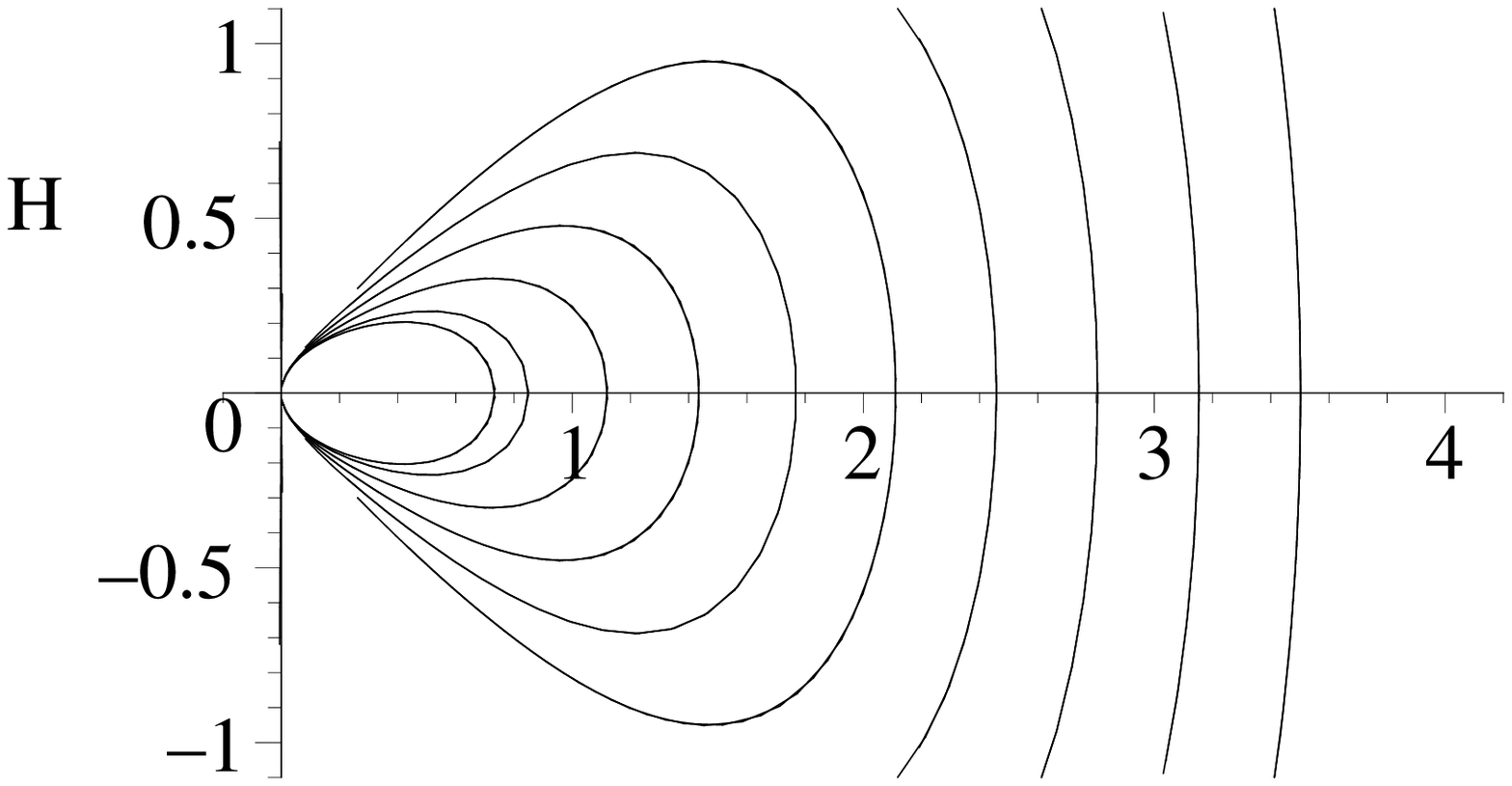} \\ h \end{tabular} &
   \begin{tabular}{c} \includegraphics[width=0.15 \textwidth]{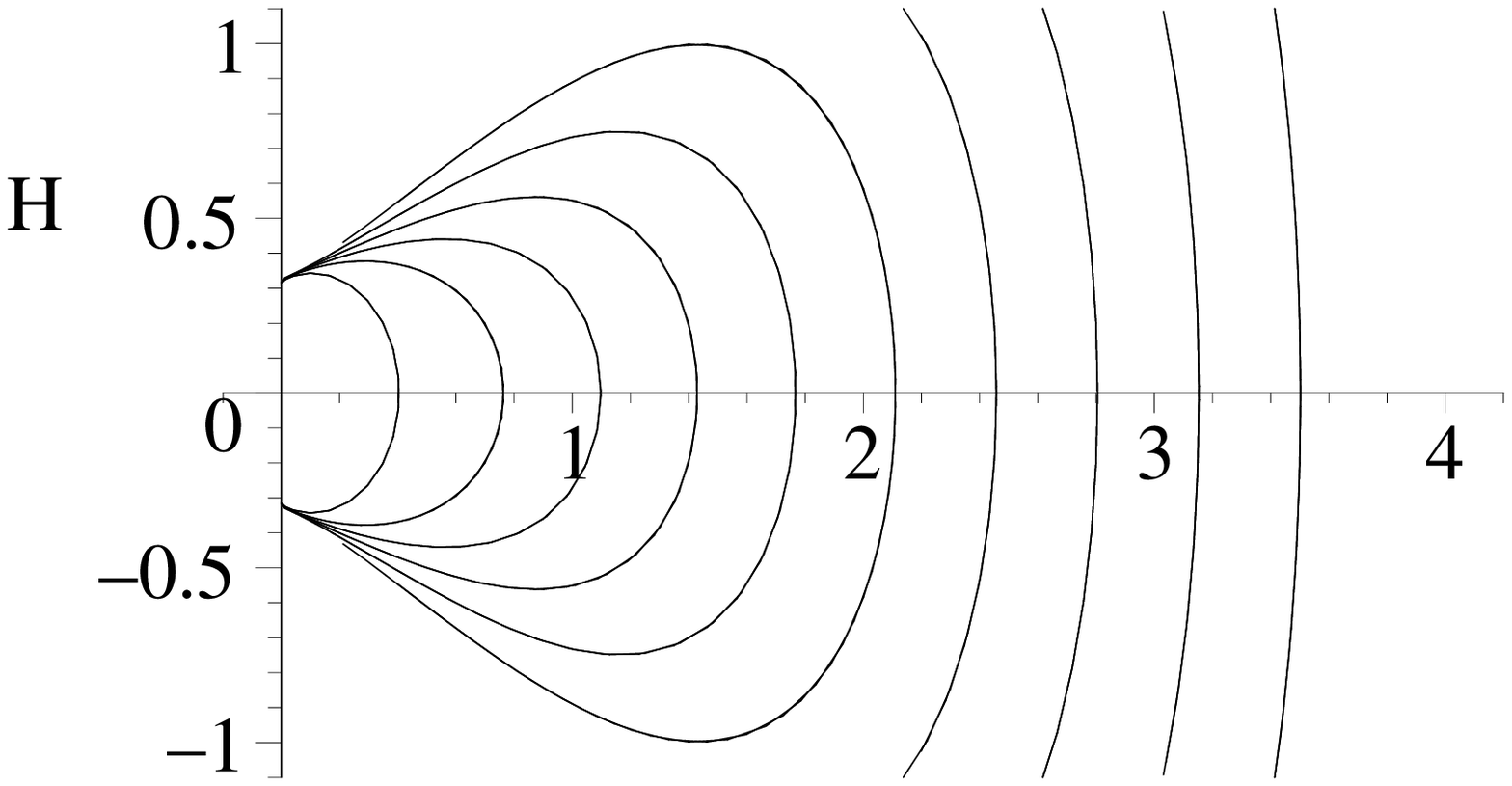} \\ i \end{tabular} \\
  \hline
\end{tabular}//
\vskip 2 mm
Table E2. Case with  $\ve=0$, $n=2$ and $\lambda=0$. \\
\vskip 2 mm

\begin{tabular}{|c|c|c|c|}
  \hline
     & $\Lambda<0$ & $\Lambda=0$ & $\Lambda>0$ \\
  \hline
   $m<0$ &
   \begin{tabular}{c} \includegraphics[width=0.15 \textwidth]{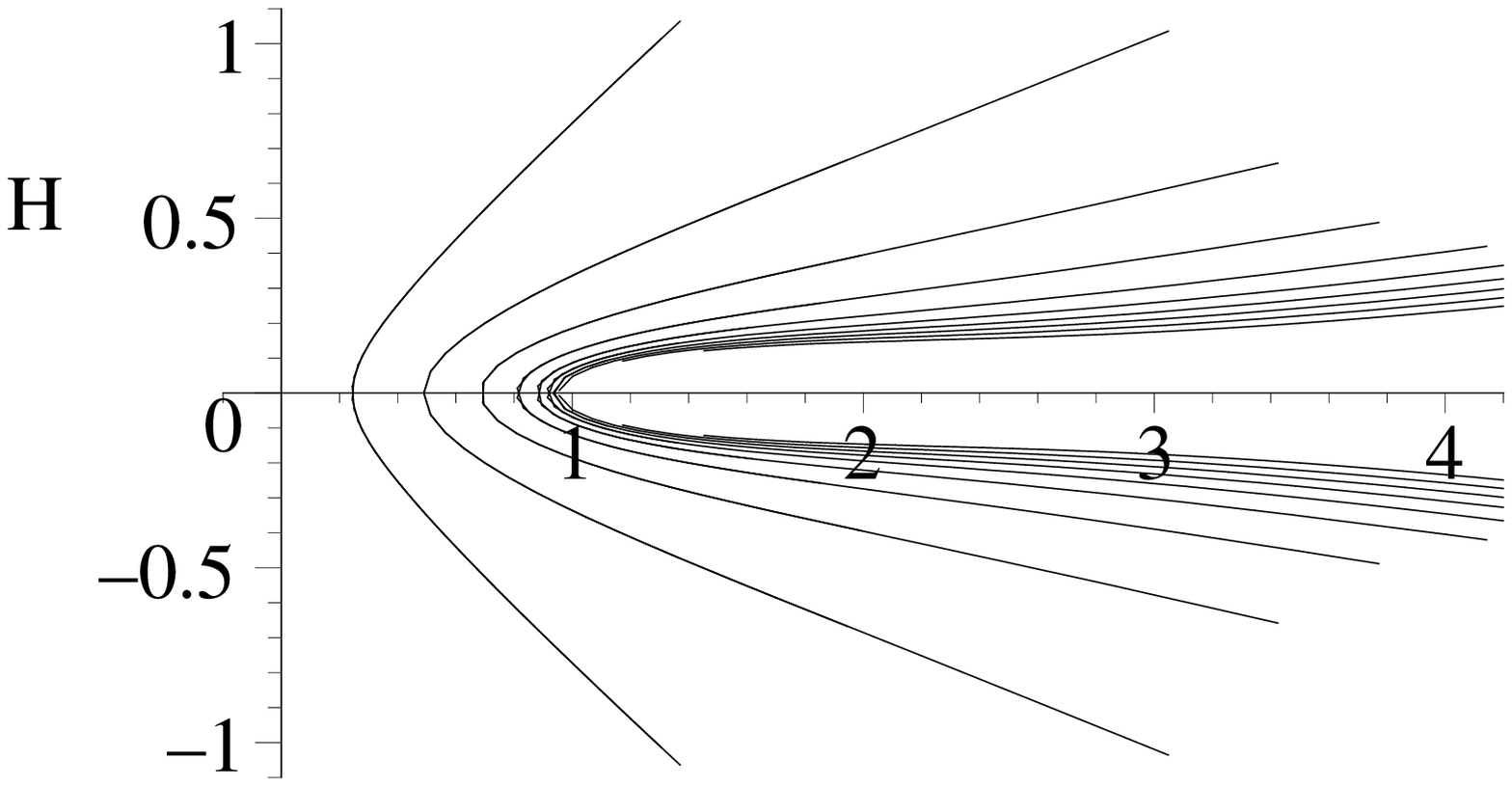} \\ a  \end{tabular} &
   \begin{tabular}{c} \includegraphics[width=0.15 \textwidth]{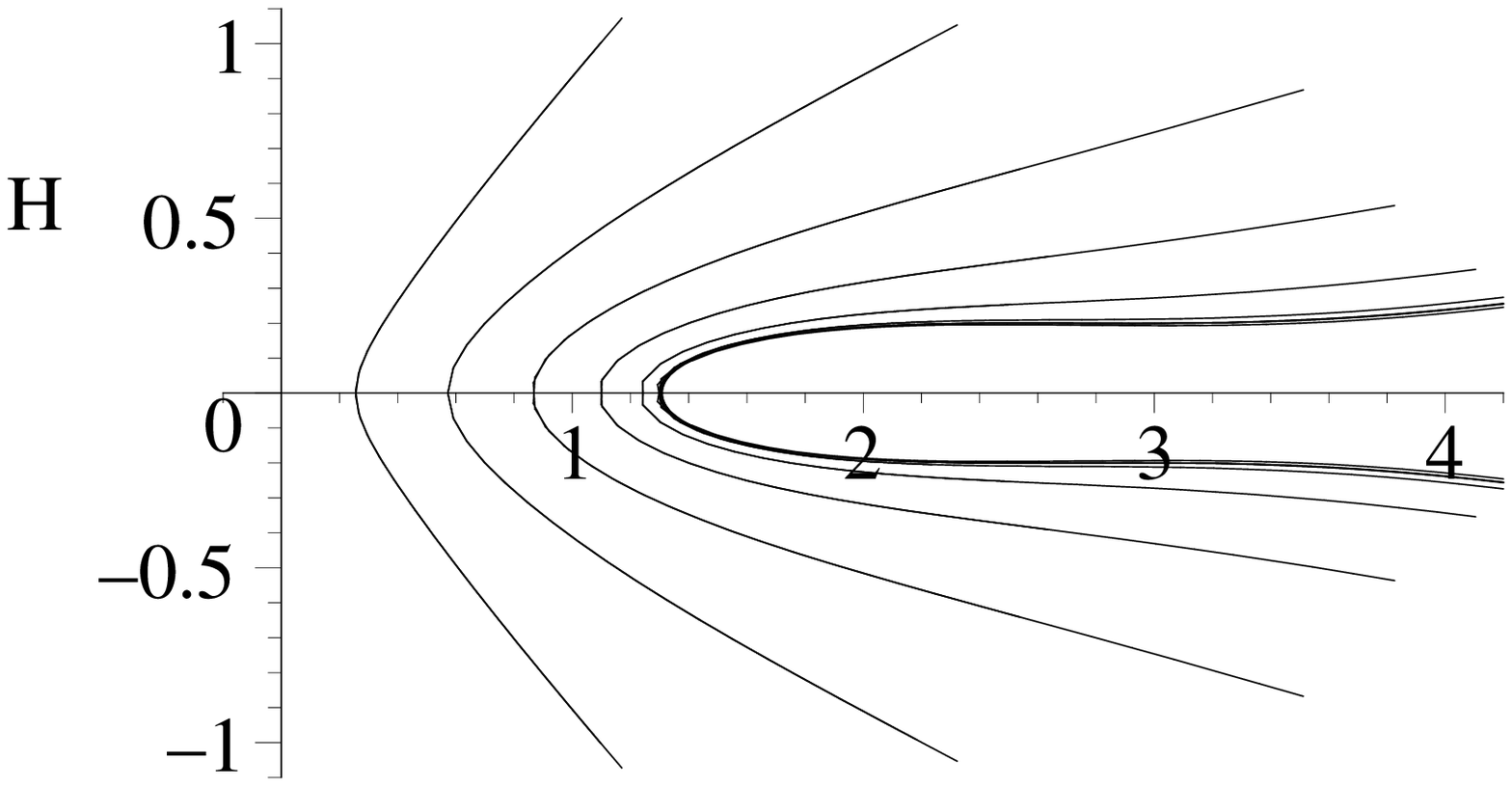} \\ b \end{tabular} &
   \begin{tabular}{c} \includegraphics[width=0.15 \textwidth]{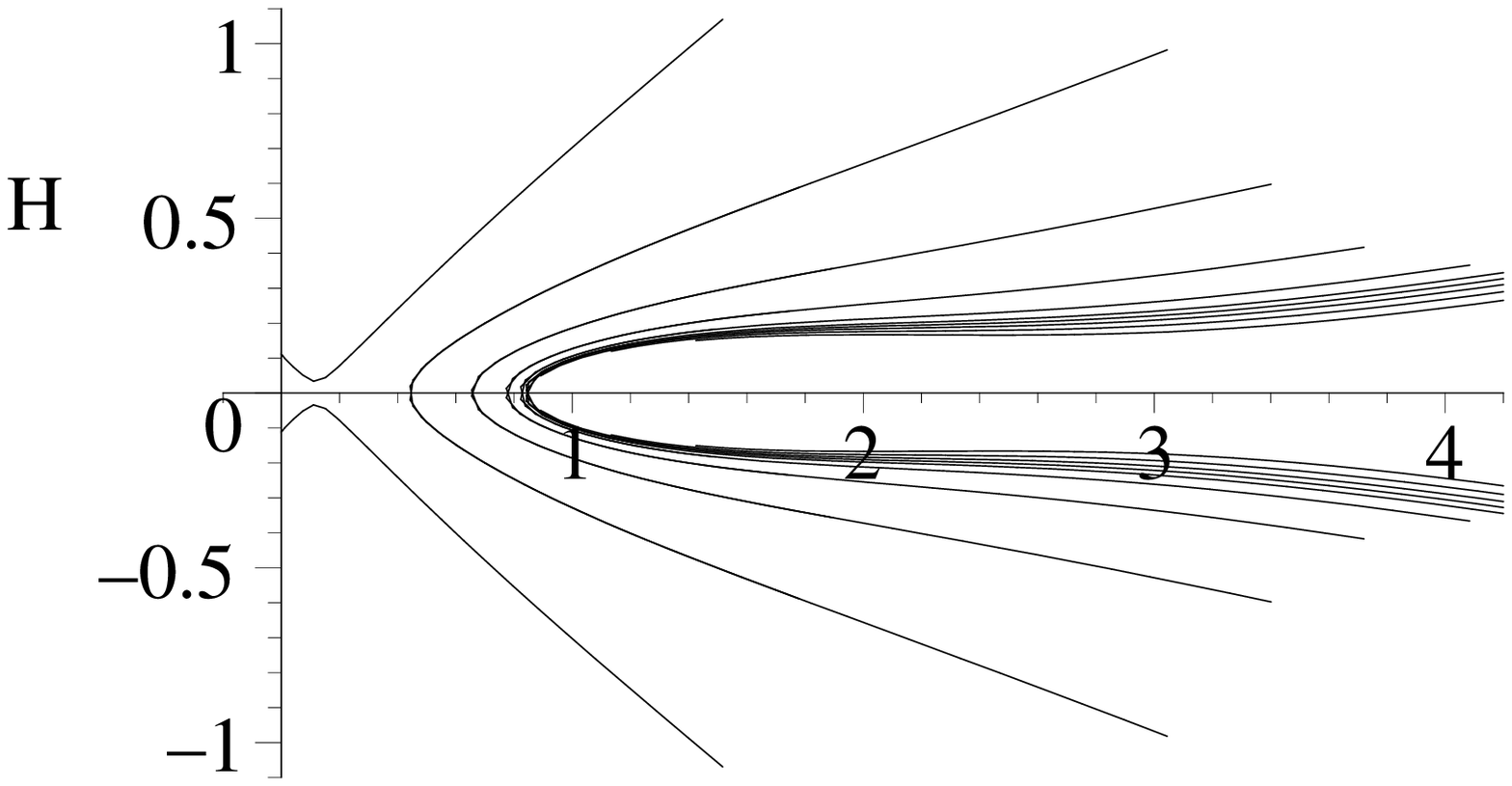} \\ c \end{tabular} \\
  \hline
   $m=0$ &
   \begin{tabular}{c} \includegraphics[width=0.15 \textwidth]{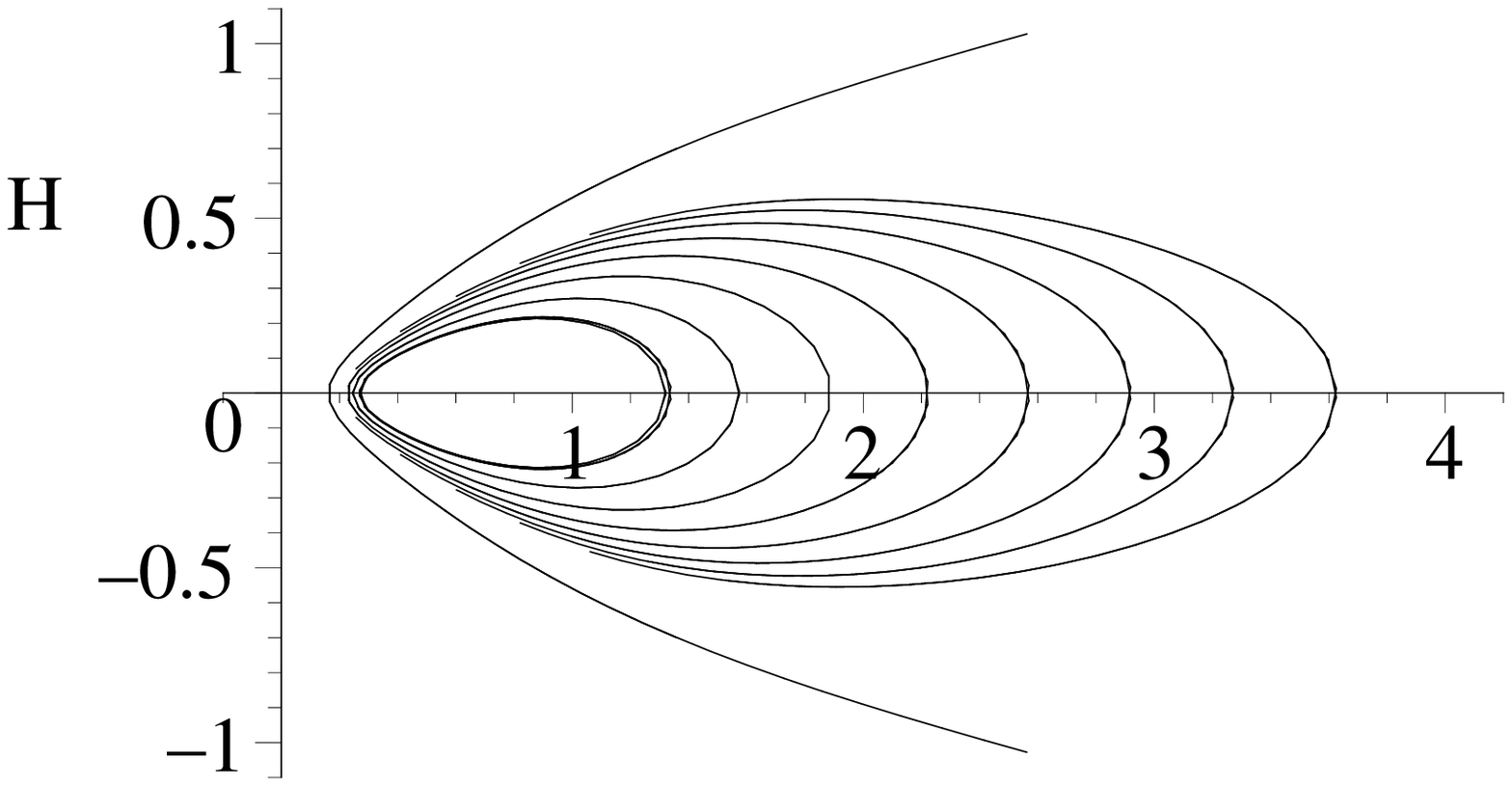} \\ d \end{tabular} &
   \begin{tabular}{c} \includegraphics[width=0.15 \textwidth]{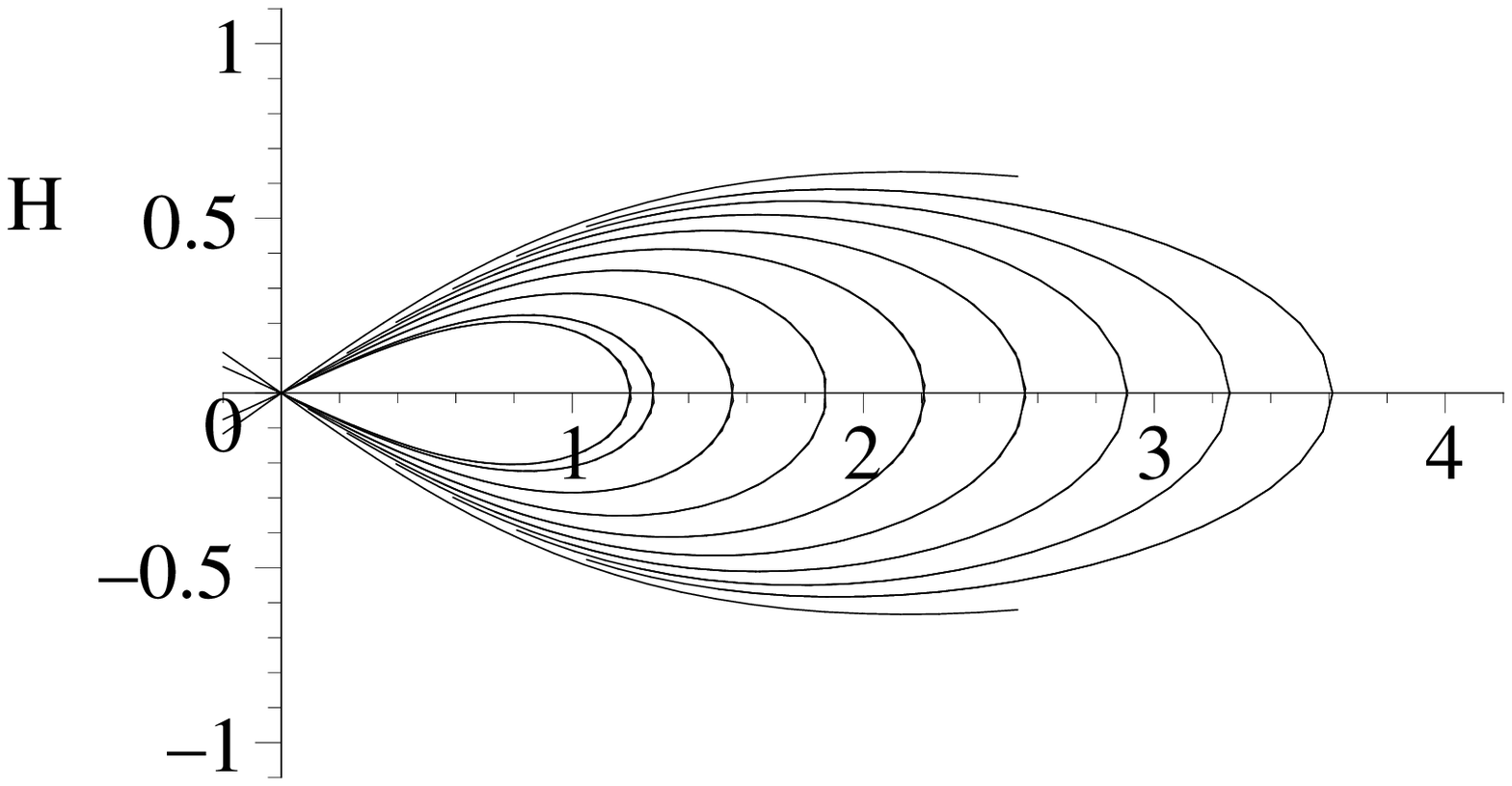} \\ e \end{tabular} &
   \begin{tabular}{c} \includegraphics[width=0.15 \textwidth]{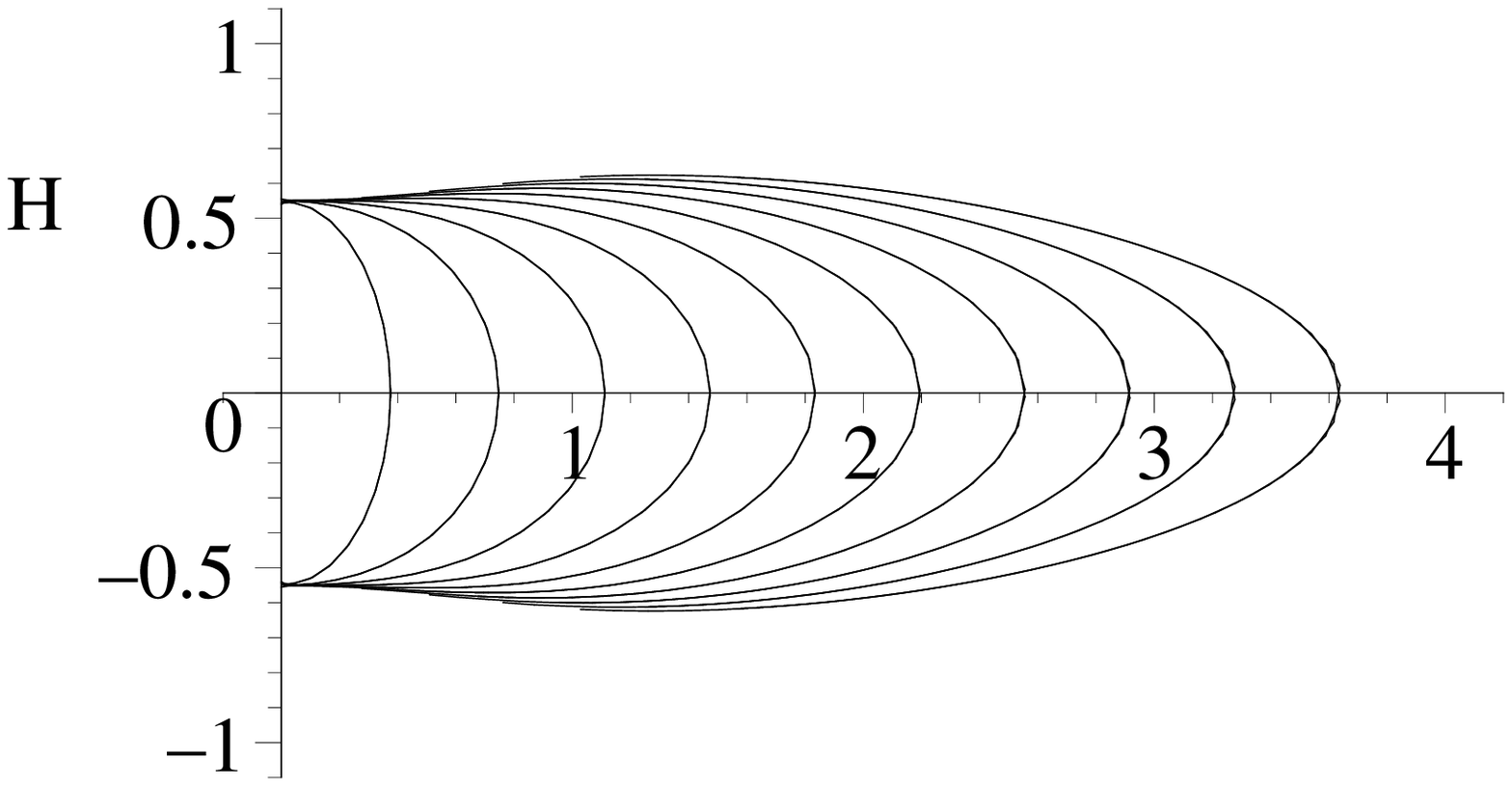} \\ f \end{tabular} \\
  \hline
   $m>0$ &
   \begin{tabular}{c} \includegraphics[width=0.15 \textwidth]{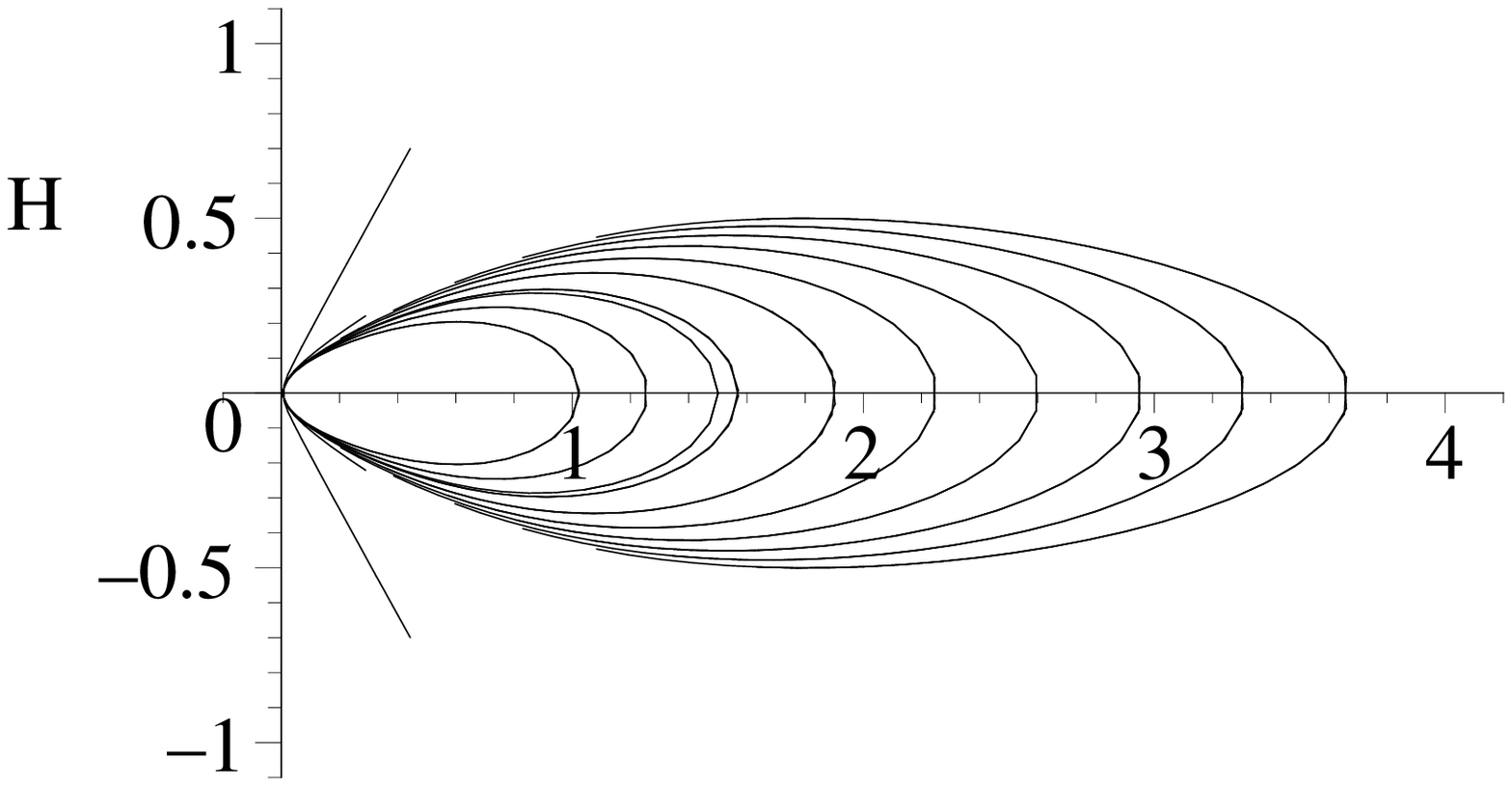} \\ g \end{tabular} &
   \begin{tabular}{c} \includegraphics[width=0.15 \textwidth]{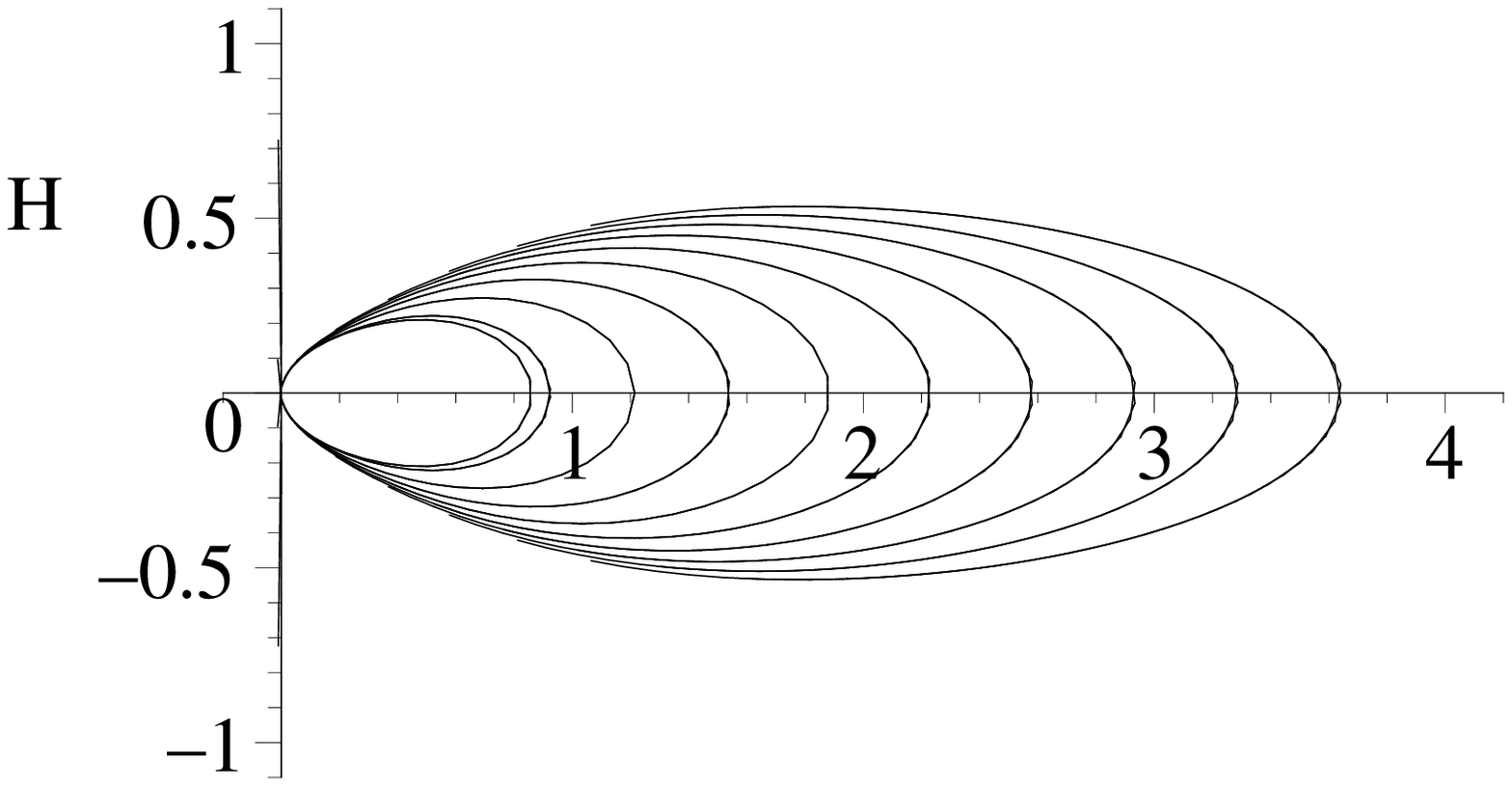} \\ h \end{tabular} &
   \begin{tabular}{c} \includegraphics[width=0.15 \textwidth]{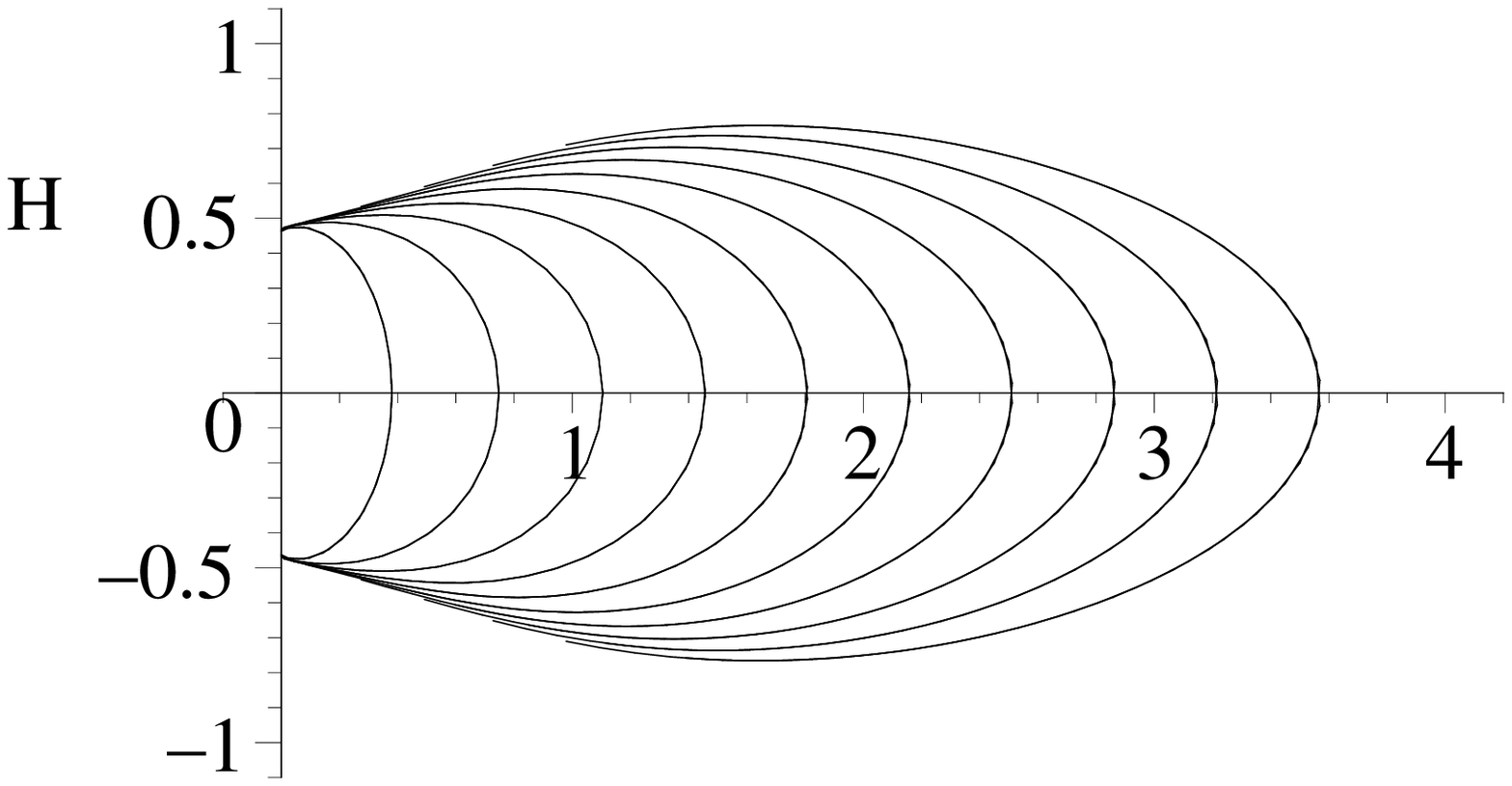} \\ i \end{tabular} \\
  \hline
\end{tabular}//
\vskip 2 mm
Table B2. Case with  $\ve=0$, $n=2$ and $\lambda>0$. \\
\vskip 2 mm

\begin{tabular}{|c|c|c|c|}
  \hline
     & $\Lambda<0$ & $\Lambda=0$ & $\Lambda>0$ \\
  \hline
   $m<0$ &
   \begin{tabular}{c} \includegraphics[width=0.15 \textwidth]{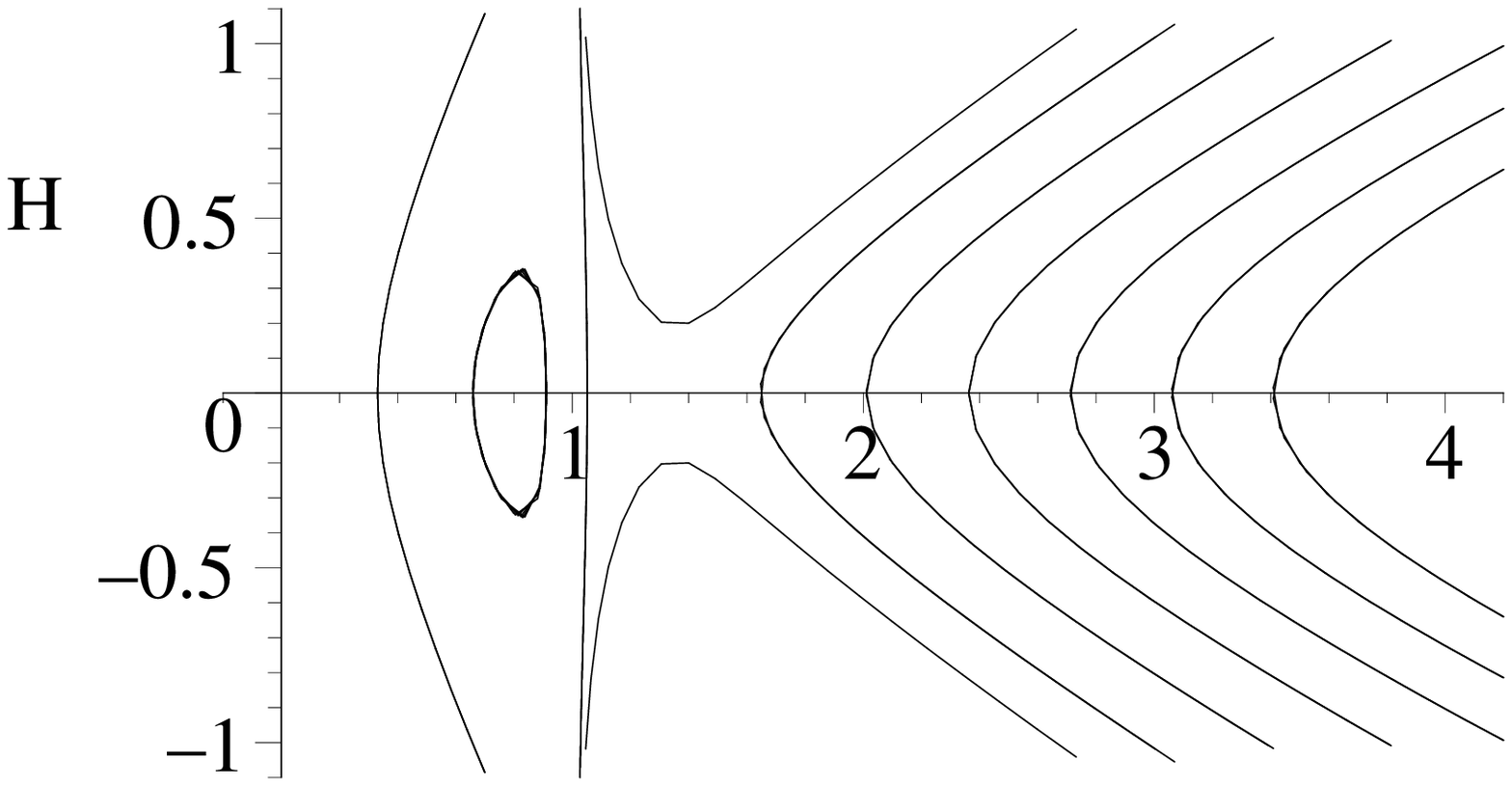} \\ a  \end{tabular} &
   \begin{tabular}{c} \includegraphics[width=0.15 \textwidth]{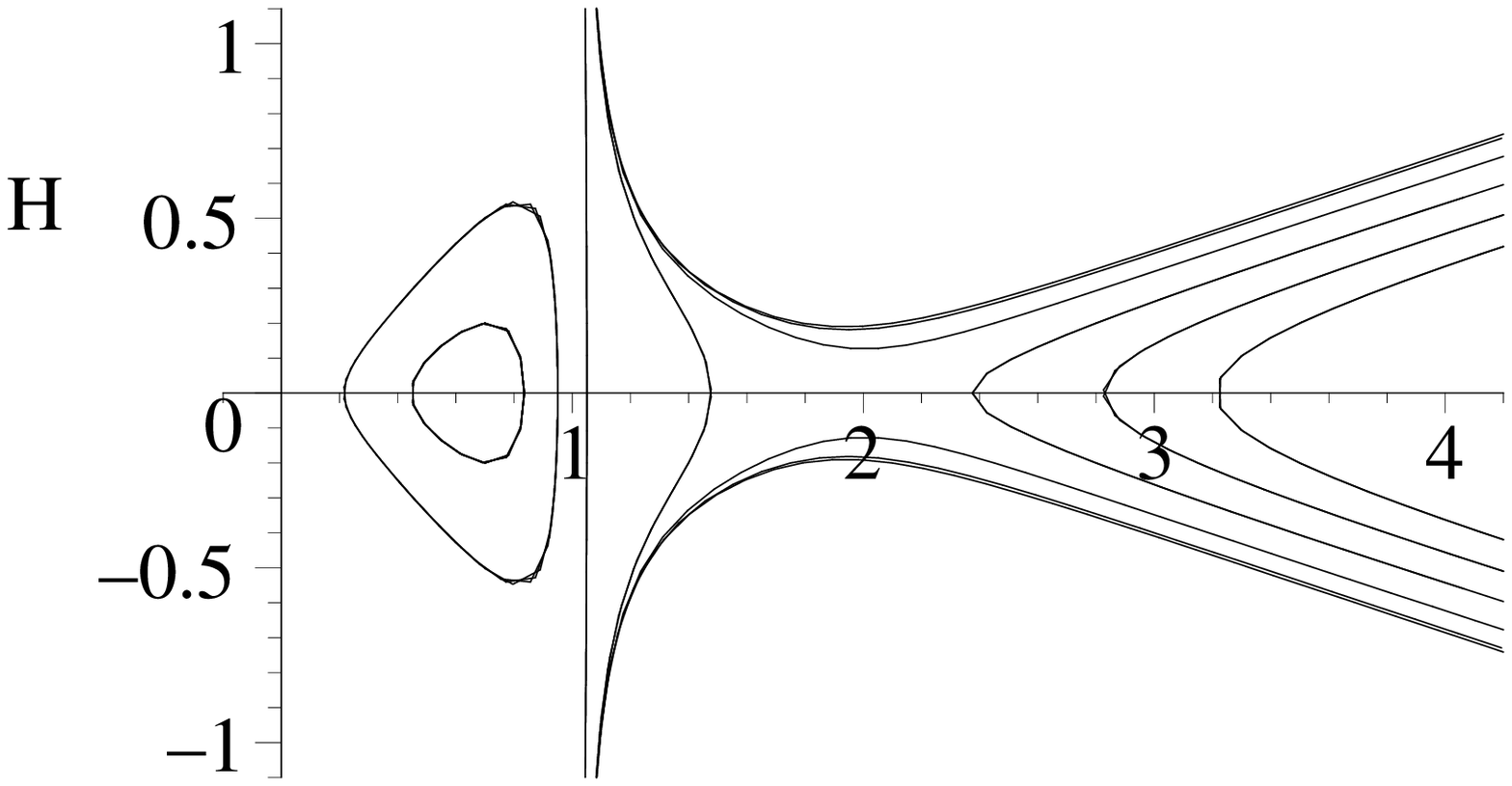} \\ b \end{tabular} &
   \begin{tabular}{c} \includegraphics[width=0.15 \textwidth]{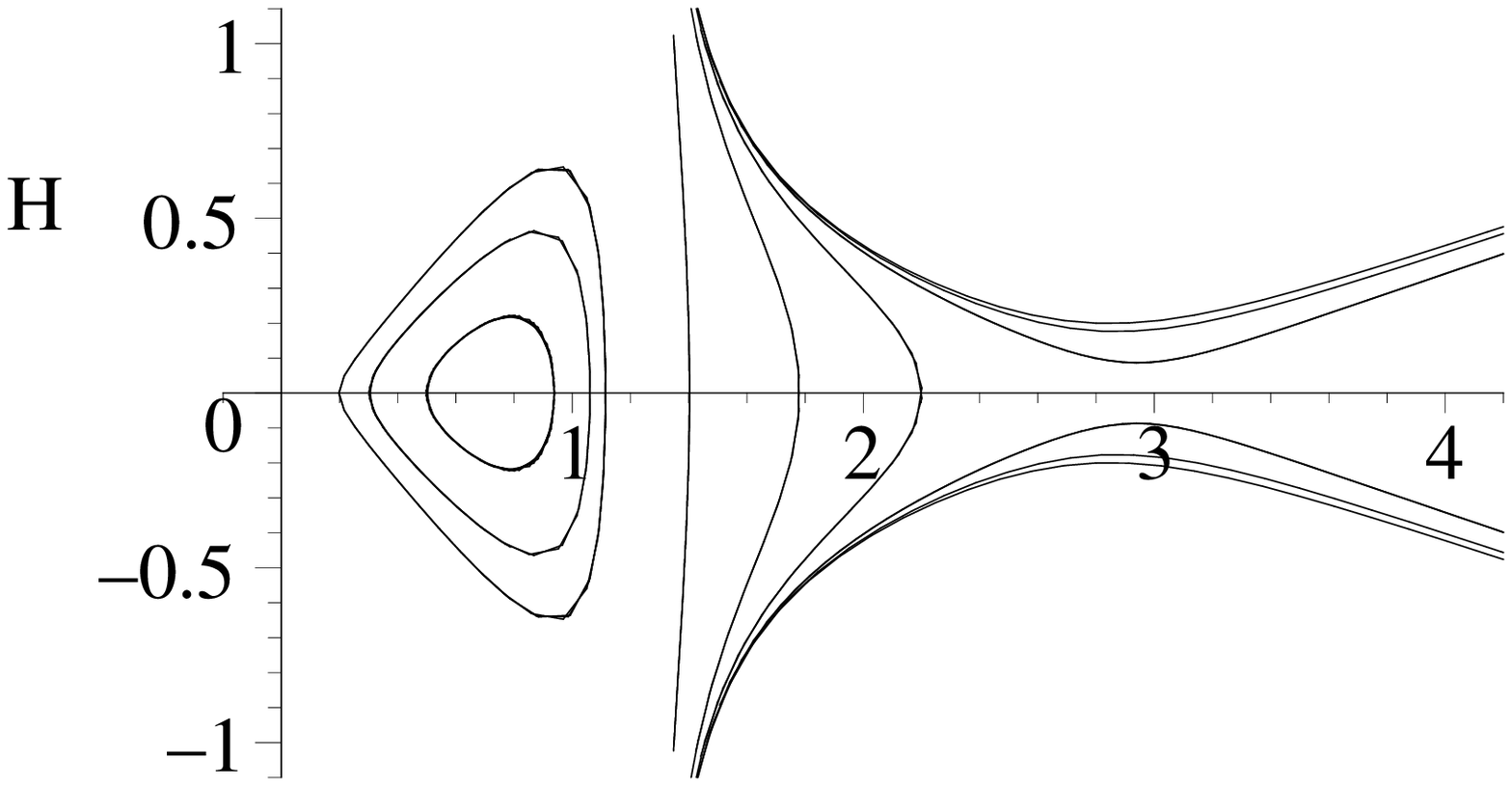} \\ c \end{tabular} \\
  \hline
   $m=0$ &
   \begin{tabular}{c} \includegraphics[width=0.15 \textwidth]{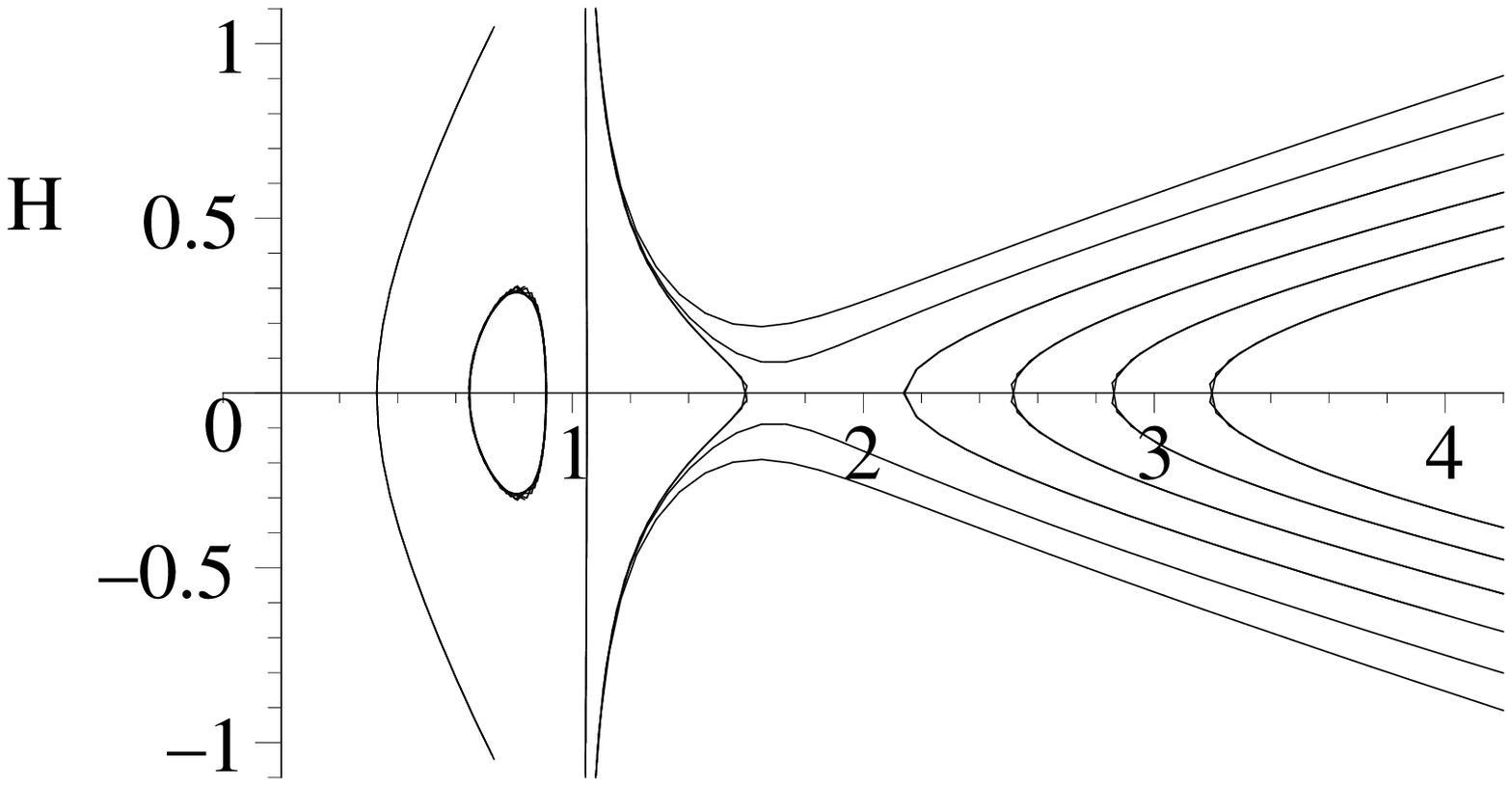} \\ d \end{tabular} &
   \begin{tabular}{c} \includegraphics[width=0.15 \textwidth]{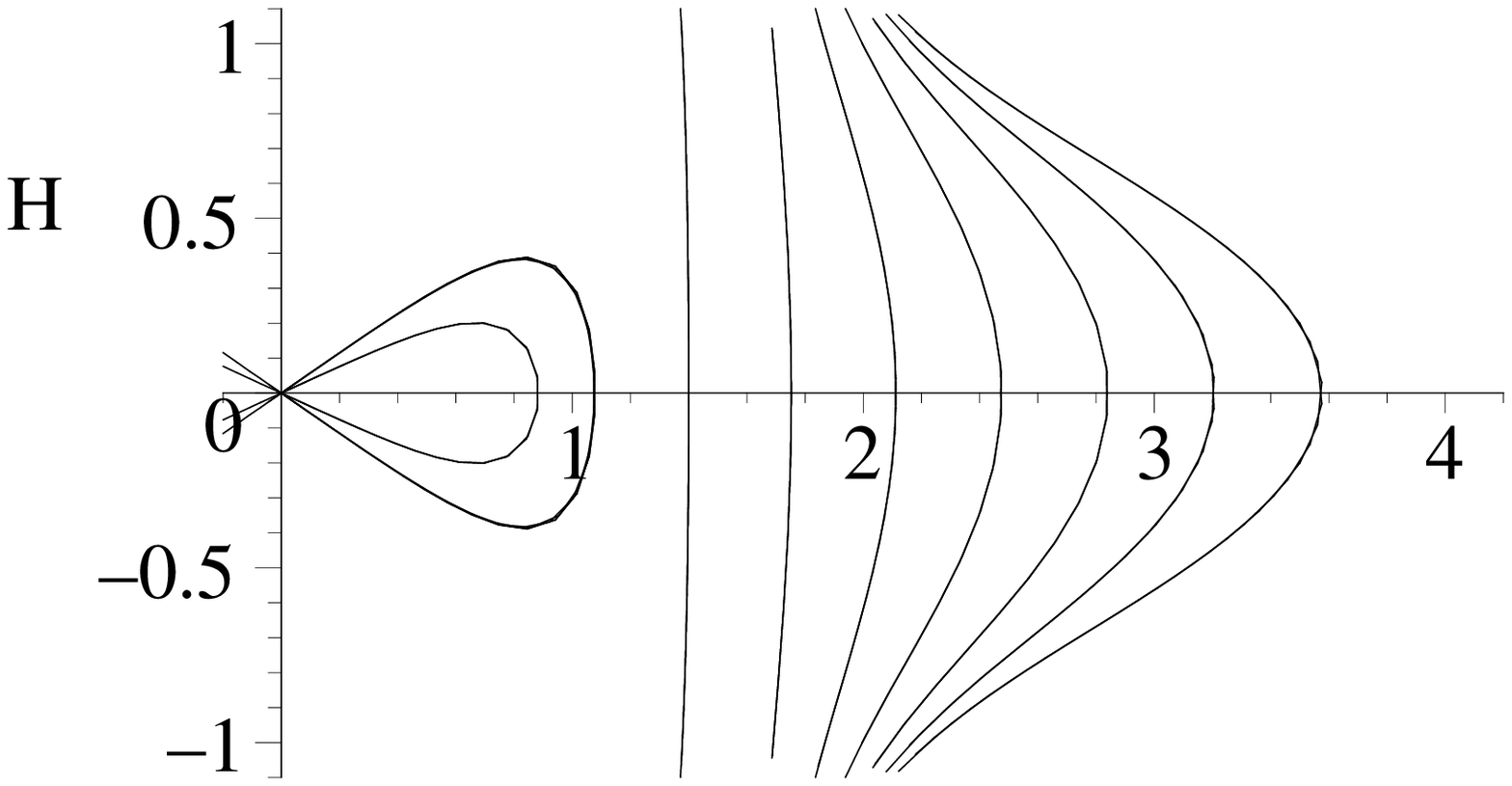} \\ e \end{tabular} &
   \begin{tabular}{c} \includegraphics[width=0.15 \textwidth]{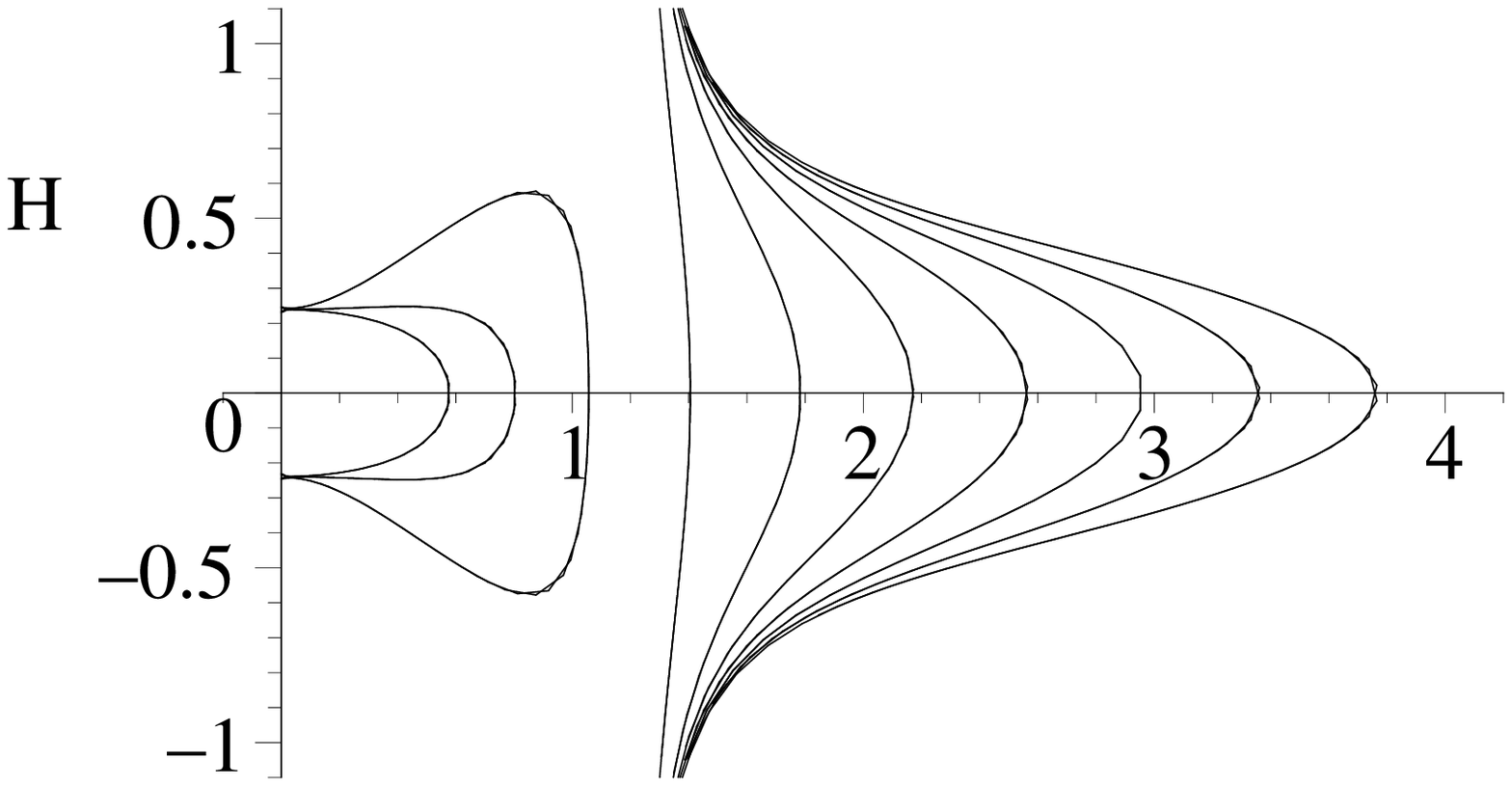} \\ f \end{tabular} \\
  \hline
   $m>0$ &
   \begin{tabular}{c} \includegraphics[width=0.15 \textwidth]{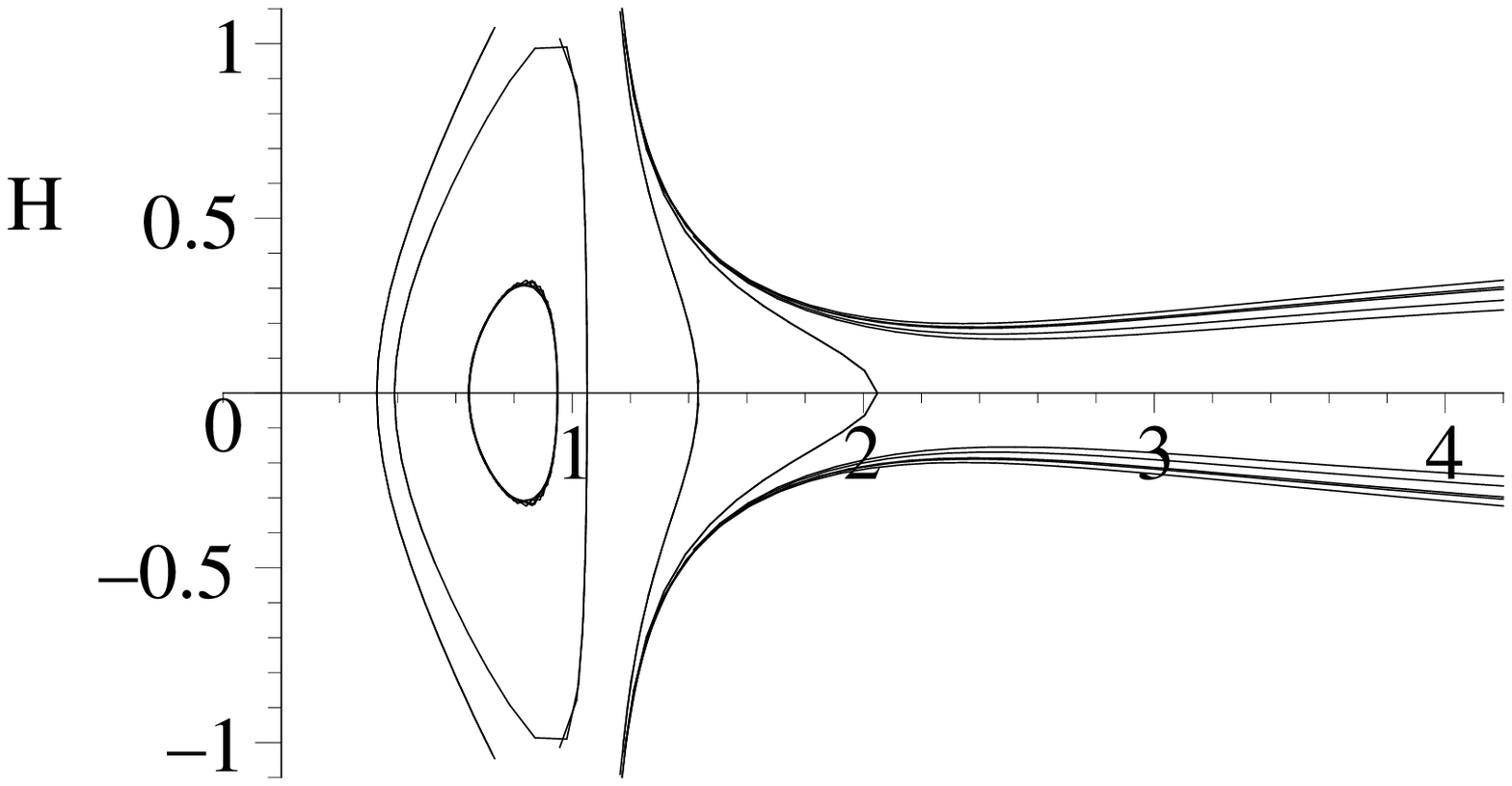} \\ g \end{tabular} &
   \begin{tabular}{c} \includegraphics[width=0.15 \textwidth]{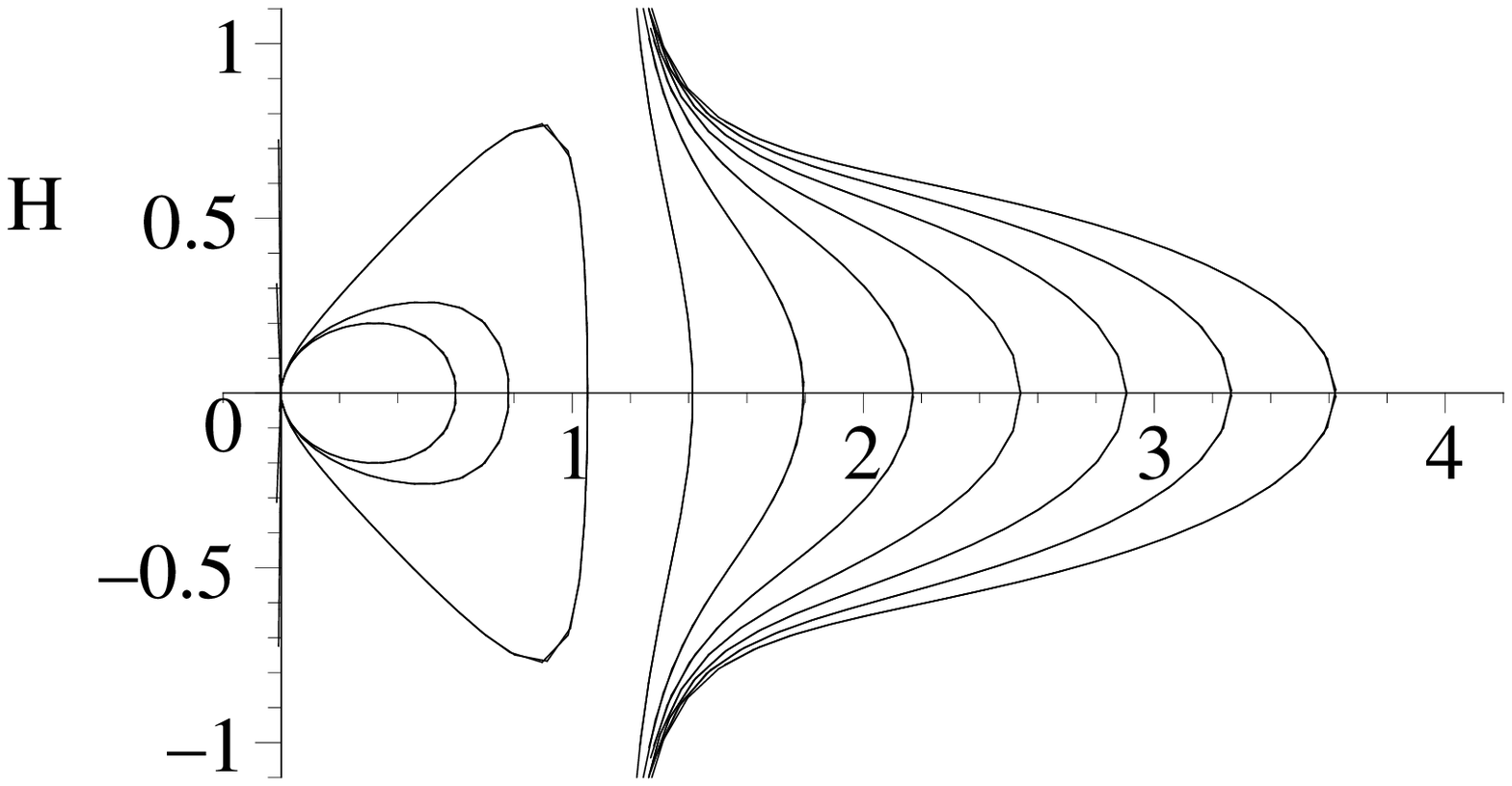} \\ h \end{tabular} &
   \begin{tabular}{c} \includegraphics[width=0.15 \textwidth]{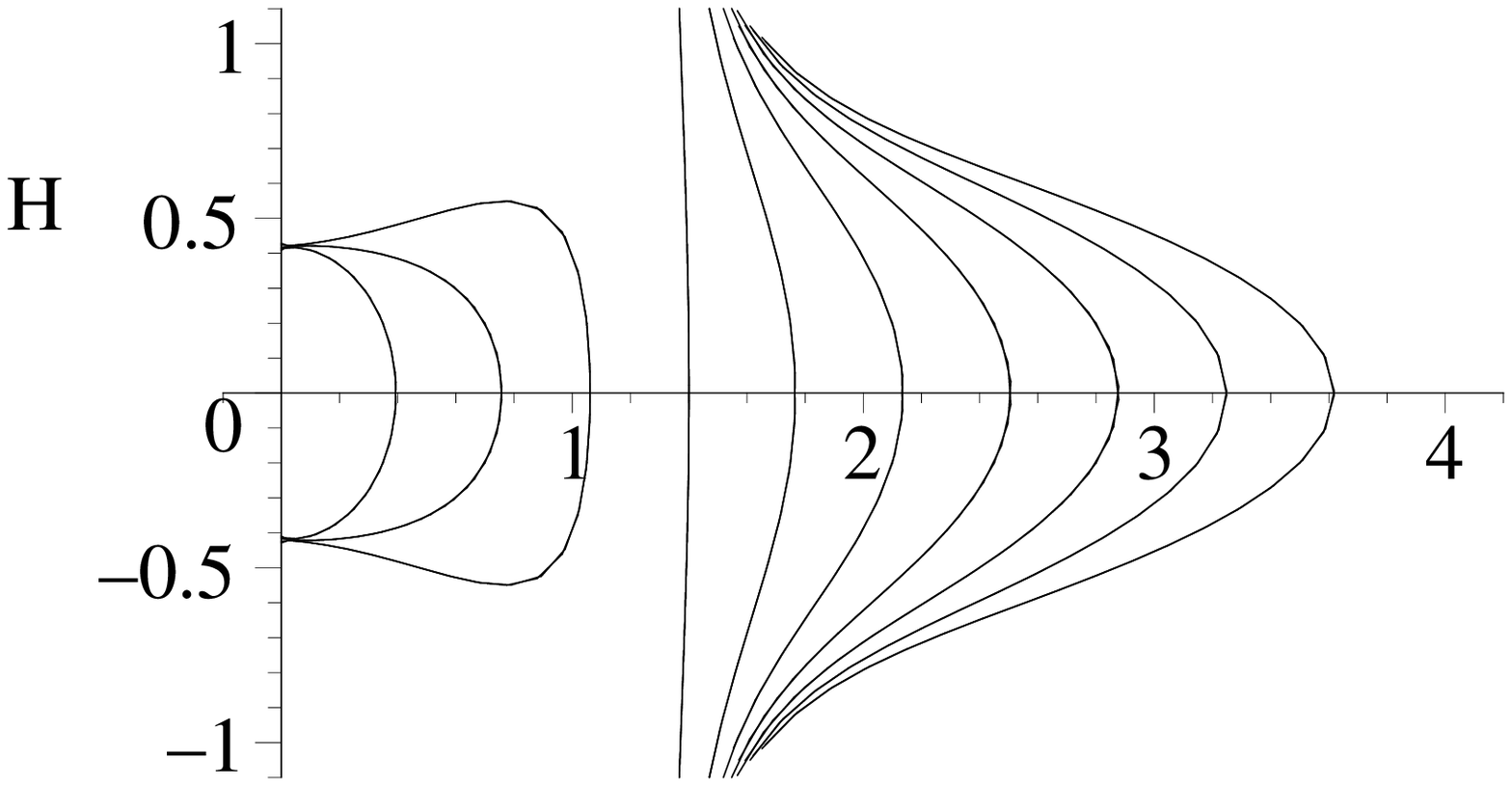} \\ i \end{tabular} \\
  \hline
\end{tabular}//
\vskip 2 mm
Table C1. Case with  $\ve=0$, $n=3$ and $\lambda<0$. \\
\vskip 2 mm

\begin{tabular}{|c|c|c|c|}
  \hline
     & $\Lambda<0$ & $\Lambda=0$ & $\Lambda>0$ \\
  \hline
   $m<0$ &
   \begin{tabular}{c} \includegraphics[width=0.15 \textwidth]{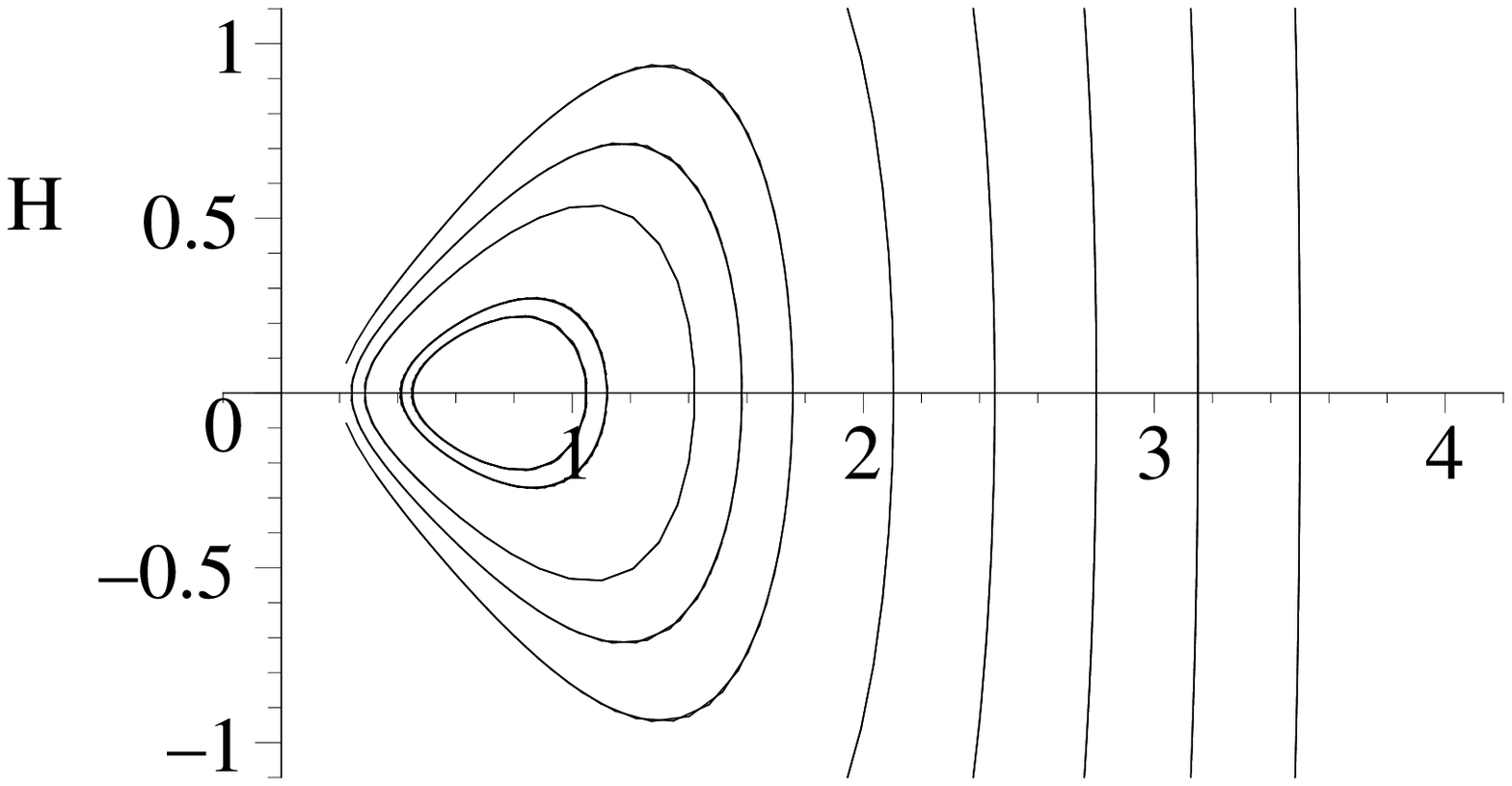} \\ a  \end{tabular} &
   \begin{tabular}{c} \includegraphics[width=0.15 \textwidth]{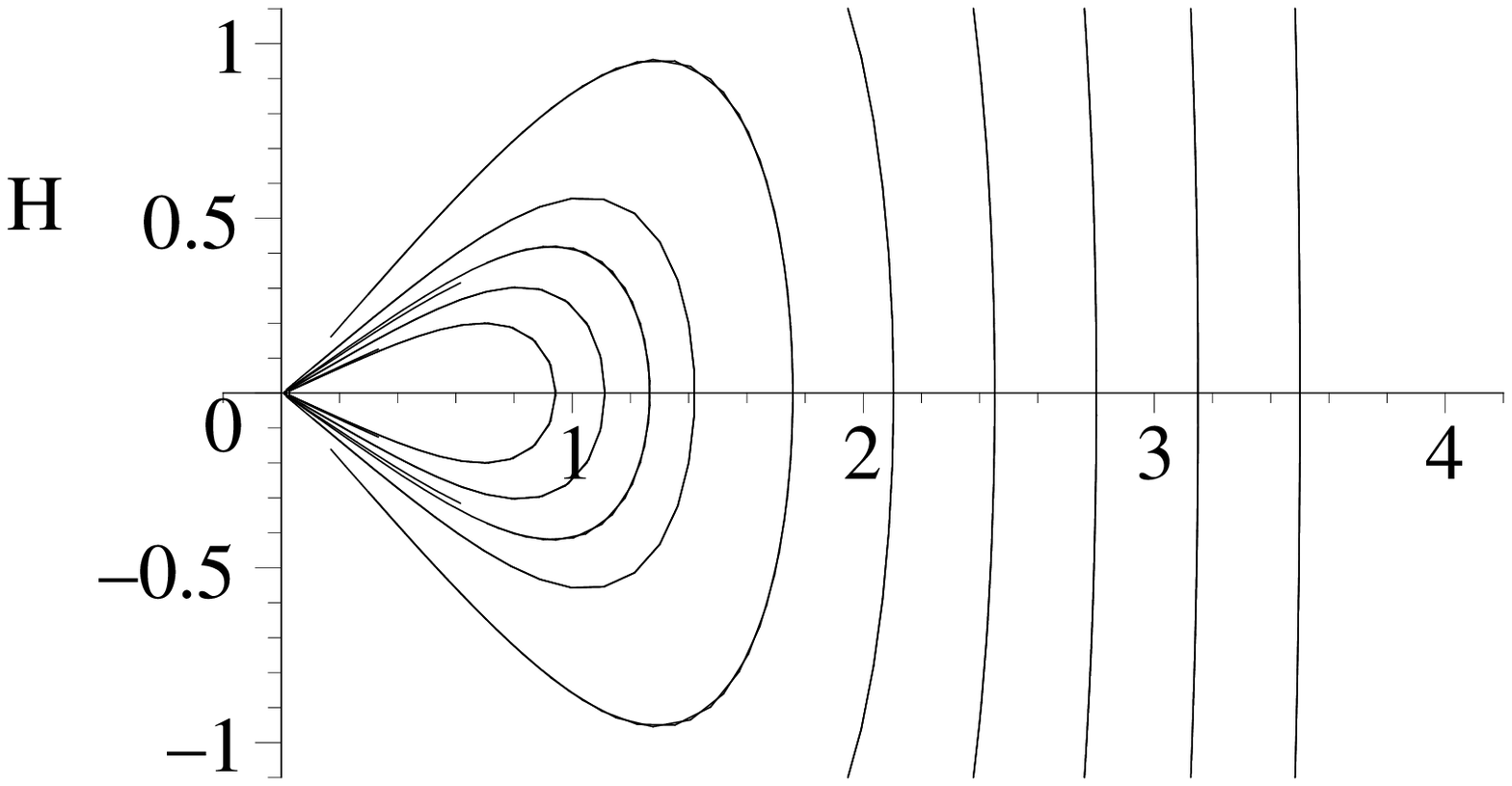} \\ b \end{tabular} &
   \begin{tabular}{c} \includegraphics[width=0.15 \textwidth]{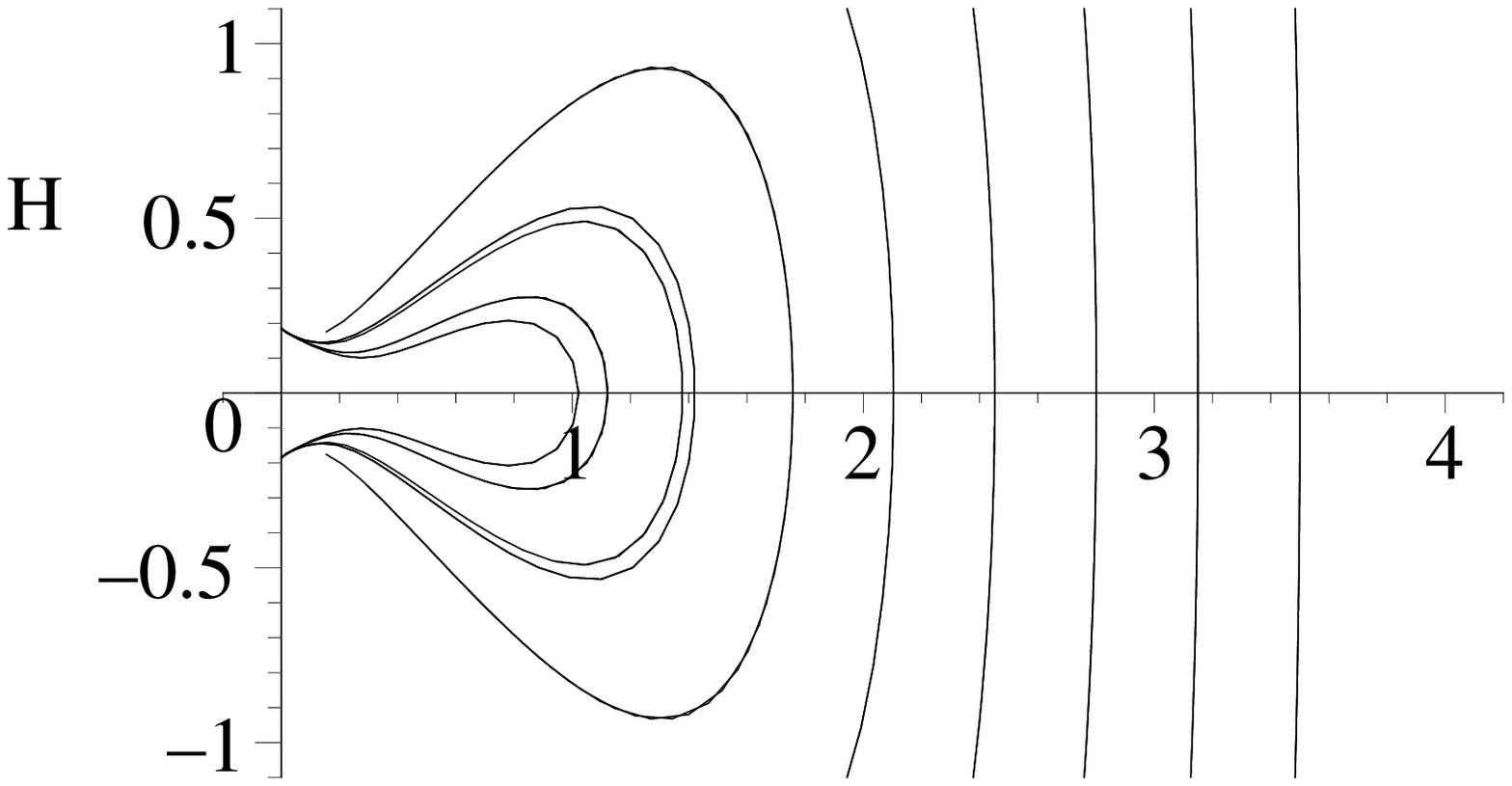} \\ c \end{tabular} \\
  \hline
   $m=0$ &
   \begin{tabular}{c} \includegraphics[width=0.15 \textwidth]{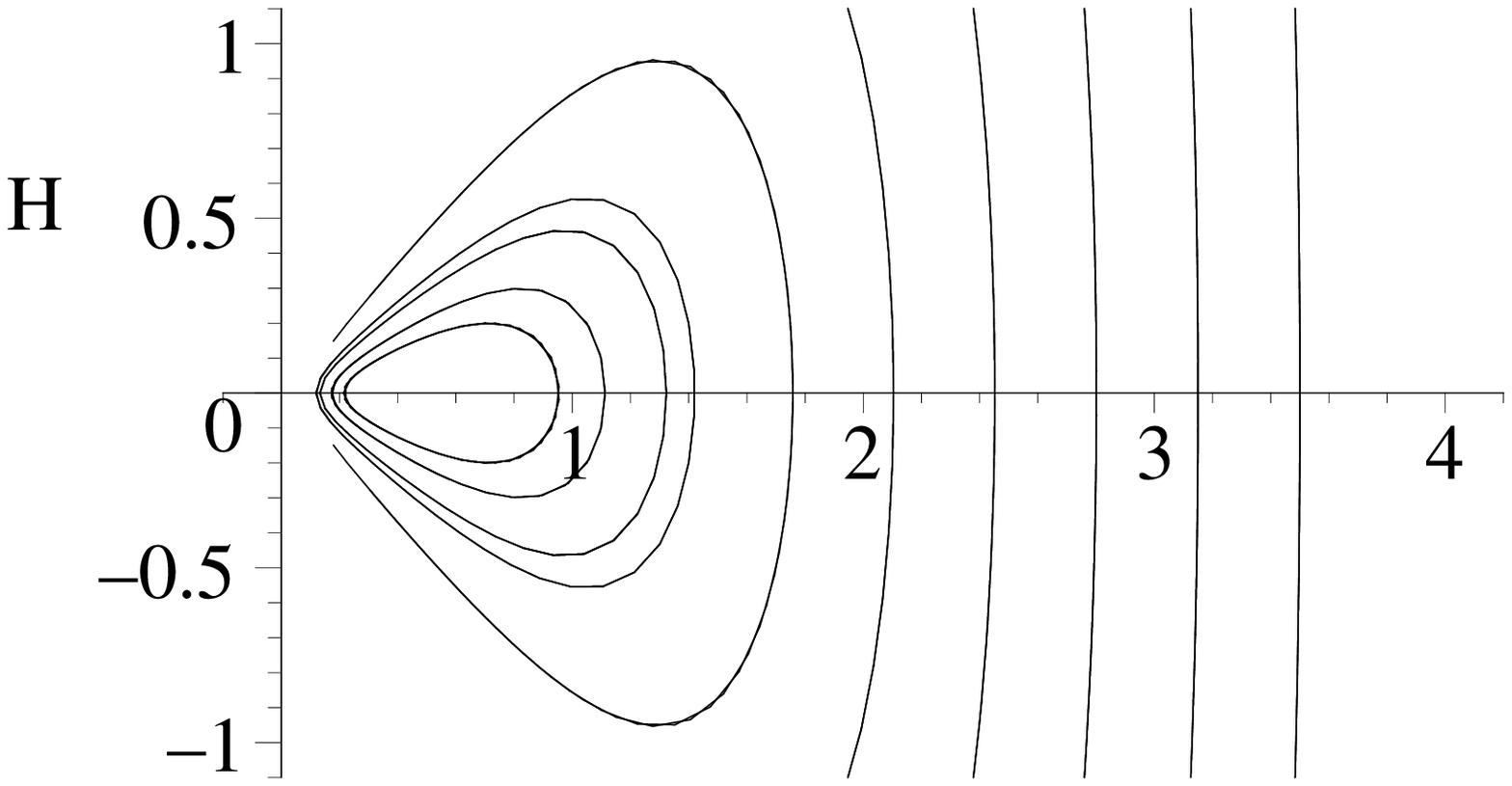} \\ d \end{tabular} &
   \begin{tabular}{c} \includegraphics[width=0.15 \textwidth]{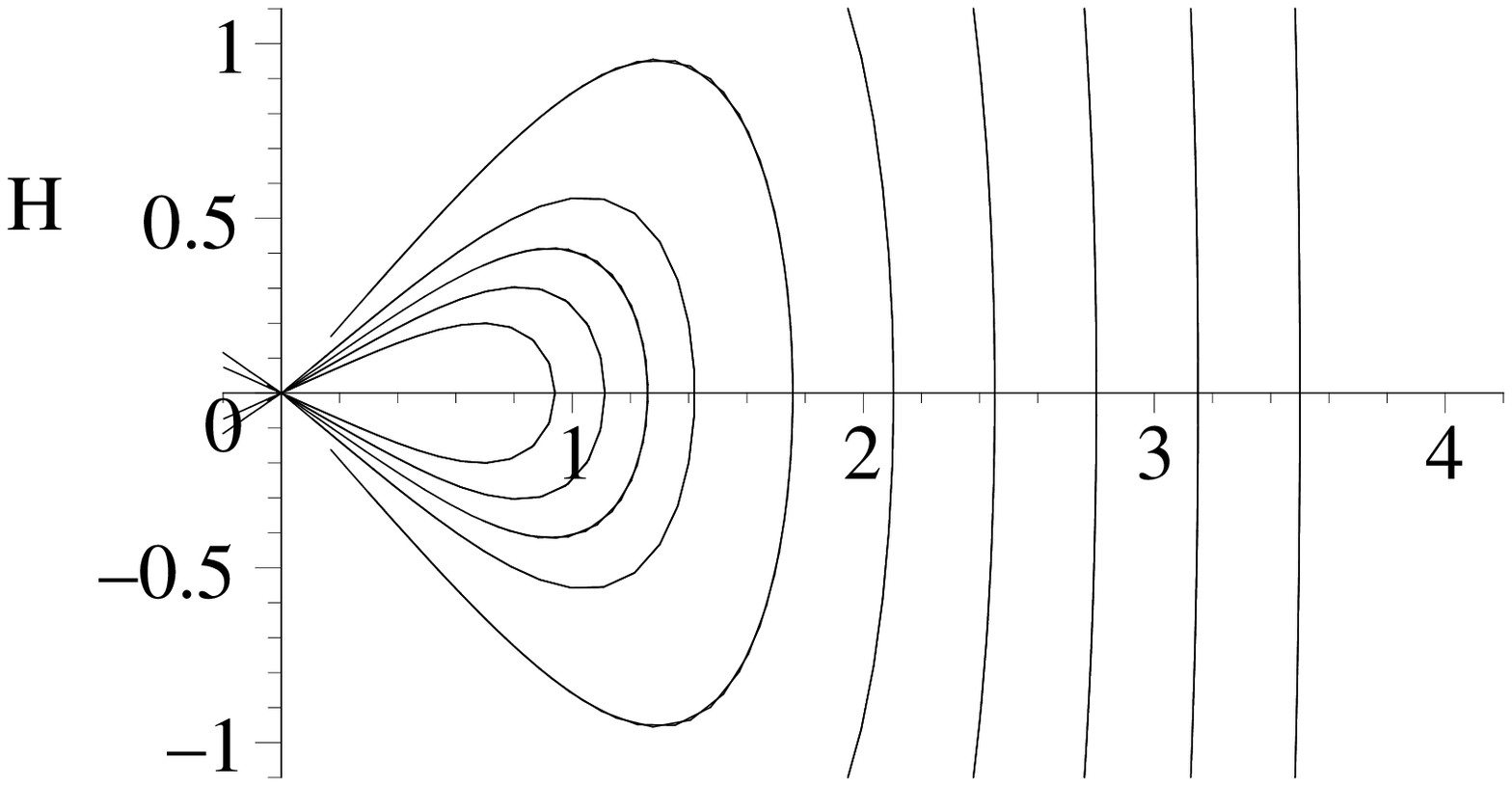} \\ e \end{tabular} &
   \begin{tabular}{c} \includegraphics[width=0.15 \textwidth]{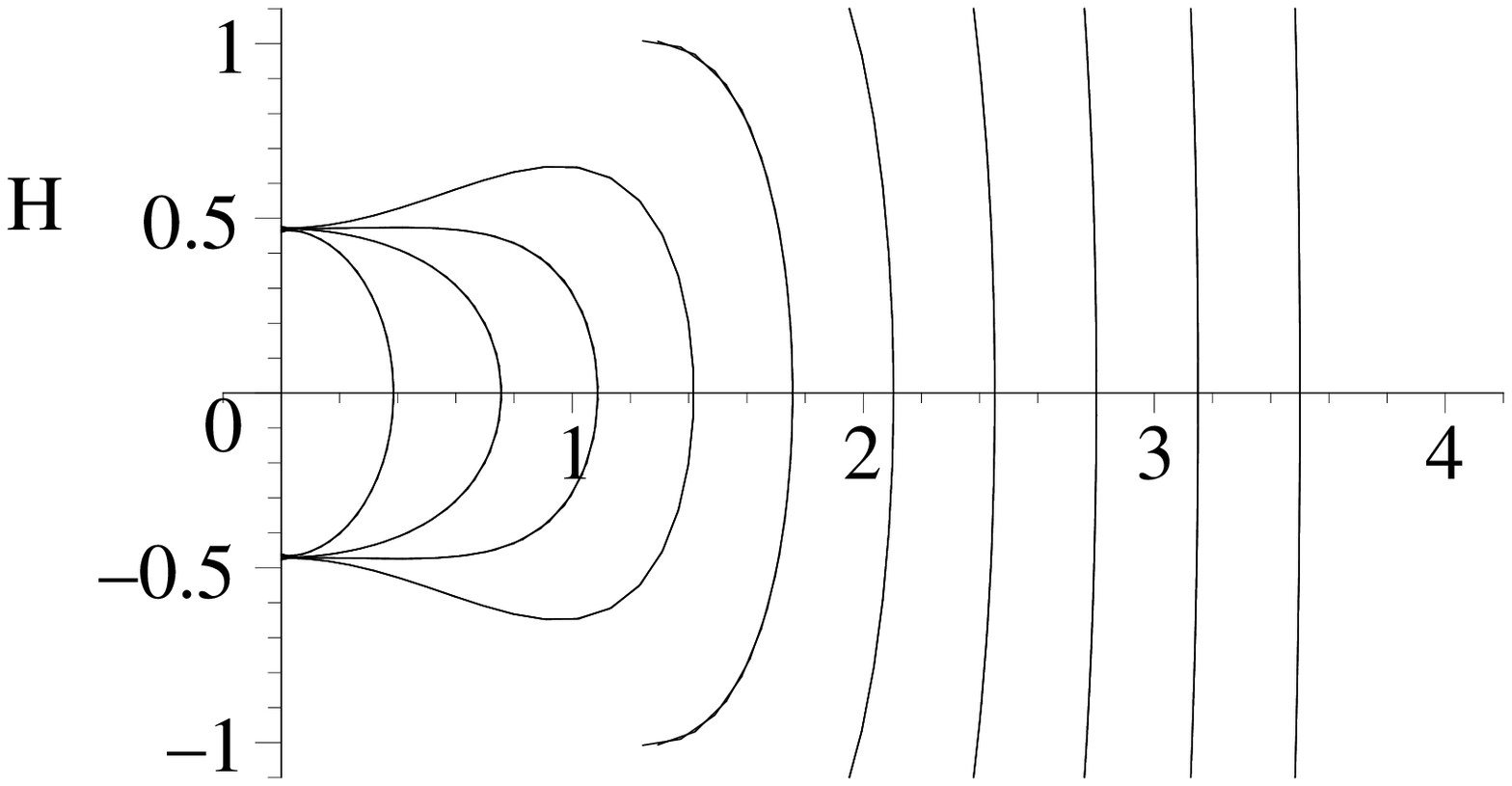} \\ f \end{tabular} \\
  \hline
   $m>0$ &
   \begin{tabular}{c} \includegraphics[width=0.15 \textwidth]{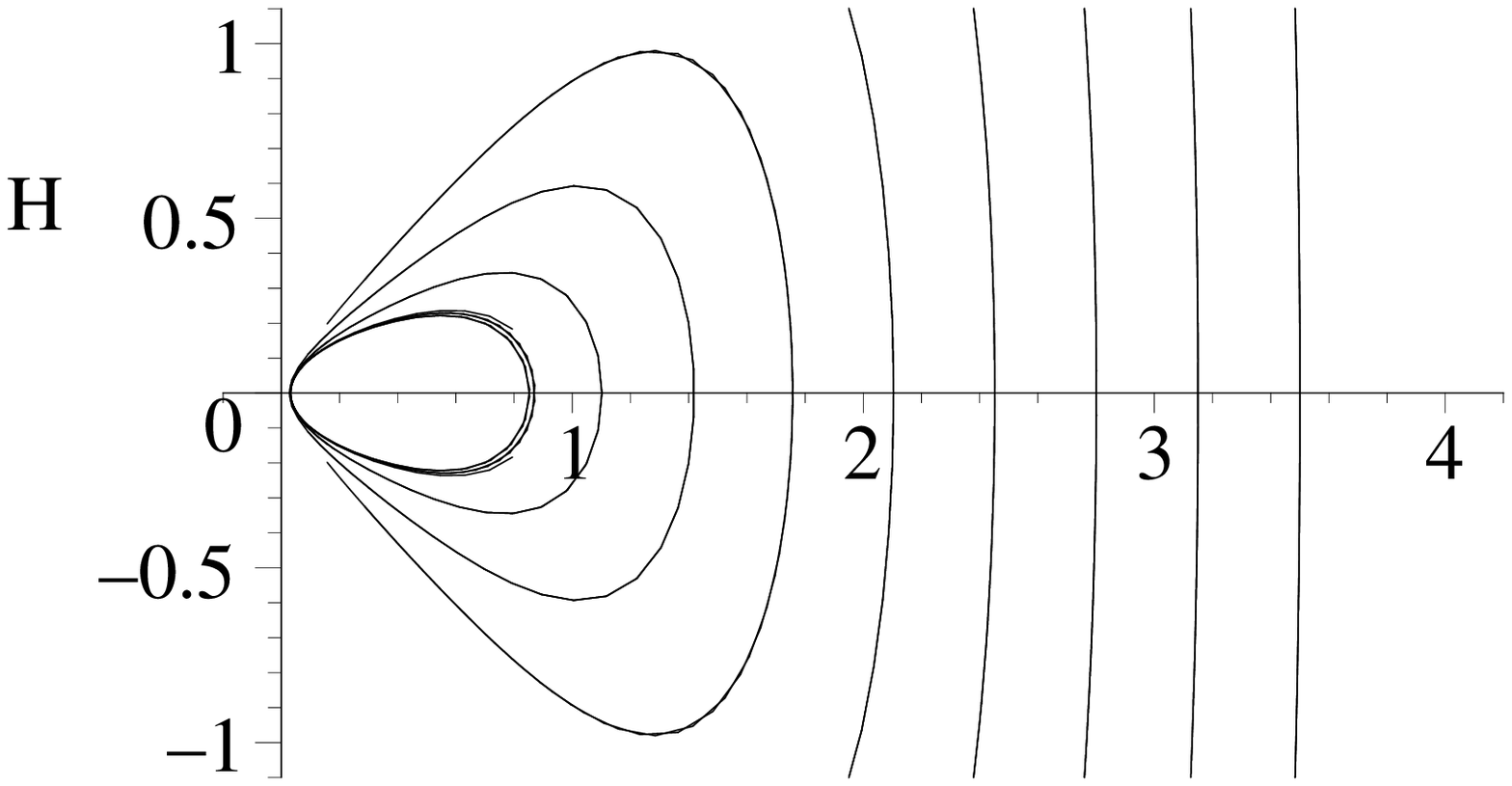} \\ g \end{tabular} &
   \begin{tabular}{c} \includegraphics[width=0.15 \textwidth]{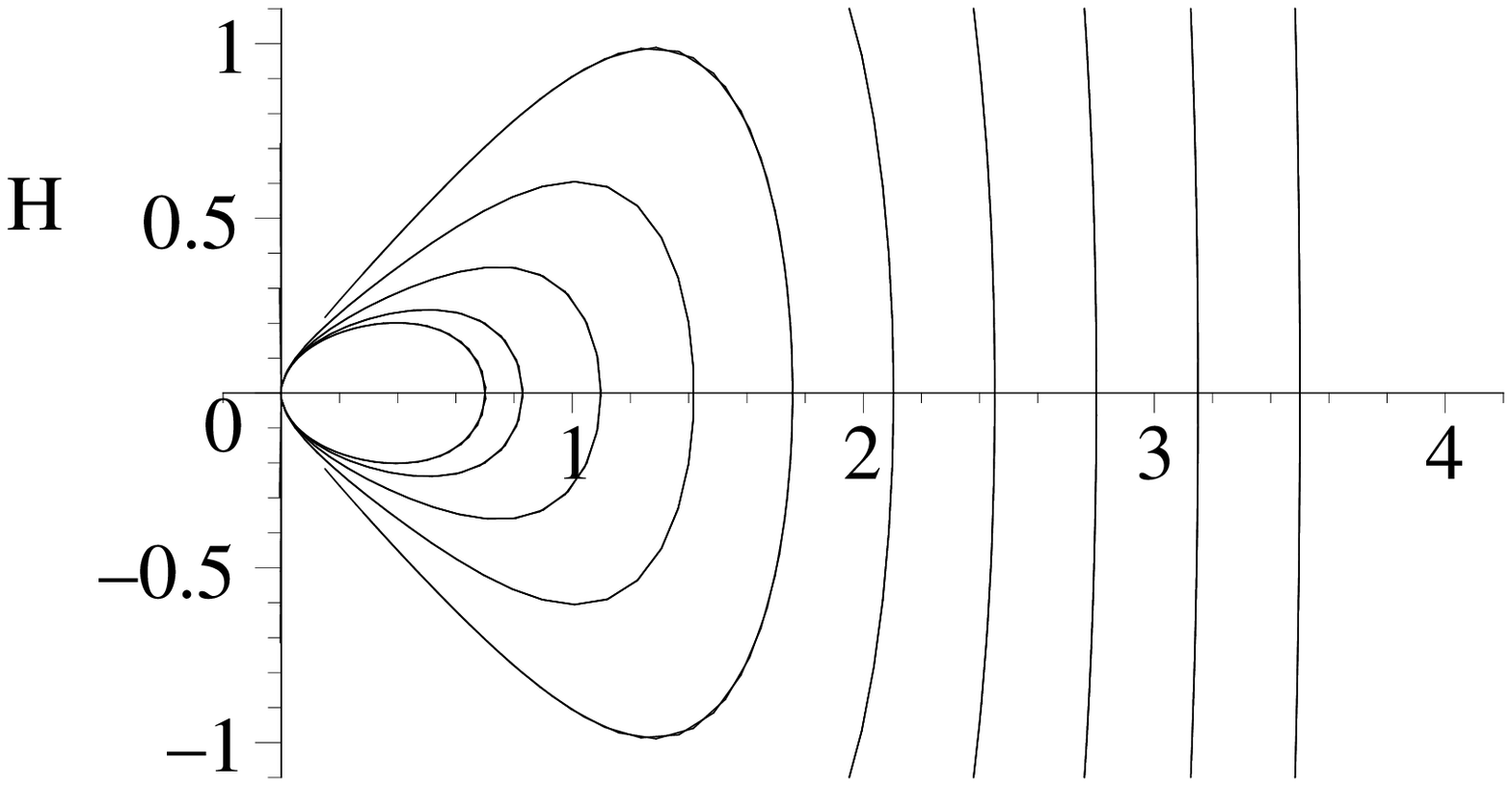} \\ h \end{tabular} &
   \begin{tabular}{c} \includegraphics[width=0.15 \textwidth]{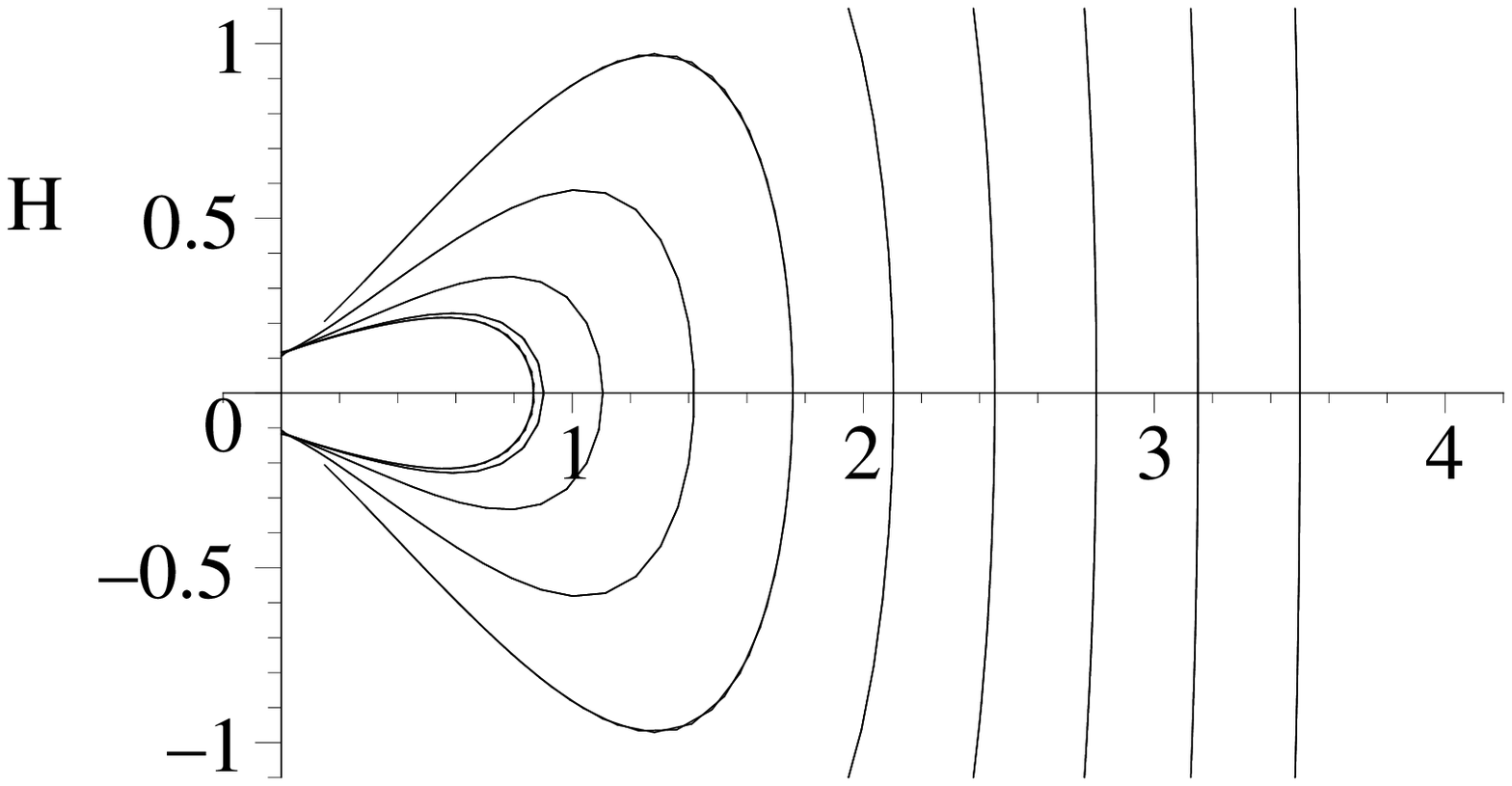} \\ i \end{tabular} \\
  \hline
\end{tabular}//
\vskip 2 mm
Table E3. Case with  $\ve=0$, $n=3$ and $\lambda=0$. \\
\vskip 2 mm

\begin{tabular}{|c|c|c|c|}
  \hline
     & $\Lambda<0$ & $\Lambda=0$ & $\Lambda>0$ \\
  \hline
   $m<0$ &
   \begin{tabular}{c} \includegraphics[width=0.15 \textwidth]{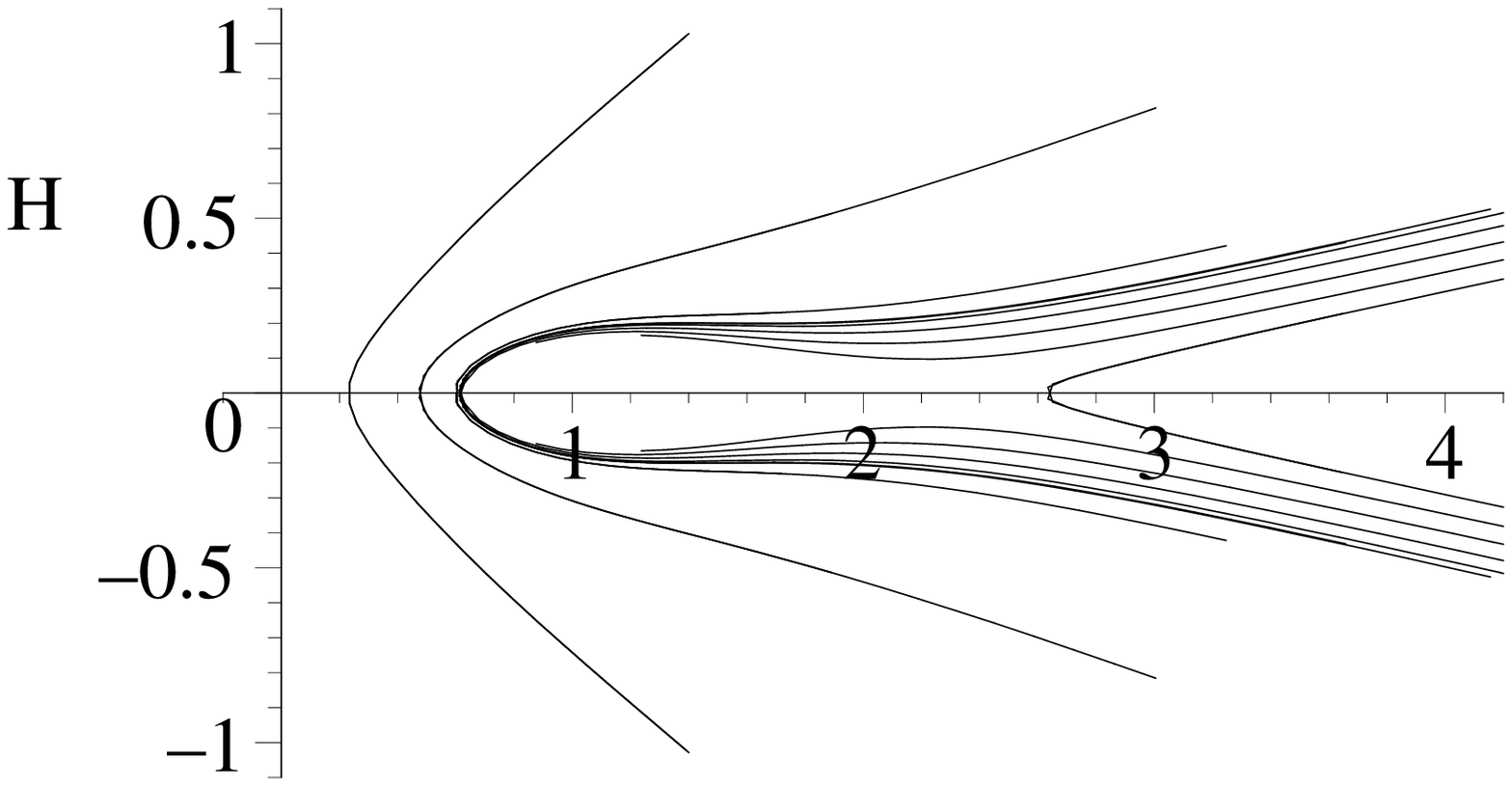} \\ a  \end{tabular} &
   \begin{tabular}{c} \includegraphics[width=0.15 \textwidth]{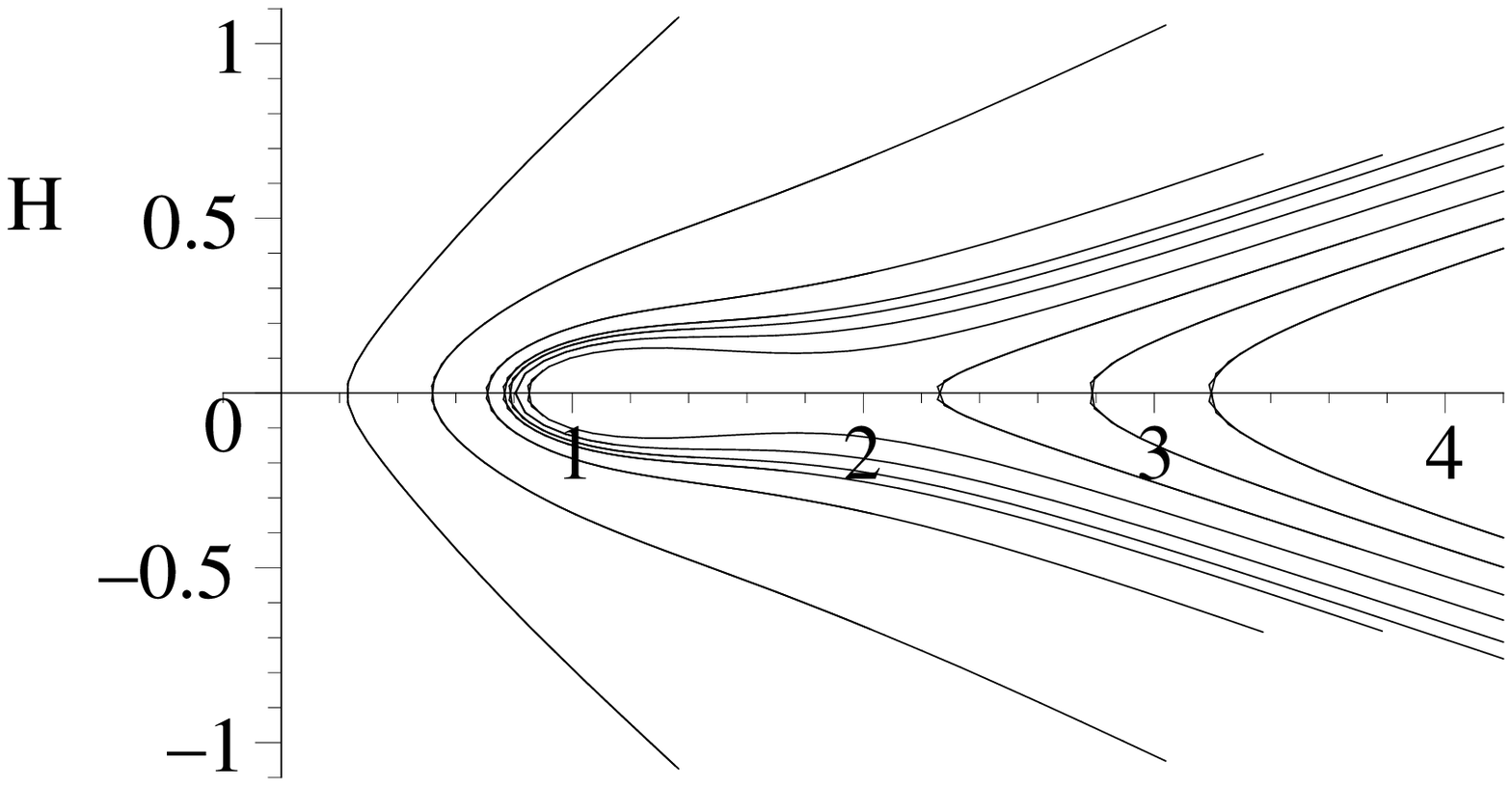} \\ b \end{tabular} &
   \begin{tabular}{c} \includegraphics[width=0.15 \textwidth]{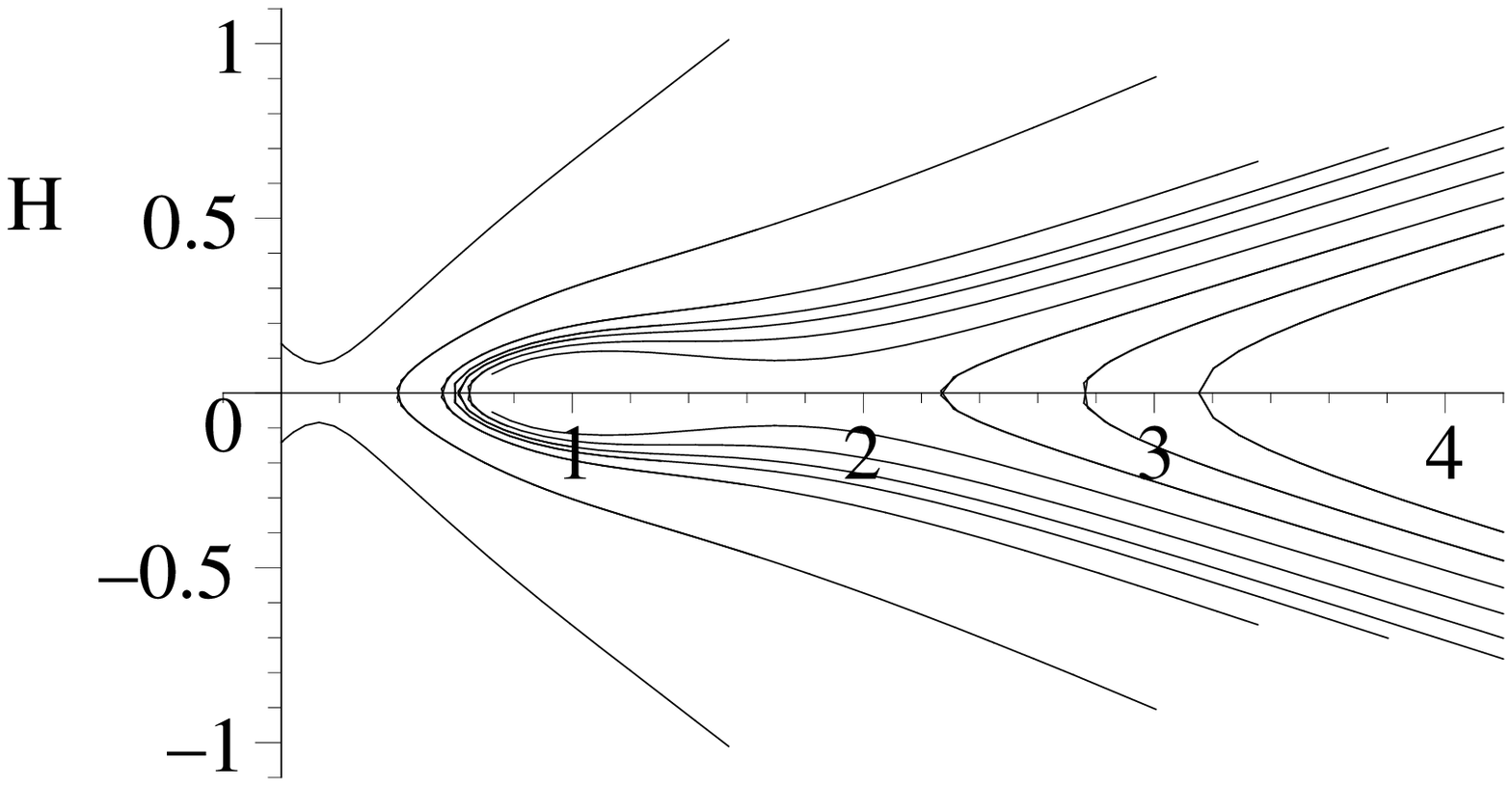} \\ c \end{tabular} \\
  \hline
   $m=0$ &
   \begin{tabular}{c} \includegraphics[width=0.15 \textwidth]{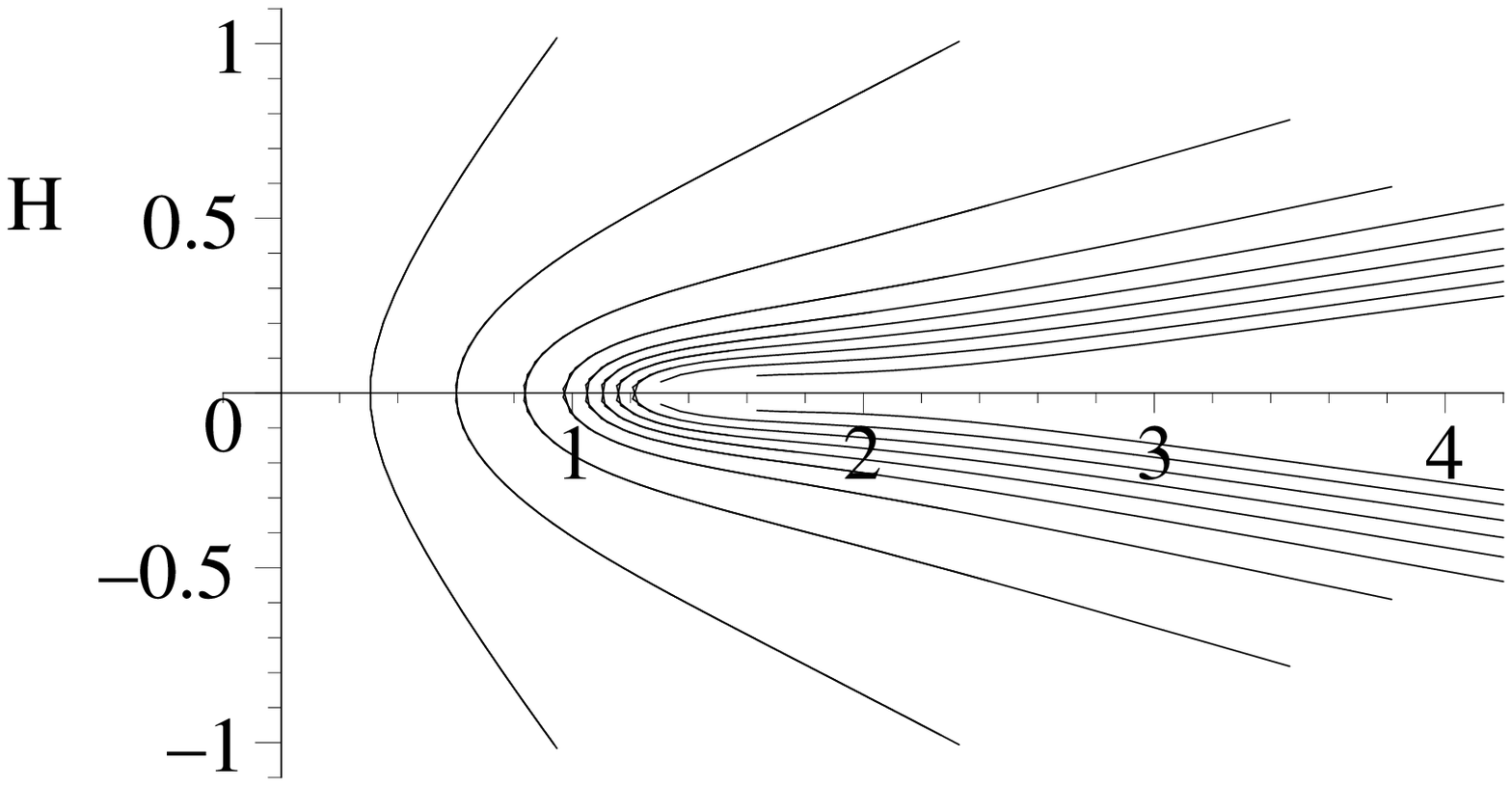} \\ d \end{tabular} &
   \begin{tabular}{c} \includegraphics[width=0.15 \textwidth]{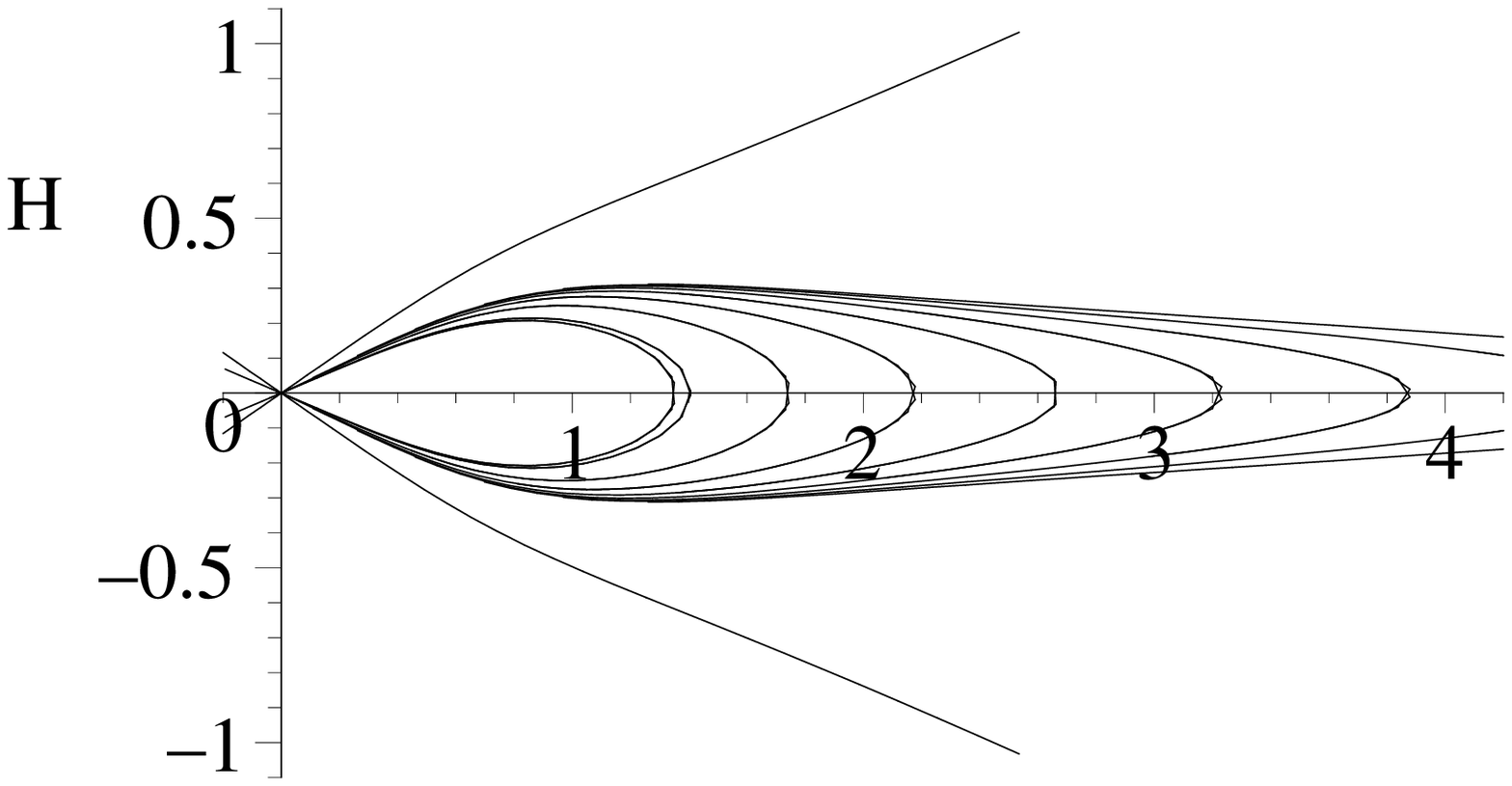} \\ e \end{tabular} &
   \begin{tabular}{c} \includegraphics[width=0.15 \textwidth]{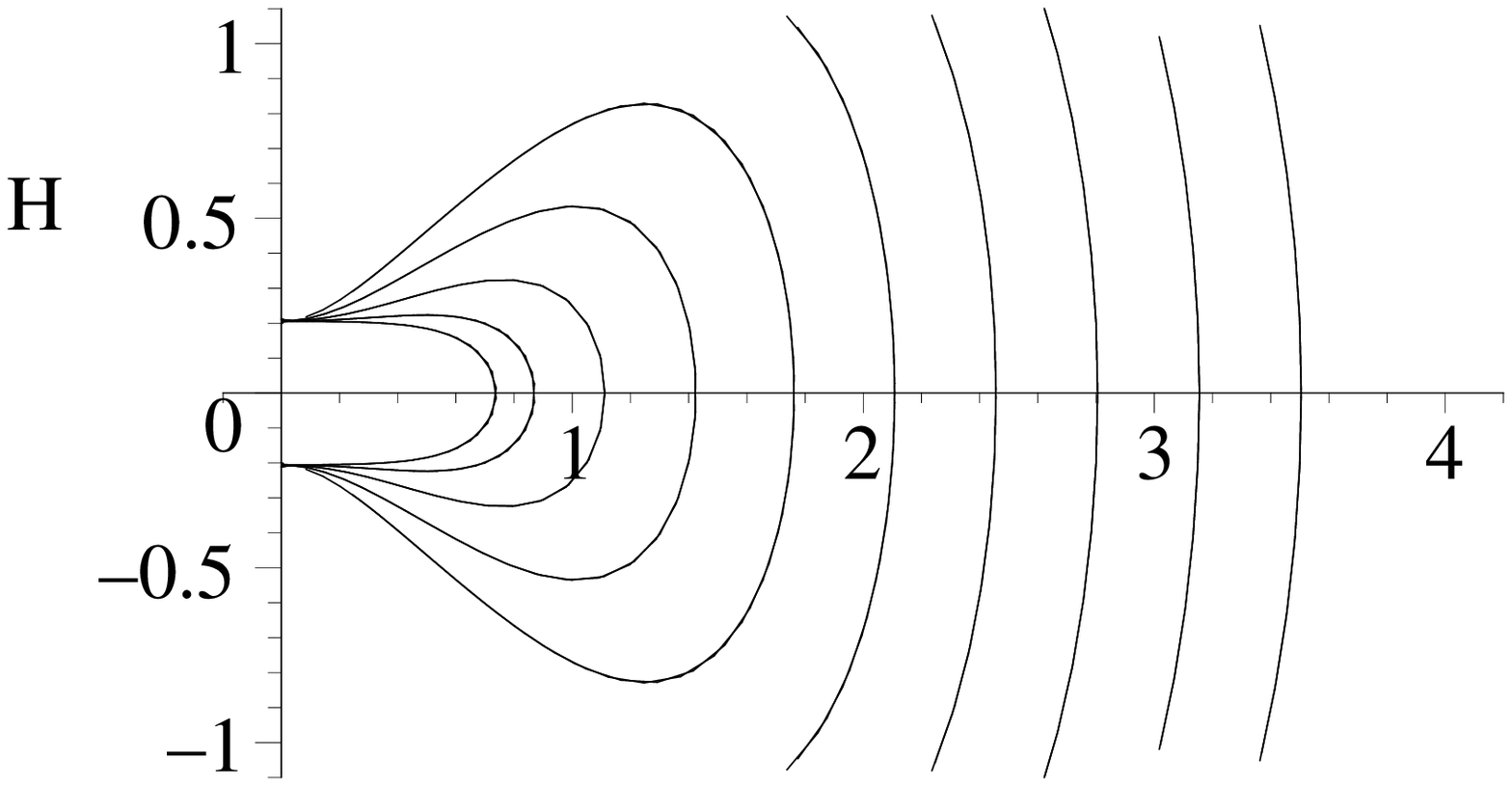} \\ f \end{tabular} \\
  \hline
   $m>0$ &
   \begin{tabular}{c} \includegraphics[width=0.15 \textwidth]{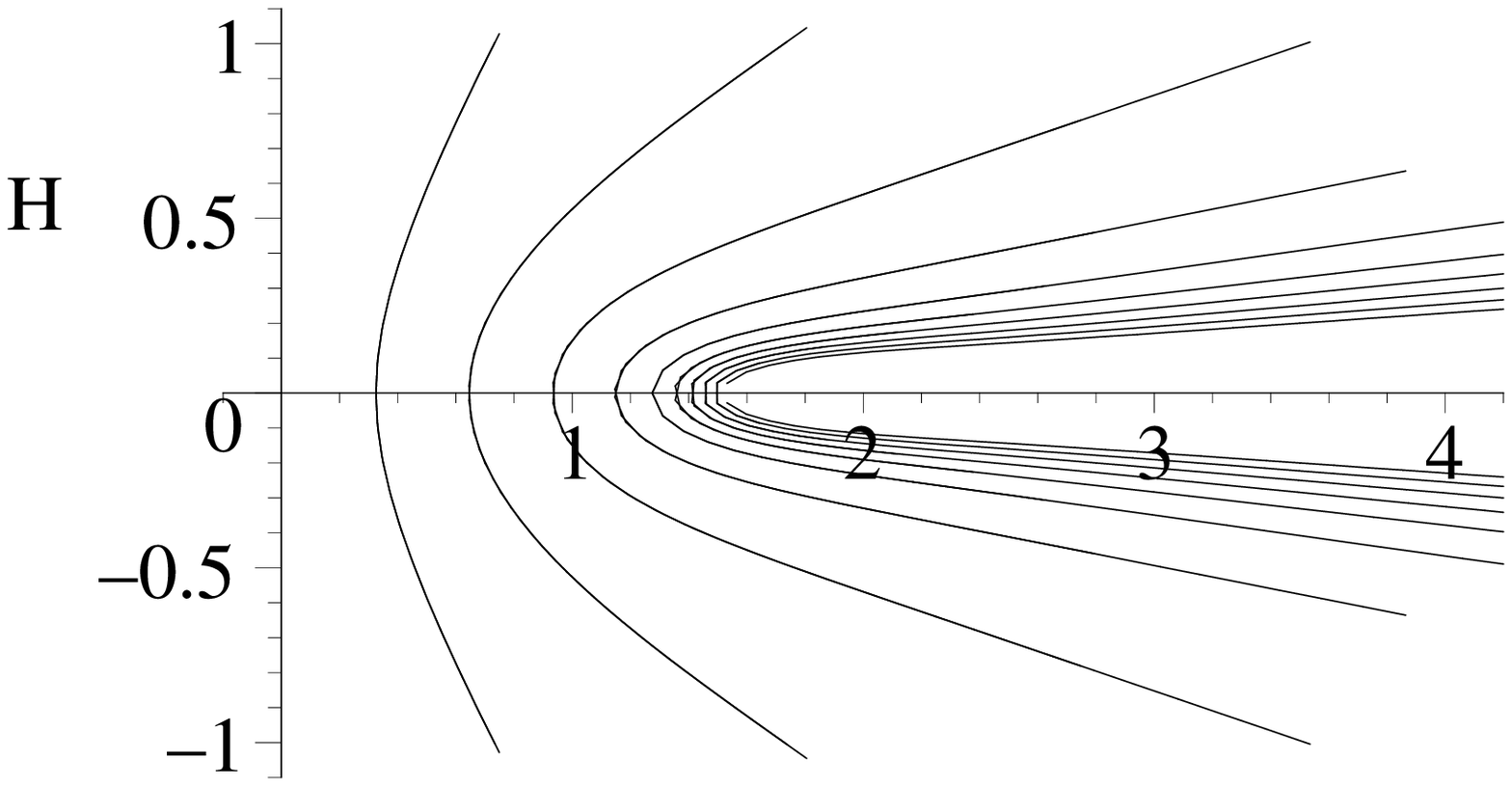} \\ g \end{tabular} &
   \begin{tabular}{c} \includegraphics[width=0.15 \textwidth]{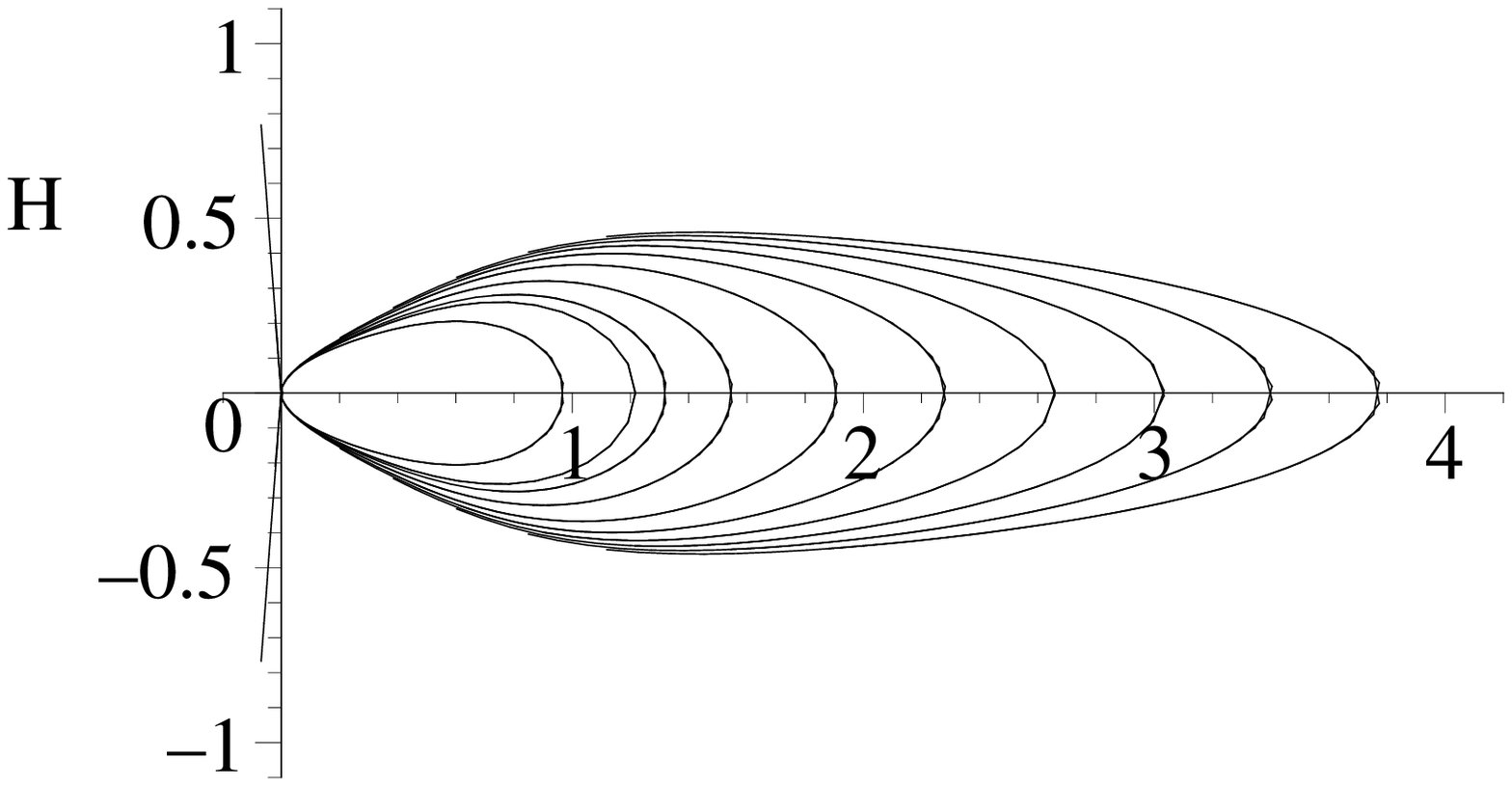} \\ h \end{tabular} &
   \begin{tabular}{c} \includegraphics[width=0.15 \textwidth]{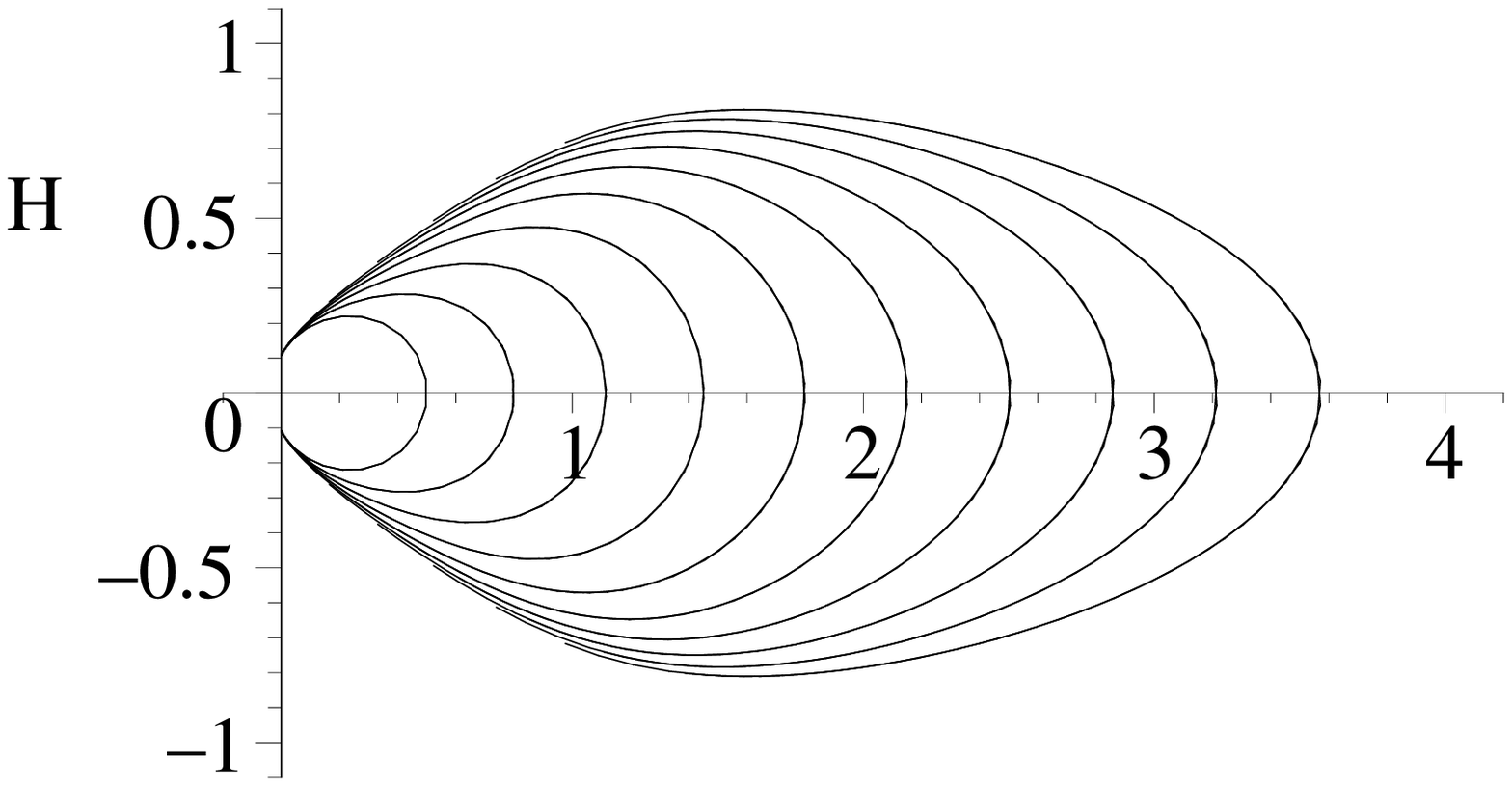} \\ i \end{tabular} \\
  \hline
\end{tabular}//
\vskip 2 mm
Table C2. Case with  $\ve=0$, $n=3$ and $\lambda>0$. \\
\vskip 2 mm

\begin{tabular}{|c|c|c|c|}
  \hline
     & $\Lambda<0$ & $\Lambda=0$ & $\Lambda>0$ \\
  \hline
   $m<0$ &
   \begin{tabular}{c} \includegraphics[width=0.15 \textwidth]{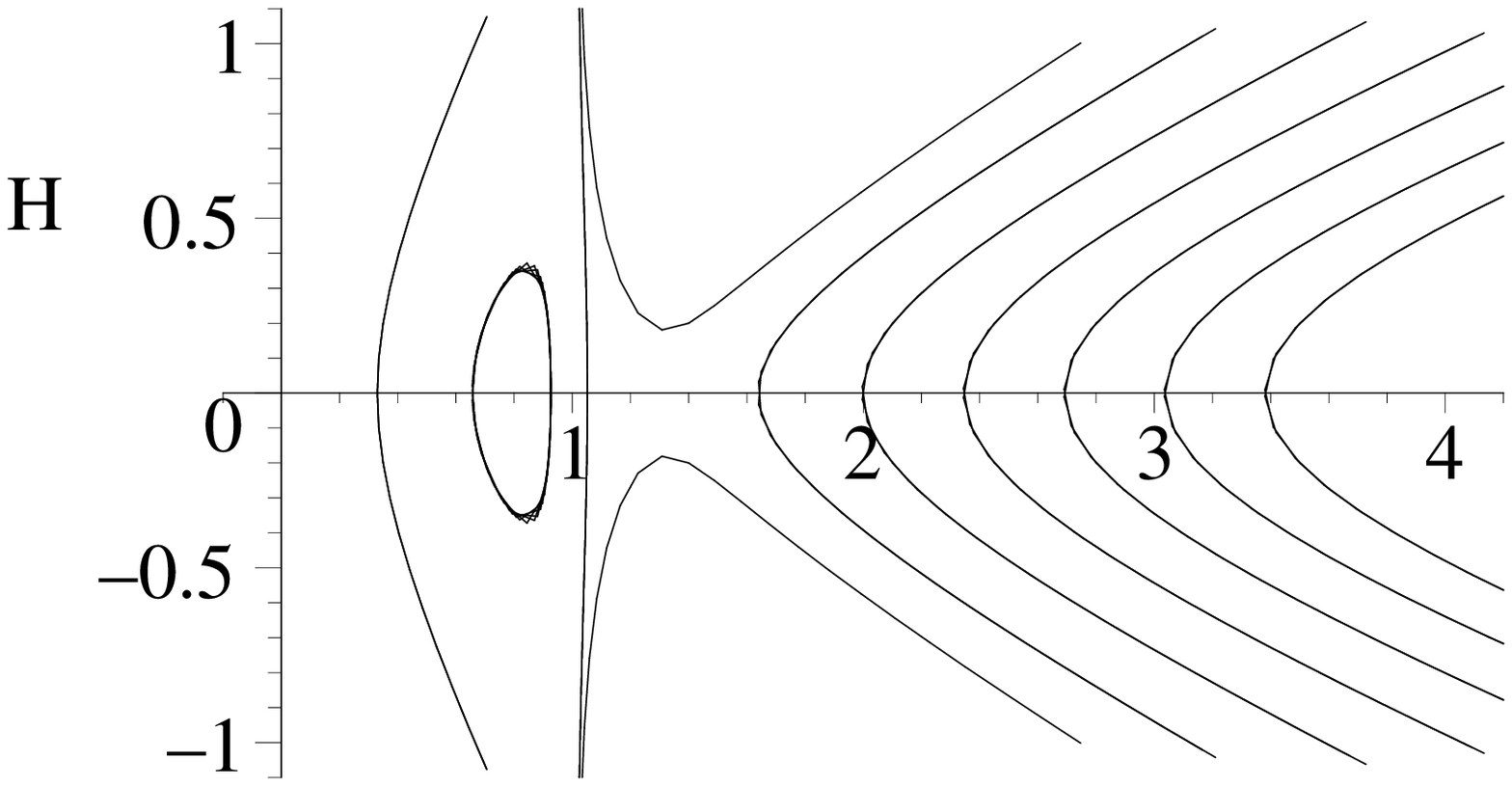} \\ a  \end{tabular} &
   \begin{tabular}{c} \includegraphics[width=0.15 \textwidth]{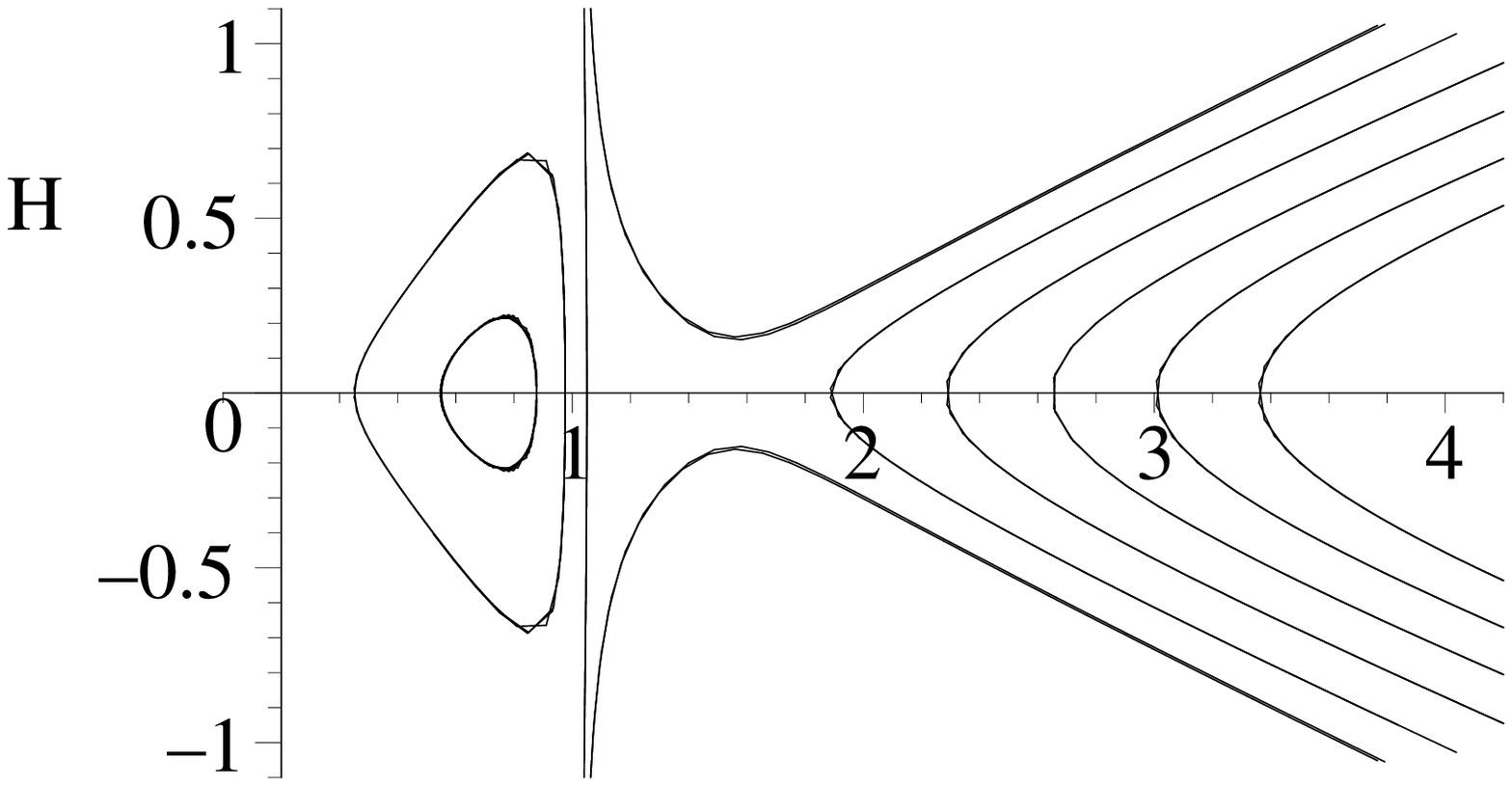} \\ b \end{tabular} &
   \begin{tabular}{c} \includegraphics[width=0.15 \textwidth]{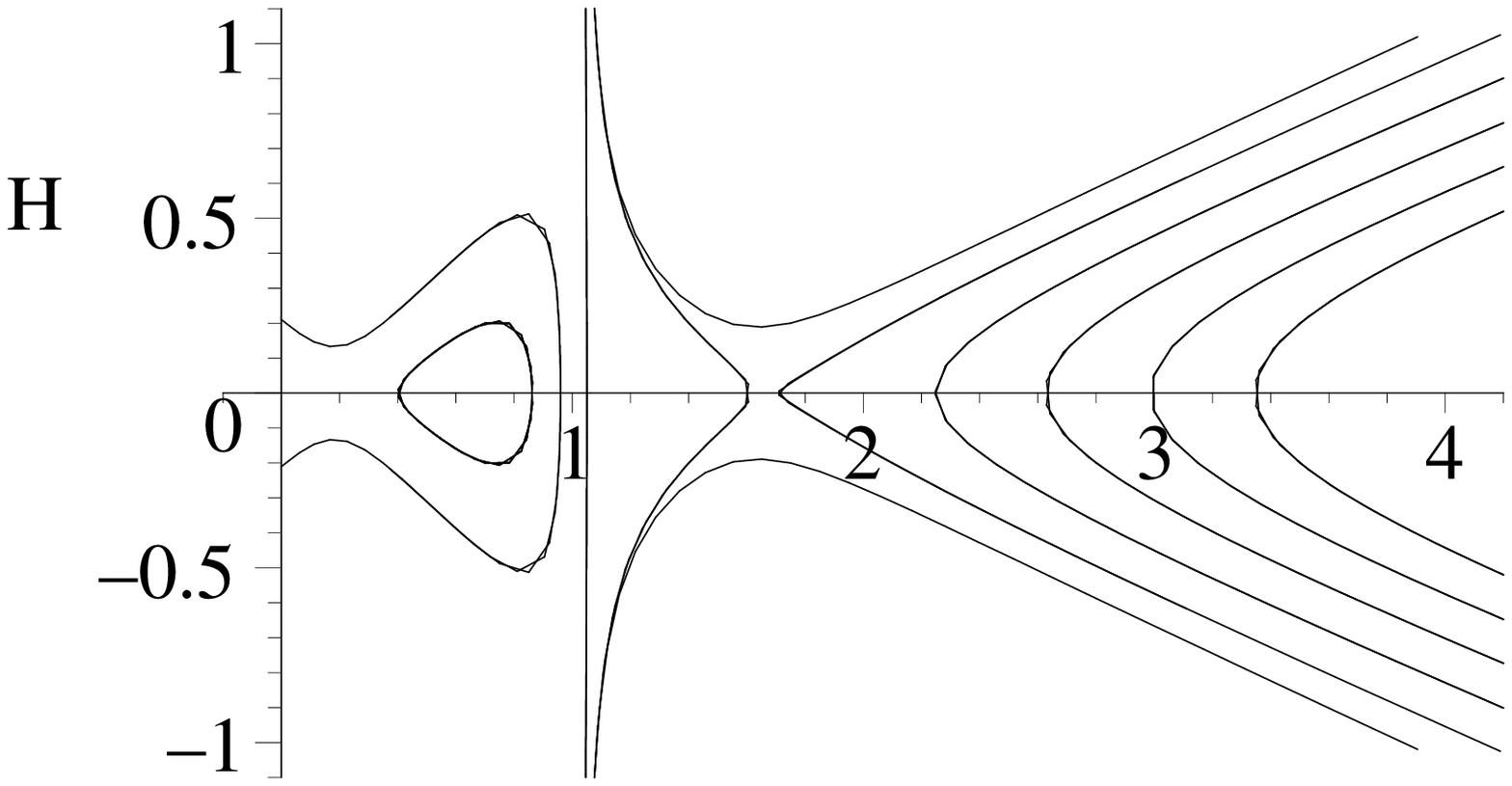} \\ c \end{tabular} \\
  \hline
   $m=0$ &
   \begin{tabular}{c} \includegraphics[width=0.15 \textwidth]{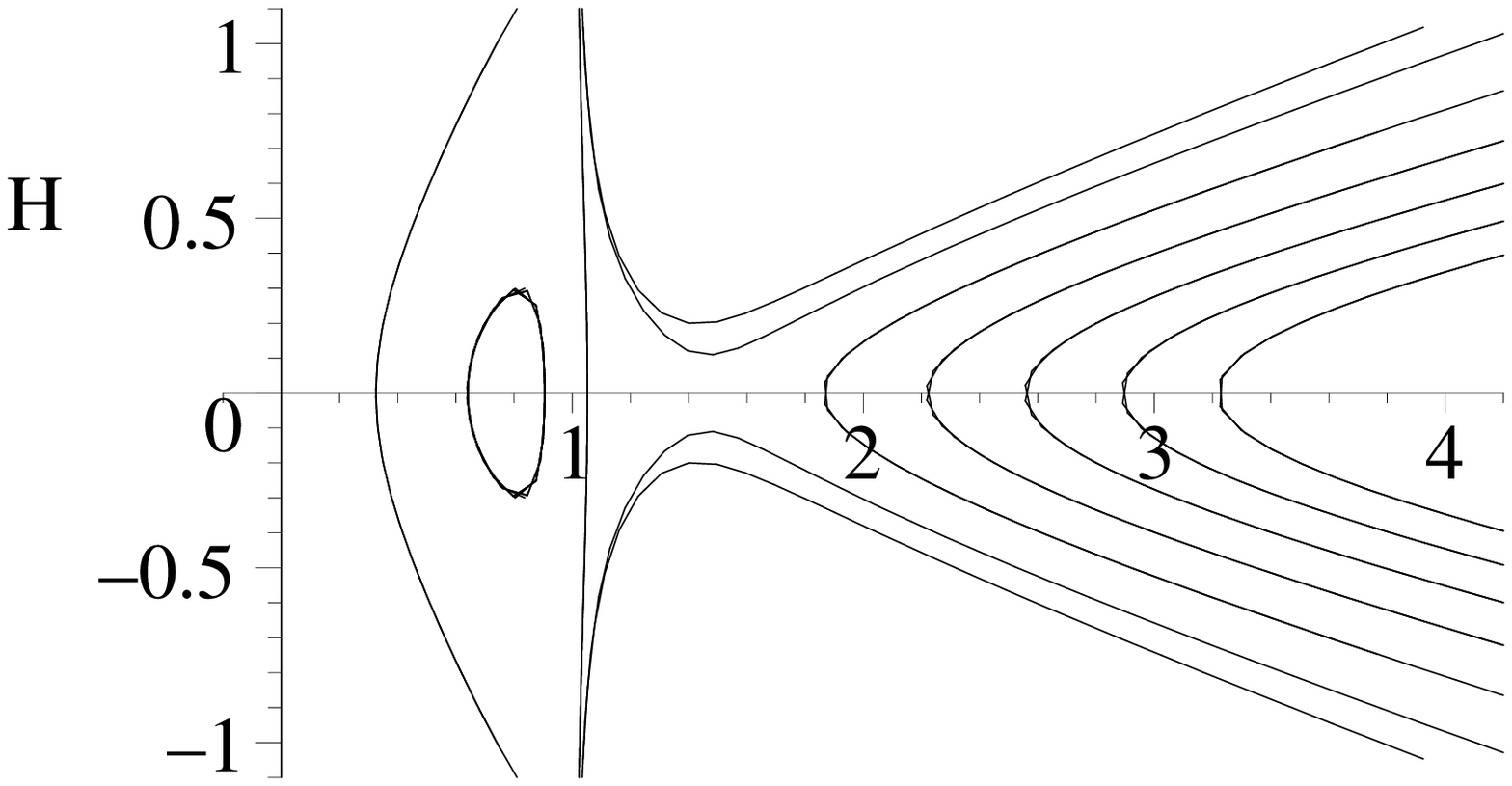} \\ d \end{tabular} &
   \begin{tabular}{c} \includegraphics[width=0.15 \textwidth]{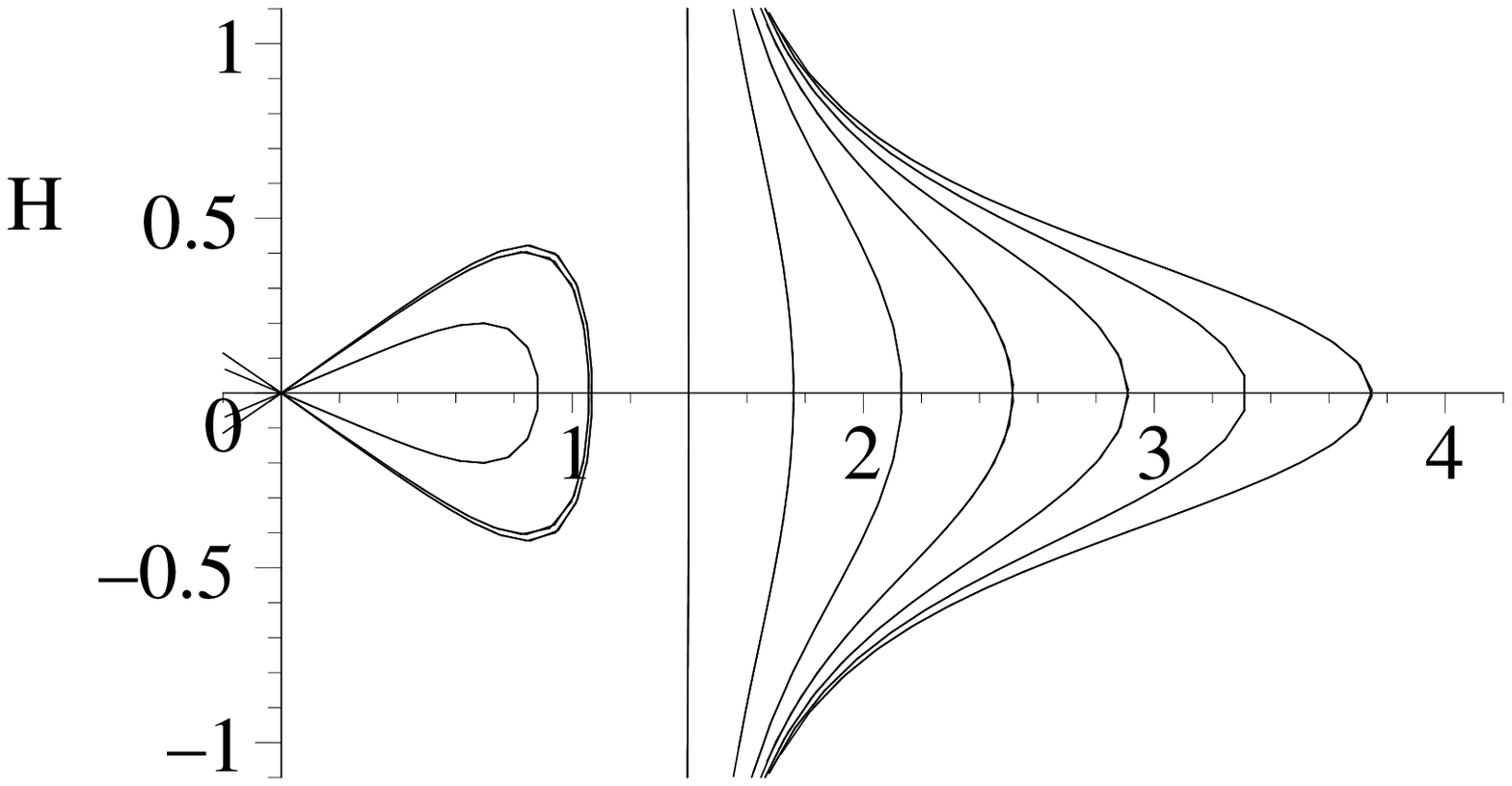} \\ e \end{tabular} &
   \begin{tabular}{c} \includegraphics[width=0.15 \textwidth]{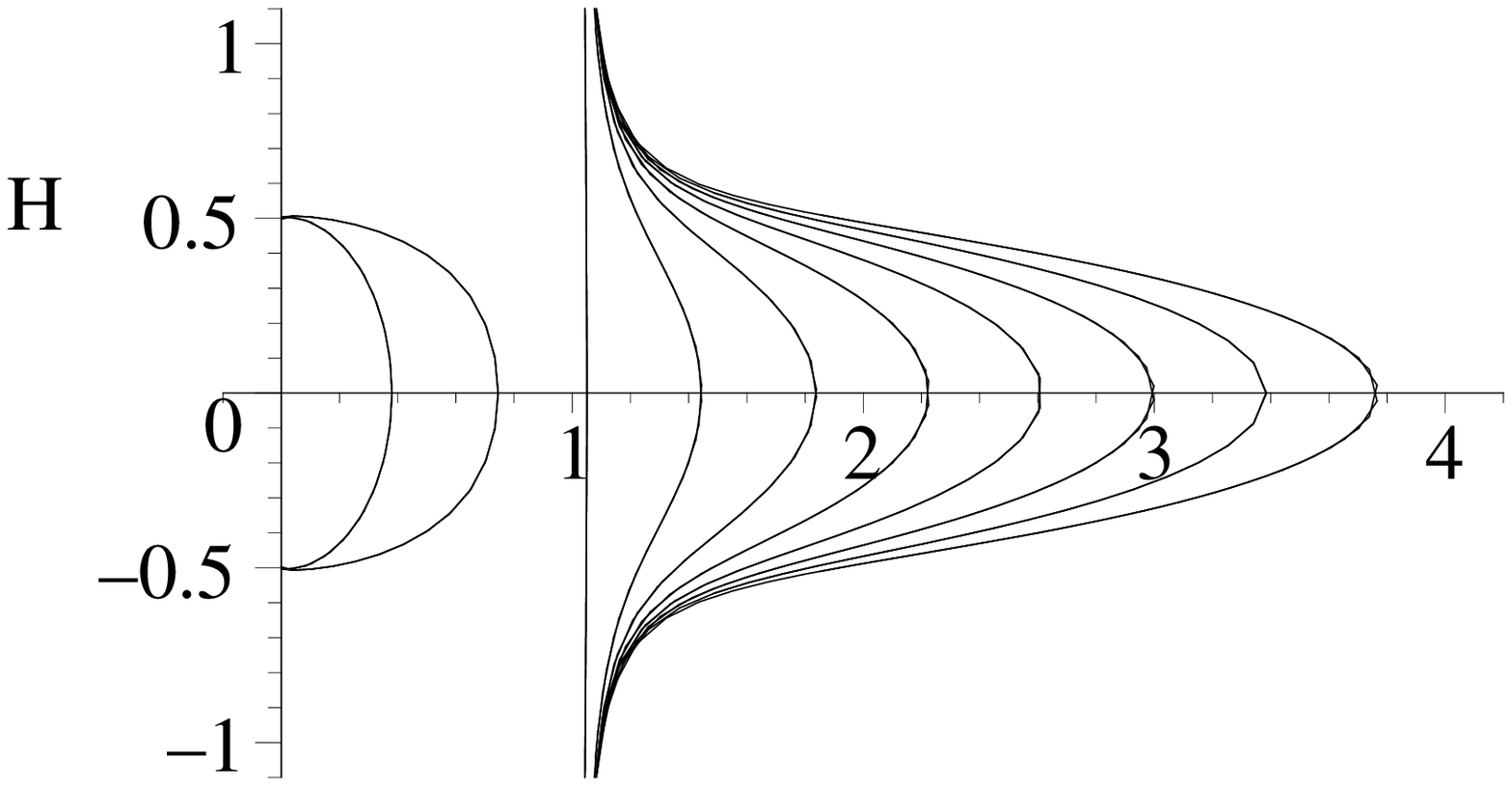} \\ f \end{tabular} \\
  \hline
   $m>0$ &
   \begin{tabular}{c} \includegraphics[width=0.15 \textwidth]{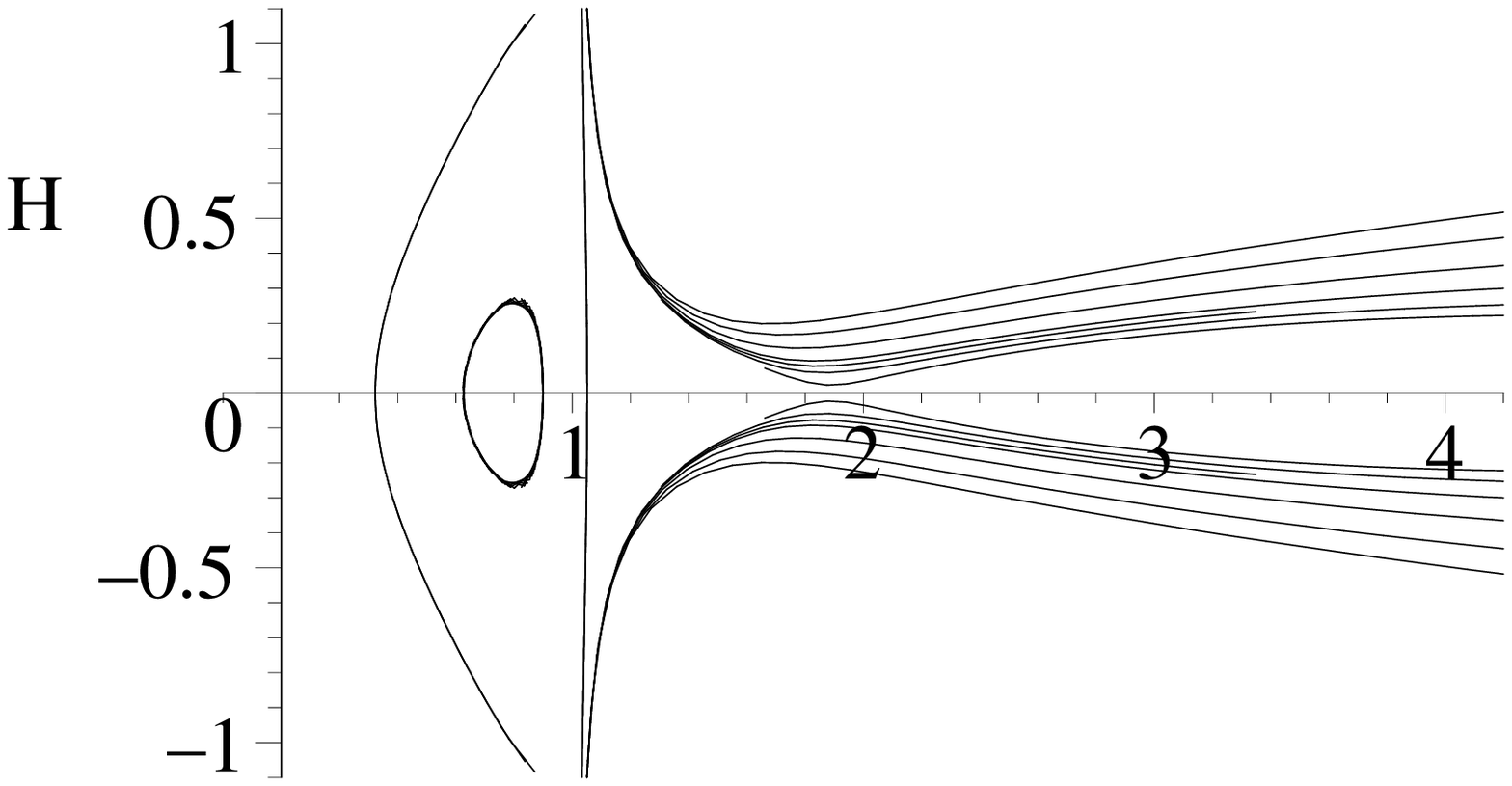} \\ g \end{tabular} &
   \begin{tabular}{c} \includegraphics[width=0.15 \textwidth]{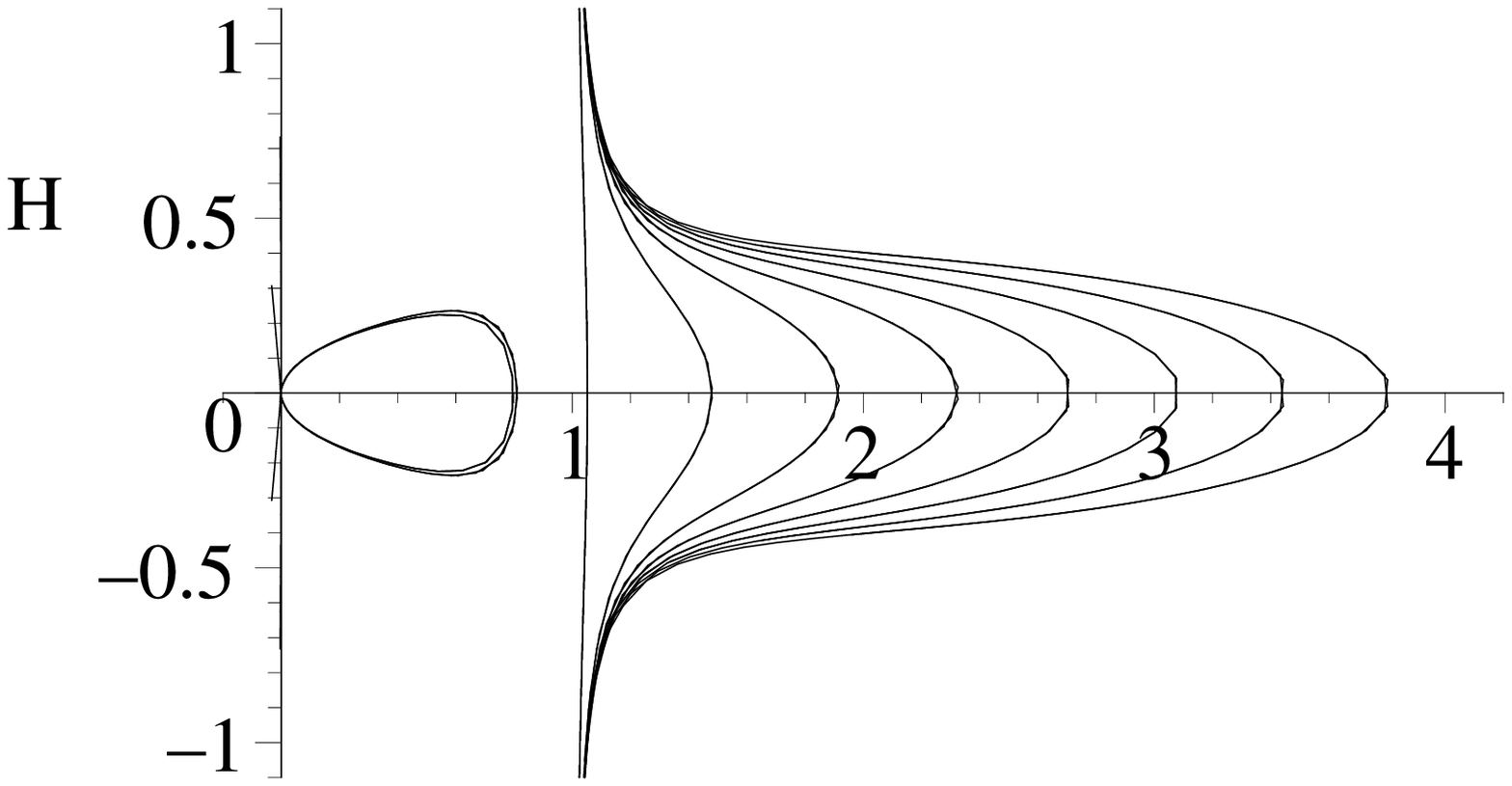} \\ h \end{tabular} &
   \begin{tabular}{c} \includegraphics[width=0.15 \textwidth]{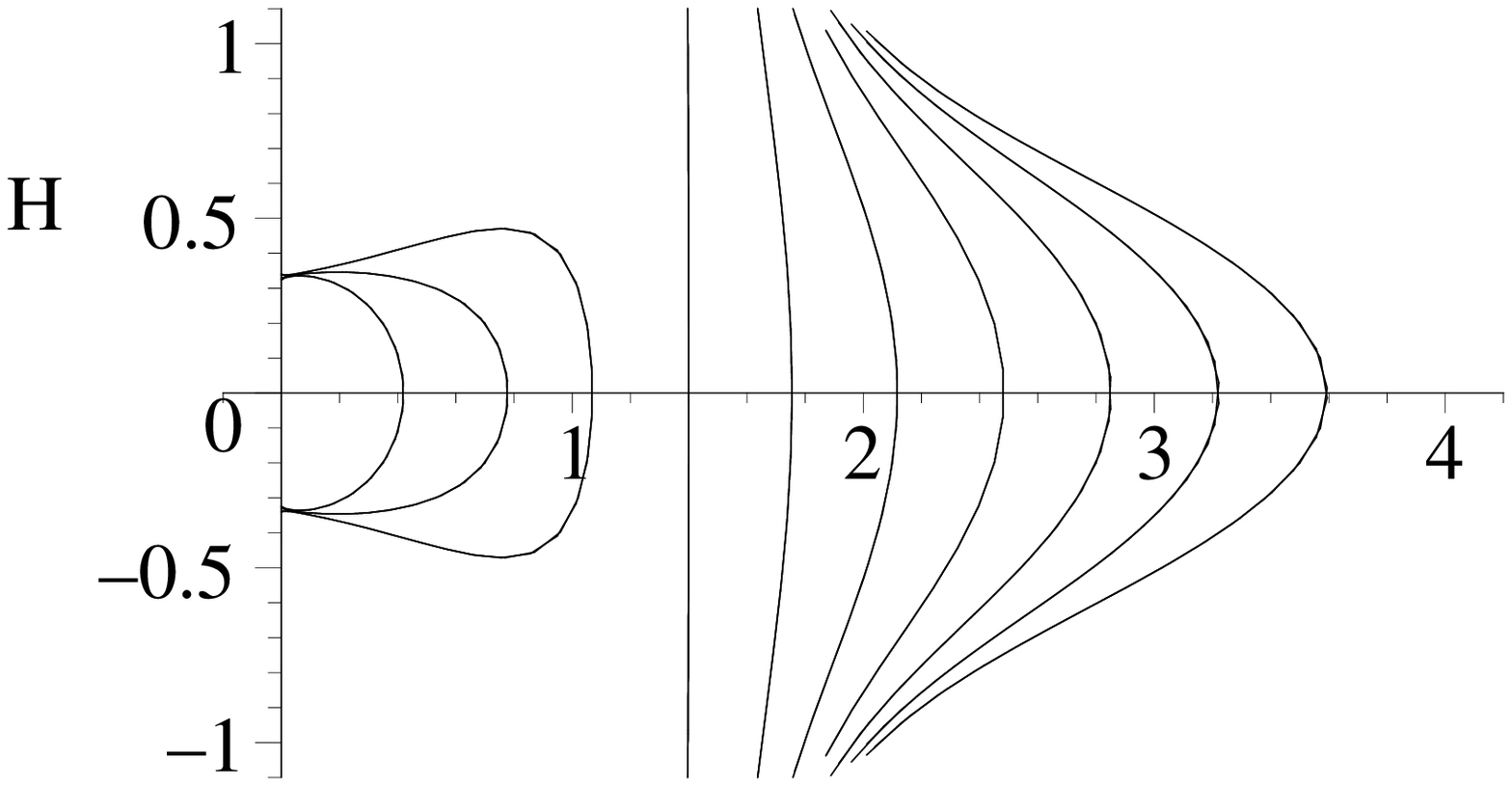} \\ i \end{tabular} \\
  \hline
\end{tabular}//
\vskip 2 mm
Table D1. Case with  $\ve=0$, $n=4$ and $\lambda<0$. \\
\vskip 2 mm

\begin{tabular}{|c|c|c|c|}
  \hline
     & $\Lambda<0$ & $\Lambda=0$ & $\Lambda>0$ \\
  \hline
   $m<0$ &
   \begin{tabular}{c} \includegraphics[width=0.15 \textwidth]{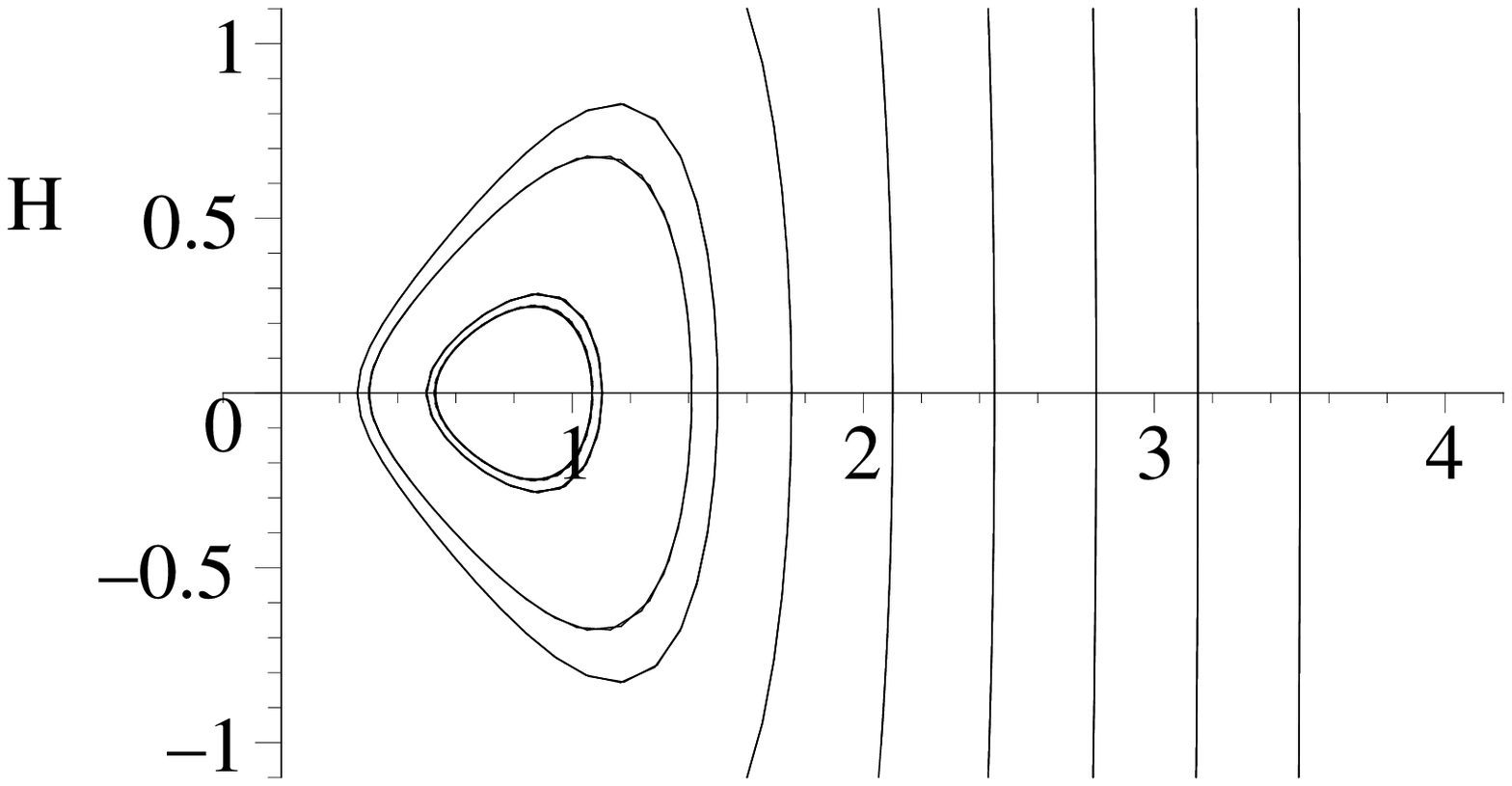} \\ a  \end{tabular} &
   \begin{tabular}{c} \includegraphics[width=0.15 \textwidth]{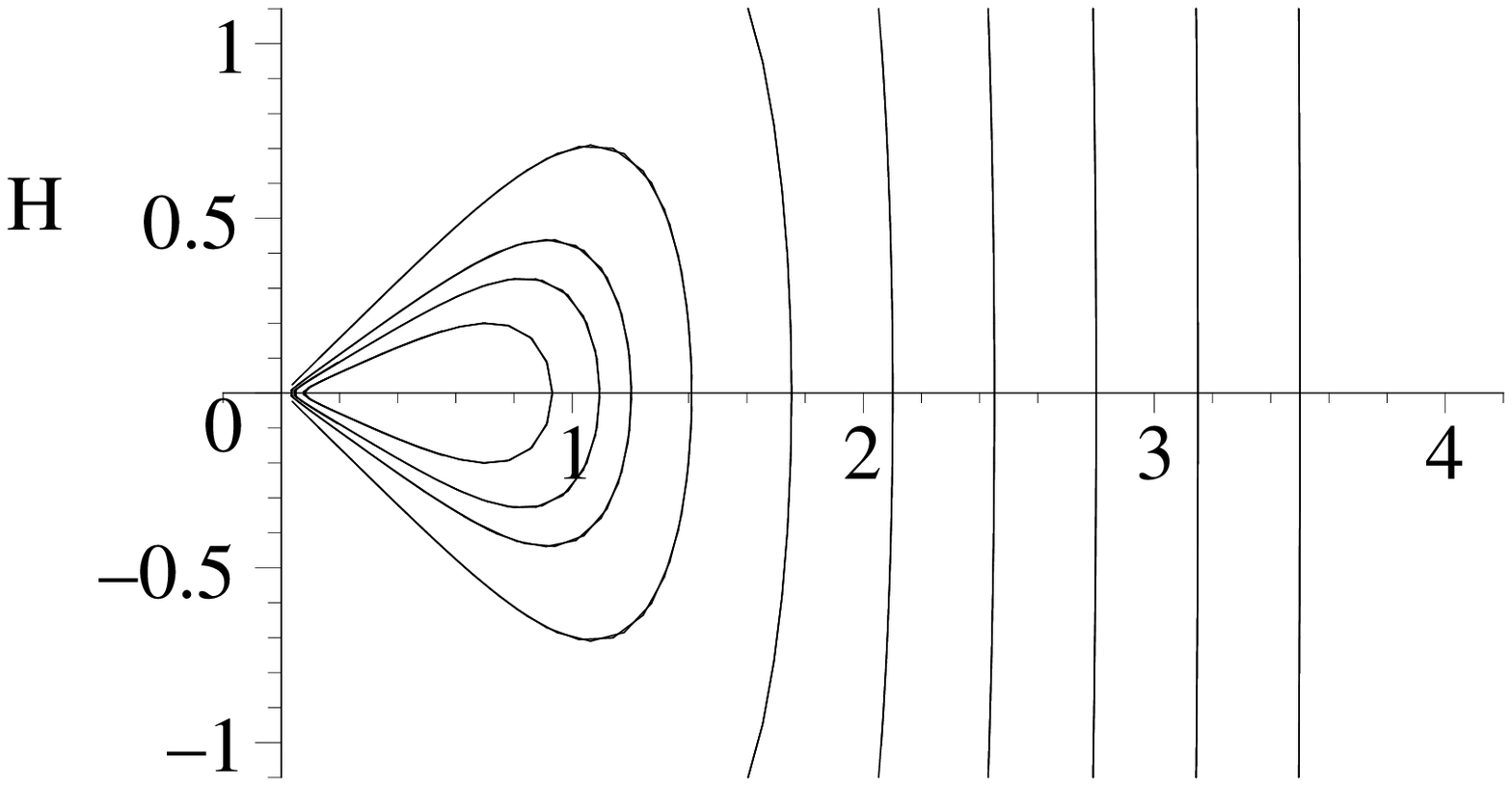} \\ b \end{tabular} &
   \begin{tabular}{c} \includegraphics[width=0.15 \textwidth]{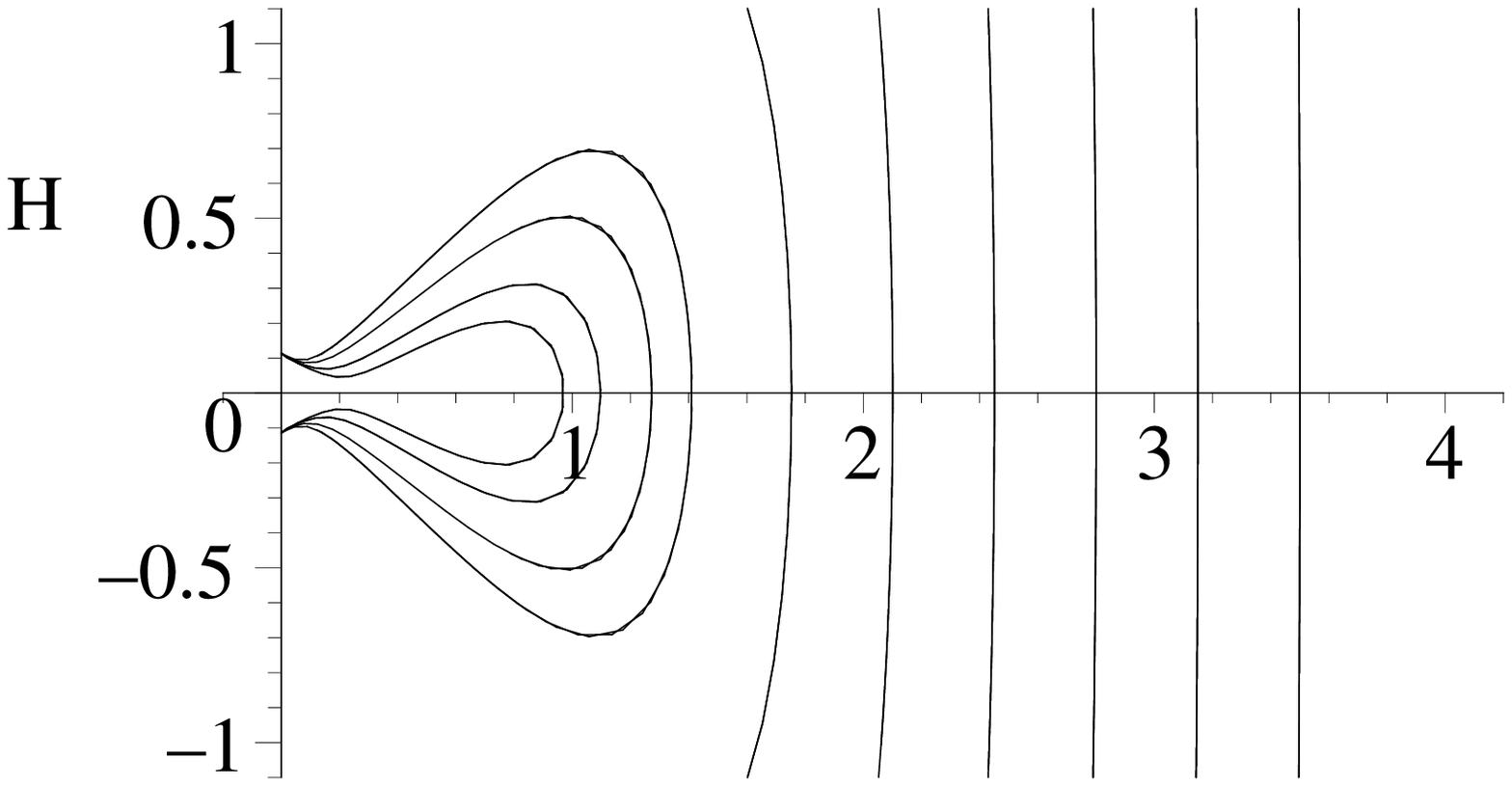} \\ c \end{tabular} \\
  \hline
   $m=0$ &
   \begin{tabular}{c} \includegraphics[width=0.15 \textwidth]{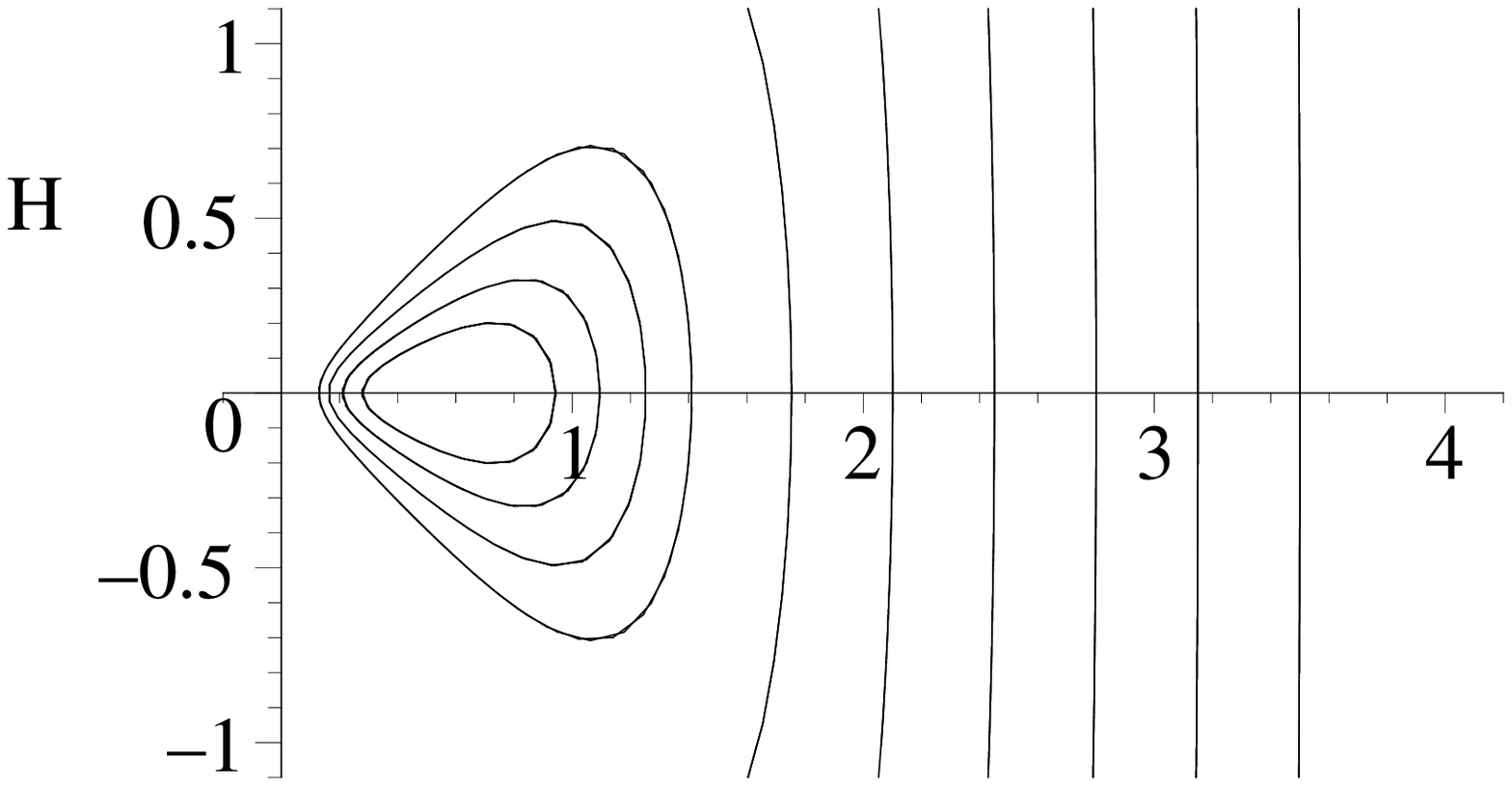} \\ d \end{tabular} &
   \begin{tabular}{c} \includegraphics[width=0.15 \textwidth]{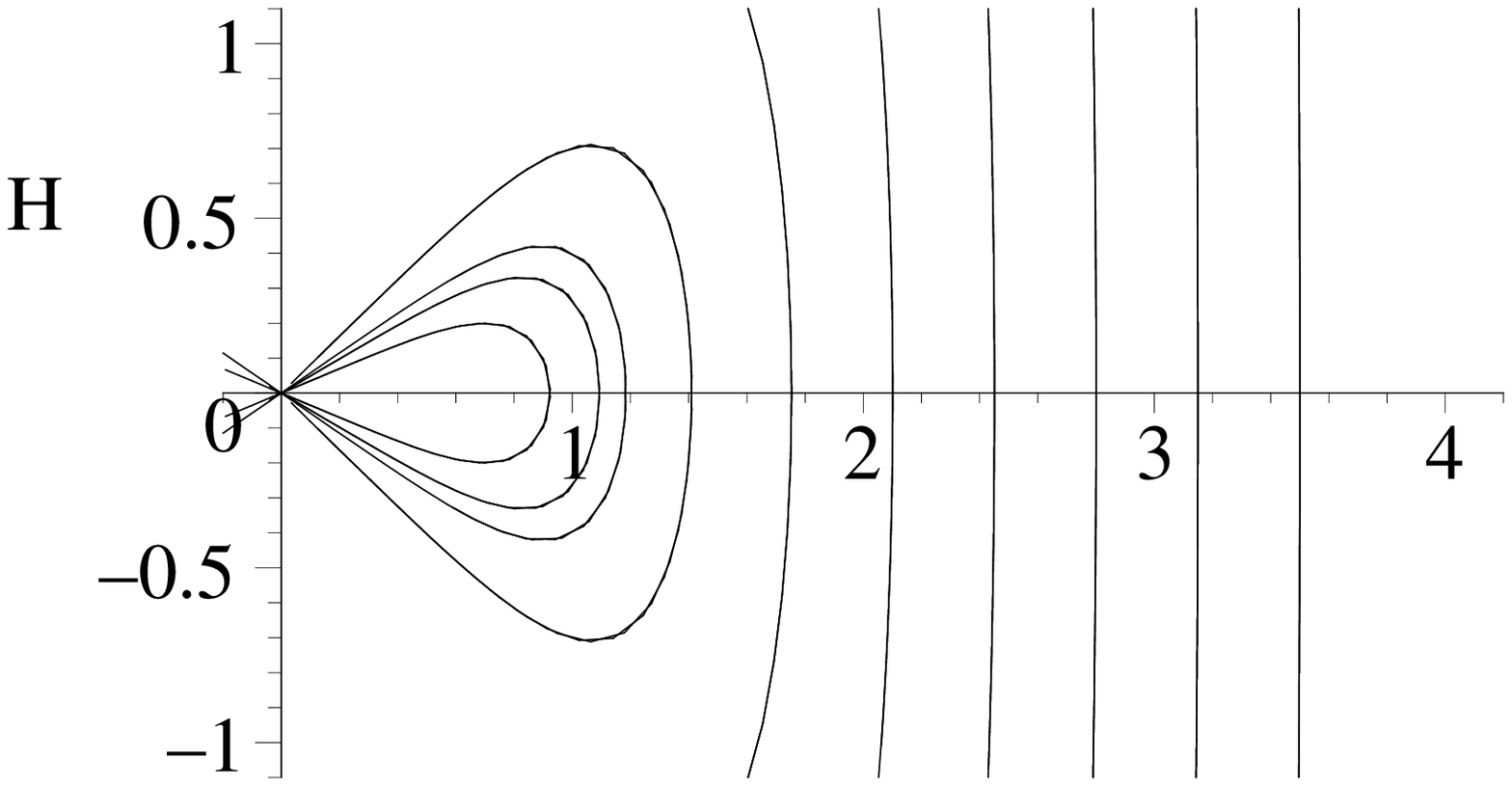} \\ e \end{tabular} &
   \begin{tabular}{c} \includegraphics[width=0.15 \textwidth]{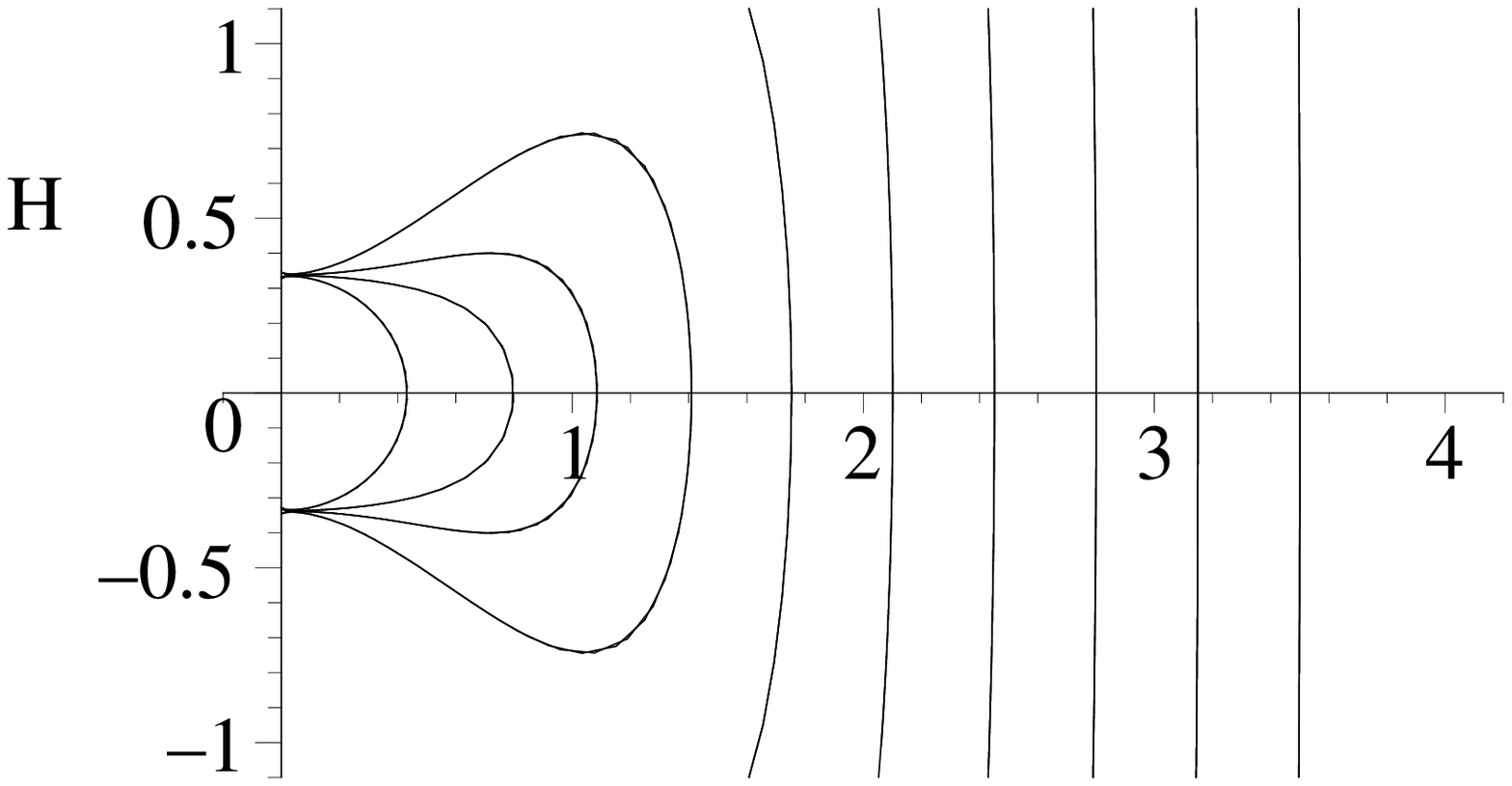} \\ f \end{tabular} \\
  \hline
   $m>0$ &
   \begin{tabular}{c} \includegraphics[width=0.15 \textwidth]{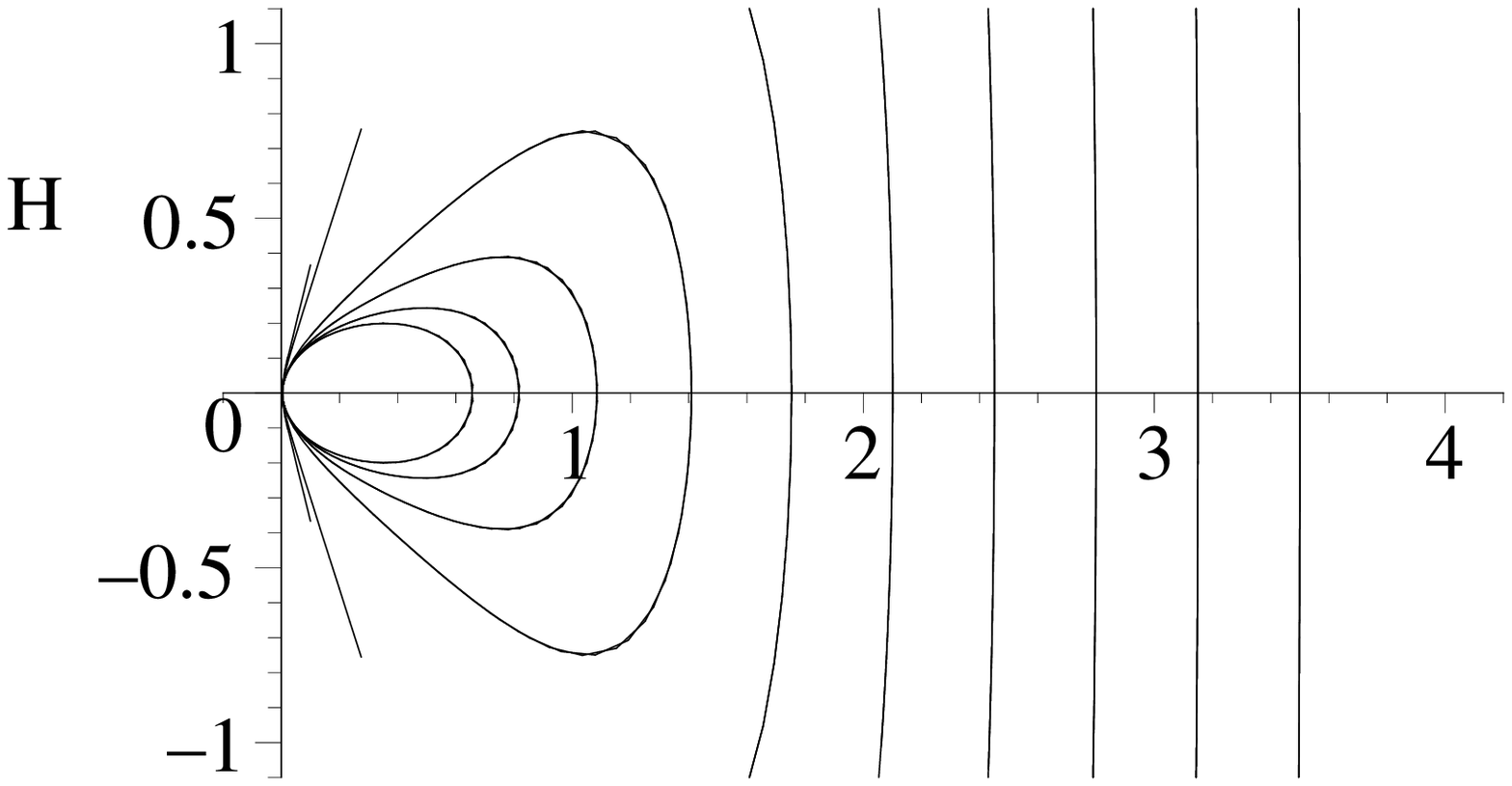} \\ g \end{tabular} &
   \begin{tabular}{c} \includegraphics[width=0.15 \textwidth]{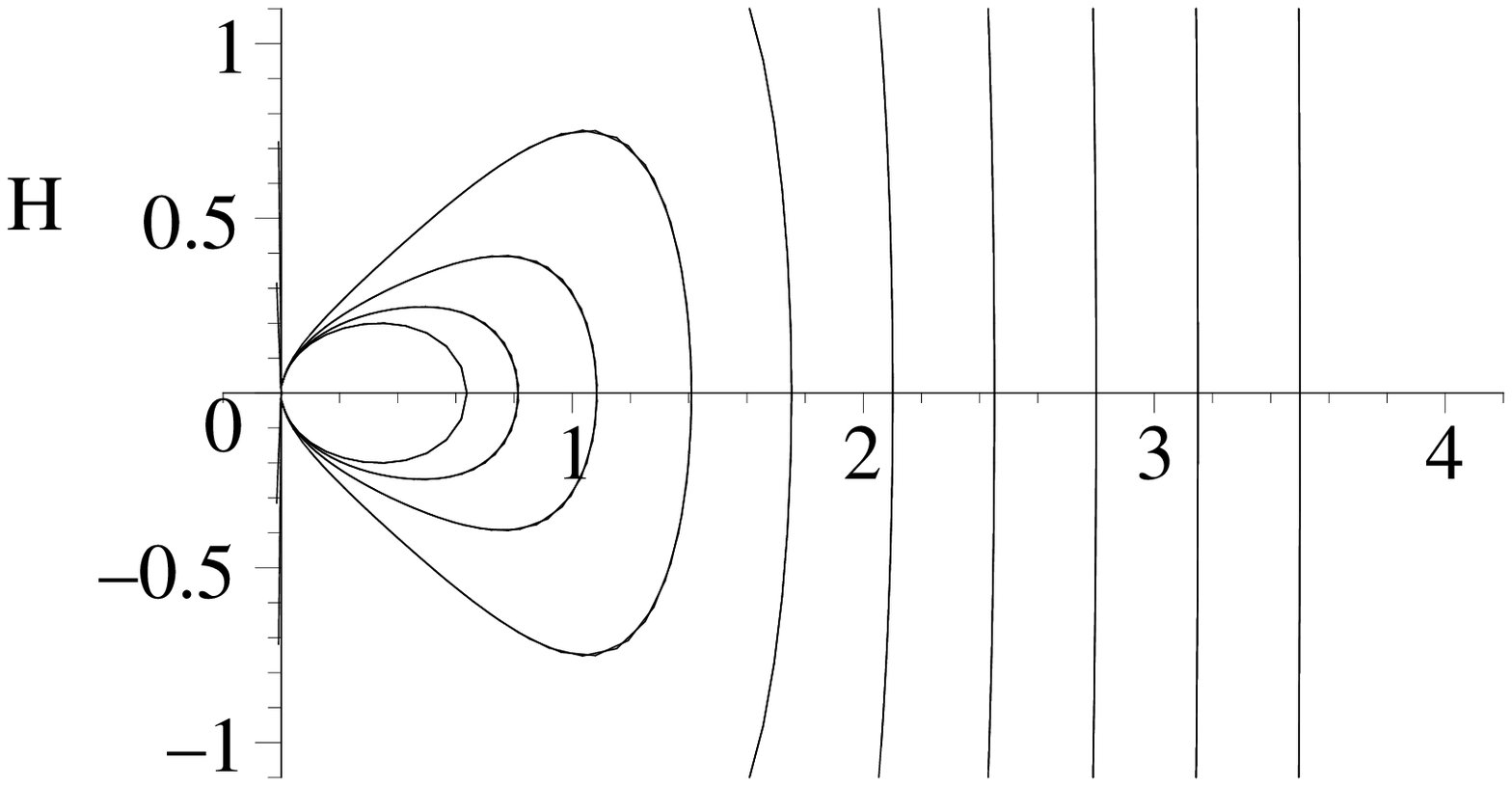} \\ h \end{tabular} &
   \begin{tabular}{c} \includegraphics[width=0.15 \textwidth]{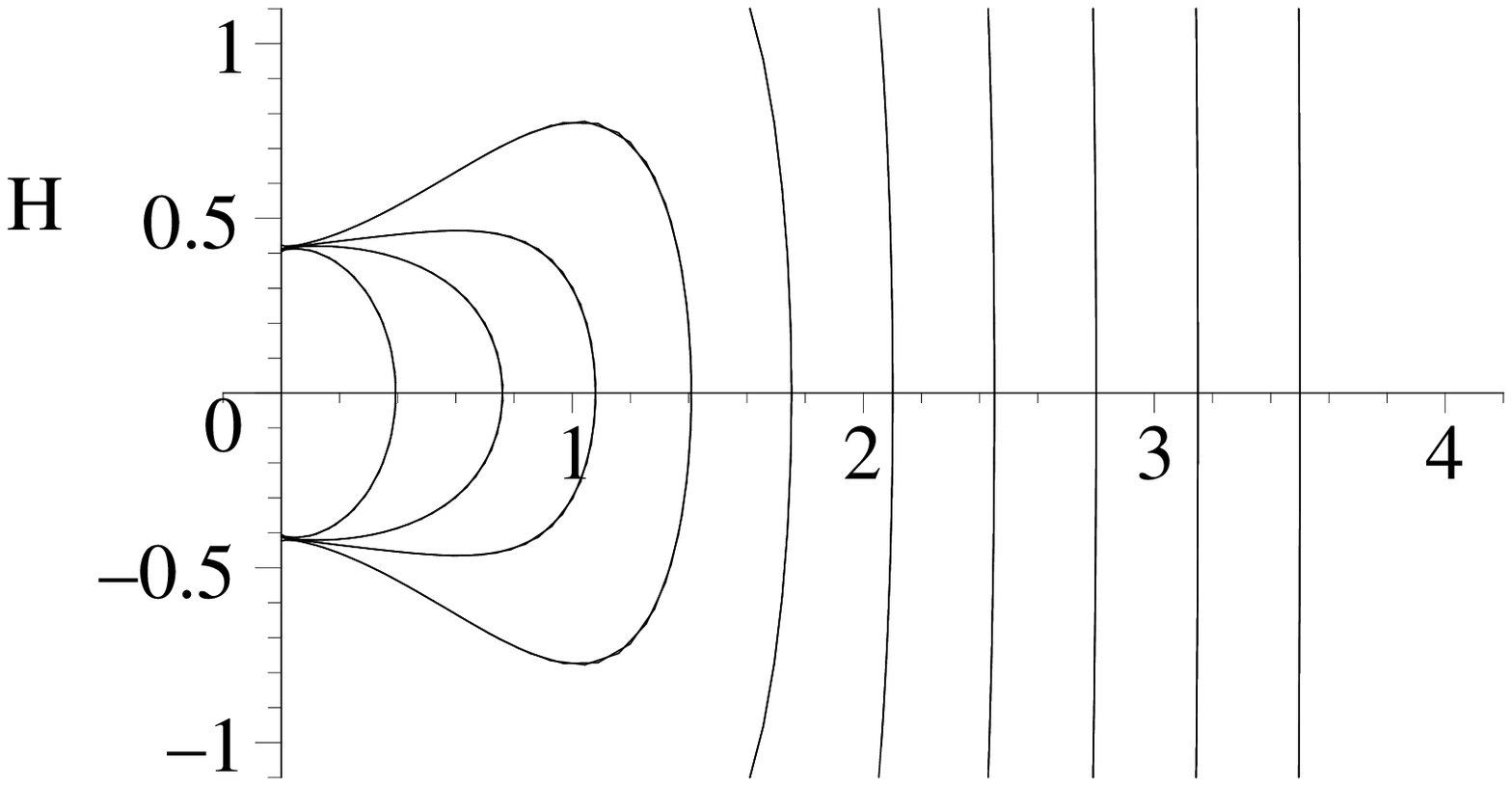} \\ i \end{tabular} \\
  \hline
\end{tabular}//
\vskip 2 mm
Table E4. Case with  $\ve=0$, $n=4$ and $\lambda=0$. \\
\vskip 2 mm

\begin{tabular}{|c|c|c|c|}
  \hline
     & $\Lambda<0$ & $\Lambda=0$ & $\Lambda>0$ \\
  \hline
   $m<0$ &
   \begin{tabular}{c} \includegraphics[width=0.15 \textwidth]{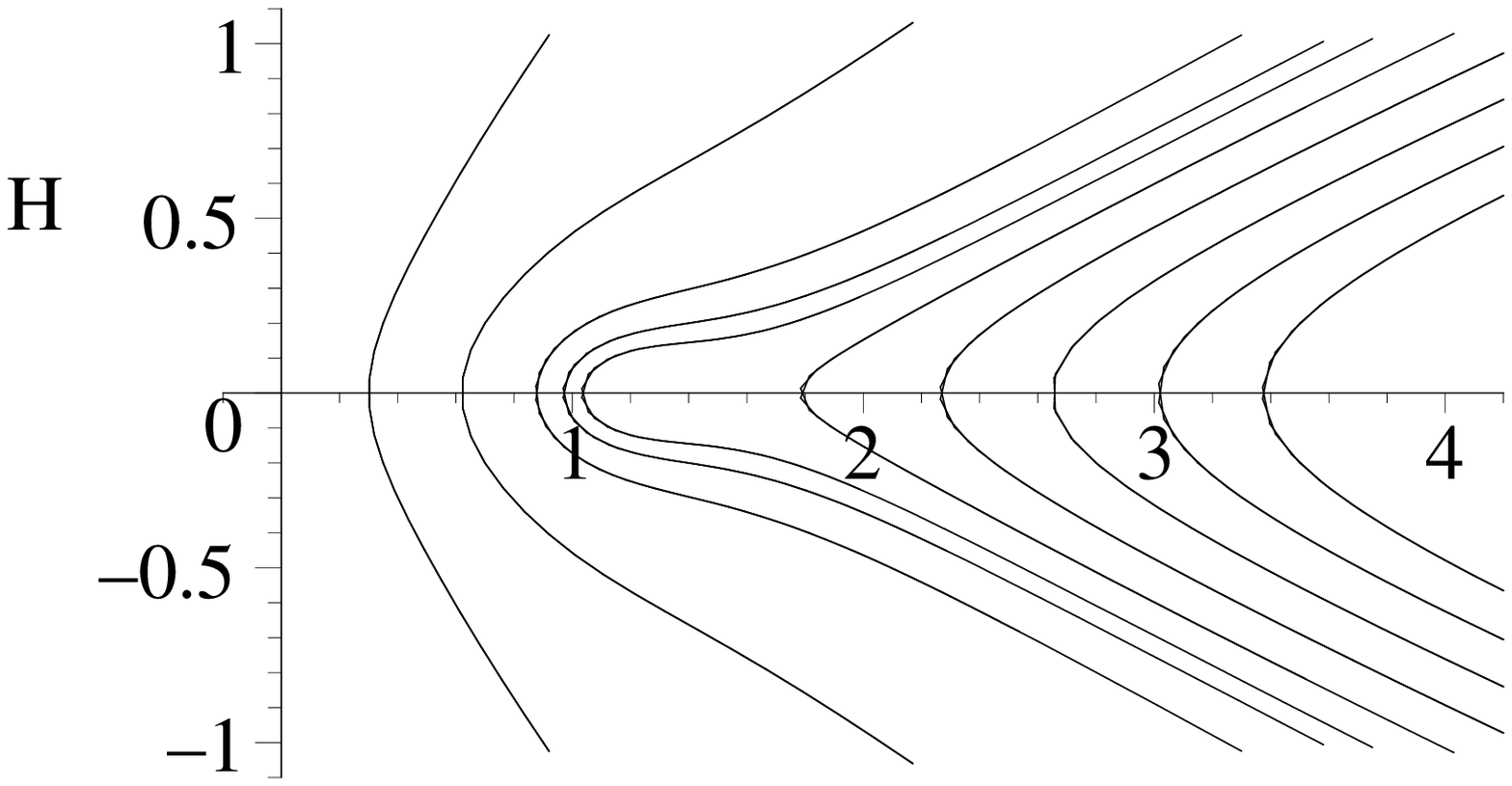} \\ a  \end{tabular} &
   \begin{tabular}{c} \includegraphics[width=0.15 \textwidth]{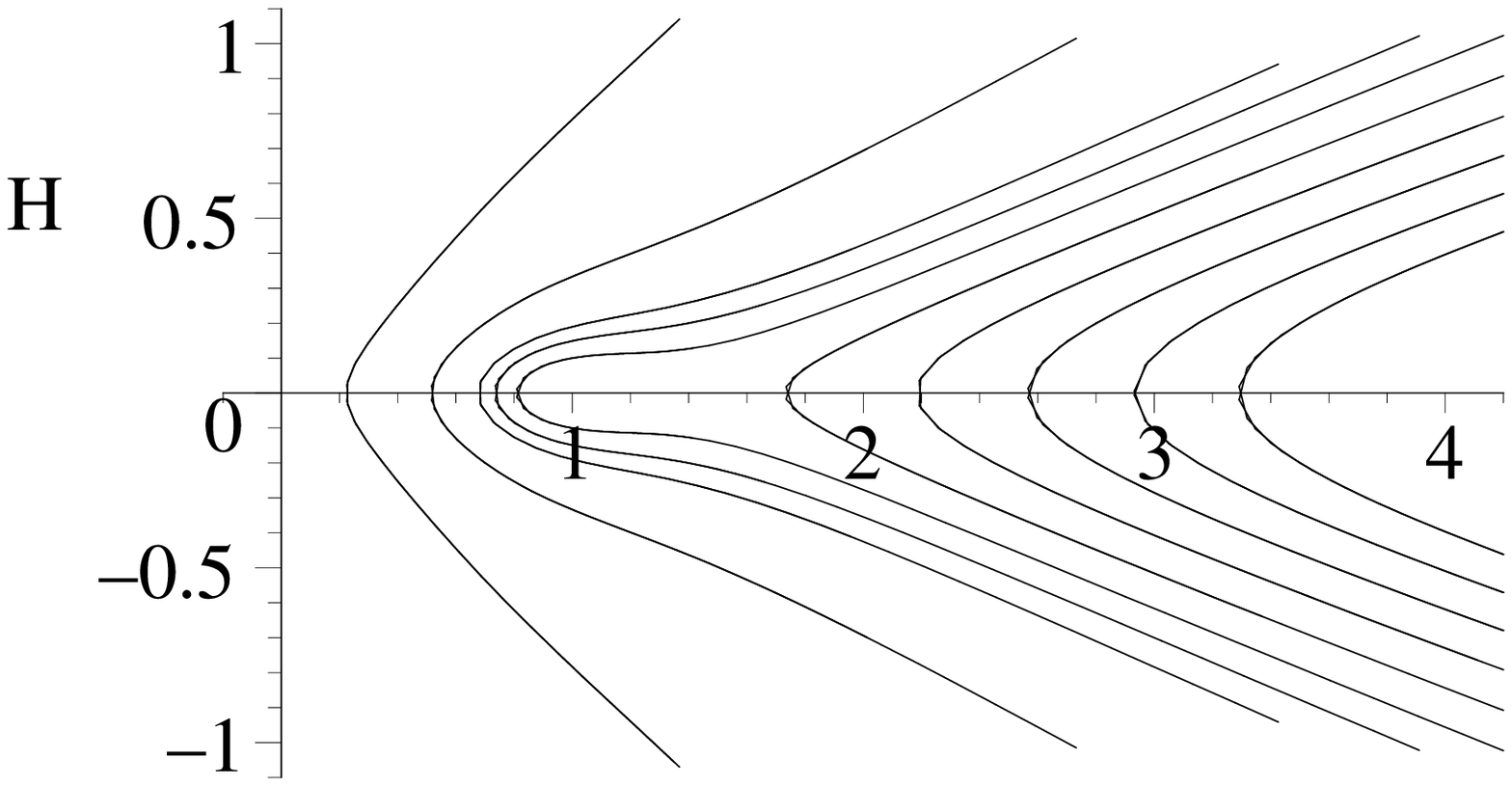} \\ b \end{tabular} &
   \begin{tabular}{c} \includegraphics[width=0.15 \textwidth]{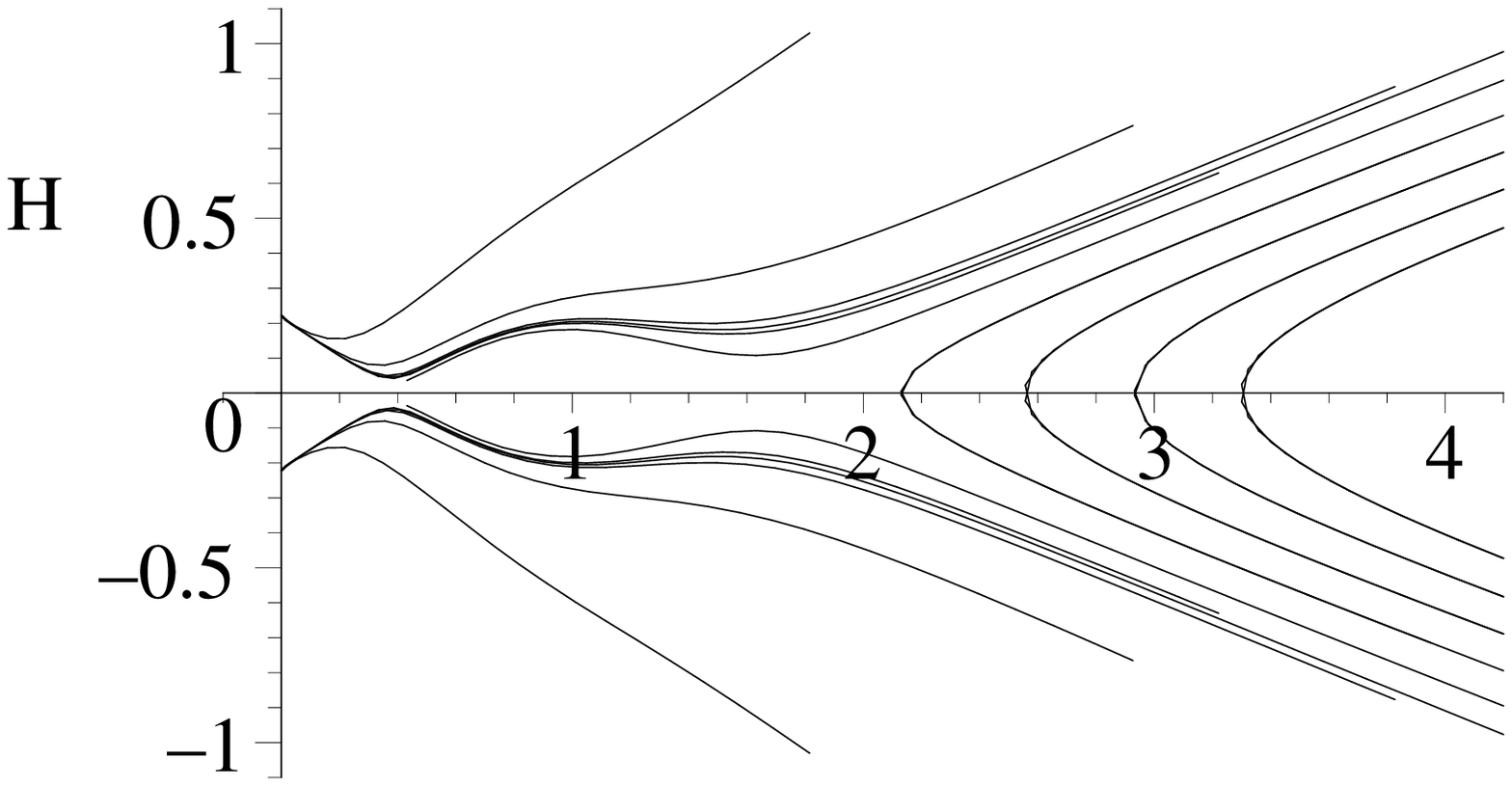} \\ c \end{tabular} \\
  \hline
   $m=0$ &
   \begin{tabular}{c} \includegraphics[width=0.15 \textwidth]{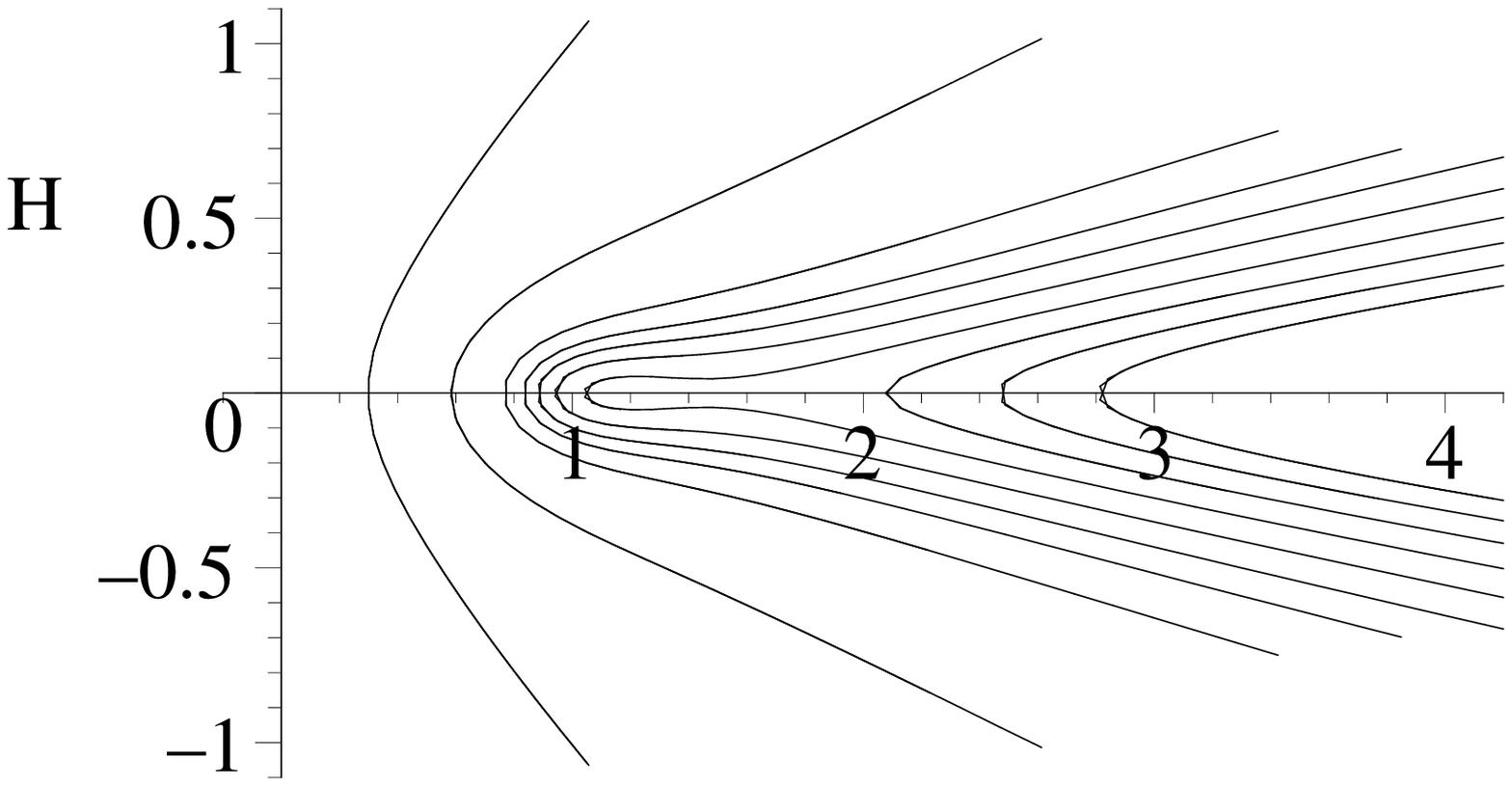} \\ d \end{tabular} &
   \begin{tabular}{c} \includegraphics[width=0.15 \textwidth]{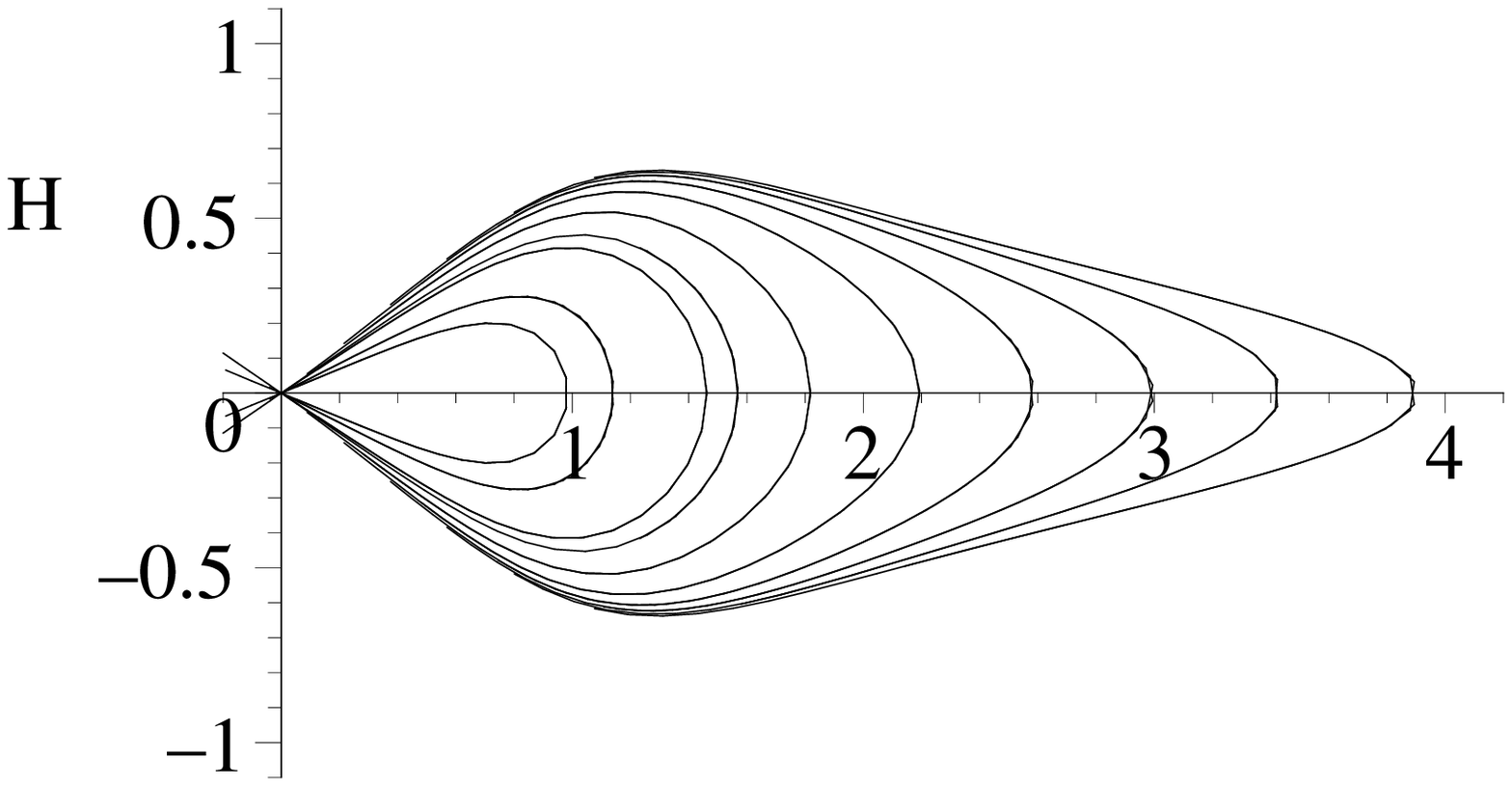} \\ e \end{tabular} &
   \begin{tabular}{c} \includegraphics[width=0.15 \textwidth]{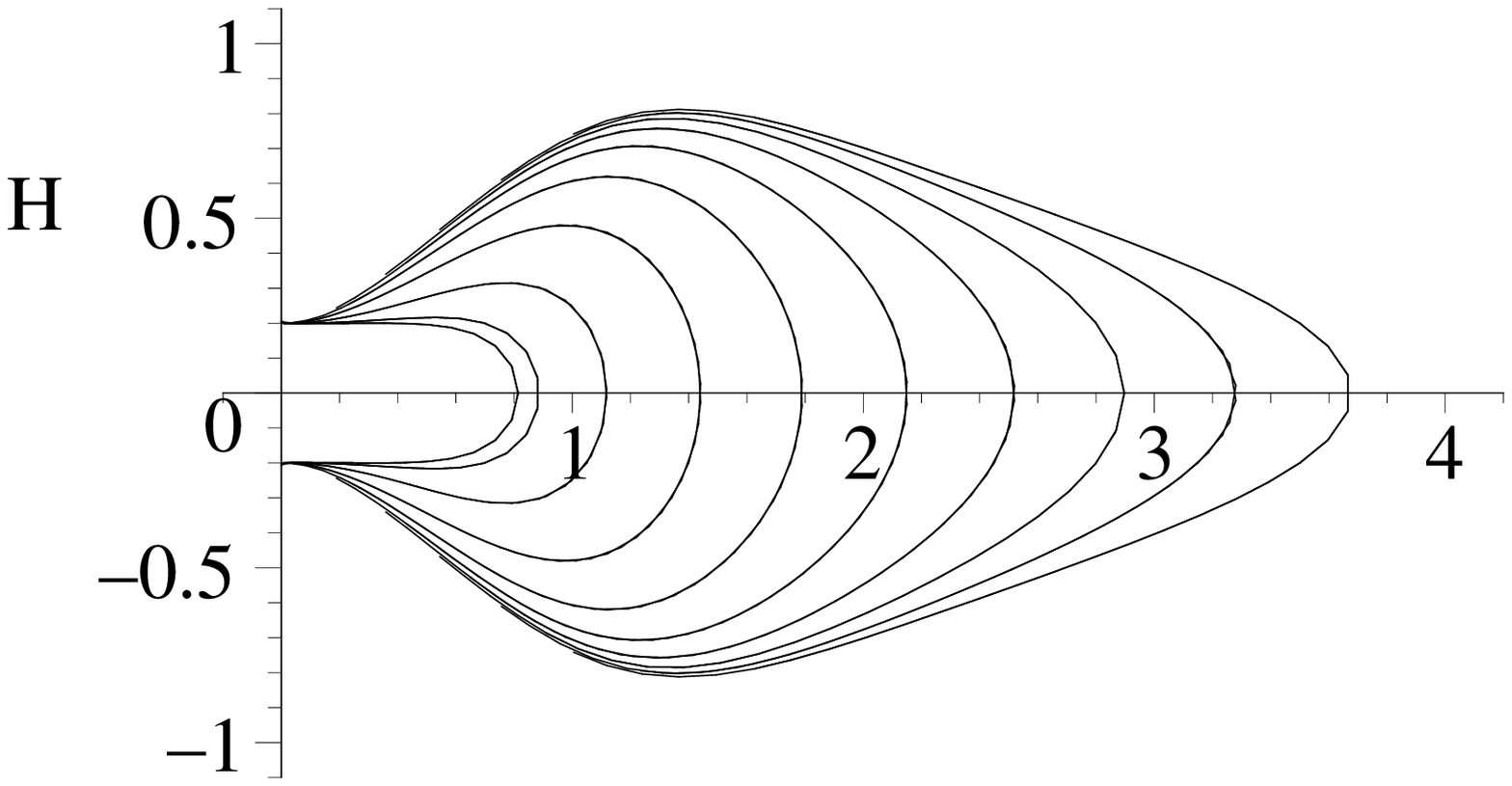} \\ f \end{tabular} \\
  \hline
   $m>0$ &
   \begin{tabular}{c} \includegraphics[width=0.15 \textwidth]{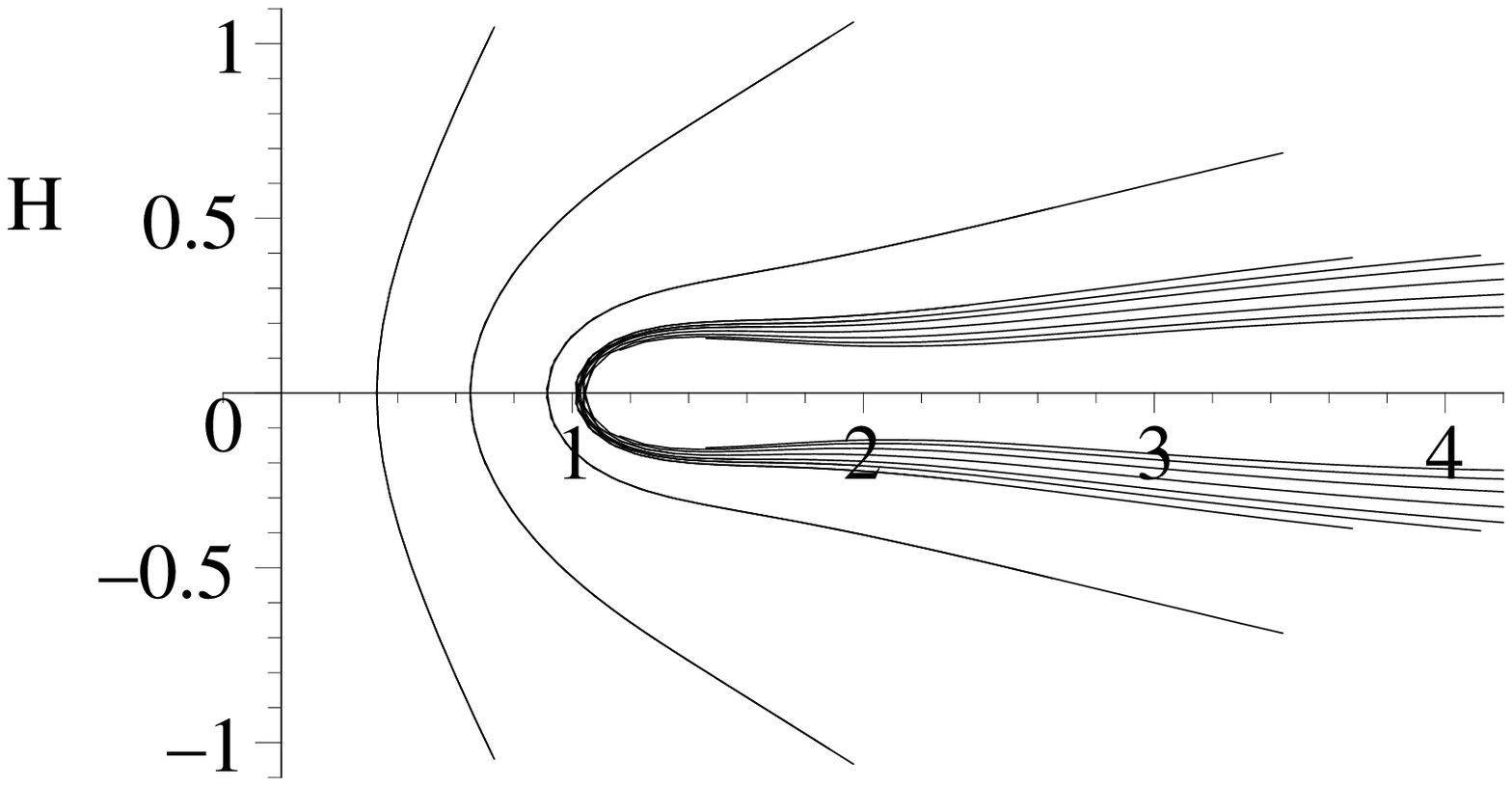} \\ g \end{tabular} &
   \begin{tabular}{c} \includegraphics[width=0.15 \textwidth]{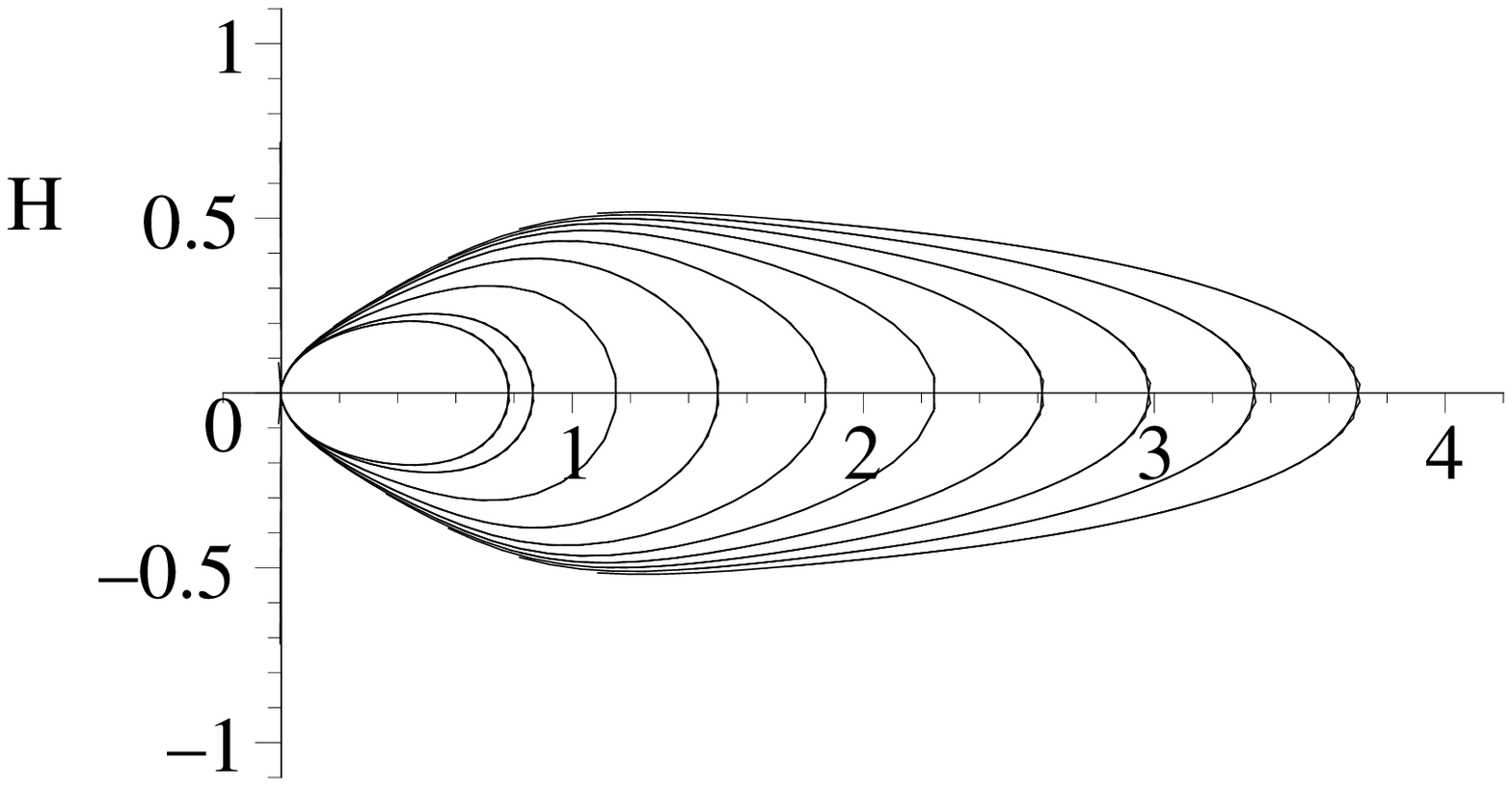} \\ h \end{tabular} &
   \begin{tabular}{c} \includegraphics[width=0.15 \textwidth]{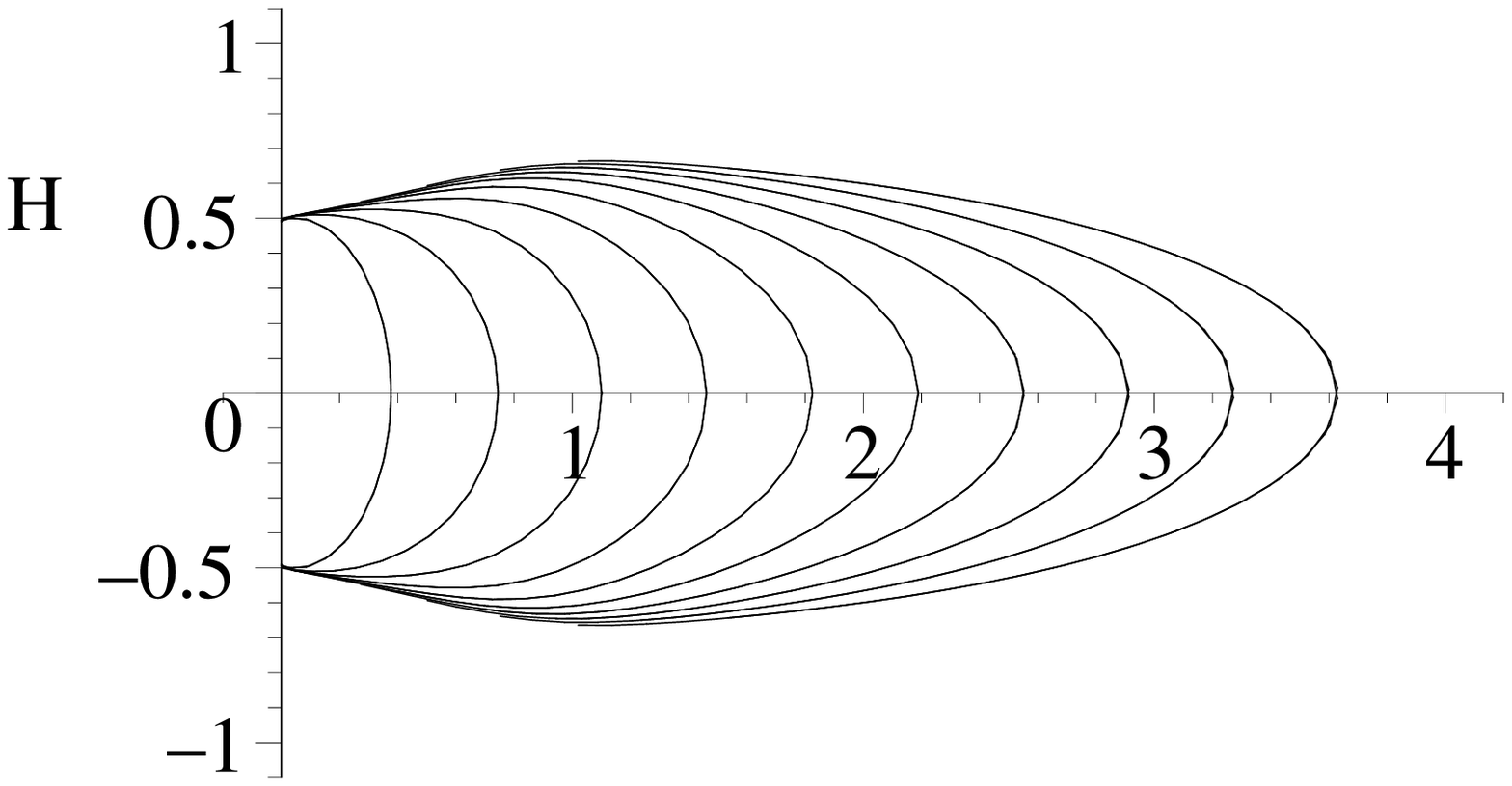} \\ i \end{tabular} \\
  \hline
\end{tabular}//
\vskip 2 mm
Table D2. Case with  $\ve=0$, $n=4$ and $\lambda>0$. \\
\vskip 2 mm
\end{center}


\subsection{Qualitative analysis of the complete system}

The system \eqref{HVespscnu} in absence of viscosity, i.e., under $\eta = 0$ and $\xi = 0$
possesses the following first integrals
\begin{subequations}
\label{F}
\begin{eqnarray}
F_1 &=& \frac{\ve}{\nu^{1+\zeta}}, \label{F1}\\
F_2 &=& \frac{(6H^2-2\ve-2\Lambda-2m\nu)}{\nu^2}-\frac{1}{\lambda(\lambda\nu^n+1)}. \label{F2}
\end{eqnarray}
\end{subequations}

The second of them \eqref{F2} remains to be the first integral even
after the introduction of bulk viscosity $\xi$. The first one, i.e.,
Eq. \eqref{F1} under $\xi \ne 0$ ceases to be the integral of motion.
Nevertheless, the introduction of bulk viscosity during the course of
time generates definite displacement of the surface given by the
formula \eqref{F1}, which allows one qualitatively, i.e., based only
on the continuity, compile the representation about the possible ways
of evolution.


\myf{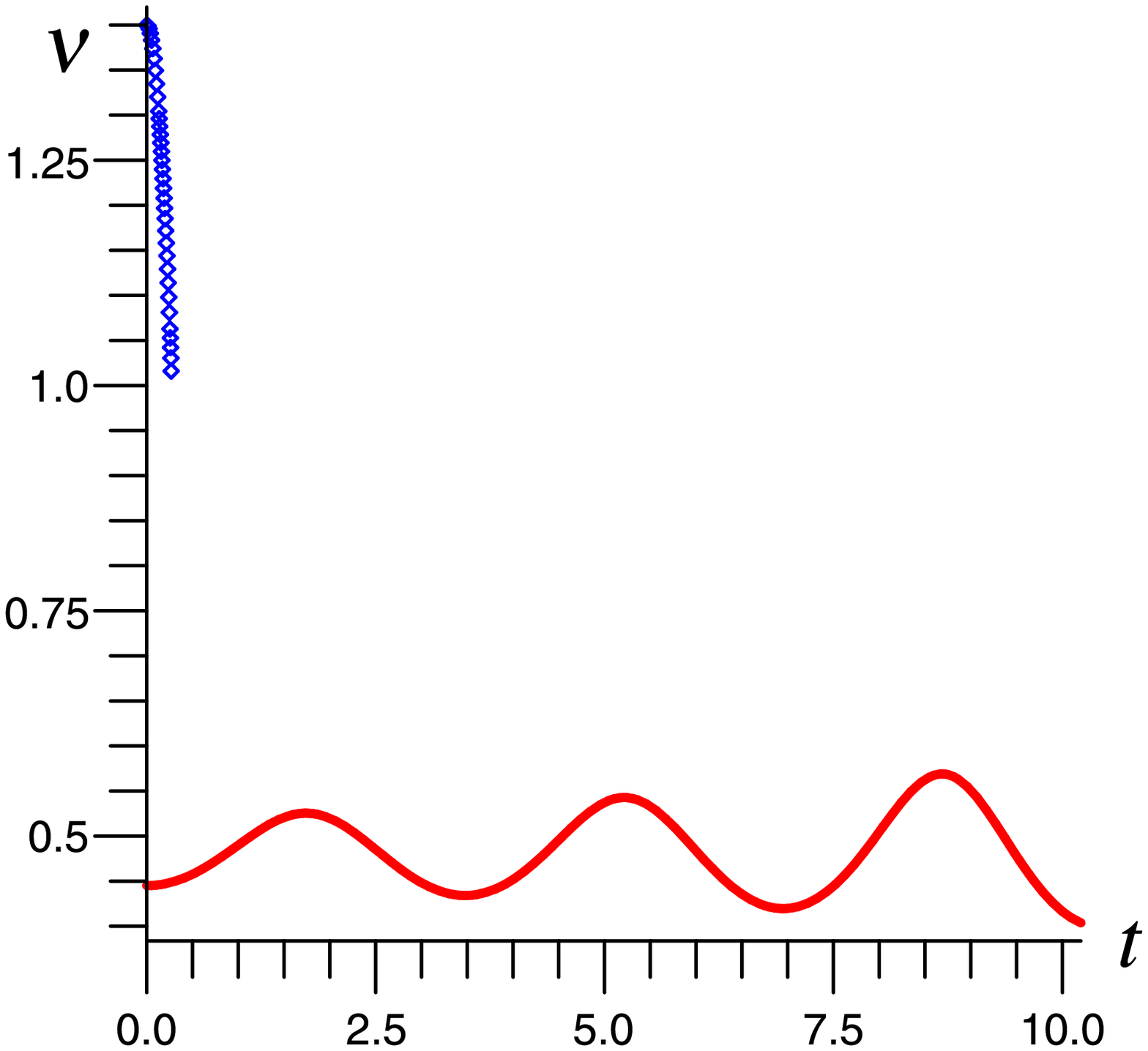}{0.25}{Evolution of function inverse to volume
scale}{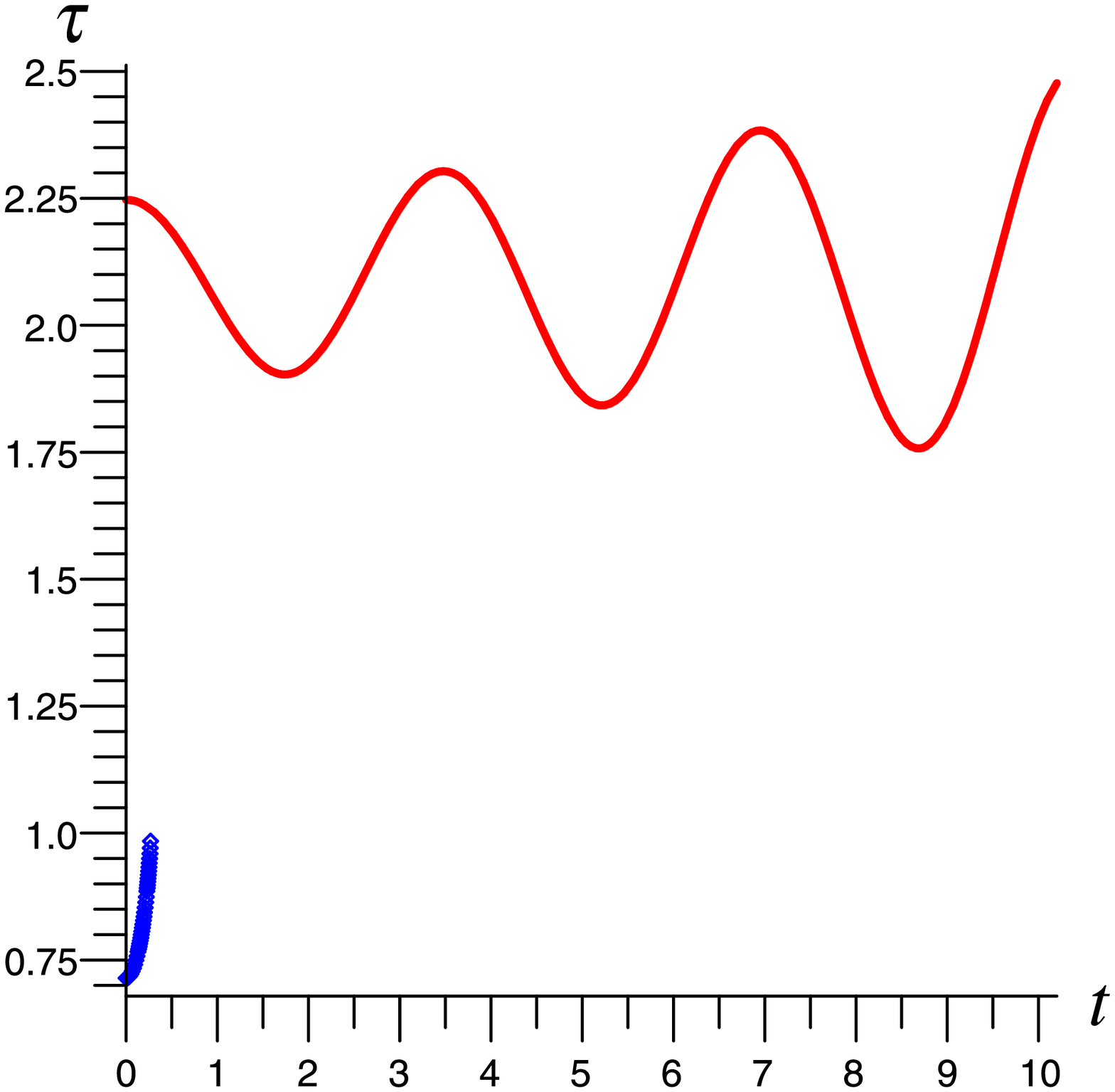}{0.20}{Evolution of volume scale}{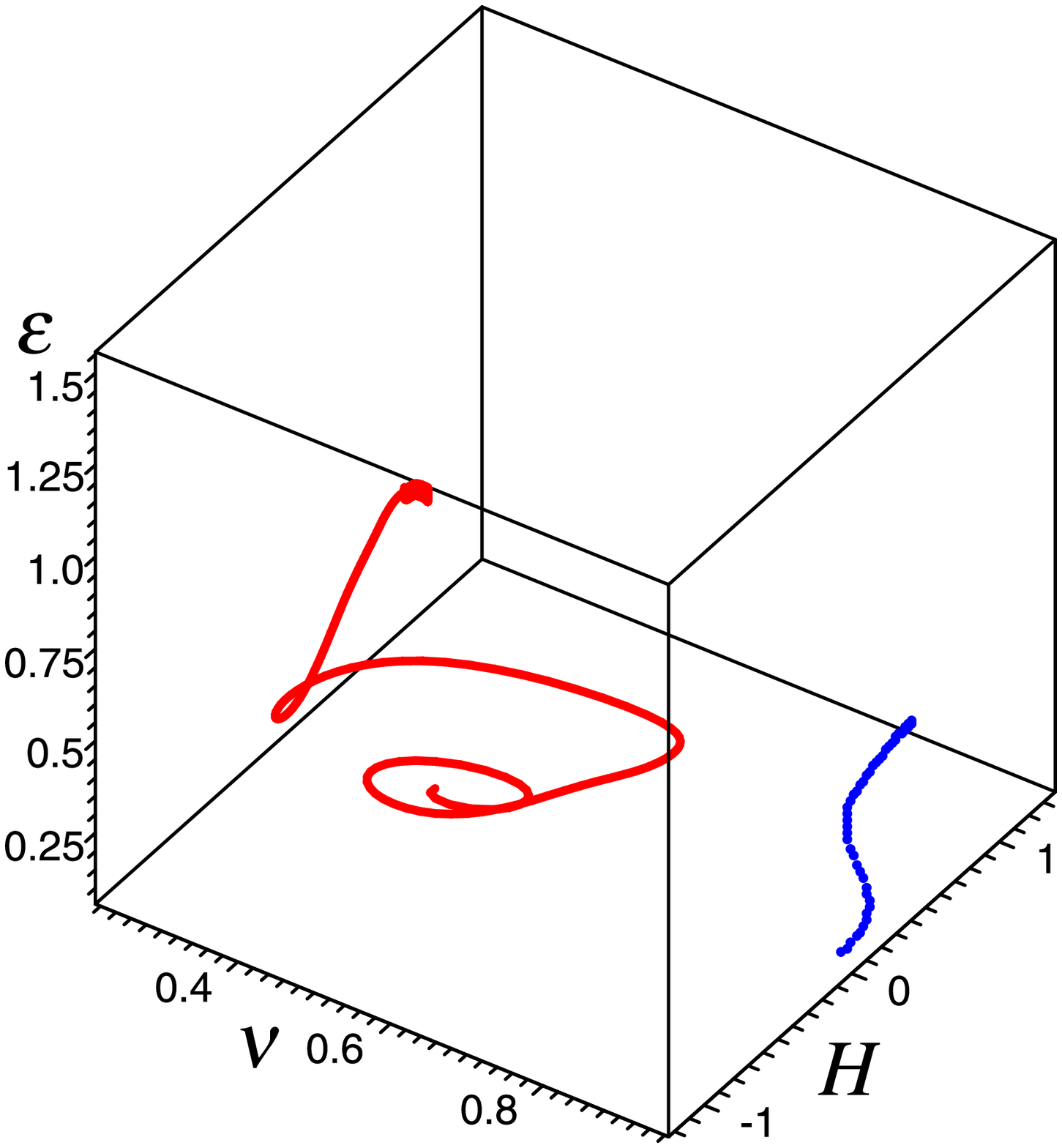}{0.20}{3D view in $\nu,H,\ve$
space}

\myf{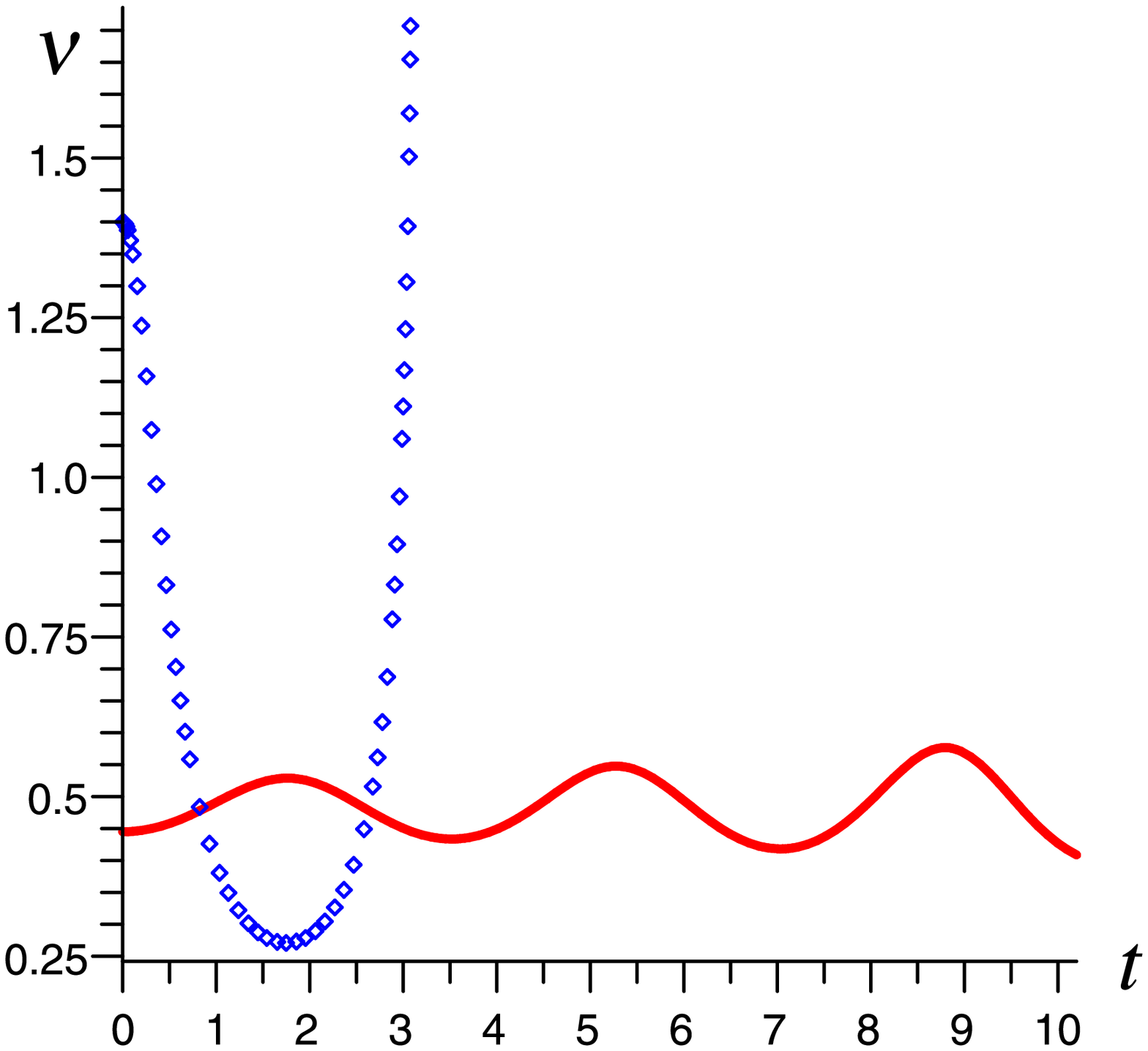}{0.25}{Evolution of function inverse to volume
scale}{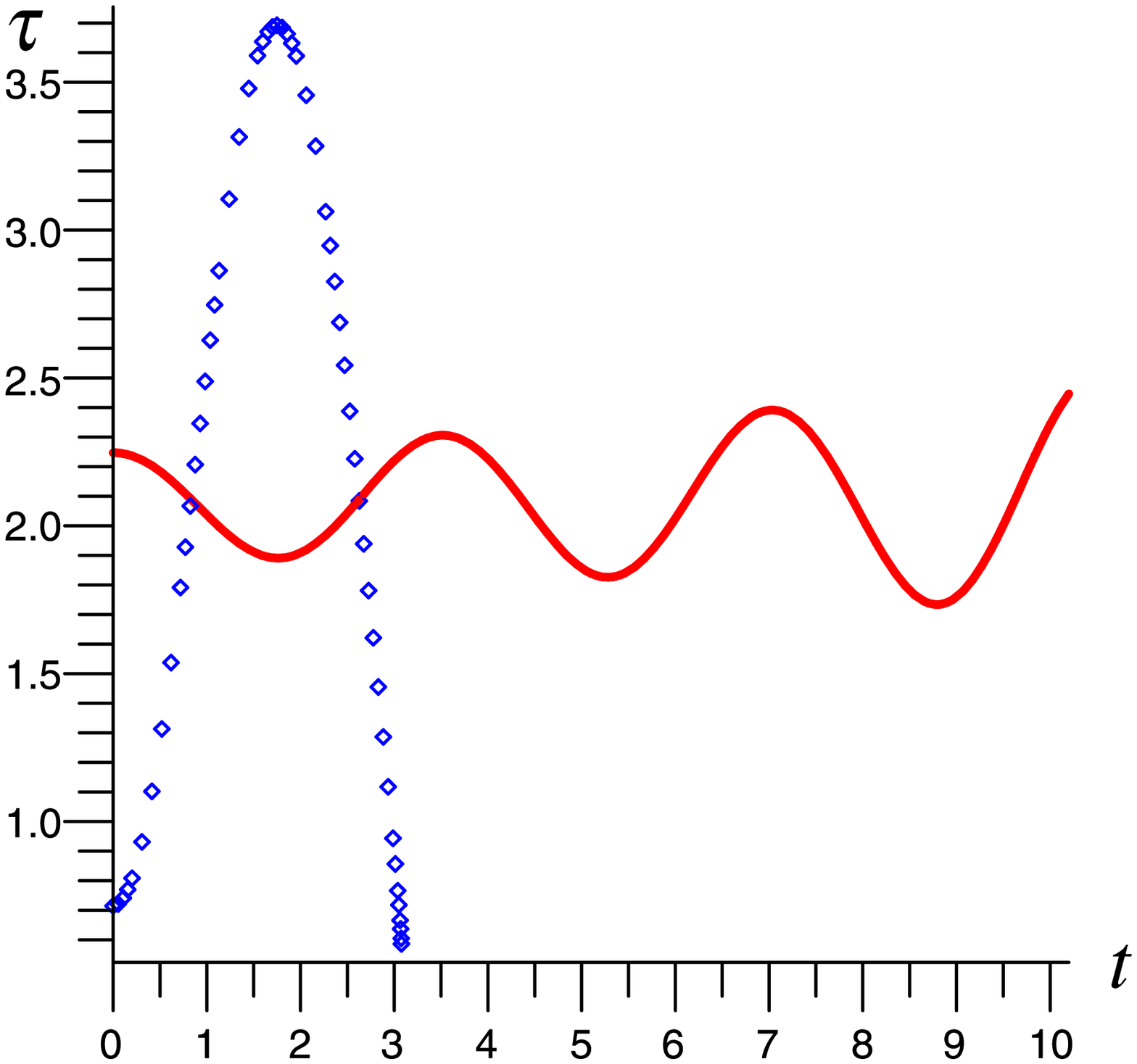}{0.20}{Evolution of volume scale}{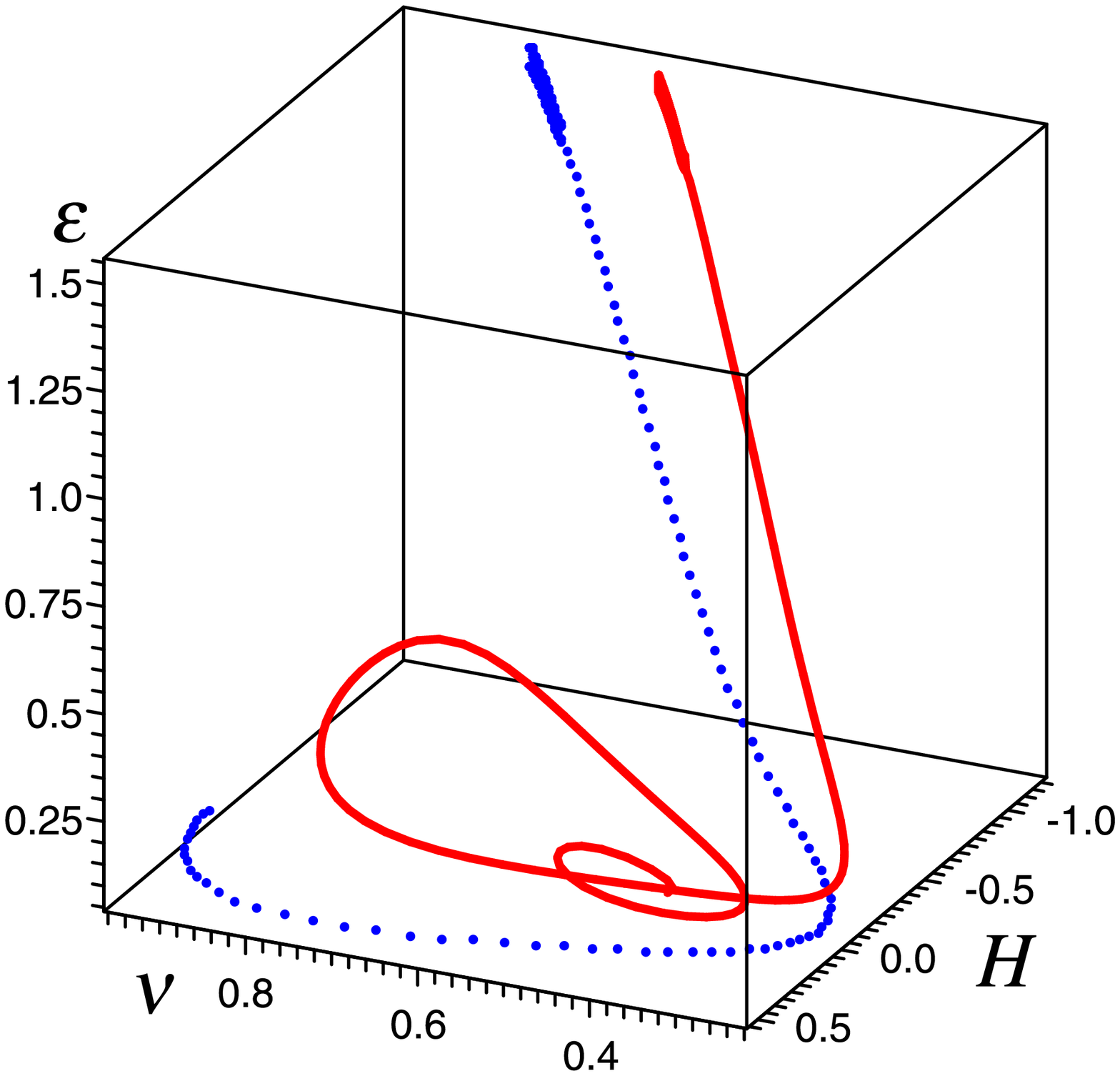}{0.20}{3D view in $\nu,H,\ve$
space}

\myf{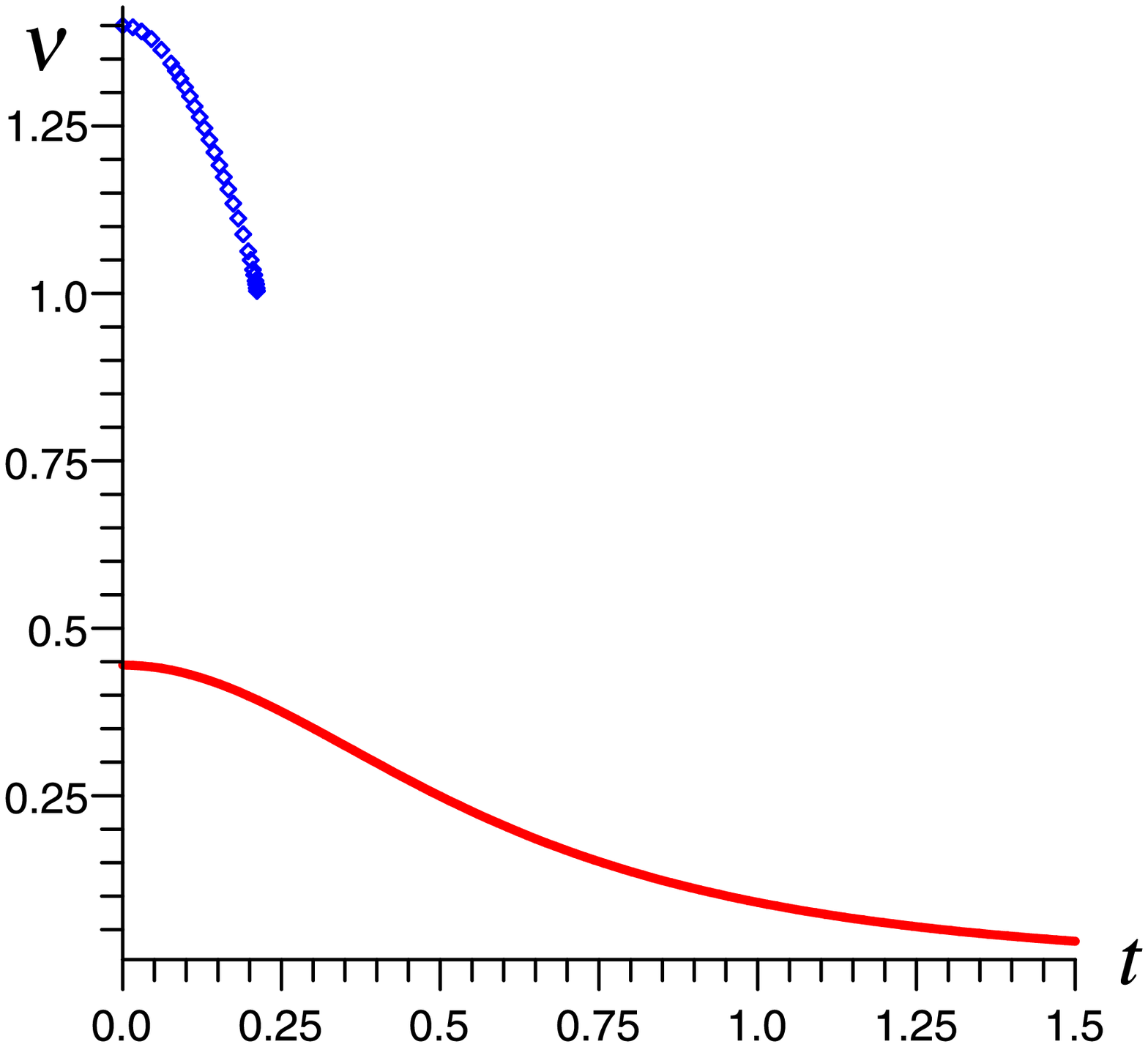}{0.25}{Evolution of function inverse to volume
scale}{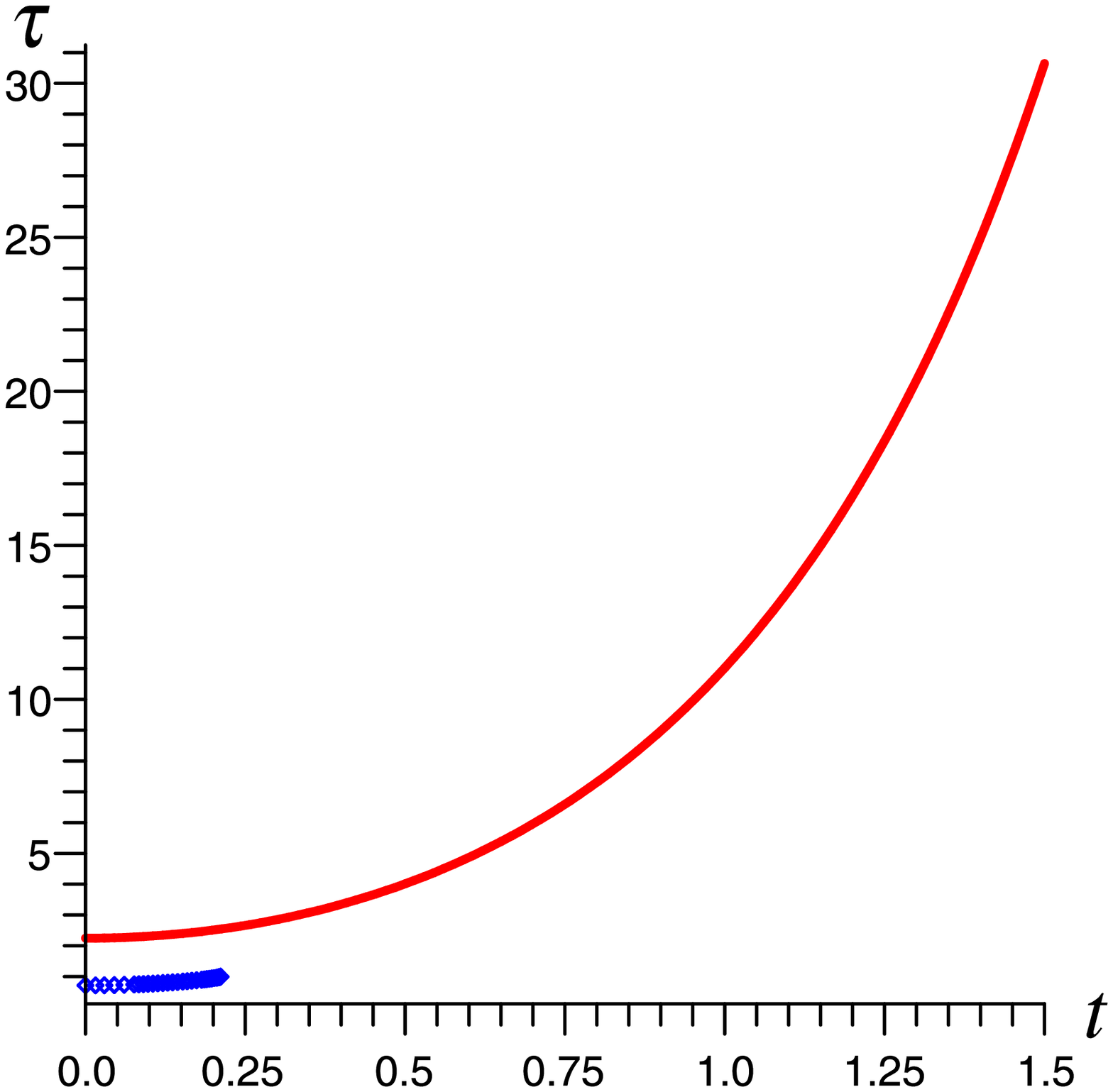}{0.20}{Evolution of volume scale}{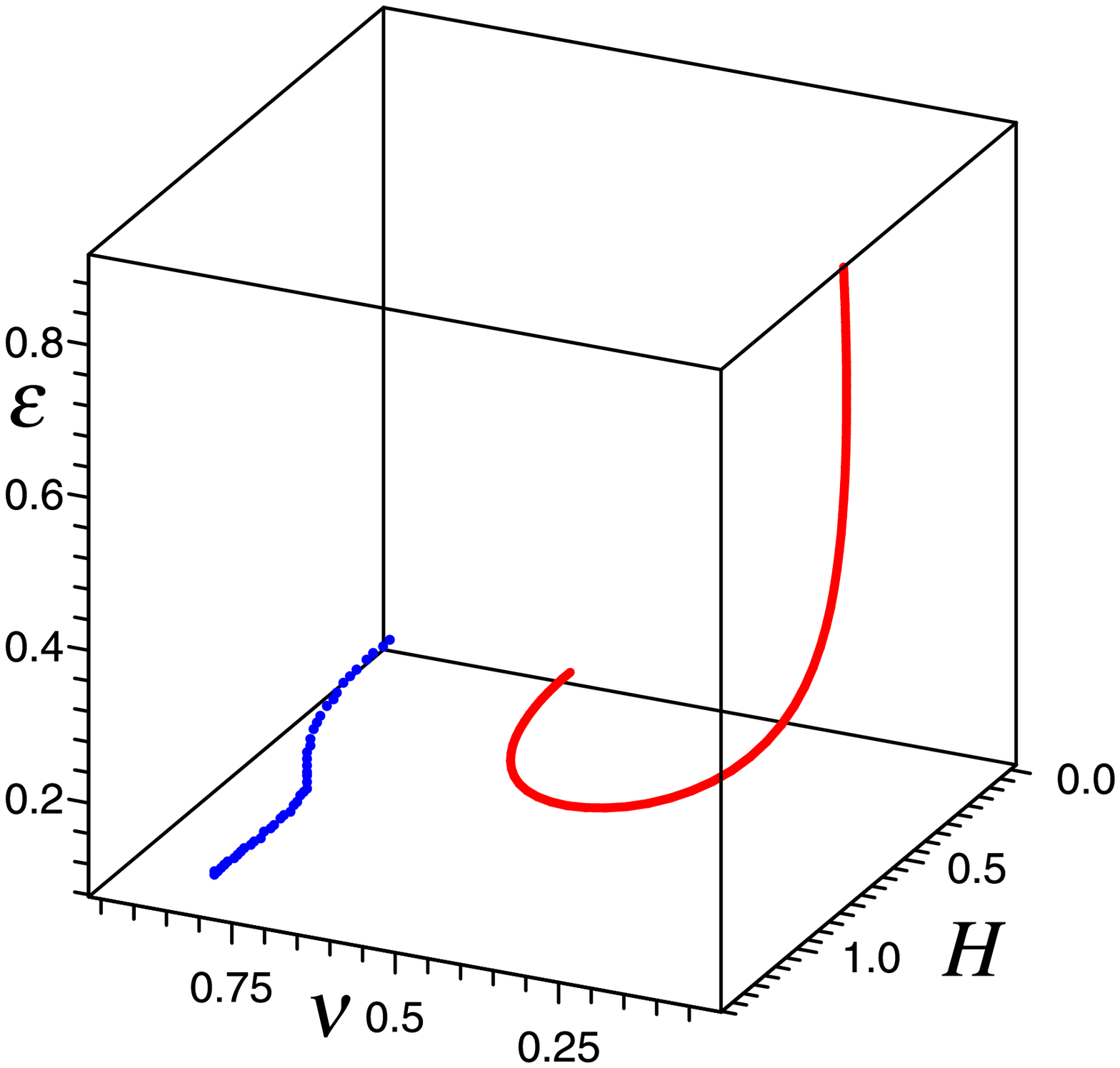}{0.20}{3D view in $\nu,H,\ve$
space}

\myf{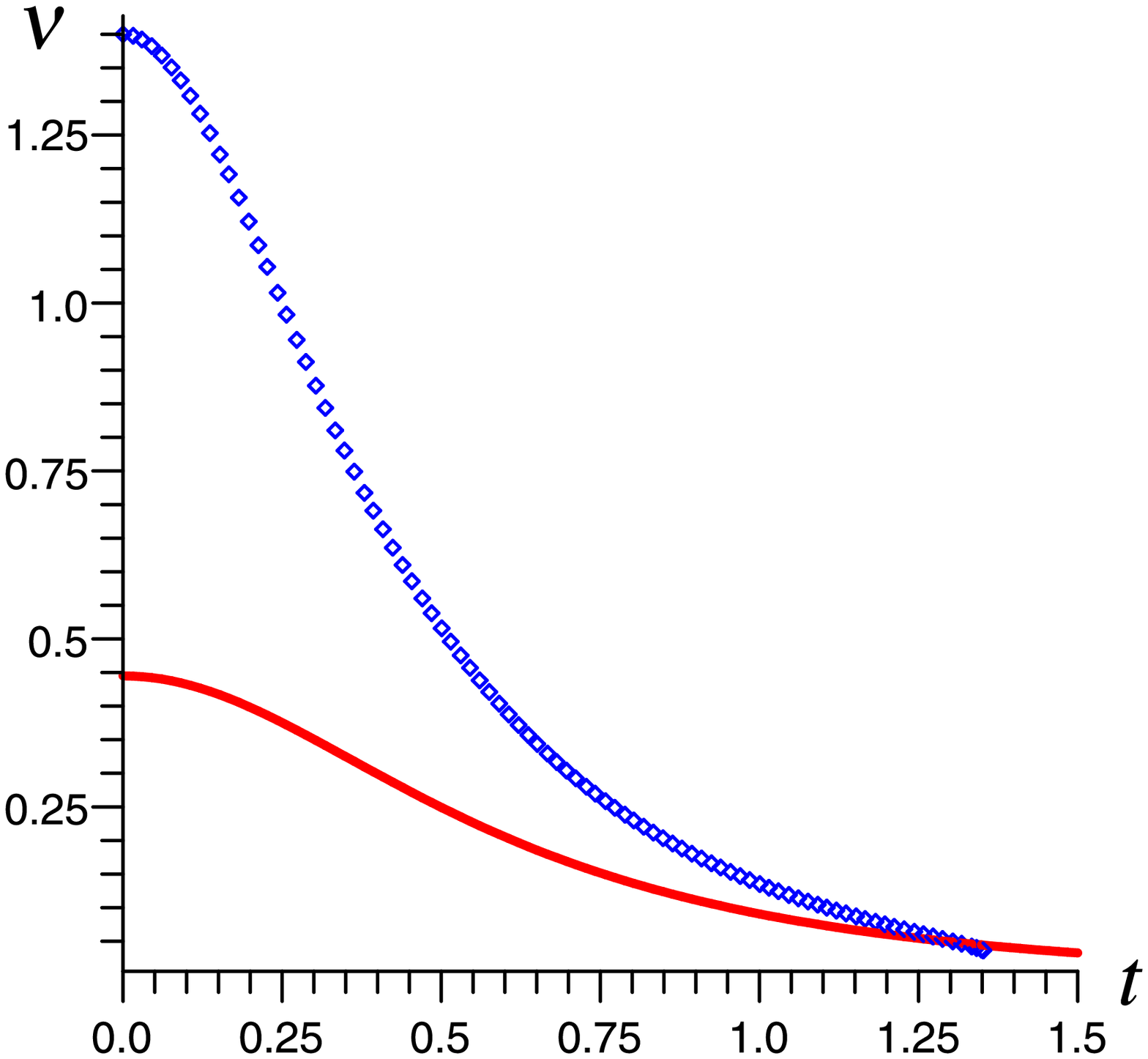}{0.25}{Evolution of function inverse to volume scale}{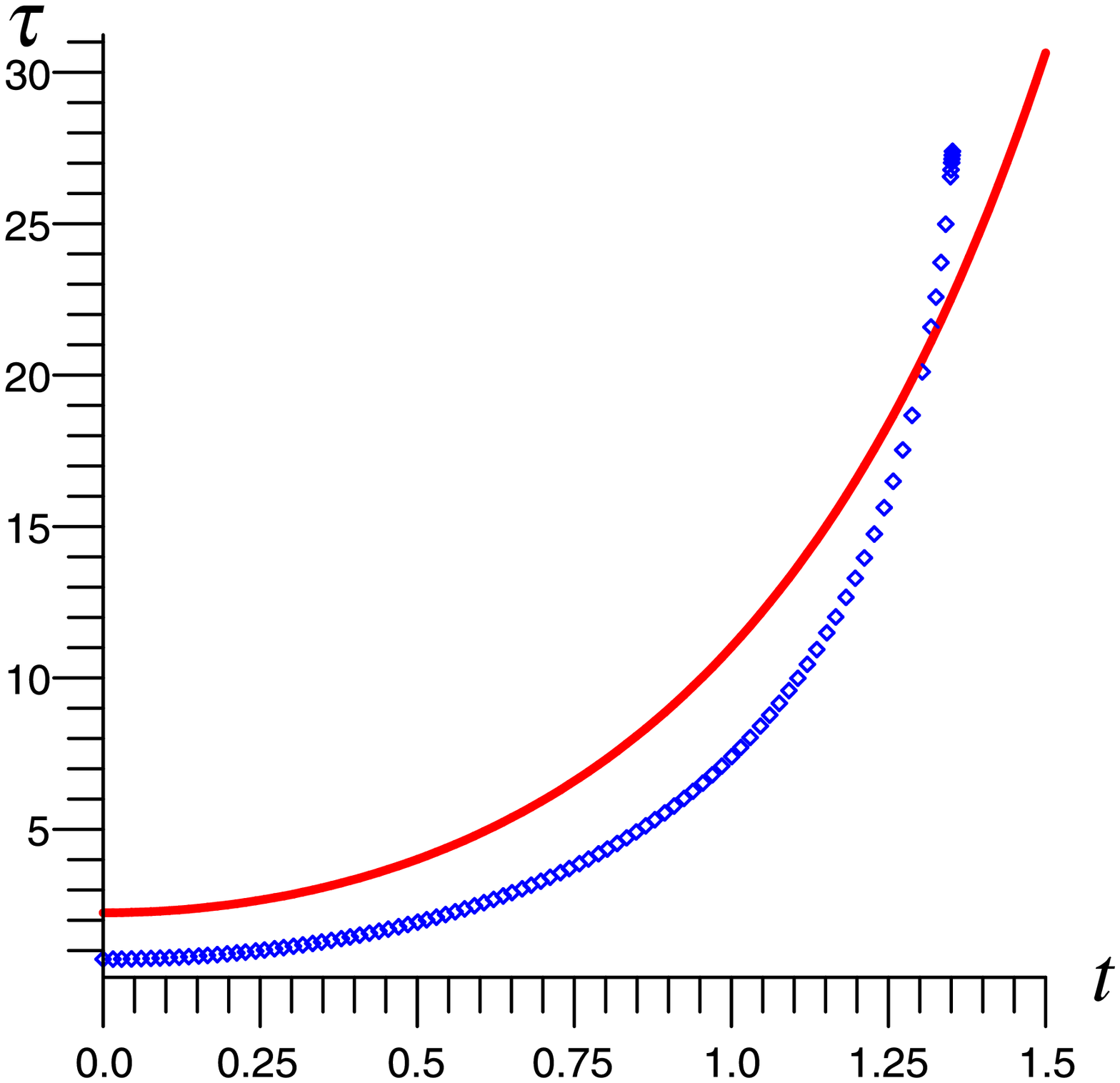}{0.20}{Evolution
of volume scale}{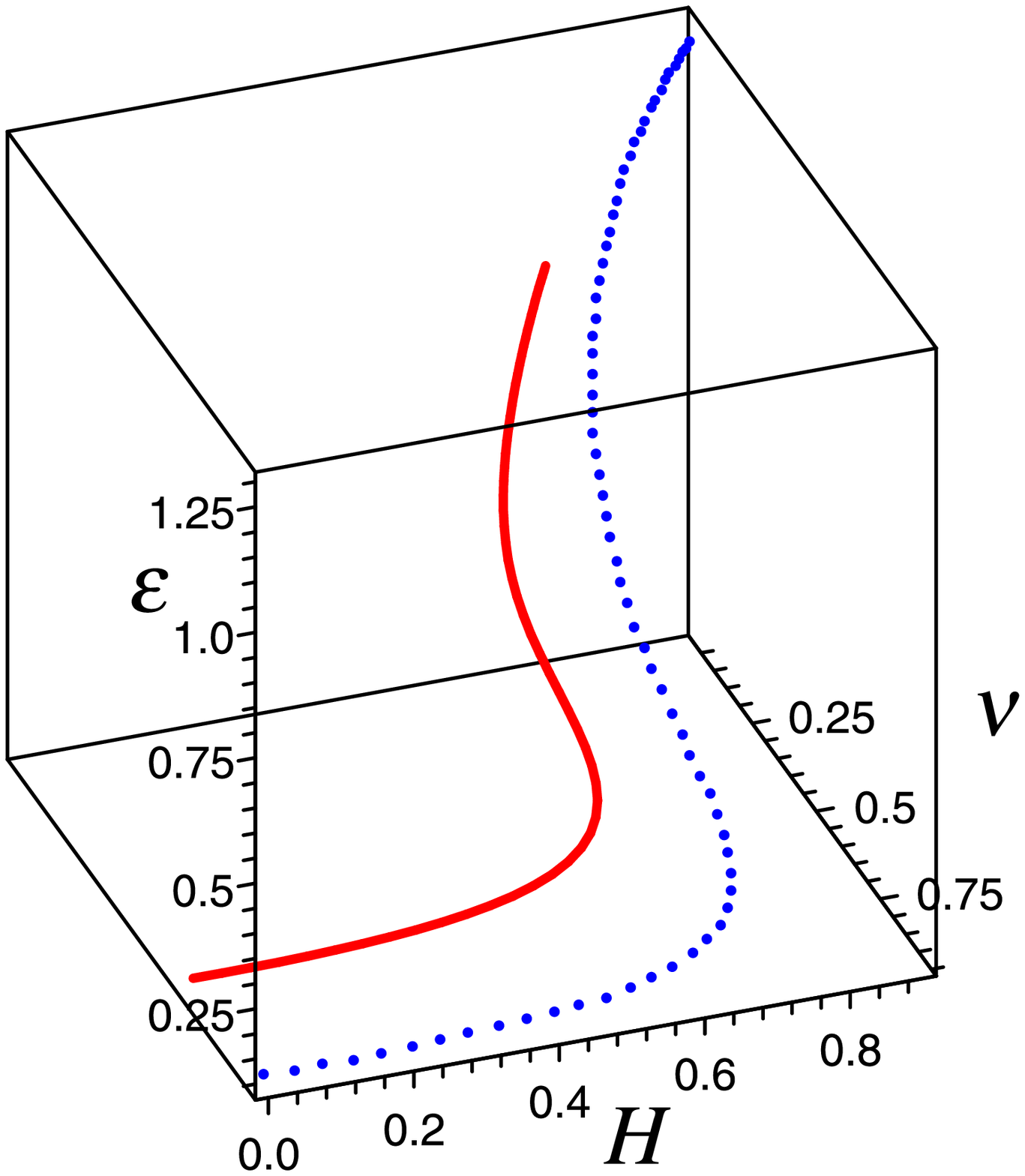}{0.20}{3D view in $\nu,H,\ve$ space}

Harnessing the Tables 1, 2 and 3, helps one to understand the 3D
phase portrait leaning on the continuous dependence of the velocity
fields of the coordinates $\nu, H, \ve$ of phase space.

In order to cover the infinite phase space completely, it is
mapped on coordinate parallelepiped with its axes being the the
arc-tangent of the corresponding coordinates. The lower horizontal
plane always represents the $\ve = 0$ plane.

It should be noted that the introduction of spinor field notably
complicates the evolution of the system. Contrary to the system in
absence of the spinor field, the initial condition with $H < 0$
already does not prevent in many cases thanks to the evolution of
volume scale entering the half-space $H > 0$ and thereupon, from
the greater value of $H$ repeats the evolution, approaching to the
$\nu = 0$ plane and displaying the classification from the table
1. In the vicinity of the borders $\ve = 0$ and $\nu = 0$ the
integral curves closely repeats the integral curves on the sides,
each time at least to some extent.

The general property of all the cases is the fact that in the
half-space $H > 0$ the velocity vectors are directed to the $\ve =
0$ plane, while in the other half opposite to it. As a result all
he invariant curves fall on $\ve = 0$, though not necessarily
reach it.

In the Figs. \ref{nu_L=1_l=1} - \ref{3d_L=-1_l=-1} we have
illustrated functions inverse to the volume scale $\nu(t)$ [Figs.
\ref{nu_L=1_l=1},\ref{nu_L=1_l=-1},\ref{nu_L=-1_l=1},\ref{nu_L=-1_l=-1}],
volume scale $\tau(t)$ [Figs.
\ref{tau_L=1_l=1},\ref{tau_L=1_l=-1},\ref{tau_L=-1_l=1},\ref{tau_L=-1_l=-1}]
and phase portrait in $\nu$, $H$, $\ve$ space
[Figs.\ref{3d_L=1_l=1},\ref{3d_L=1_l=-1},\ref{3d_L=-1_l=1},\ref{3d_L=-1_l=-1}],
for $\alpha = 4$, $\beta = 1$, $\zeta = 1/2$, $A = 1$, $B = 1$, $n =
4$, $m = 4$.

Continuous and dot lines  in the Figs. \ref{nu_L=-1_l=-1} -
\ref{3d_L=1_l=1} corresponds to two different initial conditions. For
$\Lambda  < 0$ depending on the sign of $\lambda$ there occur the
following situations: (i) for $\lambda <0 $ there exist separable
plane which does not allow the solutions with initial condition in
one part enter into the second one [cf. ╨шё.
\ref{nu_L=-1_l=-1},\ref{tau_L=-1_l=-1},\ref{3d_L=-1_l=-1}, that
corresponds to Table D1-g]. (ii) For $\lambda > 0$ there is no
separable plane in this case [cf. ╨шё.
\ref{nu_L=-1_l=1},\ref{tau_L=-1_l=1},\ref{3d_L=-1_l=1}, that
corresponds to Table D2-g]. As one sees, a negative $\Lambda$, which
is in fact the additional gravitational field, generates oscillatory
regime of expansion.

In case of $\Lambda > 0$ there are only exponential regimes of
expansion. For $\lambda <0 $ there is separable plane  [cf. ╨шё.
\ref{nu_L=1_l=-1},\ref{tau_L=1_l=-1},\ref{3d_L=1_l=-1}, that
corresponds to Table D1-i], while for $\lambda > 0$ there is no
separable plane [cf. ╨шё.
\ref{nu_L=1_l=1},\ref{tau_L=1_l=1},\ref{3d_L=1_l=1}, that corresponds
to Table D2-i].

\section{Evolution with blow up}

Studying the system of ODE let us imagine the integral curves in the
space. It is very important to know the directional field given by
this system. It is more important than the corresponding vector
field. First of all, the integral curves, by definition, are tangent
to the vector field, hence to the directional field, at those
(peculiar) points, where vector field becomes trivial with the
direction being indefinite. Secondly, like the vector field the
directional field is also continuous (excluding the peculiar points),
but it may be continuously continued at the boundary where the vector
field might be infinity.

We are interested in two aspects: how rapidly the solution can tend
to infinity at the distant boundary (simply infinity) and how does it
behave at the infinity. The way the problem is posed becomes
reasonable when the space is closed by means of infinitely remote
points in any given interpretation.

We will follow the Pensele's (яЁшэЎшя эхяЁхЁ√тэюёЄш ╧юэёхых)
principle of continuity - the properties of a system at continuous
change from one common position to another without losing generality.
We are interested in qualitative properties in solving the system of
ODE. Deforming the vector field continuously, at the same time
leaving the peculiar points unaltered, we don't change the
qualitative behavior of the integral curves with an accuracy of
topological equivalent. In this way we can simplify the analysis,
substituting the initial system by a simpler one, constructed from
convenient elements.

\subsection{Blow up}

The history of studying the regime with blow up is associated with
S.P. Kurdiumov \cite{kurdimov}. The study of the process of heat distribution in active and nonlinear medium led to an extremely distinguishing
feature, namely wave and localization. Mathematical models of
demography detects critical moments: solution to the (time dependent)
ODE may reach its limit within a finite time. The processes in the
chromosphere of the sun possess a flashing (eruptive) character, but
the mechanism of energy transference does not detect the presence of
predefined scale of time.

To illustrate the detection of a characteristic time in the system
with no explicit time-dependence, let us consider the following
example.
\begin{equation}
\dot{x}=-x^\alpha, x\in R^+
\end{equation}
It has two solutions: a) $x(t)=0$ and b)
$x(t)=[x(0)^{1-\alpha}-t(1-\alpha)]^\frac{1}{1-\alpha}$.

In case of b) the limiting value $x(t_*)=0$ is reached at a finite
time $t_*=\frac{x(0)^{1-\alpha}}{1-\alpha}$, if $\alpha<1$. Then both
solutions mix up. At moment $t_*$ the uniqueness condition
(precisely, Lipshits condition) breaks down.

The power law dependencies are typical for different types of
catastrophes: from earth quakes and flood to stock exchange collapse
and accidents in atomic power energy.

\subsection{Infinity}

The joining of infinitely remote point to the space of ODE
\begin{equation}
\dot{x}=F(x), x\in R^+
\end{equation}
we execute in the following way: let us make the change of variables
$x=\frac{s}{c}$, $s^2+c^2=1$. We call the  point
$\pm\infty=\frac{1}{0}$ infinitely remote one.

As a result we obtain a system of equations
\begin{equation}
\begin{array}{ccl}
\dot{s}c-\dot{c}s&=&c^2 F(\frac{s}{c}), \\
\dot{s}s+\dot{c}c&=&0,
\end{array}
\end{equation}
which on account of $s^2+c^2=1$ leads to
\begin{equation}
\begin{array}{ccl}
\dot{s}&=&c^3 F(\frac{s}{c}), \\
\dot{c}&=&-sc^2 F(\frac{s}{c}).
\end{array}
\end{equation}

Reducing the right hand side of the system to a common denominator in
the vicinity of the point $s=1, c=0$ (but not on it) and then
eliminating it, we do not alter the directional field. Preserving
namely this meaning, we define the direction at this point.

Let us go back to the system of equations and rewrite it in the form
\begin{subequations}
\begin{eqnarray}
\dot \nu &=& - 3 H \nu, \label{tauH1obs}\\
\dot {H} &=& \frac{1}{2}\bigl(3 \xi H - (\ve+p) \bigr) - \bigl(3 H^2
- \ve - \Lambda\bigr) + \frac{1}{2}\phi_1(\nu), \label{Hn1obs}\\
\dot {\ve} &=& 3 H\bigl(3 \xi H - (\ve+p) \bigr) + 4 \eta \bigl(3 H^2
- \ve - \Lambda \bigr)- 4 \eta \phi_2(\nu), \label{ve1obs}
\end{eqnarray}
\end{subequations}
where $\phi_1$ and $\phi_2$ are the functions of $\tau$.

In case of a spinor field only we have $\phi_1(\nu)=m\nu + \lambda
(n-2)$, $\phi_2(\nu)=m \nu - \lambda \nu^n$.

Introduction of a scalar field gives $\phi_1(\nu)=m\nu + \frac{n
\nu^{n+2}}{2(1 + \lambda \nu^n)^2}$, $\phi_2(\nu)=m\nu +
\frac{\nu^2}{2(1 + \lambda \nu^n)}$.

Near the point $\ve=\infty$ we make the following substitution
$\ve=1/\mu$. Then the system takes the form
\begin{subequations}
\begin{eqnarray}
\dot \nu &=& - 3 H \nu, \label{tau1obs}\\
\dot{H}&=&\frac{3}{2}BH\mu^{-\beta}+\frac{1}{2}(1-\zeta)\mu^{-1}-3H^2+
\Lambda+\frac{1}{2}\phi_1(\nu), \label{H1obs}\\
\dot {\mu} &=&4A\bigl(-3H^2+\Lambda+\phi_2(\nu)\bigr)\mu^{2-\alpha}
-9BH^2\mu^{2-\beta} + 4A\mu^{1-\alpha}+3H(1+\zeta)\mu. \label{mu1obs}
\end{eqnarray}
\end{subequations}

As it is seen from \eqref{mu1obs} in the absence of viscosity ($A=0$,
$B=0$) the blow up along the energy density is impossible.

The answer, whether the blow up takes place in the past or in the
future, depends on the sign of the coefficient at $\mu$ with the
lowest power.

Let $A = 0$. In order to the blow up takes place at finite $H$, it is
necessary that $\beta > 1$. In this case the singularity will be in
the future, i.e., we have Big Rip.

Now consider the case with $B = 0$. In this case the blow up takes
place in the past (Big Bang) if $\alpha > 1$.

In the figures illustrated below we plot the trajectories on which
the infinite energy density $\ve$ is achieved in a finite time. The
blue line indicates past while the red one the future.

\myfigured{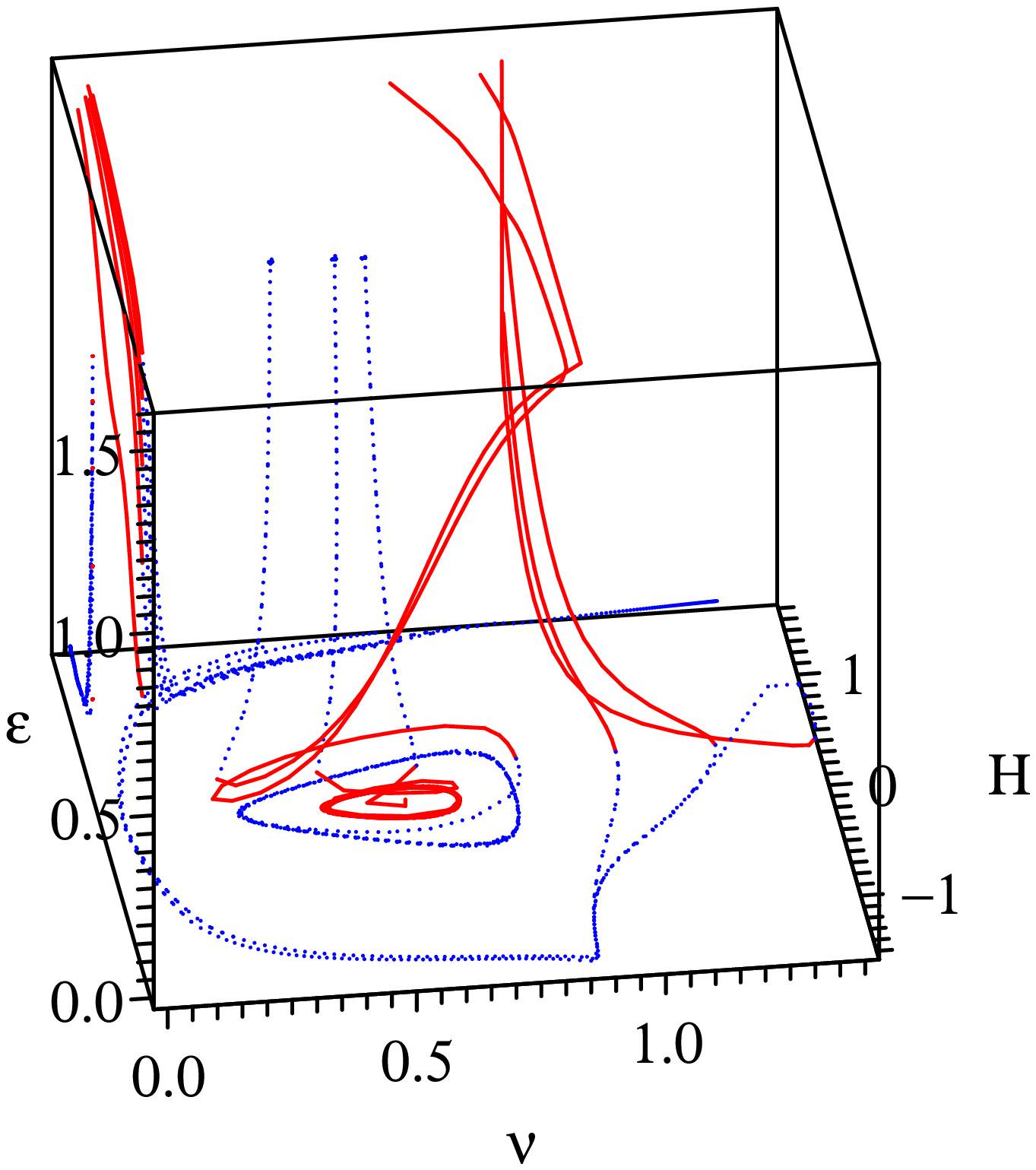}{0.6}{The trajectory of evolution in case of an
interacting spinor and scalar fields with $\alpha=4,\beta=1,
\zeta=1/2,A=1,B=1, m=4,\Lambda=-1,\lambda=-1$}{0.6}

\myfigured{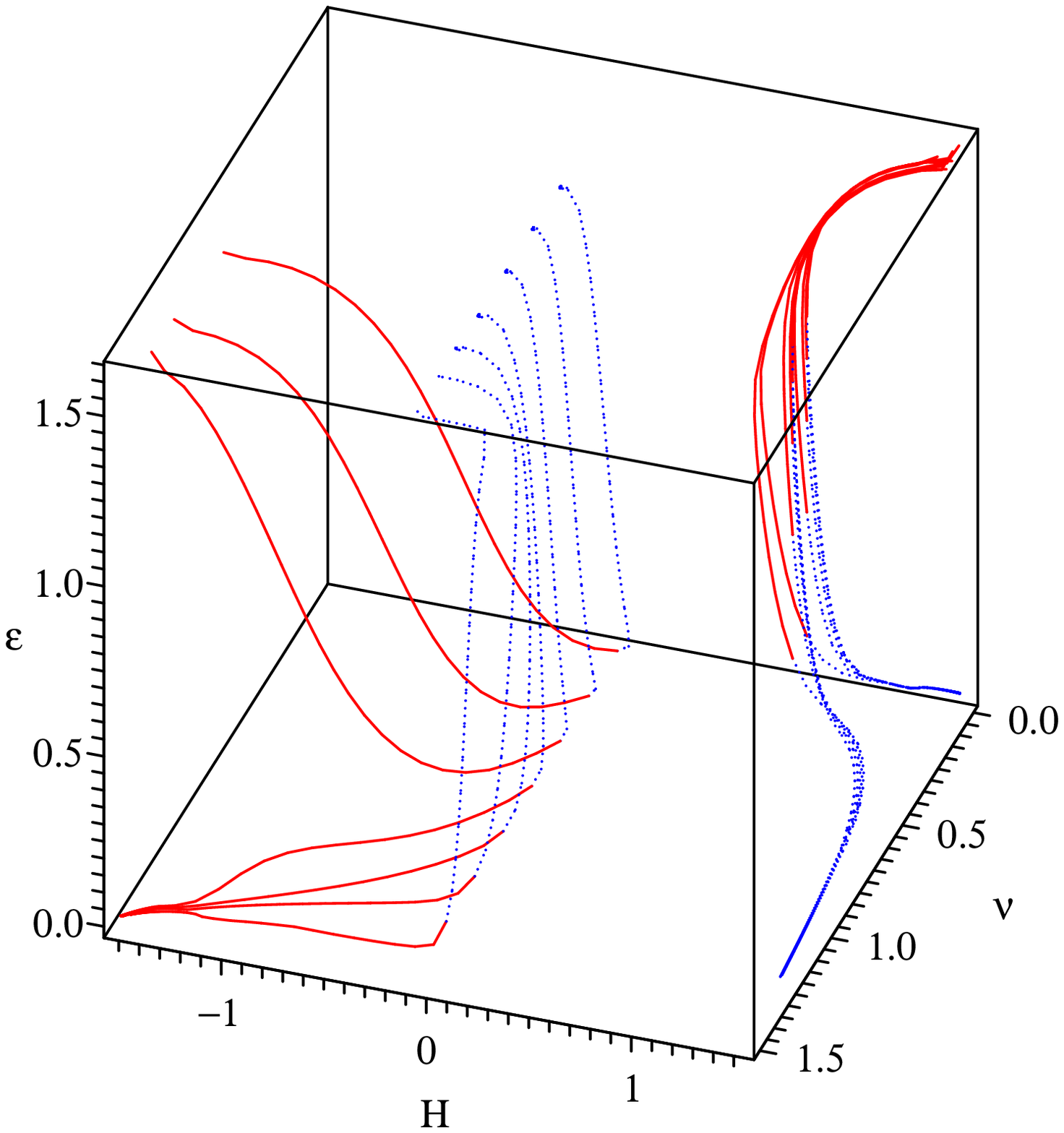}{0.6}{The trajectory of evolution in case of a
spinor field with self-action at $\alpha=4,\beta=1,
\zeta=1/2,A=1,B=1, m=4,\Lambda=-1, \lambda=-1$}{0.6}

In the figures \ref{3dspm} and \ref{3dspsclm05} we show the evolution
of $\tau$, $H$ and $\ve$ relative to each other. In both cases there
exists possibility for infinite growth of energy density at
infinitely large volume, i.e., there occurs so-called Big Rip.


               \section{Conclusion}
Recently a self consistent system of nonlinear spinor and
gravitational fields in the framework of Bianchi type-I
cosmological model filled with viscous fluid was considered by one
of the authors \cite{saharrp,visnlsrev2}. The spinor filed
nonlinearity is taken to be some power law of the invariants of
bilinear spinor forms, namely $I = S^2 = (\bp \p)^2$ and $J = P^2
= (i \bp \gamma^5 \p)^2$. Solutions to the corresponding equations
are given in terms of the volume scale of the BI space-time, i.e.,
in terms of $\tau = a b c$, with $a,b,c$ being the metric
functions. This study generates a multi-parametric system of
ordinary differential equations \cite{saharrp,visnlsrev2}. Given the
richness of the system of equations in this paper a qualitative
analysis of the system in question has been thoroughly carried
out. A complete qualitative classification of the mode of
evolution of the universe given by the corresponding dynamic
system has been illustrated. In doing so we have considered all
possible values of problem parameters independent to their
physical validity and graphically presented the most
distinguishable in our view results.

The system is studied from the view point of blow up. It has been shown
that in absence of viscosity the blow up does not occur. It should be emphasized that phenomena similar to one in question can
be observed in other discipline of physics and present enormous interest
from the point of catastrophe, demography etc.


\newcommand{\hnl}{\htmladdnormallink}

\end{document}